\documentclass{dune} %

\usepackage{subfiles}
\usepackage[numbers]{natbib}
\renewcommand{\familydefault}{\sfdefault}

\usepackage[pdftex,bookmarks,hidelinks]{hyperref}

\graphicspath{ {graphics/} }

\newif\ifdp
\newif\ifsp

\def\expshort{DUNE\xspace}

\def\explong{The Deep Underground Neutrino Experiment\xspace}

\def\thedocsubtitle{Deep Underground Neutrino Experiment (DUNE)}

\newcommand{\refsec}[2]{Volume~\csname volnumber#1\endcsname \xspace Section~#2}
\newcommand{\refch}[2]{Volume~\csname volnumber#1\endcsname \xspace Chapter~#2}
\newcommand{\refinch}[2]{#2 in Volume~\csname volnumber#1\endcsname \xspace}

\newcommand{\numu}{\ensuremath{\nu_\mu}\xspace}
\newcommand{\nue}{\ensuremath{\nu_e}\xspace}

\newcommand{\numubartonumubar}{
\ensuremath{\overline{\numu}\rightarrow\overline{\numu}}\xspace
}

\newcommand{\nominalmodsize}{\SI{10}{kt}\xspace} %
\newcommand{\fdfiducialmass}{\SI{40}{\kt}\xspace} 

\def\argon40{${}^{40}$Ar}       
\def\Ar39{$^{39}$Ar}
\def\Cl40{$^{40}$Cl}
\def\K40{$^{40}$K}
\def\B8{$^{8}$B}

\def\larmass{\SI{17.5}{\kt}\xspace} %

\def\coldbox{cold box\xspace} %

\newcommand{\efield}{E field\xspace}

\newcommand{\threed}{3D\xspace}
\newcommand{\twod}{2D\xspace}
\newcommand{\phel}{photoelectron\xspace} %
\newcommand{\frfour}{FR-4\xspace} %

\newcommand{\lsim}{{\;\raise0.3ex\hbox{$<$\kern-0.75em\raise-1.1ex\hbox{$\sim$}}\;}}
\newcommand{\gsim}{{\;\raise0.3ex\hbox{$>$\kern-0.75em\raise-1.1ex\hbox{$\sim$}}\;}}
\newcommand{\beq}{\begin{equation}}
\newcommand{\eeq}{\end{equation}}
\newcommand{\bea}{\begin{eqnarray}}
\newcommand{\eea}{\end{eqnarray}}

\mathchardef\minus="002D

\newcommand{\rrt}[1]{}

\newcommand{\hideme}[1]{{\it( #1)}}
\renewcommand{\hideme}[1]{} %
\newcommand{\nubar}{\overline{\nu}}

\DeclareSIUnit \s {\second}
\DeclareSIUnit \MB {\mega\byte}
\DeclareSIUnit \GB {\giga\byte}
\DeclareSIUnit \TB {\tera\byte}
\DeclareSIUnit \PB {\peta\byte}
\DeclareSIUnit \Mbps {\mega\bit/\s}
\DeclareSIUnit \Gbps {\giga\bit/\s}
\DeclareSIUnit \Tbps {\tera\bit/\s}
\DeclareSIUnit \Pbps {\peta\bit/\s}
\DeclareSIUnit \kton {\kilo\tonne} %
\DeclareSIUnit \kt {\kilo\tonne}
\DeclareSIUnit \Mt {\mega\tonne}
\DeclareSIUnit \eV {\electronvolt}
\DeclareSIUnit \keV {\kilo\electronvolt}
\DeclareSIUnit \MeV {\mega\electronvolt}
\DeclareSIUnit \GeV {\giga\electronvolt}
\DeclareSIUnit \m {\meter}
\DeclareSIUnit \cm {\centi\meter}
\DeclareSIUnit \in {\inchcommand}
\DeclareSIUnit \km {\kilo\meter}
\DeclareSIUnit \kV {\kilo\volt}
\DeclareSIUnit \kW {\kilo\watt}
\DeclareSIUnit \MW {\mega\watt}
\DeclareSIUnit \MHz {\mega\hertz}
\DeclareSIUnit \mrad {\milli\radian}
\DeclareSIUnit \year {year}
\DeclareSIUnit \POT {POT}
\DeclareSIUnit \sig {$\sigma$}
\DeclareSIUnit\parsec{pc}
\DeclareSIUnit\lightyear{ly}
\DeclareSIUnit\foot{ft}
\DeclareSIUnit\ft{ft}
\DeclareSIUnit \ppb{ppb}
\DeclareSIUnit \ppt{ppt}
\DeclareSIUnit \samples{S}

\newcommand{\beamreprate}{\SI[round-mode=places,round-precision=2]{0.833333333333}{\hertz}\xspace}

\usepackage[toc]{glossaries}
\makeglossaries

\newcommand{\dshort}[1]{\glsentrytext{#1}}  % doesn't provide link
\newcommand{\dfirst}[1]{\glsfirst{#1}\glsunset{#1}}

\newcommand{\dword}[1]{\gls{#1}}
\newcommand{\dwords}[1]{\glspl{#1}}
\newcommand{\Dword}[1]{\Gls{#1}}
\newcommand{\Dwords}[1]{\Glspl{#1}}

\newcommand{\newduneword}[3]{
    \newglossaryentry{#1}{
        text={#2},
        long={#2},
        name={\glsentrylong{#1}},
        first={\glsentryname{#1}},
        firstplural={\glsentrylong{#1}\glspluralsuffix},
        description={#3},
        sort={#2}
    }
}

\newcommand{\newduneabbrev}[4]{
  \newglossaryentry{#1}{
    text={#2},
    long={#3},
    shortplural={{#2}\glspluralsuffix},
    longplural={{#3}\glspluralsuffix{}},
    name={\glsentrylong{#1}{} (\glsentrytext{#1}{})},
    first={#3 (#2)},
    firstplural={#3\glspluralsuffix{} (\glsentrytext{#1}\glspluralsuffix{})},
    description={#4},
    sort={#2}
  }
}

\newcommand{\newduneabbrevs}[5]{
  \newglossaryentry{#1}{
    text={#2},
    long={#3},
    plural={#4},
    shortplural={{#2}\glspluralsuffix},
    longplural={#4},
    name={\glsentrylong{#1}{} (\glsentrytext{#1}{})},
    first={#3 (#2)},
    firstplural={#4 (\glsentrytext{#1}\glspluralsuffix{})},
    description={#5},
    sort={#2}    
  }
}

\newduneword{dword}{DUNE Word}{A term in the DUNE lexicon}

\newduneword{nasa}{NASA}{U.S. National Aereonautics and Space Administration}

\newduneabbrev{nd}{ND}{near detector}{Refers to the collection of \gls{dune} detector components %or more  generally the experimental site
 installed close to the neutrino source at \gls{fnal}; also a subproject of \gls{usproj} that  includes  installation, infrastructure, and the cryogenics systems for this detector} %updated 19 nov 21

\newduneabbrev{fd}{FD}{far detector}{The \SI{70}{kt} total (\fdfiducialmass fiducial) mass \gls{lartpc} DUNE detector, composed of four \larmass total (\nominalmodsize fiducial) mass modules,  %or more generally the experimental 
  to be installed at the far site at \gls{surf} in
  Lead, SD, USA}

\newduneabbrev{sp}{SP}{single-phase}{Distinguishes a \gls{lartpc} technology by the fact that it operates using argon in its liquid phase only; a legacy \gls{dune} term now replaced by \gls{hd} and \gls{vd}} %11 may 21 {Distinguishes one of the DUNE far detector technologies by the fact that it operates using argon in its liquid phase only}

\newduneabbrev{dp}{DP}{dual-phase}{Distinguishes a \gls{lartpc} technology by the fact that it operates using argon 
 in both gas and liquid phases; sometimes called double-phase} % 1 sept2020; Anne added `double' after SSR mentioned that EU tends to use it

\newduneabbrev{pds}{PDS}{photon detection system}{The detector 
  subsystem sensitive to light produced in the \gls{lar} }

\newduneabbrev{hvs}{HVS}{high voltage system}{The detector 
  subsystem that provides the \gls{tpc} drift field}

\newduneabbrev{tpc}{TPC}{time projection chamber}{Depending on context: (1) A type of particle detector that uses an \efield together with a sensitive volume of gas or liquid, e.g., \gls{lar}, to perform a \threed reconstruction of a particle trajectory or interaction. The activity is recorded by digitizing the waveforms of current
  induced on the anode as the distribution of ionization charge passes by
  or is collected on the electrode. (2) TPC is also used in \gls{usproj} for ``total project cost''} 

\newduneabbrev{lartpc}{LArTPC}{liquid argon time-projection chamber}{A \gls{tpc} filled with liquid argon; %A class of detector technology that this technology forms 
the basis for the \gls{dune} \gls{fd} modules} %.   It typically entails observation of ionization activity by  electrical signals and of scintillation by optical signals}

\newduneabbrevs{apa}{APA}{anode plane assembly}{anode plane assemblies}{A unit of the \gls{sphd}
  detector module containing the elements sensitive to ionization in the \gls{lar}. 
  Each anode face has three planes of wires (two induction, one collection) to provide a 3D view, and interfaces to the cold electronics and photon detection system} 

\newduneabbrev{awg}{AWG}{American wire gauge} {U.S. standard set of non-ferrous wire conductor sizes}

\newduneabbrev{ufer}{Ufer}{concrete encased electrode} {U.S. National Electrical Code grounding method refered to as Concrete Encased Electrode}

\newduneabbrev{cro}{CRO}{charge readout}{The system for detecting
  ionization charge distributions in a %\gls{dp} 
  detector module} %11 may 2021 (no more DP module) still use?

\newduneabbrev{lro}{LRO}{light readout}{The system for detecting
  scintillation photons in a \gls{lartpc} detector module}

\newduneabbrev{shv}{SHV}{safe high voltage}{Type of bayonet mount
connector used on coaxial cables that has additional insulation 
compared to standard BNC and MHV connectors that makes it safer
for handling \gls{hv} by preventing accidental contact with the
live wire connector in an unmated connector or plug}

\newduneabbrev{fe}{FE}{front-end}{The front-end refers to a point that is
  ``upstream'' of the data flow for a particular subsystem. 
 For example the \gls{sphd} front-end electronics is where the cold electronics
  meet the sense wires of the TPC and the front-end \gls{daq} is where the \gls{daq} meets the output of the electronics}

\newduneabbrev{daqrou}{DAQ RU}{DAQ readout unit}{The first element in the data flow of the \gls{daq}}

\newduneabbrev{cots}{COTS}{commercial off-the-shelf}{Items, typically hardware such as 
computers, that may be purchased whole, without any custom design or fabrication and 
thus at normal consumer prices and availability}

\newduneabbrev{i2c}{I2C}{Inter-Integrated Circuit}{I$^2$C or I2C is a synchronous, 
multi-master, multi-slave, packet switched, single-ended, serial computer bus widely used 
for attaching lower-speed peripheral ICs to processors and microcontrollers in short-distance, 
intra-board communication} %leave upper case

\newduneabbrev{spi}{SPI}{Serial Peripheral Interface}{The Serial Peripheral Interface is a 
synchronous serial communication interface specification used for short distance 
communication, primarily in embedded systems}%leave upper case

\newduneabbrev{miso}{MISO}{master in slave out}{The Master In Slave Out is a logic
signal on the \gls{spi} bus on which the data from the slave are transmitted once
a request from the master is received} %leave upper case

\newduneabbrev{mosi}{MOSI}{master out slave in}{The Master Out Slave In is a logic
signal on the \gls{spi} bus on which the data from the master is transmitted} %leave upper case

\newduneabbrev{uart}{UART}{Universal Asynchrous Receiver/Transmitter}{A universal 
asynchronous receiver-transmitter is a computer hardware device for asynchronous 
serial communication in which the data format and transmission speeds are configurable}%leave upper case

\newduneword{cr}{CR}{Capacitance-Resistance} %leave upper case

\newduneword{dc}{DC}{direct coupling} % I think these are ok lower case (anne)

\newduneword{ac}{AC}{Alternating Current; when used in the phrase ``AC coupling'' refers to a circuit element that filters out low-frequency components, such as constant offsets, leaving higher frequency signal components. The frequency filtering is determined both by a resistor and a capacitor}

\newduneabbrev{pll}{PLL}{Phase-Locked Loop}{A control system that generates an
output signal whose phase is related to the phase of an input signal}  %leave upper case

\newduneword{fifo}{FIFO}{First-In-First-Out} % leave in upper case

\newduneword{saci}{SACI}{\gls{slac} \gls{asic} Control Interface}

\newduneword{om3}{OM3}{Type of multi-mode fiber optic cable, typically capable of \SI{10}{Gbps} data transmission at lengths up to \SI{300}{m}}

\newduneword{om4}{OM4}{Type of multi-mode fiber optic cable, typically capable of \SI{10}{Gbps} data transmission at lengths up to \SI{550}{m}}

\newduneword{qfp}{QFP}{Quad Flat Package} % leave in upper case

\newduneabbrev{ams}{AMS}{analog and mixed signal}{Verilog-AMS is a derivative of the Verilog hardware description language that includes analog and mixed-signal extensions (AMS) in order to define the behavior of analog and mixed-signal systems}

\newduneabbrev{hepa}{HEPA}{High Efficiency Particulate Air}{The High Efficiency Particulate Air filters are a type of air filter that remove 99.97\% of particles that have a size greater than or equal to \SI{0.3}{\micro\meter}}  % leave in upper case

\newduneabbrev{uvm}{UVM}{universal verification methodology}{The Universal Verification Methodology is a standardized methodology for verifying integrated circuit designs}   % leave in upper case

\newduneword{lhc}{LHC}{Large Hadron Collider}

\newduneabbrev{lsb}{LSB}{least significant bit}{The bit with the lowest numerical value in a binary number}

\newduneabbrev{ldo}{LDO}{low-dropout regulator}{A low-dropout or LDO regulator is a \gls{dc} linear voltage regulator that can regulate the output voltage even when the supply voltage is very close to the output voltage}

\newduneabbrev{adc}{ADC}{analog-to-digital converter}{A sampling of a voltage
  resulting in a discrete integer count corresponding in some way to
  the input}

\newduneabbrev{inl}{INL}{integral non-linearity}{A commonly used measure of performance in \glspl{adc}. It is the deviation between the ideal input threshold value and the measured threshold level of a certain output code}

\newduneabbrev{dnl}{DNL}{differential non-linearity}{A commonly used measure of performance in \glspl{adc}. The DNL error is defined as the difference between an actual step width and the ideal value of one \gls{lsb}}

\newduneword{pnp}{PNP}{Type of bipolar junction transistor consistning of a
layer of N-doped semiconductor sandwiched between two layers of P-doped material}

\newduneabbrev{spice}{SPICE}{Simulation Program with Integrated Circuit Emphasis}{a general-purpose, 
open-source analog electronic circuit simulator. It is a program used in integrated 
circuit and board-level design to check the integrity of circuit designs and to 
predict circuit behavior} % Anne updated to abbrev oct 2022

\newduneabbrev{daq}{DAQ}{data acquisition}{The data acquisition system
  accepts data from the detector \gls{fe} electronics, buffers
  the data, performs a \gls{trigdecision}, builds events from the selected
  data and delivers the result to the offline \gls{diskbuffer}}

\newduneabbrev{iov}{IOV}{interval of validity}{Interval over which something is valid}

\newduneword{calci}{CALCI}{Calibration and Cryogenic Instrumentation}

\newduneword{detmodule}{far detector module}{The entire DUNE far detector design calls for segmentation into  four modules,  each with a total/fiducial mass of approximately \SI{17}{\kton}/\SI{10}{\kton}}
  
\newduneword{module}{module}{Many aspects of the DUNE far and near detectors are modular, so ``module'' must be understood in context.  It may refer to one of the four far detector modules,  distinct portions of a subdetector as in a ``field cage module,'' a software or electronics module,  e.g., a separate framework plug-in,  and so on}

\newduneword{detunit}{detector unit}{A portion of a \gls{detmodule} may be further partitioned into a number of similar parts.   For example, the \gls{sphd} \gls{tpc} is made up of \gls{apa}  units (and other elements)}

\newduneword{diskbuffer}{secondary DAQ buffer}{A secondary
  \gls{daq} buffer holds a small subset of the full rate as
  selected by a \gls{trigcommand}. 
  This buffer also marks the interface with the DUNE Offline}

\newduneabbrev{om}{OM}{online monitoring}{Processes that run inside
  the \gls{daq} on data ``in flight,'' specifically before landing on the
  offline disk buffer, and that provide feedback on the operation of
  the \gls{daq} itself and the general health of the data it is marshalling}

\newduneabbrev{dqm}{DQM}{data quality monitoring}{Analysis of the raw
  data to monitor the integrity of the data and the performance of the
  detectors and their electronics. This type of monitoring may be
  performed in real time, within the \gls{daq} system, or in later
  stages of processing, using disk files as input}

\newduneword{dumpbuffer}{DAQ dump buffer}{This \gls{daq} buffer
  accepts a high-rate data stream, in aggregate, from an associated
  portion of a \gls{detmodule} sufficient to collect all data likely relevant to
  a potential \gls{snb}}
\newduneabbrev{etf}{ETF}{Experiment Test Framework}{\gls{wlcg} testing middleware that runs grid jobs that actively test distributed sites' services and capabilities,  and reports back to monitoring services}

\newduneabbrev{etl}{ETL}{external trigger logic}{Trigger processing
  that consumes \gls{detmodule} level \gls{trignote} information
  and other global sources of trigger input and emits
  \gls{trigcommand} information back to the \glspl{mtl}}
\newduneabbrev{daqeti}{ETI}{external trigger interface}{Interface between \glspl{mtl} and external source and sinks of relevant trigger information}

\newduneword{trignote}{trigger notification}{Information provided by
  \gls{mtl} to \gls{etl} about \gls{trigdecision} %its 
  processing}

\newduneword{trigprimitive}{trigger primitive}{Information derived by
  the \gls{daq} \gls{fe} hardware that describes a region of space (e.g.,
  one or several neighboring channels) and time (e.g., a contiguous set
  of \gls{adc} sample ticks) associated with some activity}

\newduneword{externtrigger}{external trigger candidate}{Information
  provided to the \gls{mtl} about events external to a
  \gls{detmodule} so that it may be considered in forming
  \glspl{trigcommand}}

\newduneabbrev{daqoob}{OOB dispatcher}{out-of-band trigger command
  dispatcher}{This component is responsible for dispatching a \gls{snb} dump
  command to all \glspl{daqfer} in the \gls{detmodule}}

\newduneabbrev{mtl}{MTL}{module trigger logic}{Trigger processing
  that consumes \gls{detunit} level \gls{trigcommand} information
  and emits \glspl{trigcommand}. 
  It provides the \gls{etl} with \glspl{trignote} and receives back any
  \glspl{externtrigger}}

\newduneword{octant}{octant}{Any of the eight parts into which 4$\pi$
  is divided by three mutually perpendicular axes. 
  In particular in referencing the value for the mixing angle
  $\theta_{23}$}

\newduneword{trigcandidate}{trigger candidate}{Summary information derived
  from the full data stream and representing a contribution toward
  forming a \gls{trigdecision}}

\newduneword{trigcommand}{trigger command}{Information derived from
  one or more \glspl{trigcandidate}  that directs elements of a
  \gls{detmodule} to read out a portion of the data stream}

\newduneabbrev{tcm}{TCM}{trigger command message}{A message flowing
  down the trigger hierarchy from global to local context.  Also see \gls{tpm}}

\newduneabbrev{mlt}{MLT}{module level trigger}{The \gls{daq} component responsible for producing a \gls{trigdecision} that will be used to command the readout of a detector module}

\newduneword{trigdecision}{trigger decision}{The process by which
  \glspl{trigcandidate} are converted into \glspl{trigcommand}}

\newduneabbrev{tpm}{TPM}{trigger primitive message}{A message flowing
  up the trigger hierarchy from local to global context.  Also see \gls{tcm}}

\newduneabbrev{ipc}{IPC}{inter-process communication}{A system for software elements to exchange information between threads, local processes or across a data network.  An IPC system is typically specified in terms of protocols  composed of message types and their associated data schema}

\newduneword{daqdispre}{discovery and presence}{As used in the context of the \gls{ipc}, a system that provides mechanisms for a node on a communication network to learn of the existence of peers and their identity (discovery) as well as determine if they are currently operational or have become unresponsive (presence)}

\newduneabbrev{pubsub}{PUB/SUB}{publish-subscribe communication pattern}{An \gls{ipc} communication pattern where one element, the publisher, sends data to all connected elements, the subscribers.  Each subscriber may connect to multiple publishers.  A variant is PUB/SUB with topics where a subscriber may register an identifier, the topic, to limit the information received to just an associated subset}

\newduneabbrev{eb}{EB}{event builder}{A software agent that executes \glspl{trigcommand}  for one  \gls{detmodule} by reading out the requested data}
\newduneabbrev{daqdfo}{DFO}{data flow orchestrator}{The process by which trigger commands are executed in parallel and asynchronous manner by the back-end output subsystem of the \gls{daq}}

\newduneabbrev{daqubi}{UBI}{upstream DAQ buffer interface}{The process which provides read-only access to data residing in the upstream \gls{daq} buffers to processes on the network}

\newduneabbrev{cob}{COB}{cluster on board}{An ATCA motherboard housing four RCEs}

\newduneabbrev{rce}{RCE}{reconfigurable computing element}{Data processor located outside of the cryostat on a \gls{cob} that contains \gls{fpga}, RAM and \gls{ssd} resources, responsible for buffering data, producing trigger primitives, responding to triggered requests for data and synching \gls{snb} dumps}

\newduneabbrev{bow}{BOW}{Bump On Wire}{A working name for the front-end readout computing elements used in the nominal \gls{daq} design to interface the \gls{dp}  crates to the \gls{daq} front-end computers} %11 may 2021 (still needed? -check with Giovanna)

\newduneabbrev{atca}{ATCA}{Advanced Telecommunications Computing
  Architecture}{An advanced computer architecture specification developed for the telecommunications, military, and aerospace industries that incorporates the latest trends in  high-speed interconnect technologies, next-generation processors, and improved reliability, availability and serviceability} 

\newduneabbrev{utca}{$\mu$TCA}{Micro Telecommunications Computing Architecture}{The computer architecture specification followed by the crates that house charge and light readout electronics; used in the \gls{spvd} and \gls{dp} technologies} %11 may 2021 (still needed?)

\newduneabbrev{udp}{UDP}{user datagram protocol}{A simple,
  connectionless Internet protocol that supports data integrity
  checksums, requires no handshaking, and does not guarantee packet delivery}

\newduneabbrev{amc}{AMC}{advanced mezzanine card}{Holds digitizing
  electronics and lives in \gls{utca} crates}

\newduneabbrev{rf}{RF}{radio frequency}{Electromagnetic emissions
  that are within the (radio) frequency band of sensitivity of the detector
  electronics}

\newduneabbrev{fpga}{FPGA}{field programmable gate array}{An
integrated circuit technology that allows the hardware to be reconfigured to
execute different algorithms after its manufacture and deployment}

\newduneabbrev{fmc}{FMC}{FPGA mezzanine card}{Boards holding \glspl{fpga} and other integrated circuitry that attach to a motherboard}

\newduneabbrev{felix}{FELIX}{Front-End Link eXchange}{A
  high-throughput interface between \gls{fe} and trigger electronics
  and the standard PCIe computer bus}

\newduneword{daqpart}{DAQ partition}{A cohesive and
 coherent collection of \gls{daq} hardware and software working together to trigger and read out some portion of one detector module; it consists of an integral number of \glspl{daqfrag}. 
 Multiple \gls{daq} partitions may operate simultaneously, but each instance operates independently}

\newduneabbrev{fec}{DAQ FEC}{DAQ front-end computer}{The portion of one
  \gls{daqpart} that hosts the \gls{daqdr}, \gls{daqbuf} and
  \gls{daqds}.  It hosts the \gls{daqfer} and corresponding portion of the \gls{daqbuf}}

\newduneword{daqfrag}{DAQ front-end fragment}{The portion of one
  \gls{daqpart} relating to a single \gls{fec} and corresponding to an
  integral number of \glspl{detunit}.  See also \gls{datafrag}}

\newduneword{datafrag}{data fragment}{A block of data read out from a single \gls{daqfrag} that
span a contiguous period of time as requested by a \gls{trigcommand}}

\newduneabbrev{daqfer}{FER}{DAQ front-end readout}{The portion of a
  \gls{daqfrag} that accepts data from the detector electronics and
  provides it to the \gls{fec}}

\newduneabbrev{daqdr}{DDR}{DAQ data receiver}{The portion of the
  \gls{daqfrag} that accepts data from the \gls{daqfer}, emits
  trigger candidates produced from the input trigger primitives, and
  forwards the full data stream to the \gls{daqbuf}}

\newduneword{daqbuf}{DAQ primary buffer}{The portion
  of the \gls{daqfrag} that accepts full data stream from the
  corresponding \gls{detunit} and retains it sufficiently long for it
  to be available to produce a \gls{datafrag}}

\newduneword{daqds}{data selector}{The portion of the \gls{daqfrag}
  that accepts \glspl{trigcommand} and returns the corresponding
  \gls{datafrag}.  Not to be confused with \gls{daqdsn}}

\newduneword{daqdsn}{data selection}{The process of forming a trigger decision for selecting a subset of detector data for output by the \gls{daq} from the content of the detector data itself.  Not to be confused with \gls{daqds}}

\newduneabbrev{daqros}{DAQ RO}{DAQ readout subsystem}{The subsystem of the \gls{daq} for accepting and buffering data input from detector electronics}

\newduneabbrev{daqdss}{DAQ DS}{DAQ data selection subsystem}{The subsystem of the \gls{daq} responsible for forming a trigger decision based on a portion of the input data stream.  The majority subset of the \gls{daqtrs}}

\newduneabbrev{daqtrs}{DAQ TS}{DAQ trigger subsystem}{The subsystem of the \gls{daq} responsible for forming a trigger decision}

\newduneabbrev{daqbes}{DAQ BE}{DAQ back-end subsystem}{The portion of the \gls{daq} that is generally toward its output end.  It is responsible for accepting and executing trigger commands and marshaling the data they address to output storage buffers}

\newduneabbrev{daqtss}{DAQ TSS}{DAQ timing and synchronization subsystem}{The portion of the \gls{daq} that provides for timing and synchronization to various components}

\newduneabbrev{femb}{FEMB}{front-end mother board}{Refers to a unit of
  the \gls{sp} \gls{ce} that contains the \gls{fe} amplifier
  and \gls{adc} \glspl{asic} covering 128 channels}

\newduneword{asic}{ASIC}{application-specific integrated circuit}

\newduneword{lv}{LV}{low voltage}

\newduneabbrev{iceberg}{ICEBERG}{ICEBERG R\&D cryostat and electronics}{Integrated Cryostat and Electronics Built for Experimental Research Goals: a double-walled cryostat built and installed at \gls{fnal}  
for liquid argon detector R\&D and for testing of DUNE detector components}

\newduneword{coldadc}{ColdADC}{A newly developed 16-channels \gls{asic} providing analog to digital conversion}

\newduneword{coldata}{COLDATA}{A 64-channel control and communications \gls{asic}}
\newduneword{cryo}{CRYO}{(1) Integrated ASIC including \gls{fe} circuitry providing signal amplification and pulse shaping, analog to digital conversion, and control and communication functionalities for 64 channels; (2) acryonym for cryogenic systems and cryostat work scopes in \gls{lbnf}}

\newduneword{larasic}{LArASIC}{A 16-channel \gls{fe} \gls{asic} that provides signal amplification and pulse shaping}

\newduneword{cmos}{CMOS}{Complementary metal-oxide-semiconductor}

\newduneabbrev{enc}{ENC}{equivalent noise charge}{The equivalent noise charge is the input charge that corresponds to a \gls{s/n}$=1$}
\newduneword{sar}{SAR}{successive approximation register}

\newduneword{protodune}{ProtoDUNE}{Either of the two initial DUNE prototype detectors constructed at \gls{cern}. % and operated in  a CERN test beam (expected fall 2018). 
  One prototype implemented \gls{sp} technology and the other \gls{dp}}
  
\newduneword{protodune2}{ProtoDUNE-II}{The second run of a \gls{protodune} detector}  % #33

\newduneword{pdsp}{ProtoDUNE-SP}{The \gls{sphd} \gls{protodune} detector constructed at \gls{cern}  in \gls{np04}}

\newduneword{pddp}{ProtoDUNE-DP}{The \gls{dp} \gls{protodune} detector constructed at \gls{cern} in \gls{np02}}

\newduneword{wa105}{WA105 DP demonstrator}{The %\SI[product-units=power]{3x1x1}{m} 
3m$\times$1m$\times$1m WA105 \gls{dp} prototype detector at \gls{cern}}

\newduneword{rawevent}{DAQ event block}{The unit of data output by the
  \gls{daq}.  
  It contains trigger and detector data spanning a unique, contiguous
  time period and a subset of the detector channels}

\newduneabbrev{ssd}{SSD}{solid-state disk}{Any storage device that
  may provide sufficient write throughput to receive, both collectively and
  distributed, the sustained full rate of data from a \gls{detmodule}
  for many seconds}
\newduneabbrev{nvme}{NVMe}{Non-volatile memory express}{A specification for an interface to storage media attached via PCIe}

\newduneabbrev{hlt}{HLT}{high-level trigger}{This is actually a filter applied to data that has been triggered and aggregated in order to further reduce or characterize it}

\newduneabbrev{pid}{PID}{particle ID}{Particle identification}

\newduneword{readout window}{readout window}{A fixed, atomic and
  continuous period of time over which data from a \gls{detmodule}, in
  whole or in part, is recorded. 
  This period may differ based on the trigger that initiated the
  readout}

\newduneabbrev{zs}{ZS}{zero-suppression}{Used to delete some portion of a
  data stream that does not significantly deviate from zero or
  intrinsic noise levels. 
  It may be applied at different granularity from per-channel to per
  \gls{detunit}}

\newduneword{rc}{RC}{Depending on context, one of (1) resistive-capacitive (circuit), (2) run control, the system for configuring, starting and terminating the \gls{daq}, or (3) resource coordinator, a member of the \gls{dune} management team responsible for coordinating the financial resources of the project} % #33

\newduneabbrev{daqccm}{CCM}{DAQ control, configuration and monitoring subsystem}{A system for controlling, configuring and monitoring other systems in particular those that make up the \gls{daq} where the CCM encompasses \gls{rc}}

\newduneword{daqrun}{DAQ run}{A period of time over which relevant data taking conditions and \gls{daq} configuration are asserted to be unchanged. 
  Multiple \gls{daq} runs may occur simultaneously when multiple \glspl{daqpart} are active. 
  This term should not be confused with DUNE experiment or beam ``runs'' that typically span many \gls{daq} runs}
\newduneword{daqrunnum}{DAQ run number}{A monotonically increasing count that uniquely and globally identifies a \gls{daqrun}}

\newduneabbrev{snb}{SNB}{supernova neutrino burst}{A prompt 
  increase in the flux of low-energy neutrinos emitted in the first few seconds of a core-collapse supernova.  It can also refer to a trigger command type that may be due to this phenomenon,
  or detector conditions that mimic its interaction signature}

\newduneabbrev{snble}{SNB/LE}{supernova neutrino burst and low
  energy}{Supernova neutrino burst and low-energy physics program}

\newduneabbrev{snews}{SNEWS}{SuperNova Early Warning System}{A global
  supernova neutrino burst trigger formed by a coincidence of \gls{snb} 
  triggers collected from participating experiments}

\newduneabbrev{pps}{1PPS signal}{one-pulse-per-second signal}{An
  electrical signal with a fast rise time and that arrives in real
  time with a precise period of one second}

\newduneabbrev{sls}{SLS}{spill location system}{A system residing at
  the DUNE far detector site that provides information, possibly
  predictive, indicating periods of time when neutrinos are being
  produced by the \gls{fnal} Main Injector beam spills}

\newduneabbrev{wib}{WIB}{warm interface board}{Digital electronics
  situated just outside a FD cryostat that receives digital data
  from the \glspl{femb} (part of \gls{ce}) over cold copper connections and sends it to the \gls{rce}
  \gls{fe} readout hardware}

\newduneabbrev{gps}{GPS}{Global Positioning System}{A satellite-based system that provides a highly accurate \gls{pps} that may be used to synchronize clocks and determine location}

\newduneabbrev{ntp}{NTP}{Network Time Protocol}{A networking protocol that allows synchronizing of clocks to within a few \si{\milli\second} of a time standard on a local network and within a few tens of \si{\milli\second} over the Internet} 

\newduneword{ptp}{PTP}{Depending on context, either p-terphenyl, a \gls{wls} material, or Precision Time Protocol, a networking protocol that allows synchronizing of clocks to within a few \si{\micro\second} of a time standard on a local network}  % #33

\newduneabbrev{irig}{IRIG}{inter-range instrumentation group}{A standards body that defined a time-code standard for transferring timing information}

\newduneabbrev{nic}{NIC}{network interface controller}{Hardware for controlling the interface to a communication network.  Typically, one that obeys the Ethernet protocol}

\newduneabbrev{wiec}{WIEC}{warm interface electronics crate}{Crates mounted on the signal flanges that contain the \glspl{wib}}

\newduneabbrev{ptc}{PTC}{power and timing card}{Cards that provide further processing and distribution of the signals entering and exiting the \gls{sp} cryostat}

\newduneabbrev{ptb}{PTB}{power and timing backplane}{Backplane used to connect the \glspl{wib} and the \glspl{ptc} on the \gls{wiec}. Also connects the \gls{ce} flange on the cryostat penetration}

\newduneabbrev{sipm}{SiPM}{silicon photomultiplier}{A solid-state
  avalanche photodiode sensitive to single \phel signals}

\newduneabbrev{cisc}{CISC}{cryogenic instrumentation and slow controls}{Includes equipment to monitor all detector  components and  \gls{lar} quality and behavior, and provides a control system for many of the detector components}

\newduneword{fte}{FTE}{full-time equivalent. A unit of labor
  for the project. One year of work from one person}

\newduneword{art}{art}{A software framework implementing an
  event-based execution paradigm} %http://art.fnal.gov/

\newduneabbrev{sam}{SAM}{sequential
  access via metadata}{A data-handling system to store and retrieve
  files and associated metadata, including a complete record of the
  processing that has used the files}

\newduneword{artdaq}{artdaq}{A data acquisition toolkit for data transfer, aggregation and processing}

\newduneword{beamline}{beamline}{A sequence of control and monitoring devices used for the formation of a directed collection of particles; also subproject within \gls{usproj}} % #36

\newduneword{cdr}{CDR}{Depending on context, either ``conceptual design report,'' a formal project  document  that describes the experiment at a conceptual level, or ``conceptual design review,'' a formal review of the conceptual design of the experiment or of a component}  % #33

\newduneabbrev{cf}{CF}{conventional facilities}{Pertaining to
  construction and operation of buildings and conventional infrastructure, and includes  cavern excavation}

\newduneabbrev{cp}{CP}{charge conjugation and parity}{Product of charge conjugation and parity
  transformations} % updated to include `conjugation' 12/14/20 per Steve Manly

\newduneabbrev{cpt}{CPT}{charge, parity, and time reversal symmetry}{product of charge, parity
  and time-reversal transformations}

\newduneabbrev{cpv}{CPV}{charge-parity symmetry violation}{Lack of
  symmetry in a system before and after charge and parity
  transformations are applied. 
  For \gls{cp} symmetry to hold,  a particle turns into its
 corresponding antiparticle under a charge transformation,  and a parity
transformation inverts its space coordinates,  i.e. produces the mirror image}

\newduneword{doe}{DOE}{U.S. Department of Energy}

\newduneabbrev{fra}{FRA}{Fermi Research Alliance}{A joint partnership of the University of Chicago and the Universities Research Association (URA) that manages and operates \gls{fnal} on behalf of the \gls{doe}}

\newduneabbrev{dune}{DUNE}{Deep Underground Neutrino Experiment}{A leading-edge, international experiment for neutrino science and proton decay studies; refers to the entire international experiment and collaboration} %updated 11 may 21

\newduneabbrev{esh}{ES\&H}{environment, safety and health}{A discipline and specialty that studies and implements practical aspects of environmental protection and safety at work} % The LBNF/DUNE ES\&H program complies with applicable standards and local, state, and federal legal requirements through the Fermilab ``work smart'' set of standards and the contract between Fermi Research Alliance and the DOE Office of Science (FRA-DOE)}

\newduneabbrev{ppe}{PPE}{personnel protective equipment}{Equipment worn to minimize exposure to hazards that cause serious workplace injuries and illnesses}

\newduneabbrev{odh}{ODH}{oxygen deficiency hazard}{a hazard that occurs when inert gases such as nitrogen, helium, or argon displace room air and thus reduce the percentage of oxygen below the level required for human life}

\newduneabbrev{feshm}{FESHM}{Fermilab Environment, Safety and Health Manual}{The document that contains \gls{fnal} 's policies and procedures designed to manage environment, safety, and health in all its programs}

\newduneabbrev{fscf}{FSCF}{Far Site Conventional Facilities}{The \gls{cf} at the DUNE far detector site, \gls{surf}, including all detector caverns and support infrastructure}%updated 11 may 21
  
\newduneabbrev{nscf}{NSCF}{Near Site Conventional Facilities}{The \gls{cf} at the DUNE near detector site, \gls{fnal}}

\newduneabbrev{fd1c}{FD1+C}{far detector module 1 + cryogenics}{The first far detector module to be built at \gls{surf}, including integration and installation, and all cryogenics infrastructure to support FD1 and \gls{fd2}; also a subproject of \gls{usproj}} %new 11 may 21

\newduneabbrev{fd2}{FD2}{far detector module 2}{The second DUNE far detector module to be built at \gls{surf}} %new 11 may 21

\newduneabbrevs{gut}{GUT}{grand unified theory}{grand unified theories}{A class of theories that unifies the electroweak and strong forces}

\newduneabbrev{lar}{LAr}{liquid argon}{Argon in its liquid phase; it is a cryogenic liquid with a boiling point of \SI{87}{K} and density of \SI{1.4}{g/ml}}

\newduneabbrev{lbl}{LBL}{long-baseline}{Refers to the distance between the 
  neutrino source  and the \gls{fd}.  It can also refer to the distance between the near and far detectors. 
  The ``long'' designation is an approximate and relative distinction. For DUNE, this distance  (between \gls{fnal} and \gls{surf}) is approximately \SI{1300}{km}}

\newduneabbrev{lbnf}{LBNF}{Long-Baseline Neutrino Facility}{Long-Baseline Neutrino Facility; refers to the facilities that support the experiment including in-kind contributions under the line-item project. The portion of \gls{usproj} responsible for developing the neutrino beam, the far site cryostats,  and far and near site cryogenics systems, and the conventional facilities, including the excavations } % #36, updated 11 may 2021
  
\newduneabbrev{lbnf-dune}{LBNF/DUNE}{LBNF and DUNE enterprise}{Long-Baseline Neutrino Facility/Deep Underground Neutrino Experiment; refers to the overall enterprise or program including \gls{usproj}, participating international projects, and the \gls{dune} experiment and collaboration} %updated 11 may 2021 (#35 09nov20)

\newduneword{usproj}{LBNF/DUNE-US}{Long-Baseline Neutrino Facility/Deep Underground Neutrino Experiment - United States; project to design and build the conventional and beamline facilities and the \gls{doe} contributions to the detectors. It is organized as a \gls{doe}/\gls{fnal} project and incorporates contributions to the facilities from international partners. It also acts as host for the installation and integration of the DUNE detectors} % new oct 2021

\newduneword{duneus}{DUNE-US}{Deep Underground Neutrino Experiment - United States; refers to the U.S. contribution to DUNE under the line-item \gls{usproj} project} % new 11 may 2021

\newduneabbrev{lbnc}{LBNC}{Long-Baseline Neutrino Committee}{The committee, composed of internationally prominent scientists with relevant expertise, charged by the \gls{fnal} director to review the scientific, technical, and managerial progress, plans and decisions associated with \gls{dune}}

\newduneabbrev{ncg}{NCG}{Neutrino Cost Group}{A group of internationally prominent scientists with relevant experience that is charged by the \gls{fnal} director to review the cost, schedule, and associated risks for the \gls{dune} experiment}

\newduneabbrev{mh}{MH}{mass hierarchy}{Describes the separation
  between the mass squared differences related to the solar and
  atmospheric neutrino problems (also written as \gls{mo})}

\newduneabbrev{mo}{MO}{mass ordering}{See \gls{mh}}

\newduneabbrev{mi}{MI}{Fermilab Main Injector}{An accelerator at
  \gls{fnal} that provides a beam of high-energy protons to the the \gls{usproj}  beamline} 

\newduneabbrev{pot}{POT}{protons on target}{Typically used as a unit
  of normalization for the number of protons striking the neutrino
  production target}

\newduneabbrev{qa}{QA}{quality assurance}{The process of ensuring that %The set of actions taken to provide confidence that 
the quality of each element meets requirements during design and development, and to detect and correct poor results prior to production} % #33

\newduneabbrev{qc}{QC}{quality control}{The process (e.g., inspection, testing, measurements) %An aggregate of activities (such as design analysis and inspection for defects) performed 
of ensuring that each manufactured element meets its quality requirements prior to assembly or installation} %adequate quality in manufactured products}  % #33

\newduneabbrev{sm}{SM}{Standard Model}{Refers to a theory describing
  the interaction of elementary particles}

\newduneword{tdr}{TDR}{Depending on context, either ``technical design report,'' a formal project  document  that describes the experiment at a technical level, or ``technical design review,'' a formal review of the technical design of the experiment or of a component}  % #33

\newduneabbrev{tp}{IDR}{interim design report}{An intermediate
milestone on the path to a full \gls{tdr}} % changed from ``technical proposal'' 6/6/2018

\newduneabbrev{ckm}{CKM matrix}{Cabibbo-Kobayashi-Maskawa
  matrix}{Refers to the matrix describing the mixing between mass and
  weak eigenstates of quarks}

\newduneabbrev{cl}{CL}{confidence level}{Refers to a probability
  used to determine the value of a random variable given its
  distribution}

\newduneabbrev{pmns}{PMNS}{Pontecorvo-Maki-Nakagawa-Sakata}{A type of matrix that describes the mixing between mass and weak eigenstates of
  the neutrino}

\newduneword{hnl}{HNL}{heavy neutral lepton} %  #37, 21 may 21
\newduneabbrevs{cpa}{CPA}{cathode plane assembly}{cathode plane assemblies}{The component of the \gls{sphd} detector module that provides the drift HV cathode}

\newduneword{fc}{field cage}{The component of a \gls{lartpc} that contains and shapes the applied \efield}

\newduneword{cpafc}{CPA/FC}{A pair of \gls{cpa} panels and the top and bottom \gls{fc} portions that attach to the pair; an intermediate assembly for installation into the \gls{spmod} }

\newduneabbrev{topfc}{top FC}{top field cage}{The horizontal portions of the \gls{sphd} \gls{fc}   on the top of the \gls{tpc}}

\newduneabbrev{botfc}{bottom FC}{bottom field cage}{The horizontal portions of the \gls{sphd} \gls{fc} on the bottom of the \gls{tpc}}

\newduneabbrev{ewfc}{endwall FC}{endwall field cage}{The vertical portions of the \gls{fc} near the end walls}

\newduneword{gp}{ground plane}{An electrode held electrically neutral relative to Earth ground voltage; it is mounted on the \gls{fc} to protect the cryostat wall}

\newduneword{gg}{ground grid}{An electrode held electrically neutral relative to Earth ground voltage; it is installed between the cathode and the \glspl{pd} in a \gls{dpmod} to protect the \glspl{pmt}, maintaining high transparency to light} %11 may 2021 (no more DP module) still used? Ask Jae Yu

\newduneabbrev{alara}{ALARA}{as low as reasonably
  achievable}{Typically used with regard management of radiation
  exposure but may be used more generally. It means making every
  reasonable effort to maintain e.g., exposures, to as far below the
  limits as practical, consistent with the purpose for that the
  activity is undertaken}

\newduneabbrev{ecal}{ECAL}{electromagnetic calorimeter}{A detector
  component that measures energy deposition of traversing particles (in the \gls{dune} near detector design)}

\newduneabbrev{hv}{HV}{high voltage}{Generally describes a voltage
  applied to drive the motion of free electrons through some media, e.g., LAr}

\newduneword{spmod}{SP module}{single-phase DUNE \gls{fd} module}  % to change? 11 may 2021 
\newduneword{vdmod}{vertical drift module}{vertical drift DUNE \gls{fd} module} %11 may 2021
\newduneword{dpmod}{DP module}{dual-phase DUNE \gls{fd} module} %11 may 2021 (no more DP module)
\newduneword{dsp}{DUNE-SP}{a single-phase DUNE far detector module} % 33 to change? 11 may 2021 (no more DP module)
\newduneabbrev{tcoord}{TC}{technical coordinator}{A member of the \gls{dune} management team responsible for organizing the technical aspects of the project effort; is head of \gls{tc}}

\newduneabbrev{tc}{TCN}{technical coordination}{The DUNE organization responsible for overall integration of the detector elements and successful execution of the detector construction project; areas of responsibility include general project oversight, systems engineering, \gls{qa} and safety}  % #33

\newduneabbrev{exb}{EB}{executive board}{The highest level DUNE
  decision-making body for the collaboration}

\newduneabbrev{tb}{TB}{technical board}{The DUNE organization responsible for
  evaluating technical decisions}

\newduneabbrev{rrb}{RRB}{Resources Review Board}{A part of \gls{dune}'s international project governance structure, composed of representatives of all funding agencies that sponsor the project, and of  \gls{fnal} management, established to provide coordination among funding partners and oversight of \gls{dune}}

\newduneabbrev{inc}{INC}{International Neutrino Council}{A highest-level international advisory body to the U.S. \gls{doe} and the  \gls{fnal} directorate on matters related to the  \gls{lbnf} and the  \gls{pip2} projects. This council is composed of representatives from the international funding agencies and  \gls{cern} that make major contributions the infrastructure}

\newduneabbrev{cc}{CC}{charged current}{Refers to an interaction
  between elementary particles where a charged weak force carrier
  ($W^+$ or $W^-$) is exchanged}

\newduneabbrev{dis}{DIS}{deep inelastic scattering}{Refers to interaction between
  elementary particles and a nucleus in an energy range where the
  interaction can be modeled as occurring between constituent quarks
  of one nucleon and resulting in no bulk recoil of the resulting
  nucleus} % 1sep2020 ArturA: swapped with {qe}

\newduneabbrev{fsi}{FSI}{final-state interactions}{Refers to
  interactions between elementary or composite particles subsequent to
  the initial, fundamental particle interaction, such as may occur as
  the products exit a nucleus}
  
\newduneabbrev{fsint}{FSI}{far site integration}{The scope of work at the \gls{fs} for the \gls{integoff}} % #36 - can't put it together with prev item

\newduneword{geant4}{Geant4}{A
  software toolkit for the simulation of the passage of particles
  through matter using \gls{mc} methods}

\newduneabbrev{genie}{GENIE}{Generates Events for Neutrino Interaction
  Experiments}{Software providing an object-oriented neutrino
  interaction simulation resulting in kinematics of the products of
  the interaction}

\newduneabbrev{mc}{MC}{Monte Carlo}{Refers to a method of numerical
  integration that entails the statistical sampling of the integrand
  function. 
  Forms the basis for some types of detector and physics simulations}

\newduneabbrev{qe}{QE}{quasi-elastic}{Refers to the 
  interaction of an elementary charged particle with a nucleus in an
  energy range where the interaction can be modeled as taking place with
  individual nucleons} % 1sep2020 ArturA: swapped with {dis}

\newduneabbrev{mou}{MoU}{memorandum of understanding}{A project management methodology that  \gls{usproj} uses to document agreement,s,  e.g., between \gls{fnal} and the Project, for how \gls{fnal}  will support the Project. More generally, a document
  summarizing an agreement between two or more parties} % #36 elaine

\newduneabbrev{pip2}{PIP-II}{Proton Improvement Plan II}{A \gls{fnal} project for
  improving the protons on target delivered delivered by the \gls{lbnf} neutrino production beam. 
  This is version two of this plan and it is planned to be followed by a PIP-III}
  
\newduneabbrev{sdsta}{SDSTA}{South Dakota Science and Technology
  Authority}{The legal entity that manages \gls{surf}, in Lead, S.D}
  
\newduneabbrev{sdsd}{SDSD}{Fermilab South Dakota Services Division}{A \gls{fnal}  division responsible providing host laboratory functions at SURF in South Dakota}

\newduneabbrev{firus}{FIRUS}{Facility Information Reporting Utility System}
 {Facility incident reporting systems, one at \gls{fnal} and at \gls{surf}, that monitors and reports the status of various fire, security and utility sensors} % #36

\newduneabbrev{bsi}{BSI}{building and site infrastructure}
 {The work package for outfitting of the \gls{lbnf} underground infrastructure}

\newduneabbrev{wbs}{WBS}{work breakdown structure}{An organizational
  project management tool by which the tasks to be performed are
  partitioned in a hierarchical manner}

\newduneabbrev{br}{BR}{branching ratio}{A fractional probability for a
  decay of a composite particle to occur into some specified set or
  sets of products}
\newduneword{bsm}{BSM}{beyond the Standard Model}

\newduneabbrev{dm}{DM}{dark matter}{The term given to the unknown
  matter or force that explains measurements of galaxy motion % motion of galaxies
  that are otherwise inconsistent with the amount of mass associated
  with the observed amount of photon production}
  
\newduneabbrev{bdm}{BDM}{boosted dark matter}{A new model that describes a relativistic dark matter particle boosted by the annihilation of heavier dark matter particles in the galactic center or the sun}

\newduneabbrev{cern}{CERN}{European Laboratory for Particle Physics}{The leading particle physics laboratory in Europe and home to the \glspl{protodune} and other prototypes and demonstrators, including the \glspl{mod0}} % per Steve Manly/Hiro Tanaka Dec 2020 (In French, the Organisation Europ\'{e}enne pour la Recherche Nucl\'{e}aire, derived from Conseil Europ\'{e}en pour la Recherche Nucl\'{e}aire)}

\newduneabbrev{dsnb}{DSNB}{diffuse supernova neutrino background}{The
  term describing the pervasive, constant flux of neutrinos due to all
  past supernova neutrino bursts}

\newduneabbrev{espp}{ESPP}{European Strategy for Particle Physics}{The
cornerstone of Europe's
decision-making process for the long-term future of the
field. Mandated by the \gls{cern} Council, it is formed through a broad
consultation of the grass-roots particle physics community, it
actively solicits the opinions of physicists from around the world,
and it is developed in close coordination with similar processes in
the USA and Japan in order to ensure coordination between regions and
optimal use of resources globally}

\newduneabbrev{gar}{GAr}{gaseous argon}{argon in its gas phase}
\newduneabbrev{gartpc}{GArTPC}{gaseous argon time-projection chamber}{A \gls{tpc} filled with gaseous argon} %; a possible technology choice for the \gls{nd}} AH rmvd 12/2020

\newduneabbrev{globes}{GLoBES}{General Long-Baseline Experiment
  Simulator}{A software package for simulating energy spectra of
  neutrino flux, interactions, and energy spectra measured after application of some
  model of a detector response)}

\newduneabbrev{snowglobes}{SNOwGLoBES}{SuperNova
Observatories with GLoBES} {From the official description: SNOwGLoBES is public software for computing interaction rates and distributions of observed quantities for \gls{snb} neutrinos in common detector materials} % (Anne thinks this was too long.)
\newduneword{l/e}{L/E}{length-to-energy ratio}
\newduneword{lri}{LRI}{long-range interactions}
\newduneabbrev{nc}{NC}{neutral current}{Refers to an interaction
  between elementary particles where a neutrally charged weak force carrier
  ($Z^0$) is exchanged}

\newduneabbrev{nh}{NH}{normal hierarchy}{Refers to the neutrino mass
  eigenstate ordering whereby the sign of the mass squared difference
  associated with the atmospheric neutrino problem is positive}

\newduneabbrev{ih}{IH}{inverted hierarchy}{Refers to the neutrino mass
  eigenstate ordering whereby the sign of the mass squared difference
  associated with the atmospheric neutrino problem is negative}

\newduneabbrev{no}{NO}{normal ordering}{Refers to the neutrino mass
  eigenstate ordering whereby the sign of the mass squared difference
  associated with the atmospheric neutrino problem is positive}

\newduneabbrev{io}{IO}{inverted ordering}{Refers to the neutrino mass
  eigenstate ordering whereby the sign of the mass squared difference
  associated with the atmospheric neutrino problem is negative}

\newduneabbrev{msw}{MSW}{Mikheyev-Smirnov-Wolfenstein effect}{Explains
  the oscillatory behavior of neutrinos produced inside the sun as
  they traverse the solar matter}

\newduneabbrev{nsi}{NSI}{nonstandard interaction}{A general class of
  theory of elementary particles other than the Standard Model}

\newduneabbrev{pfive}{P5}{Particle Physics Project Prioritization
Panel}{The Particle Physics Project Prioritization Panel (P5) was a
subpanel of the High Energy Physics Advisory Panel (HEPAP). It completed
its Report, a ten-year strategic plan for high energy physics in the
U.S., in 2014. This report included a recommendation that ``host a world-leading neutrino
program that will have an optimized set of short- and long-baseline neutrino oscillation experiments, and its long-term focus
is a reformulated venture referred to here as the Long Baseline
Neutrino Facility (LBNF)''}

\newduneabbrev{sme}{SME}{standard-model extension}{an effective field theory that contains the \gls{sm}, general relativity, and all possible operators that break Lorentz symmetry (Wikipedia)}

\newduneabbrev{susy}{SUSY}{supersymmetry}{Theoretical symmetry between a fermion and a boson}

\newduneabbrev{wimp}{WIMP}{weakly-interacting massive particle}{A
  hypothesized particle that may be a component of dark matter}

\newduneabbrev{ce}{CE}{cold electronics}{Analog and digital readout electronics that operate at cryogenic temperatures}

\newduneabbrev{crp}{CRP}{charge-readout plane}{An anode technology using a stack of perforated \glspl{pcb} with etched electrode strips to provide \gls{cro} in 3D; it has two induction layers and one collection layer; it is used in the \gls{spvd} \gls{fd} and \gls{dp} designs} % with projecting electrodes immersed in \dword{lar}}
\newduneabbrev{dram}{DRAM}{dynamic random access memory}{A computer memory technology}

\newduneabbrev{fnal}{Fermilab}
{Fermi National Accelerator Laboratory}{U.S. national laboratory in Batavia, IL.  It is the laboratory that hosts \gls{lbnf} and \gls{dune}, and serves as the experiment's near site}

\newduneabbrev{bnl}{BNL}{Brookhaven National Laboratory}{US national laboratory in Upton, NY}

\newduneabbrev{slac}{SLAC}{SLAC National Accelerator Laboratory}{US national laboratory in Menlo Park, CA}

\newduneabbrev{lbnl}{LBNL}{Lawrence Berkeley National Laboratory}{US national laboratory in Berkeley, CA}

\newduneabbrev{anl}{ANL}{Argonne National Laboratory}{US national laboratory in Lemont, IL}

\newduneabbrev{lanl}{LANL}{Los Alamos National Laboratory}{US national laboratory in Los Alamos, NM}

\newduneword{fs}{FS}{Depending on context, one of (1) the far site, \gls{surf}, where the DUNE far detector is located; (2) ``Full Stream'' relates to a data stream that has not undergone selection, compression or other form of reduction} % #36 (see change from item above)

\newduneabbrev{lem}{LEM}{large electron multiplier}{A micro-pattern detector suitable for use in ultra-pure argon vapor; LEMs consist of copper-clad PCB boards with sub-millimeter-size holes through which electrons undergo amplification}

\newduneabbrev{lng}{LNG}{liquefied natural gas}{Pertaining to natural gas in its liquid phase}

\newduneabbrev{mip}{MIP}{minimum ionizing particle}{Refers to a
  particle traversing some medium such that the particle's mean energy loss is  
  near the minimum}
\newduneabbrev{pd}{PD}{photon detector}{The detector
  elements involved in measurement of the number and arrival times of
  optical photons produced in a detector module} 

\newduneabbrev{pmt}{PMT}{photomultiplier tube}{A device that makes use
  of the photoelectric effect to produce an electrical signal from the
  arrival of optical photons}

\newduneabbrev{ppm}{ppm}{parts per million}{A concentration equal to one part in $10^{6}$}
\newduneabbrev{ppb}{ppb}{parts per billion}{A concentration equal to one part in $10^{9}$}
\newduneabbrev{ppt}{ppt}{parts per trillion}{A concentration equal to one part in $10^{12}$}
\newduneword{rio}{RIO}{reconfigurable input output}

\newduneabbrev{s/n}{S/N}{signal-to-noise}{signal-to-noise ratio}
\newduneword{ssp}{SSP}{\gls{sipm} signal processor}

\newduneabbrev{sbn}{SBN}{Short-Baseline Neutrino}{A \gls{fnal} program consisting of three collaborations, \gls{microboone}, \gls{sbnd}, and \gls{icarus}, to perform sensitive searches for $\nue$ appearance and $\numu$ disappearance in the Booster Neutrino Beam}

\newduneword{stt}{STT}{straw tube tracker}

\newduneword{wire board}{wire board}{At the head end of the APA in the \gls{sphd} \gls{tpc}, stacks of electronics boards referred to as ``wire boards'' are arrayed to anchor the wires.  They also provide the connection between the wires and the cold electronics} %?? long for a word. ??

\newduneabbrev{wls}{WLS}{wavelength-shifting}{A material or process by
  which incident photons are absorbed by a material and photons are
  emitted at a different, typically longer, wavelength}
  
\newduneabbrev{tpb}{TPB}{tetra-phenyl butadiene}{A \gls{wls} material}

\newduneabbrev{sft}{SFT}{signal feedthrough}{A cryostat penetration allowing for the passage of cables or other extended parts}

\newduneabbrev{sftchimney}{SFT chimney}{signal feedthrough chimney}{A volume above the cryostat penetration used for a signal feedthrough} %19 nov 2021  

\newduneabbrev{catiroc}{CATIROC}{charge and time integrated readout chip}{A complete read-out chip manufactured in AustriaMicroSystem designed to read arrays of 16 photomultipliers}

\newduneabbrev{wr}{WR}{White Rabbit}{A component of the timing system that forwards clock signal and time-of-day reference data to the master timing unit}

\newduneabbrev{mch}{MCH}{MicroTCA Carrier Hub}{A network switching device}

\newduneabbrev{wrmch}{WR-MCH}{White Rabbit \gls{utca} Carrier Hub}{A card mounted in \gls{utca} crate that recieves time syncronization information and trigger data packets over \gls{wr} network and disributes them to the \gls{amc} over \gls{utca} backplane} 

\newduneabbrev{wrtsn}{WR-TSN}{White Rabbit TimeStamping Node}{A unit on the \gls{wr} network that timestamps the trigger signals and sends out trigger data packets to \gls{wrmch}}

\newduneword{qap}{QAP}{quality assurance plan} %{A project management device for planning \gls{qa}}
\newduneword{ieshp}{IESHP}{integrated environmental, safety and health plan}%{Refers to the LBNF/DUNE project planning instrument}
\newduneword{dmp}{DMP}{data management plan} %{A project management device to state how the experimental data will be managed}
\newduneword{qam}{QAM}{quality assurance manager} %{The manager of \gls{qa} for the LBNF/DUNE project}

\newduneabbrev{dss}{DSS}{detector support system}{The system of rails suspended from the cryostat ceiling in a \gls{spmod} used to support the \gls{apa}s, \gls{cpa}s, and the \glspl{ewfc}} % update from MV 1sept2020 

\newduneabbrev{ddss}{DDSS}{DUNE detector safety system}{The hardware system responsible for the safety of the detector, implemented either via a \gls{plc} or via custom hardware protections} % update from MV 1sept2020 

\newduneabbrev{tco}{TCO}{temporary construction opening}{An opening in the side of a cryostat through which detector elements are brought into the cryostat; utilized during construction and installation}

\newduneabbrev{surf}{SURF}{Sanford Underground Research Facility}{The laboratory in Lead, SD, USA where the \gls{dune} \gls{fd} will be installed and operated; also where the \gls{usproj} \gls{fscf} and the \gls{fs} cryostat and cryogenics systems will be constructed} %  36

\newduneabbrev{uit}{UIT}{underground installation team}{An organizational unit responsible for installation in the underground area at the \gls{surf} site}

\newduneabbrev{cmgc}{CMGC}{construction manager/general contractor}{The contracted company hired to manage overall construction, used by \gls{lbnf} at the \gls{surf} site for the \gls{fscf} construction} % #36 old: The organizational unit responsible for management of the construction of conventional facilities at the underground area at the \gls{surf} site}

\newduneword{pdr}{PDR}{Depending on context, either ``preliminary design report,'' a formal project document  that describes the experiment at a preliminary level, or ``preliminary design review,'' a formal review of the preliminary design of the experiment or of a component} % #33

\newduneword{fdr}{FDR}{Depending on context, either ``final design report,'' a formal project document  that describes the experiment at a final level, or ``final design review,'' a formal review of the final design of the experiment or of a component} %  per #33, not sure if we need a `final design report' - yes per #36

\newduneabbrev{prr}{PRR}{production readiness review}{A project management mechanism by which the production readiness is reviewed}  % #33

\newduneabbrev{irr}{IRR}{installation readiness review}{A project management mechanism by which the plan for installation is reviewed}  % #33
\newduneabbrev{orr}{ORR}{operational readiness review}{A project management mechanism by which the operational readiness is reviewed}  % #33

\newduneabbrev{ppr}{PPR}{production progress review}{A project management mechanism by which the progress of production is reviewed}  % #33

\newduneabbrev{edms}{EDMS}{engineering document management system}{A computerized document management system developed and supported at the \gls{cern} in which some \gls{dune} project and collaboration documents, drawings and engineering models are managed} % #36  add `lbnf' before dune
\newduneabbrev{docdb}{DocDB}{Document DataBase}{A computerized document management system developed and supported at \gls{fnal} in which most \gls{lbnf-dune} documents are managed (docs.dunescience.org); some documents are maintained in \gls{edms}}

\newduneword{wrgm}{WR grandmaster}{White Rabbit grandmaster}

\newduneabbrev{larsoft}{LArSoft}{Liquid Argon Software}{A shared base of physics software across \gls{lartpc} experiments}
\newduneword{nova}{NOvA}{The \gls{nova} off-axis neutrino oscillation experiment at \gls{fnal}}
\newduneword{minerva}{MINERvA}{Neutrino cross sections experiment at \gls{fnal},  shut down in 2019}
\newduneword{microboone}{MicroBooNE}{A \gls{lartpc} neutrino oscillation experiment at \gls{fnal}}
\newduneword{sbnd}{SBND}{The Short-Baseline Near Detector experiment at  \gls{fnal}}
\newduneabbrev{nexo}{nEXO}{Enriched Xenon Observatory}{Experiment at Lawrence Livermore National Laboratory (U.S. national lab in Livermore, CA) searching for new physics with neutrinoless double-beta decay}
\newduneword{argoneut}{ArgoNeuT}{The ArgoNeuT test-beam experiment and \gls{lartpc} prototype at  \gls{fnal}}
\newduneword{icarus}{ICARUS}{A neutrino experiment that was located at the Laboratori Nazionali del Gran Sasso (LNGS) in Italy, then refurbished at \gls{cern} for re-use in the same neutrino beam from \gls{fnal} used by the \gls{miniboone} , \gls{microboone} and \gls{sbnd} experiments at \gls{fnal}}
\newduneword{atlas}{ATLAS}{One of two general-purpose detectors at the \gls{lhc}. It investigates a wide range of physics, from the measurements of the Higgs boson properties to searches for extra dimensions and particles that could make up \gls{dm}}

\newduneword{lbne}{LBNE}{Long Baseline Neutrino Experiment; (1) a terminated U.S. experiment that was reformulated in 2014 under the auspices of the new \gls{dune} collaboration, an internationally coordinated and internationally funded program, with \gls{fnal} as host; and (2) the former incarnation of \gls{usproj} } % #36

\newduneabbrev{lbno}{LBNO}{Long Baseline Neutrino Observatory} {A terminated European project that, during its six-year duration, assessed the feasibility of a next-generation deep underground neutrino observatory in Europe)}
\newduneword{wirecell}{Wire-Cell}{A tomographic automated \threed neutrino event reconstruction method for \glspl{lartpc}}
\newduneabbrev{wct}{WCT}{Wire-Cell Toolkit}{A software toolkit with data flow processing components for \gls{lartpc} noise and signal simulation, noise filtering, signal processing, and tomographic \threed ionization activity imaging}
\newduneword{ftslite}{F-FTS-lite}{Light-weight version of the \gls{fnal} File Transfer system used for rapid data transfers out of the online systems}
\newduneabbrev{fts}{FTS}{File Transfer System}{A file transfer system developed at \gls{fnal} to catalog and move data to permanent storage}

\newduneword{35t}{35 ton prototype}{A prototype cryostat and \gls{sp} detector built at \gls{fnal} before the \gls{protodune} detectors}

\newduneabbrev{mcr}{MCR}{main communications room}{Space at the \gls{fd} site for cyber infrastructure}

\newduneabbrev{cuc}{CUC}{central utility cavern}{The utility cavern at the 4850L of \gls{surf} located between the two detector caverns. It contains utilities such as central cryogenics and other systems, and the underground data center and control room}

\newduneabbrev{cfd}{CFD}{computational fluid dynamics}{High performance computer-assisted modeling of fluid dynamical systems}
\newduneword{vuv}{VUV}{vacuum ultra-violet}
\newduneword{tallbo}{TallBo}{A cylindrical cryostat at \gls{fnal} primarily used for developing scintillation light collection technologies for \gls{lartpc} detectors}

\newduneword{root}{ROOT}{A modular scientific software toolkit. It provides all the functionalities needed to deal with big data processing,  statistical analysis, visualization, and storage.  It is mainly written in C++ but integrated with other languages such as Python and R}

\newduneabbrev{eos}{EOS}{EOS}{The XRootD-based distributed file system developed by CERN}
\newduneabbrev{ehn1}{EHN1}{Experiment Hall North One}{Location at \gls{cern} of the \gls{np02} and \gls{np04} areas used for the \glspl{protodune} and for other test and prototyping activities for DUNE} 

\newduneword{led}{LED}{Light-emitting diode}
\newduneabbrev{rtd}{RTD}{resistance temperature detector}{A temperature sensor consisting of a material with an accurate and reproducible resistance/temperature relationship}
\newduneword{swc}{SWC}{Software \& Computing}
\newduneabbrev{las}{LAS}{LEM-anode Sandwich}{In the \gls{dp} technology, a \gls{lem} and its corresponding anode are mounted together in a module called a LEM-anode sandwich}

\newduneword{roi}{ROI}{region of interest}
\newduneabbrev{hpc}{HPC}{high-performance computing}{high-performance computing facilities; generally computing facilities emphasizing parallel computing with aggregate power of more than a teraflop}

\newduneword{comfund}{common fund}{The shared resources of the collaboration}
\newduneabbrev{ims}{IMS}{integrated master schedule}{A project management device consisting of linked tasks and milestones}

\newduneword{hvdb}{HVDB}{HV divider board}

\newduneword{hvft}{HVFT}{HV feedthrough}  % #33

\newduneword{sas}{SAS}{Another term for the materials airlock; a pass-through chamber used to ensure safe transfer of materials into a clean room, avoiding contamination in both directions}

\newduneabbrev{fea}{FEA}{finite element analysis}{Simulation of a physical phenomenon using the numerical technique called Finite Element Method (FEM), a numerical method for solving problems of engineering and mathematical physics}

\newduneword{fss}{FSS}{field shaping strips}
\newduneword{lvds}{LVDS}{low-voltage differential signaling}

\newduneword{esd}{ESD}{electrostatic discharge}%{ESD is the sudden flow of electricity between two electrically charged objects caused by contact, an electrical short, or dielectric breakdown. ESD can cause failure of electronic components such as integrated circuits}

\newduneabbrev{rp}{RP}{resistive panel}{Resistive panels form the constant potential surfaces for a \gls{spmod} \gls{cpa}; they are composed of a thin layer of carbon-impregnated Kapton and laminated to both sides of a \frfour sheet}

\newduneword{uhmwpe}{UHMWPE}{ultra-high molecular weight polyethylene}

\newduneword{cts}{CTS}{Cryogenic Test System}

\newduneabbrev{plc}{PLC}{programmable logic controller}{An industrial digital computer that has been ruggedized and adapted for the control of manufacturing or other processes that require high reliability, ease of programming, and process fault diagnosis} % update from Marco V 1sep2020

\newduneword{mppc}{MPPC}{\SI{6}{mm}$\times$\SI{6}{mm} Multi-Pixel Photon Counters produced by Hamamatsu\texttrademark{} Photonics K.K}

\newduneabbrev{sfp}{SFP}{small form-factor pluggable}{a particular standard for optical transceivers}

\newduneabbrev{minipod}{MiniPOD}{miniature parallel optical device}{a family of types of multi-channel optical transceivers}

\newduneword{ccc}{CCC}{configuration change command}
\newduneword{ccondc}{CCC}{code of conduct committee}

\newduneword{act}{ACT}{activation time stamp}
\newduneword{lcm}{LCM}{light calibration module}
\newduneword{lpm}{LPM}{light pulser module}
\newduneword{dac}{DAC}{digital-to-analog converter}
\newduneword{arapuca}{ARAPUCA}{A \gls{pds} design that consists of a light trap that captures wavelength-shifted photons inside boxes with highly reflective internal surfaces until they are eventually detected by \gls{sipm} detectors or are lost}
\newduneword{sarapu}{S-ARAPUCA}{Standard \gls{arapuca} design with different \gls{wls} coatings on both faces of the dichroic filter window(s) of the cell}
\newduneword{xarapu}{X-ARAPUCA}{Extended \gls{arapuca} design with \gls{wls} coating on only the external face of the dichroic filter window(s) but with a \gls{wls} doped plate inside the cell}
\newduneword{feb}{FEB}{front-end board}

\newduneabbrev{lsnd}{LSND}{Liquid Scintilator Neutrino Detector}{A scintillation detector and associated experiment located at Los Alamos National Laboratory}

\newduneabbrev{cvn}{CVN}{convolutional visual network}{An algorithm for identifying neutrino interactions based on their topology and without the need for detailed reconstruction algorithms}

\newduneword{pandora}{Pandora}{The Pandora multi-algorithm approach to pattern recognition} 

\newduneabbrev{pma}{PMA}{Projection Matching Algorithm}{A reconstruction algorithm that combines \twod reconstructed objects to form a \threed representation}
\newduneabbrev{bdt}{BDT}{boosted decision tree}{A method of multivariate analysis}
\newduneabbrev{cnn}{CNN}{convolutional neural network}{A deep learning technique most commonly applied to analyzing visual imagery}
\newduneword{pdg}{PDG}{Particle Data Group}

\newduneword{pci}{PCI}{Peripheral Component Interconnect}

\newduneword{labview}{LabVIEW}{Laboratory Virtual Instrument Engineering Workbench is a system-design platform and development environment for a visual programming language from National Instruments}

\newduneword{pcb}{PCB}{printed circuit board}

\newduneword{crio}{cRIO}{Compact Reconfigurable Input Output}

\newduneword{dcs}{DCS}{Distributed Communications System}

\newduneword{opc-ua}{OPC-UA}{OPC  Unified Architecture is a machine to machine communication protocol for industrial automation developed by the OPC Foundation. OPC stands for Object Linking and Embedding for Process Control}

\newduneword{cabangle}{Cabibbo angle}{A quark mixing parameter that governs the coupling of up quarks to strange quarks}
\newduneword{valor}{VALOR}{A neutrino oscillation fitting framework that is used by \gls{t2k}; the name stands for VALencia-Oxford-Rutherford, the original three institutions that developed it}
\newduneword{cafana}{CAFAna}{Common Analysis File Analysis}
\newduneabbrev{pca}{PCA}{principal component analysis}{A statistical procedure that uses an orthogonal transformation to convert a set of observations of possibly correlated variables into a set of values of linearly uncorrelated variables called principal components (Wikipedia)}
\newduneword{numi}{NuMI}{a set of facilities at \gls{fnal}, collectively called ``Neutrinos at the Main Injector.''  The NuMI neutrino beamline target system converts an intense proton beam into a focused neutrino beam}
\newduneword{gibuu}{GiBUU}{Giessen Boltzmann-Uehling-Uhlenback Project; a unified theory and transport framework in the MeV and GeV energy regimes for elementary reactions on nuclei }
\newduneabbrev{rpa}{RPA}{random phase approximation} {an approximation method commonly used for describing the dynamic linear electronic response of electron systems (Wikipedia)}%(from nu-osc 05)}
\newduneword{t2k}{T2K}{T2K (Tokai to Kamioka) is a long-baseline neutrino experiment in Japan studying neutrino oscillations}
\newduneword{mptdet}{MPT detector}{multipurpose tracking detector}

\newduneword{lariat}{LArIAT}{The repurposed ArgoNeuT \gls{lartpc}, modified for use in a charged particle beam, dedicated to the calibration and precise characterization of the output response of these detectors}

\newduneword{captain}{CAPTAIN}{Experimental program sited at \gls{lanl} that is designed to make measurements of scientific importance to \gls{lbl} neutrino physics and physics topics that will be explored by large underground detectors}

\newduneword{dayabay}{Daya Bay}{a neutrino-oscillation experiment in Daya Bay, China, designed to measure the mixing angle $\Theta_{13}$  using antineutrinos produced by the reactors of the Daya Bay and Ling Ao nuclear power plants}

\newduneword{nuwro}{NuWro}{neutrino interaction generator}

\newduneabbrev{neut}{NEUT}{neutrino interaction generator}{A neutrino interaction simulation program library for the studies of atmospheric and accelerator neutrinos}

\newduneword{minos}{MINOS}{A long-baseline neutrino experiment, with a near detector at \gls{fnal} and a far detector in the Soudan mine in Minnesota,  designed to observe the phenomena of neutrino oscillations (ended data runs in 2012)}

\newduneabbrev{efig}{EFIG}{Experimental Facilities Interface Group}{The body that provides a means to coordinate and discuss issues related to the interfaces within and between the facility and the DUNE detectors at both the \gls{fnal} and \gls{surf} sites} %The body responsible for the required high-level coordination between the project and collaboration in \gls{lbnf-dune}}

\newduneword{ashriver}{Ash River}{The Ash River, Minnesota, USA \gls{nova} experiment far site, used as an assembly test site for \gls{dune}} 

\newduneword{ipd}{project integration director}{Responsible for integration and installation of DUNE detector deliverables. Manages the integration project} %#35 09nov20 Responsible for integration and installation of \gls{lbnf} and \gls{dune} deliverables in South Dakota. Manages the \gls{integoff}}

\newduneabbrev{jpo}{JPO}{Joint Project Office}{The framework used to facilitate close and coherent coordination across all elements with many shared management resources; 
JPO functions include systems engineering, procurement, \gls{esh}, \gls{qa}, finance, project controls, risk management, compliance, internal reviews, partner agreement management, document management, and administrative support} % #35 09nov20 global project configuration and integration, installation planning and coordination, scheduling, safety assurance, technical review planning and oversight, development of partner agreements, and financial reporting}

\newduneword{ifbeam}{IFbeam}{Database that stores beamline information indexed by timestamp}

\newduneabbrev{marley}{MARLEY}{Model of Argon Reaction Low Energy Yields}{Developed at UC Davis, MARLEY is the first realistic model of neutrino electron interactions on argon for enegies less than \SI{50}{MeV}. This includes the energy range important for \gls{snb} neutrinos and also solar 8--boron neutrinos}

\newduneabbrev{es}{ES}{elastic scattering}{Events in which a neutrino
elastically scatters off of another particle}

\newduneabbrev{cno}{CNO}{carbon nitrogen oxygen}{The CNO cycle (for carbon-nitrogen-oxygen) is one of the two known sets of fusion reactions by which stars convert hydrogen to helium, the other being the proton-proton chain reaction (pp-chain reaction). In the CNO cycle, four protons fuse, using carbon, nitrogen, and oxygen isotopes as catalysts, to produce one alpha particle, two positrons and two electron neutrinos}

\newduneabbrev{sdwf}{SDWF}{South Dakota Warehouse Facility}{Warehousing operations in South Dakota responsible for receiving LBNF and DUNE goods and coordinating shipments to the access shaft (Ross Shaft) at \gls{surf}}

\newduneabbrev{wms}{WMS}{warehouse management system}{Commercial software package used to track shipments and interface to freight forwarders. This includes a database for shipping}

\newduneabbrev{dcdb}{DCDB}{DUNE construction database}{Database used by DUNE to track the history and testing of all parts of each \gls{detmodule}}

\newduneabbrev{aup}{AUP}{acceptance for use and possession}{Required for beneficial occupancy of the underground areas at \gls{surf} for \gls{lbnf} and \gls{dune}}

\newduneabbrev{bms}{BMS}{building management system}{A system provided by the \gls{cf} to manage the utility (cooling, ventilation, power, etc.) and fire/life safety systems. Separate systems are provided at \gls{surf} and at \gls{fnal}} % #36 Elaine  old: Part of the safety system at \gls{surf} that includes the fire and life safety system}

\newduneabbrev{fls}{FLS}{fire and life safety system}{Fire and life safety; systems designed with \gls{cf} to meet building/safety code compliance for safe facilities at \gls{surf} and at \gls{fnal}} % #36 

\newduneabbrev{sno}{SNO}{Sudbury Neutrino Observatory}{The Sudbury Neutrino Observatory was a detector built 6800 feet under ground, in INCO's Creighton mine near Sudbury, Ontario, Canada. SNO was a heavy-water Cherenkov detector designed to detect neutrinos produced by fusion reactions in the sun}

\newduneword{sk}{Super-Kamiokande}{Experiment sited in the Kamioka-mine, Hida-city, Gifu, Japan that uses a large water Cherenkov detector to study neutrino properties through the observation of solar neutrinos, atmospheric neutrinos and man-made neutrinos}

\newduneabbrev{id}{ID}{inner diameter}{Inner diameter of a tube}

\newduneabbrev{od}{OD}{outer diameter}{Outer diameter of a tube}

\newduneabbrev{rms}{RMS}{root mean square}{The square root of the arithmetic mean of the squares of a set of values, used as a measure of the typical magnitude of a set of numbers, regardless of their sign}

\newduneabbrev{orc}{ORC}{operational readiness clearance}{Final safety approval prior to the start of operation}

\newduneabbrev{gsc}{GSC group}{global safety coordination group}{DUNE group that evaluates applicable codes and standards, including international code equivalency, for the design, assembly, and installation of the \gls{fd}}

\newduneabbrev{ha}{HA}{hazard analysis}{A first step in a process to assess risk; the result of hazard analysis is the identification of the hazards present for a task or process}%different types of hazards}(anne)
\newduneword{har}{HAR}{hazard analysis report}

\newduneabbrev{tap}{TAP}{trip action plan}{A document required for any trip by a worker to the underground area at \gls{surf}, per that site's access control program; it describes the work to be accomplished during the trip} % ask Jim Stewart to check

\newduneword{em}{EM}{emergency management}
\newduneword{ert}{ERT}{emergency response team}

\newduneabbrev{ndk}{NDK}{nucleon decay}{The hypothetical, baryon number violating decay of a proton or a bound neutron into lighter particles}

\newduneabbrev{emi}{EMI}{electromagnetic interference}{Disturbance generated by an external source that affects an electrical circuit by electromagnetic induction, electrostatic coupling, or conduction}

\newduneabbrev{pe}{PE}{photoelectron}{An electron ejected from the surface of a material by the photoelectric effect}

\newduneabbrev{spe}{SPE}{single photoelectron}{A single photoelectron}

\newduneabbrev{fwhm}{FWHM}{full width at half maximum}{Width of a distribution measured between those points at which the distribution is equal to half of its maximum amplitude}

\newduneabbrev{gdml}{GDML}{geometry description markup language}{An application-independent, geometry-description format based on XML}

\newduneabbrev{xml}{XML}{extensible markup language}{A markup language that defines a set of rules for encoding documents in a format that is both human-readable and machine-readable}

\newduneabbrev{crt}{CRT}{cosmic-ray tagger}{Detector external to the TPC designed to tag TPC-traversing cosmic ray particles}

\newduneabbrev{sn}{SN}{supernova}{Event that occurs upon the death of certain types of stars}

\newduneabbrev{wg}{WG}{working group}{A group of persons working together to achieve specified goals}

\newduneabbrev{ctsf}{CTSF}{coating, testing and storage facility}{A facility where the the \gls{dp} photon detectors will be coated, tested, and stored} %11 may 2021 (no more DP module)- will other DP be tested? Is it at CERN?

\newduneword{rucio}{Rucio}{Data management system originally developed
by \gls{atlas} but now open-source and shared across \gls{hep}}
\newduneabbrev{doma}{DOMA}{data organization, management, and access}{data organization, management, and access efforts through the \gls{hsf}}

\newduneabbrev{hsf}{HSF}{HEP Software Foundation}{A foundation that facilitates cooperation and common efforts in high energy physics software and computing internationally} %updated 3/4/22 AH per P. Laycock

\newduneabbrev{wlcg}{WLCG}{Worldwide LHC Computing Grid}{Worldwide LHC
Computing Grid}
\newduneabbrev{osg}{OSG}{Open Science Grid}{Open Science Grid}
\newduneabbrev{sci}{SCI}{Scientific Computing Infrastructure}{Proposed
extension of the infrastructure component of \gls{wlcg} to other
experiments}
\newduneabbrev{csc}{CSC}{computing and software consortium}{DUNE
computing and software consortium}

\newduneword{dirac}{DIRAC}{Computing workflow management designed for LHCb and now used by many HEP experiments}

\newduneword{frp}{FRP}{fiber-reinforced plastic}
\newduneabbrev{hdpe}{HDPE}{high-density polyethylene}{A thermoplastic polymer made from petroleum commonly used to make plastic bottles}
\newduneword{hvps}{HVPS}{\gls{hv} power supply}
\newduneword{aisi}{AISI}{American Iron and Steel Institute}
\newduneword{ific}{IFIC}{Instituto de Fisica Corpuscular (in Valencia, Spain)}
\newduneabbrev{rsds}{RSDS}{radioactive source deployment system}{Proposed calibration system based on the deployment of
radioactive sources inside the \gls{dune} cryostat}
\newduneword{2p2h}{2p2h}{two particle, two hole}
\newduneabbrev{duneprism}{DUNE-PRISM}{\gls{dune} Precision Reaction-Independent Spectrum Measurement}{a mobile near detector that can perform measurements over a range of angles off-axis from the neutrino beam direction in order to sample many different neutrino energy distributions}
\newduneword{arcube}{ArgonCube}{The name of the core part of the \gls{dune} \gls{nd}, a \gls{lartpc}}

\newduneabbrev{citf}{CITF}{cryogenic instrumentation test facility}{A facility at \gls{fnal} with small ($<$\SI{1}{ton}) to intermediate ($\sim$\SI{1}{ton}) volumes of instrumented, purified TPC-grade \gls{lar}, used for testing devices intended for use in \gls{dune}}

\newduneabbrev{3dst}{3DST}{3D scintillator tracker}{The core part of the \threed projection scintillator tracker spectrometer in the near detector conceptual design}
\newduneabbrev{3dsts}{3DST-S}{3D scintillator tracker spectrometer}{The \threed projection scintillator tracker spectrometer  in the near detector conceptual design}
\newduneabbrev{mpd}{MPD}{multi-purpose detector}{A component of the near detector conceptual design; it is a magnetized system consisting of a \gls{hpgtpc} and a surrounding \gls{ecal}}
\newduneabbrev{hpg}{HPG}{high-pressure gas}{gas at high pressure to be used in a \gls{hpgtpc}} 
\newduneabbrev{hpgtpc}{HPgTPC}{high-pressure gaseous argon TPC}{A \gls{tpc} filled with gaseous argon; a possible component of the \gls{dune} \gls{nd}}

\newduneword{src}{SRC}{short-range correlated nucleon-nucleon interactions}
\newduneword{larpix}{LArPix}{ \gls{asic} pixelated charge readout for a \gls{tpc} }
\newduneword{arclt}{ArCLight}{a light detector for the \gls{arcube} effort}
\newduneword{fhc}{FHC}{forward horn current ($\numu$ mode)}
\newduneword{rhc}{RHC}{reverse horn current (\numubartonumubar mode)}
\newduneword{mwpc}{MWPC}{multi-wire proportional chamber}
\newduneword{na61}{NA61}{CERN hadron production experiment}
\newduneword{pdnd}{ProtoDUNE-ND}{a prototype \gls{dune} \gls{nd}}
\newduneabbrev{roc}{ROC}{readout chamber}{readout chamber for gaseous argon \gls{tpc}}
\newduneabbrev{iroc}{IROC}{inner readout chamber}{inner (radial) readout chamber for gaseous argon \gls{tpc}}
\newduneabbrev{oroc}{OROC}{outer readout chamber}{outer (radial) readout chamber for gaseous argon \gls{tpc}}

\newduneword{lux}{LUX}{Large Underground Xenon (LUX) dark matter detector at \gls{surf} }

\newduneword{mjdemo}{Majorana Demonstrator}{Experiment sited at \gls{surf} that  seeks to determine whether neutrinos are their own antiparticles}

\newduneword{lz}{LZ}{Experiment sited at \gls{surf} that  seeks to detect faint interactions between galactic dark matter and regular matter}

\newduneword{mu2e}{Mu2e}{An experiment sited at \gls{fnal} that searches for charged-lepton flavor violation and seeks to discover physics beyond the \gls{sm}}

\newduneword{pdsp2}{ProtoDUNE-SP-II}{A second test run in the single-phase ProtoDUNE test stand at CERN, acting as a validation of the final
single-phase detector design}  % #33 11 may 21-  may need an update

\newduneword{osha}{OSHA}{Occupational Safety and Health Administration (USA Department of Labor) formed by the Occupational Safety and Health Act of 1970}
\newduneabbrev{pns}{PNS}{pulsed neutron source}{Calibration system based
on neutron capture gamma showers spread out in the whole detector}

\newduneabbrev{fv}{FV}{fiducial volume}{The detector volume within the \gls{tpc} that is selected for physics analysis through cuts on reconstructed event position}

\newduneword{p6}{P6}{software framework used to plan and status the resource-loaded schedule of activities associated with \gls{usproj}}

\newduneabbrev{evms}{EVMS}{earned value management system}{Earned Value Management is a systematic approach to the integration and measurement of cost, schedule, and technical (scope) accomplishments on a project or task. It provides both the government and contractors the ability to examine detailed schedule information, critical program and technical milestones, and cost data (text from the US DOE); the EVMS is a system that implements this approach}

\newduneword{core}{CORE}{CORE contributions are in either monetary units or labor hours. They can be technical components for the facility or experiment and the effort of the staff needed to produce, install, and test them;  major facilities for the experiment; or other products and services relevant for the completion of the facility or experiment} % 15 May - gina needs to `bless' this

\newduneabbrev{ahj}{AHJ}{Authority Having Jurisdiction}{An organization, office, or individual responsible for enforcing the requirements of a code or standard, or for approving equipment, materials, an installation, or a procedure (\gls{osha})}
\newduneword{cte}{CTE}{coefficient of thermal expansion}

\newduneabbrev{opc}{OPC}{open platform communications}{Open platform communications is a series of standards and specifications for industrial telecommunication} 
\newduneword{scada}{SCADA}{supervisory control and data acquisition}
\newduneword{ln}{LN$_2$}{liquid nitrogen}
\newduneabbrev{lapd}{LAPD}{Liquid Argon Purity Demonstrator}{Cryostat at \gls{fnal}  for long-term studies requiring a large volume of argon}

\newduneabbrev{pab}{PAB}{Proton Assembly Building}{Home of several \gls{lar} facilities at \gls{fnal} }
\newduneword{hep}{HEP}{high energy physics}
\newduneword{cms}{CMS}{Compact Muon Solenoid experiment; one of two general-purpose detectors at the \gls{lhc}. }
\newduneword{alice}{ALICE}{A Large Ion Collider Experiment, at CERN}
\newduneword{gpib}{GPIB}{general purpose interface bus}

\newduneabbrev{pfparticle}{PFParticle}{particle flow particle}{Each of the individual reconstructed particles in the hierarchy (or particle flow) describing the reconstructed event interaction}

\newduneabbrev{mcparticle}{MCParticle}{Monte Carlo Particle}{Individual true simulated particle}
\newduneword{au}{AU}{astronomical unit}
\newduneword{nufit}{NuFIT 4.0}{The NuFIT 4.0 global fit to neutrino oscillation data}

\newduneword{uhv}{UHV}{ultra high vacuum}
\newduneword{lps}{LPS}{laser positioning system}

\newduneword{unicamp}{UNICAMP}{University of Campinas, Sao Paulo, Brazil}
 
\newduneabbrev{fbk}{FBK}{Fondazione Bruno Kessler}{FBK is a research non-profit entity in Trento, Italy that partners in the development of technology with applications in various fields including High Energy Physics}

\newduneabbrev{enob}{ENOB}{effective number of bits}{The effective number of bits is a measure of the dynamic range of an \gls{adc} and its associated circuitry. The resolution of an \gls{adc} is specified by the number of bits used to represent the analog value, in principle giving 2N signal levels for an N-bit signal. However, all real \gls{adc} circuits introduce noise and distortion. ENOB specifies the resolution of an ideal \gls{adc} circuit that would have the same resolution as the circuit under consideration}
\newduneabbrev{sndr}{SNDR}{signal to noise and distortion ratio}{Also known as SINAD. Ratio of the \gls{rms} signal amplitude to the mean value of the root-sum-square of all other spectral components, including harmonics, but excluding \gls{dc} levels. It is a good indication of the overall dynamic performance of an \gls{adc} because it includes all components which make up noise and distortion}
\newduneabbrev{sfdr}{SFDR}{spurious free dynamic range}{Spurious free dynamic range is the ratio of the \gls{rms} value of the signal to the \gls{rms} value of the worst spurious signal regardless of where it falls in the frequency spectrum. The worst spur may or may not be a harmonic of the original signal}
\newduneabbrev{thd}{THD}{total harmonic distortion}{Total harmonic distortion is the ratio of the \gls{rms} value of the fundamental signal to the mean value of the root-sum-square of its harmonics} 
\newduneword{tvs}{TVS}{transient voltage suppression}

\newduneword{riskprob}{risk probabilities}{The risk probability, after taking into account the planned mitigation activities, is ranked as 
 L (low $<\,$\SI{10}{\%}), 
M (medium \SIrange{10}{25}{\%}), or 
H (high $>\,$\SI{25}{\%}). 
The cost and schedule impacts are ranked as 
L (cost increase $<\,$\SI{5}{\%}, schedule delay $<\,$2 months), 
M (\SIrange{5}{25}{\%} and 2--6 months, respectively) and 
H ($>\,$\SI{20}{\%} and $>\,$2 months, respectively)}

\newduneabbrev{lbls}{LBLS}{laser beam location system}
{Auxiliary calibration system providing an independent location measurement of the ionization laser beams direction}

\newduneabbrev{lsst}{LSST}{Large Synoptic Survey Telescope}{8.4 m telescope with 3.2G-pixel camera that will start taking data in 2023}
\newduneabbrev{ska}{SKA}{Square Kilometer Array}{International radio telescope array planned to start data-taking in 2027}
\newduneabbrev{hyperk}{HyperK}{Hyper Kamiokande}{260 kt water Cerenkov neutrino detector to begin construction at Kamiokande in 2020}
\newduneword{lhcb}{LHCb}{LHC experiment dedicated to forward physics}
\newduneword{belleii}{Belle II}{B-factory experiment now running at KEK}

\newduneabbrev{ldm}{LDM}{light-mass dark matter}{Refers to dark matter particles with mass values much lower than the electroweak scale, specifically below the 1~GeV level}
 
\newduneabbrev{bnv}{BNV}{baryon-number violating}{Describing an interaction where \gls{baryonnumber} is not conserved}

\newduneword{bugey}{Bugey}{Neutrino experiment that operated at the Bugey nuclear power plant in France}

\newduneword{minosplus}{MINOS$+$}{The successor to the \gls{minos} experiment, utilizing the same detectors and beam line, but operating at higher beam energy tune than \gls{minos}, parasitic with \gls{nova}}

\newduneword{baryonnumber}{baryon number}{A quantity expressing the total number of baryons in a system minus the number of antibaryons}

\newduneword{np04}{NP04}{%Experiment 
The CERN North Area in \gls{ehn1} intersected by the \gls{h4} hadron beamline,  the location of  \gls{pdsp} and \gls{pdsp2}; also used to refer to the 800\,t cryostat in this area}

\newduneword{np02}{NP02}{%Experiment 
The CERN North Area in \gls{ehn1} intersected by the \gls{h2} hadron beamline, the location of  the 800\,t cryostat used for \gls{pddp} and for \gls{spvd} tests and prototypes; also used to refer to the 800\,t cryostat in this area}

\newduneword{h4}{H4}{CERN North Area hadron beamline used for \gls{pdsp} and \gls{pdsp2}}  % #33

\newduneword{h2}{H2}{CERN North Area hadron beamline used for \gls{pddp} and \gls{spvd} prototypes and demonstrators}  % #33

\newduneword{ua1}{UA1}{UA1 (Underground Area 1) was a particle detector at \gls{cern}'s  Super Proton Synchrotron (SPS). It ran from 1981 until 1990, when the SPS was used as a proton-antiproton collider, searching for traces of W and Z particles in collisions. (CERN) The UA1 dipole magnet was reused in the NOMAD experiment and currently provides the magnetic field for the \gls{t2k} ND280 detector}

\newduneword{ssc}{SSC}{The Superconducting Super Collider was to be a huge underground ring complex beneath the area near Waxahachie, Texas, USA, that would have been the world’s most energetic particle accelerator. It was begun in 1990, but canceled by the U.S. Congress in 1993 (scientificamerican.com Oct 2013)}

\newduneword{daphne}{DAPHNE}{Detector electronics for Acquiring PHotons from NEutrinos is a custom-developed warm front-end waveform digitizing electronics module derived from the readout system developed at \gls{fnal}  for the Mu2e experiment}
 
\newduneword{nersc}{NERSC}{National Energy Research Computing Facility at \gls{lbnl}}

\newduneword{integoff}{integration project}{The \gls{doe} project element that organizes the onsite teams responsible for coordinating far detector installation and detector-facility integration activities at \gls{surf} as well as near detector installation activities at \gls{fnal}.  The integration project office includes overall LBNF/DUNE systems engineering, compliance and review offices} % #35

\newduneabbrev{sma}{SMA}{SubMiniature version A}{Connector interface for coaxial cables
with a screw-type coupling mechanism}

\newduneword{kloe}{KLOE}{KLOE is a $e^+ e^-$ collider detector spectrometer operated at DAFNE,  the $\phi$-meson factory at Frascati, Rome.  In DUNE it will consist of a \SI{26}{cm} Pb+scintillating fiber \gls{ecal} surrounding a cylindrical open detector region that is  \SI{4.00}{m} in diameter and \SI{4.30}{m} long.  The \gls{ecal} and detector region are embedded in a \SI{0.6}{T} magnetic field created by a \SI{4.86}{m} diameter superconducting coil and a \SI{475}{tonne} iron yoke}

\newduneabbrev{ro}{RO}{review office}{An office within the \gls{integoff} that organizes reviews} % #33
\newduneabbrev{doecd}{CD}{critical decision}{The U.S. DOE's Order 413.3B outlines a series of staged project approvals, each of which is referred to as a critical decision (CD)}

\newduneabbrev{lbnfspac}{LBNF/DUNE-US SPAC}{LBNF / DUNE-US Strategic Project Advisory Committee}{A committee charged by the host laboratory director to provide expert, independent advice on significant issues and strategies related to \gls{usproj} project organization, management, and risks} % #36

\newduneabbrev{sand}{SAND}{System for on-Axis Neutrino Detection}{The beam monitor component of the near detector that remains on-axis at all times and serves as a dedicated neutrino spectrum monitor}

\newduneword{4850l}{4850L}{The depth in feet (1480 m) of the access level for the DUNE underground area at SURF; called the ``4850 level''} % #33

\newduneword{apb}{APB}{authorship and publications board}
\newduneword{crb}{CRB}{collaboration resources board}
\newduneword{cube}{CuBe}{beryllium copper, used to make \gls{sphd} \gls{apa} readout planes}
\newduneword{drift}{drift}{(1) refers to electron drift under the influence of an electric field in a \gls{tpc}; (2) an excavated horizontal corridors (tunnels) in the underground areas at \gls{surf}}
\newduneword{exposure}{exposure}{The integrated detector fiducial mass times beam intensity; it is proportional to the number of interactions and is used to normalize cross sections in a data sample}
\newduneword{fr4}{FR-4}{Flame-retardant fiberglass-reinforced epoxy resin laminate used in making PCBs and other detector components}
\newduneword{g10}{G-10}{Non-flame-retardant fiberglass-reinforced epoxy resin laminate used in making PCBs}
\newduneword{kapton}{Kapton}{A polyimide plastic film that is stable over a broad range of temperatures and is resistant to radiation damage}
\newduneword{shaft}{shaft}{A vertical excavation at \gls{surf} connecting with the surface}
\newduneword{winze}{winze}{A vertical excavation at \gls{surf} connecting two drifts, not connecting to the surface}
\newduneword{ib}{IB}{institutional board; all institutions participating in DUNE are represented on this board}
\newduneword{irb}{IBR}{institutional board representative}
\newduneword{sc}{SC}{depending on context, either speakers committee or scientific computing} % #33
\newduneword{sac}{SAC}{spokespersons advisory committee}
\newduneword{eoc}{EOC}{education and outreach committee}
\newduneword{indico}{Indico}{Web-based meeting organization tool}
\newduneabbrev{htc}{HTC}{High Throughput Computing}{Computing facilities typically consisting of large numbers of commodity servers as opposed to a single large machine. Best suited for running large numbers of independent jobs in parallel, these facilities are what is usually meant by ``grid computing''}
\newduneword{dcache}{dCache}{A distributed, highly scalable (multi-PB) storage system, usable as both a standalone system and as a high-speed frontend to a tape storage system (such as \gls{pnfs} at \gls{fnal} )}
\newduneabbrev{ifdhc}{IFDHC}{Intensity Frontier Data Handling Client}{A multi-protocol tool for data transfer and file delivery in jobs. It is able to automatically select transfer protocols based on source and destination characteristics}
\newduneabbrev{ifdh}{ifdh}{Intensity Frontier Data Handling}{The actual command invoked when using \gls{ifdhc}, on the command line, e.g. ifdh cp source\_file dest\_file}
\newduneabbrev{pnfs}{PNFS}{Pseudo Network File System}{A file system often used in large storage systems. Typically interaction is very similar to a regular NFS volume, but there can be some subtle and important differences}

\newduneword{nde}{NDE}{non-destructive evaluation} % 33
\newduneword{psv}{PSV}{pressure safety valve}
\newduneword{pickling}{pickling}{steel pickling and oiling is a metal surface treatment finishing process used to remove surface impurities such as rust and carbon scale from hot rolled carbon steel}

\newduneabbrev{tof}{ToF}{time of flight}{The time a particle takes to fly between two visible interactions observed in the detector. If combined with the distance traveled by the particle, for example a neutron, it can be used for energy reconstruction}

\newduneword{pep4}{PEP-4}{TPC for the Positron Electron Project 4 Collider at Stanford}

\newduneword{ndlar}{ND-LAr}{\gls{lartpc} component of the near detector based on \gls{arcube} technology}

\newduneword{ndgar}{ND-GAr}{component of the near detector with a core gaseous argon \gls{tpc} surrounded by an \gls{ecal} and a magnet}

\newduneword{sfgd}{SuperFGD}{Super Fine-Grained Detector (SuperFGD) is a 3D granular plastic scintillator detector that adopts the same technology as \dword{3dst}. It will be installed in the \dword{t2k} \dword{nd280} system \cite{Abe:2019whr}. The \dword{3dst} design will inherit in large part from the SuperFGD detector}

\newduneword{nd280}{ND280}{Near Detector 280, is the \dword{t2k} magnetized near detector \cite{Abe:2011ks}}

\newduneword{fee}{FEE}{front-end electronics}

\newduneword{mpgd}{MPGD}{MicroPattern Gas Detectors}

\newduneword{rmm}{RMM}{Resistive MicroMegas}

\newduneabbrev{stv}{STV}{Single Transverse Variables}{Kinematical variables obtained by projecting the neutrino interaction onto the transverse plane}

\newduneabbrev{ccqe}{CCQE}{charged current quasielastic interaction} {An interaction where a neutrino scatters from a nucleon, producing a charged lepton and converting a neutron to a proton or vice versa}

\newduneword{sis}{SIS}{shallow inelastic scattering}

\newduneabbrev{res}{RES}{resonant scattering}{The mode of scattering where the target nucleon is excited to a resonant state and decays, typically producing one or more pions}

\newduneabbrev{hadw}{W}{invariant mass of the hadronic system}{Refers to the invariant mass of the hadronic system formed during the neutrino scatter}

\newduneabbrev{agky}{AGKY}{Andreopoulos-Gallagher-Kehayias-Yang}{A model for hadronization of non-resonant inelastic neutrino reactions used in \gls{genie}. At low invariant hadronic masses, typically less than 2.3\,GeV/c$^2$, it is a KNO-inspired empirical model anchored on several bubble chamber measurements of neutrino-induced shower characteristics. For invariant hadronic masses between 2.3 and 3.0\,GeV/c$^2$, the model transitions linearly to a \gls{genie}-tuned version of PYTHIA, which is also used for the simulation of events at higher invariant masses} % update from Costas Andreopoulos sept2020
\newduneabbrev{clas}{CLAS}{CEBAF Large Acceptance Spectrometer}{A nuclear and particle physics detector located in the experimental Hall B at Jefferson Laboratory in Newport News, Virginia, United States. It is used to study the properties of the nuclear matter by the collaboration of over 200 physicists. Of particular relevance is its study of electron interactions with nuclei, including argon} 

\newduneabbrev{e4nu}{e4nu}{Electrons for Neutrinos}{A collaboration dedicated to using JLab's electron-scattering data to deliver improved neutrino-nucleus cross sections} 

\newduneabbrev{ldmx}{LDMX}{Light Dark Matter eXperiment}{The LDMX detector concept consists of a small precision tracker, and electromagnetic and hadronic calorimeters, all with near $2\pi$ azimuthal acceptance from the forward beam axis out to $\sim40^\circ$ angle. This detector would be capable of measuring correlations among electrons, pions, protons, and neutrons in electron-nucleus scattering at exactly the energies relevant for DUNE physics} 

\newduneabbrev{mec}{MEC}{meson-exchange currents} {An nuclear effect wherein pairs or larger groups of nucleons within a nucleus are bound together through the exchange of pions or other mesons. Neutrinos and other particles can scatter from these correlated pairs}

\newduneword{imt}{IMT}{Intranuclear momentum transfer}

\newduneword{miniboone}{MiniBooNE}{The Mini Booster Neutrino Experiment,  at \gls{fnal} , was designed to fully explore the LSND result}

\newduneabbrev{tms}{TMS}{The Muon Spectrometer}{A muon spectrometer for the Near Detector that will be installed for the initial running period of DUNE, before the \gls{mpd} detector component is ready} % #34 09nov20  08Apr22 Temporary -> The

\newduneabbrev{cvmfs}{CVMFS}{CERN VM File System}{A distributed file system designed for scalable, high-performance distribution of software to interactive and batch computers} %from tjunk sept2020 #21

\newduneword{kerberos}{Kerberos}{A strong authentication system used by the computing resources at \gls{fnal} and \gls{cern}} %from tjunk sept2020 #23

\newduneabbrev{mrb}{MRB}{Multi Repository Build System}{A \gls{fnal}-developed build system based on \dword{cmake} that allows development and builds of code from multiple repositories} %heidi sept2020

\newduneword{cmake}{Cmake}{CMake is an open-source, cross-platform family of tools designed to build, test and package software}%heidi sept2020

\newduneabbrev{ups}{UPS}{UNIX product support}{A software tool that sets up a consistent environment of versions of pre-installed products and their dependencies on UNIX-like platforms} % tjunk sept2020 #20

\newduneabbrev{upd}{UPD}{UNIX product distribution}{A tool for uploading and downloading pre-built software products between local systems and centralized software distribution servers.  UPD is not frequently used on DUNE because newer tools are more convenient} % tjunk sept2020 #22

\newduneabbrev{sso}{SSO}{single sign-on}{Used at \gls{fnal} to indicate that a group of services,  such as DocDB or the DUNE Wiki share common sign-in credentials and active sessions.  \gls{fnal}  services that say "Sign in with SSO username and password" mean to use your \gls{fnal} Services or federated username and password} % tjunk sept2020 #24

\newduneabbrev{vo}{VO}{virtual organization}{A database containing a list of member names, certificate distinguishing information, and a list of permissions members have to access computing grid and data resources} % tjunk sept2020 #27

\newduneabbrev{nas}{NAS}{network attached storage}{Disk storage that is available on computers but shared between them.  Relies on \gls{nfs} mounts rather than authenticated file transfer protocols.  Usually found on interactive servers to provide space for home directories, app and data storage} % tjunk sept2020 #25

\newduneabbrev{nfs}{NFS}{network file system}{Industry-standard mechanism for mounting disks over a network.  Provides regular UNIX file and directory access} % tjunk sept2020 #26

\newduneword{recombination}{recombination}{Electrons freed from Argon atoms will sometimes recombine with the positive argon ions, either the same ones from which they came or nearby ones.  Sometimes called ``quenching"} % tjunk sept2020 #29

\newduneword{elife}{electron lifetime}{The attachment of electrons drifting through liquid argon to impurity molecules such as oxygen or water is parameterized by an exponential as a function of time with a time constant called the electron lifetime} % tjunk sept2020 #30

\newduneword{gplane}{grid plane}{The uninstrumented plane of wires or electrodes on an anode plane %in an \gls{apa} 
facing the drift volume.   It shapes the signals and provides \gls{esd} protection} % tjunk sept2020 #31

\newduneword{gmesh}{grounding mesh}{A metal mesh attached to the SP \gls{apa} frame between the collection-plane wires and the space inside the frame where the \gls{pd} modules are installed.  It provides electric field uniformity so the collection-plane wires all have similar fields around them} % tjunk sept2020 #32

\newduneabbrev{bcr}{BCR}{baseline change request}{A DOE project change, part of the change management system process}

\newduneword{ncav}{North Cavern}{the location of two of the planned four DUNE far detector modules at \gls{surf}}

\newduneword{scav}{South Cavern}{the location of two of the planned four DUNE far detector modules at \gls{surf}}

\newduneword{semp}{SEMP}{systems engineering management plan}  % #36 replaced cmp 

\newduneabbrev{tpcost}{TPC}{total project cost}{The DOE terminology for the total budget and contingency for the entire \gls{usproj} project from CD-0 to CD-4}  % #36 replaced cmp 
\newduneabbrev{nsint}{NSI}{near site integration}{The scope of work at the near site for the \gls{integoff}} % #36 - can't put it together with other nsi

\newduneabbrev{opcost}{OPC}{other project costs}{The DOE project costs that support conceptual design, pre-operations commissioning, technical coordination, and power}

\newduneabbrev{moa}{MOA}{memorandum of agreement}{A project management methodology that documents an agreement between \gls{fnal} and the \gls{usproj}  Project for how \gls{fnal}  will support the project} %%%  MISSED THIS IN 36!!! ADDED AFTER
\newduneabbrev{croc}{CROC}{central readout chamber}{central (radial) readout chamber for the ND \gls{gartpc}} % from SManley 12/2/20 via email

\newduneabbrev{tki}{TKI}{transverse kinematic imbalance}{The imbalance among final-state particle momenta in the transverse plane to the neutrino direction; different aspects of the imbalance are sensitive to the detail of the nuclear effects in neutrino-nucleus interactions} % from SManley 25feb2021 via email

\newduneabbrev{iandi}{I\&I}{Integration and Installation}{One of the three project areas in the LBNF/DUNE-US \gls{doe} project, along with LBNF and DUNE-US} 

\newduneword{hmi}{HMI}{human-machine interface}

\newduneword{lhe}{LHe}{liquid helium}

\newduneword{echain}{energy chain}{mechanical machine elements used to carry and guide power to moving parts of machines or structures, as required for \gls{duneprism} to carry power, data, and utilities to and from each movable near detector component at any arbitrary position along its travel path}

\newduneabbrev{sphd}{FD1-HD}{horizontal drift technology}{LArTPC design in which electrons drift horizontally to wire plane anodes (\glspl{apa}) that along with the front-end electronics are immersed in LAr} % updated 21 May 21

\newduneabbrev{spvd}{FD2-VD}{vertical drift technology}{LArTPC design in which electrons drift vertically to PCB-based anodes at the top and bottom of the LAr volume,  with a cathode in the middle} % updated 21 May 21

\newduneabbrev{cru}{CRU}{charge-readout unit}{In the \gls{spvd} design an assembly of the \glspl{pcbp} plus adapter boards; two to a \gls{crp}} % 4 aug 2021 four --> two

\newduneword{mpv}{MPV}{most probable value}

\newduneword{pcbp}{PCB panel}{In the \gls{spvd} design, one of four \glspl{pcb} of size  %\SI{1.5x1.7}{m} 
1.5 $\times$ 1.7\.,m assembled into a \gls{cru}}

\newduneword{anodepln}{anode plane}{a planar array of charge readout devices covering an entire face of a detector module}

\newduneword{msps}{MSPS}{megasamples per second}

\newduneword{gpsdo}{GPSDO}{\gls{gps} disciplined oscillator}
\newduneword{tdaq}{TDAQ}{trigger and DAQ system} % 25feb2021 SPVD DAQ

\newduneword{nios}{NIOS}{network identity operating system} 

\newduneabbrev{corsika}{CORSIKA}{COsmic Ray SImulations for KAscade}{a program for detailed simulation of extensive air showers initiated by high-energy cosmic ray particles}

\newduneabbrev{scd}{SCD}{scientific computing division}{\gls{fnal}'s Scientific Computing Division}

\newduneabbrev{garsoft}{GArSoft}{Gaseous Argon Software}{A software toolkit similar to \gls{larsoft}, but targeted at the gaseous argon time projection chamber and calorimeter of \gls{ndgar}}

\newduneword{ndgarlite}{ND-GAr-Lite}{a temporary muon spectrometer consisting of the magnet and steel flux return of \gls{ndgar}, but with a simplified tracking chamber made with scintillating bars}

\newduneword{github}{GitHub}{a commercial web service providing code version management, storage,  and browsing via \gls{git}}

\newduneword{git}{git}{a distributed version-control system, commonly used to manage software}

\newduneword{xrootd}{xrootd}{a high-performance data system widely used in \gls{hep} to store and to distribute data to jobs.  It allows streaming of data}

\newduneabbrev{gpuaas}{GPUAAS}{GPU As A Service}{a technique that allows many non-GPU-enabled compute nodes to share a GPU resource by sending it work over the network and waiting for results to be returned}
\newduneword{sce}{SCE}{space charge effect}

\newduneabbrev{nest}{NEST}{noble element simulation technique}{Comprehensive simulation code modeling the excitation, ionization, and corresponding scintillation and electroluminescence processes in liquid noble elements}

\newduneword{bgr}{BGR}{A bandgap voltage reference is a circuit block in ASIC for providing stable reference voltages} %in electronics chapter

\newduneword{comsol}{COMSOL}{General-purpose simulation software based on advanced numerical methods (comsol.com)}

\newduneword{greyrm}{grey room}{ISO-8 clean room with a cleanliness level of 3.5M particles of 0.5 micron or less per cubic meter volume. ISO-8 clean rooms are referred to as grey rooms because at this level of cleanliness most standard clean room attire is not required.}

\newduneabbrev{pof}{PoF}{power-over-fiber}{a technology in which a fiber optic cable carries optical power, which is used as an energy source rather than, or as well as, carrying data; this allows a device to be remotely powered, while providing electrical isolation between the device and the power supply} %in prototype chap; got def from wikipedia (Anne)

\newduneword{ppc}{PPC}{photovoltaic power converter}% in PD
\newduneword{eelaser}{edge-emitting laser}{a laser in which  light is emitted from the edge of the substrate}% in PD
\newduneword{peek}{PEEK}{Polyether ether ketone, a colorless organic thermoplastic polymer}

\newduneword{fsm}{Finite State Machine}{a mathematical model of computation; it is an abstract machine that can be in exactly one of a finite number of states at any given time} % in DAQ
\newduneword{icecube}{IceCube}{South Pole Neutrino Observatory}
\newduneword{pde}{PDE}{photo detection efficiency} % in PD
\newduneword{pmos}{PMOS}{(see \gls{cmos}; PMOS is constructed with the p-type source and drain and an n-type substrate}% in PD
\newduneabbrev{poe}{PoE}{power-over-Ethernet}{systems that pass electric power along with data on twisted-pair Ethernet cabling} % in PD

\newduneabbrev{daqdts}{DUNE DTS}{DUNE timing and synchronization subsystem}{The portion of the \gls{daq} that provides for timing and synchronization to various detector systems}

\newduneword{tai}{TAI}{International Atomic Time}% in DAQ

\newduneword{larzic}{LARZIC}{The cryogenic amplifier \dword{asic} that is the principal component of the \gls{spvd} top drift \dword{fe} analog cards}

\newduneword{dpdfd}{DPDFD}{Deputy Project Director for far detectors}
\newduneword{bsws}{BSWS}{bearing sensor wire compression seal}

\newduneabbrev{ly}{LY}{light yield}{detected photons per unit deposited energy}

\newduneword{mtbf}{MTBF}{mean time between failures}

\newduneword{ingaas}{InGaAs}{Indium gallium arsenide is a room-temperature semiconductor commonly used as a high-speed, high-sensitivity photodetector for optical fiber telecommunications}

\newduneword{intlproj}{LBNF/DUNE Construction Project}{The international project to design and build the facilities and detectors for the \gls{lbnf-dune}; it includes the \gls{usproj} and projects at multiple international partners to manage the contributions from non-US institutions and funding agencies to design, build, and install the detector components}% new oct 2021

\newduneword{ddmp}{DUNE Management Plan}{DUNE Management Plan}
  
\newduneword{pd2hd}{ProtoDUNE-II-HD}{The second \gls{protodune} for the \gls{sphd} design, using \gls{np04}, to test installation, integration, and detector component performance}

\newduneword{pd2vd}{ProtoDUNE-II-VD}{The \gls{protodune} for the \gls{spvd} design, using \gls{np02}, to test installation, integration, and detector component performance}

\newduneword{mod0}{Module 0}{The final pre-production instance of a detector; for the DUNE \glspl{detmodule}, these will use the 800\,t cryostats in \gls{np02} and \gls{np04}}

\newduneword{fsii}{FSII}{far site integration and installation}
\newduneword{fdc}{FDC}{Far Detector and Cryogenics Subproject}

\newduneabbrev{co}{CO}{compliance office}{a team of engineers from multiple partners that provides clear direction for designing and constructing components that will be used during integration}

\newduneword{fscfbsi}{FSCF-BSI}{\gls{usproj} subproject for far site conventional facilities, building and site infrastructure}

\newduneabbrev{poms}{POMS}{Production Operations Management System}{A workflow management system available for all DUNE users to submit and monitor grid jobs as well as view job log files}
\newduneabbrev{fifemon}{FIFEMON}{FabrIc for Frontier Experiments MONitoring}{Comprehensive suite of job and storage monitoring information available for most \gls{fnal} experiments, including DUNE}

\newduneword{larg4}{LArG4}{LArG4 is the replacement for LArsim/LArG4. LArG4 is based on \gls{artg4tk}}

\newduneword{artg4tk}{artg4tk}{artg4tk provides a general interface between \gls{geant4} and \gls{art}}

\newduneword{tde}{TDE}{top detector electronics}

\newduneword{bde}{BDE}{bottom detector electronics}

\newduneword{sigproc}{Signal Processing}{The goal of TPC signal processing is to reconstruct the distribution of ionization electrons arriving at wire planes from the digitized TPC waveform}
\newduneword{mcnd}{MCND}{More Capable Near Detector} % for ND TDR 9/26/22, Steve M

\newduneabbrev{compass}{COMPASS}{
Common Muon and Proton Apparatus for Structure and Spectroscopy }{a multipurpose experiment at \gls{cern}’s Super Proton Synchrotron (SPS)} %

\newduneword{vd}{VD}{vertical drift \gls{tpc} technology} %

\newduneword{hd}{HD}{horizontal drift \gls{tpc} technology} %

\newduneabbrev{esnet}{ESnet}{Energy Sciences Network}{The \gls{doe}'s dedicated science network} %

\newduneword{project.py}{project.py}{XML-based job configuration system developed by the \gls{microboone} collaboration} %

\newduneabbrev{ppfx}{PPFX}{Package to Predict the FluX}{\gls{fnal}-supported package that implements hadron production corrections to geant4 simulations and propagates uncertainties for the NuMI  and LBNF beam lines} %

\newduneabbrev{ml}{ML}{Machine Learning}{Machine Learning}

\newduneword{datalake}{data lake}{The not-yet-realized concept of a storage service with multiple levels of quality of service in which the end user can access data without knowing the data's source location} % from where the data are served}

\newduneabbrev{metacat}{MetaCat}{MetaCat}{Metadata Catalog, a modern replacement for the file description portion of the sam metadata catalog} %

\newduneword{g4lbnf}{g4lbnf}{LBNF neutrino beamline simulation program\cite{g4lbnf}} %

\newduneword{Geant4Reweight}{Geant4Reweight}{Framework for evaluating and propagating hadronic interaction uncertainties in Geant4\cite{Calcutt:2021zck}} %

\newduneword{geantnet}{G\'EANT}{G\'EANT interconnects Europe's national research and education networking (NREN) organizations with the high bandwidth, high speed and highly resilient pan-European backbone.
}

\newduneabbrev{nren}{NREN}{National Research and Education Network}{National level research computing network infrastructure} %

\newduneabbrev{vrf}{VRF}{Virtual Routing and Forwarding}{Networking overlays that provide  a logical routing infrastructure  that allows flexible traffic engineering} %

\newduneword{samvalue}{value}{A generic quantity describing a file in the \dword{sam} data catalog} %

\newduneword{samparameter}{parameter}{A user or experiment described quantity describing a file in the \dword{sam} data catalog} %

\newduneword{samproject}{SAM-project}{A server process running centrally that maintains a predefined list of files and delivers information about  their locations when asked by distributed processes. The project tracks success and failure of file processing} %

\newduneword{samconsumer}{SAM-consumer}{A client process that requests information about file locations from a \dword{samproject}, process the file and reports success or failure} %

\newduneword{samdataset}{SAM-dataset}{ A dynamic collection of files defined by queries to the \dword{sam} data catalog} %
\newduneword{metacatdataset}{MetaCat-dataset}{ A fixed but mutable collection of files defined by queries to the \dword{metacat} data catalog} %

\newduneword{samsnapshot}{SAM-snapshot}{A fixed collection of files corresponding to a \dword{sam} \dword{samdataset} at a particular point in time} %

\newduneabbrev{mql}{MQL}{\dword{metacat} Query Language} {A query language which supports queries of the \dword{metacat} data catalog, including parentage and logical functions such as union, join and subtraction} %

\newduneword{pdhd}{ProtoDUNE-HD}{\gls{protodune} with horizontal drift technology.  This refers to the \gls{sp} \gls{apa}-based prototype to run in \gls{np04} (in the  \gls{pd2} phase)} 

\newduneword{pdvd}{ProtoDUNE-VD}{ProtoDUNE with vertical drift technology.  This refers to the \gls{crp}-based prototype to run in \gls{np02}  (in the  \gls{pd2} phase)} 

\newduneword{pd2}{ProtoDUNE-II}{\gls{protodune} test runs at CERN in 2022-2023; includes \gls{pdhd} and \gls{pdvd}}

\newduneabbrev{ucondb}{uconDB}{Unstructured Conditions Database}{Unstructured conditions database developed for \gls{fnal} fixed target experiments\cite{bib:ucondb}} 

\newduneabbrev{json}{JSON}{JavaScript Object Notation}{Open standard data interchange format that uses pair-value pairs and maps well onto python data formats such as tuples and lists} 

\newduneabbrev{dqmdb}{DQMDB}{Data Quality and Monitoring Database}{Database storing the results of data-quality monitoring} % removed .

\newduneword{db}{DB}{database}

\newduneabbrev{vm}{VM}{virtual machine}{Emulator of a physical computer that allows multiple users to configure different operating systems  while sharing physical hardware}

\newduneword{enstore}{Enstore}{A mass storage system developed by \gls{fnal} that provides distributed access and management of data stored on tapes} % \cite{Litvintsev:2012dw}}

\newduneword{stashcache}{StashCache}{A distributed caching federation that enables opportunistic users to utilize nearby opportunistic storage\cite{bib:stashcache}}

\newduneabbrev{hdf5}{HDF5}{Hierarchical Data Format}{Data format\cite{bib:hdf5} widely used in \gls{ml}} 

\newduneword{glideinwms}{GlideinWMS}{A system of submitting pilot jobs to grid computing sites, inside of which user jobs run, presenting a uniform setup across many different sites} % KH 2022-10-15  \cite{sfiligoi2009pilot

\newduneword{mars}{MARS}{The MARS code system is a set of \gls{mc} programs for detailed simulation of coupled hadronic and electromagnetic cascades, with heavy ion, muon and neutrino production and interactions} % removed .

\newduneabbrev{fife}{FIFE}{Fabric for Intensity Frontier Experiments}{\gls{fnal} computing infrastructure for \gls{if} Experiments} % \cite{box2014fife}} \newduneword{postgres}{Postgres}{also known as PostgreSQL, Postgres is a free and open-source relational database management system used extensively for databases in \gls{hep}}

\newduneword{tr}{trigger record}{A data record produced by the \gls{dune} \gls{daq} system.  A Trigger Record can contain multiple interaction ``events'' or none} 

\newduneword{postgres}{Postgres}{also known as PostgreSQL, Postgres is a free and open-source relational database management system used extensively for databases in \gls{hep}}

\newduneword{fhicl}{FHICL}{Fermilab Hierarchical Configuration Language; a standard configuration language for the storage, communication, and manipulation of scientific parameter sets}

\newduneword{api}{API}{Application Programming Interface}

\newduneword{hwdb}{HWDB}{hardware database}   

\newduneword{s3}{S3}{The Amazon cloud-based commercial storage service} % removed .  %anne added

\newduneword{openstack}{OpenStack}{An open source cloud software used to deploy instances of containers}

\newduneabbrev{cric}{CRIC}{Computing Resource Information System}{a framework providing a centralized (and flexible) way to describe which resources are being used by the experiment and how}

\newduneabbrev{ccb}{CCB}{Computing Contributions Board}{a board made up of institutional representatives for larger countries and laboratories. It meets annually to negotiate collaboration contributions to computing infrastructure} 

\newduneword{garfield}{Garfield}{A simulation program\cite{Veenhof:1998tt} developed at \gls{cern} for gaseous detectors } 

\newduneword{datatier}{data tier}{Differing data types produced in a processing sequence, for example, raw data, reconstructed, derived analysis sample, histograms} 

\newduneabbrev{gpu}{GPU}{Graphical Processing Unit}{Specialized computing hardware optimized for image processing} 

\newduneabbrev{gpvm}{dunegpvm}{DUNE General Purpose Virtual Machine}{Centrally managed virtual Linux systems at \gls{fnal} with access to network attached and \gls{pnfs} storage. Used for small-scale data analysis and algorithm development} 

\newduneword{madx}{MAD-X}{framework that provides the de facto standard scripting language to describe particle accelerators, simulate beam dynamics, and optimize beam optics at \gls{cern}}

\newduneabbrev{cpu}{CPU}{Central Processing Unit}{A computing processor, when used as a unit of processing; generally refers to a single core} 

\newduneabbrev{fft}{FFT}{Fast Fourier Transform}{An algorithm that calculates the frequency components of a time-domain waveform in a computationally efficient manner} 

\newduneword{grain}{GRAIN}{In the \gls{sand} detector, a small cryostat containing \gls{lar} installed upstream of the straw-tube tracker inside the \gls{ecal}}

\newduneword{voms}{VOMS}{Virtual Organization Membership Service (in grid computing)}

\newduneword{cry}{CRY}{A cosmic ray shower library and software tool}

\newduneword{jwt}{JWT}{\gls{json} web token}

\newduneabbrev{ferry}{FERRY}{Frontier Experiments RegistRY}{a central database that keeps track of all scientific computing users at \gls{fnal}, the experiments and groups of which they are members, and the various capabilities they are allowed}

\newduneabbrev{perfsonar}{perfSONAR}{performance Service-Oriented Network monitoring ARchitecture}{a network measurement toolkit designed to provide federated coverage of paths and to help establish end-to-end usage expectations}

\newduneword{ipv6}{IPv6}{the most recent version of the Internet communications protocol that provides an identification and location system for computers on networks and routes traffic across the Internet}

\newduneword{monit}{MONIT}{a suite of tools using open source technology used in the monitoring of the \gls{cern} IT data center and \gls{wlcg} infrastructure. }

\newduneword{garana}{GArAna}{provides a facility for making an analysis ntuple from information stored in \gls{garsoft} data products}

\newduneword{ci}{CI}{continuous integration}

\newduneword{datadis}{Data Dispatcher}{communicates between running processing jobs and the data delivery systems.  It provides file location information and basic bookkeeping on file access and transfers} % removed .

\newduneabbrev{castor}{CASTOR}{CERN Advanced STORage manager}{a hierarchical storage system with tape and disk developed at \gls{cern}. It is being replaced by \gls{cta}} 

\newduneabbrev{cta}{CTA}{CERN Tape Archive}{a hierarchical storage system with tape and disk developed at \gls{cern}. It is replacing \gls{castor}} 

\newduneabbrev{sonic}{SONIC}{Services for Optimized
Network Inference on Coprocessors}{Framework for implementing machine learning algorithms on co-processors developed by \gls{fnal}} 

\newduneword{jobsub}{JobSub}{\cite{box2014fife} the \gls{fnal} \gls{if} job submission client that supports user submission of complex workflows to HT-Condor} 

\newduneword{hepcloud}{HEPCloud}{routes jobs to local or remote computing resources based on the policy for a particular experiment, workflow requirements, and cost and efficiency of accessing the various resources. It expands the resources available to include \gls{hpc} centers and commercial cloud resources}

\newduneword{wmsmon}{WMS monitoring}{Workflow Management System Monitoring}

\newduneword{redmine}{Redmine}{an open source code repository and issue tracking tool that was historically used by many of the \gls{fnal} general computing projects and \gls{if} experiments} 

\newduneabbrev{if}{IF}{Intensity Frontier}{refers to \gls{hep} experiments, particularly in the U.S., that rely on high luminosity instead of high energy for discovery.  Includes B-physics, neutrino, and muon experiments}

\newduneword{slack}{Slack}{a commercial business communication platform} 

\newduneword{servicenow}{ServiceNow}{a commercial enterprise workflow system used by \gls{fnal} for formal issue tracking and IT workflows such as user account preparation} 

\newduneabbrev{rse}{RSE}{Rucio Storage Element}{A storage element that is known to the DUNE \gls{rucio} instance} 

\newduneabbrev{coffea}{COFFEA}{Columnar Object Framework For Effective Analysis}{Columnar data analysis frame\-work\cite{Smith:2020pxs} developed at \gls{fnal}} 

\newduneword{nuisance}{Nuisance}{An open source C++ framework for studying neutrino interaction cross sections\cite{Stowell:2016jfr}} 

 \newduneabbrev{highland}{HighLAND}{High Level Analysis at a Neutrino Detector}{Analysis framework developed by the \gls{t2k} collaboration} 
 
 \newduneabbrev{raii}{RAII}{Resource Acquisition Is Initialization}{a C++ programming technique that binds the lifecycle of a resource that must be acquired before use (e.g., allocated heap memory, thread of execution, open socket, open file, locked mutex, disk space, database connection—anything that exists in limited supply) to the lifetime of an object}  
 
 \newduneword{fts3}{FTS3}{File Transfer Service version 3, developed by \gls{cern}, and distinct from, the \gls{fnal} File Transfer Service} 
 
 \newduneword{wfms}{WFMS}{Workflow Management System}
 
  \newduneword{condmeta}{conditions metadata}{defined as the information necessary to understand the context of physics data, e.g., beam data or calibrations} 
 
 \newduneword{cdb}{conditions DB}{The Conditions Database stores the \gls{condmeta} needed for data processing and analysis} 
 
 \newduneword{rntuple}{RNtuple}{The next generation of the \gls{root} I/O system\cite{Blomer:2020usr, ROOTTeam:2020jal}}
 
 \newduneabbrev{lan}{LAN}{Local Area Network}{Computing network confined to a relatively small geographic area} 
 
 \newduneabbrev{wan}{WAN}{Wide Area Network}{Computing network that interconnects \glspl{lan} across relatively broad geographic areas} 

 \newduneword{layer3}{Layer-3}{Networking protocol  {https://en.wikipedia.org/wiki/Network\_layer}}
 
 \newduneabbrev{rest}{REST}{Representational State Transfer}{Standard for web interfaces} %  \href{https://en.wikipedia.org/wiki/Representational\_state\_transfer}{https://en.wikipedia.org/wiki/Representational_state_transfer}}
 
 \newduneabbrev{faq}{FAQ}{Frequently Asked Questions}{A software system for collecting and answering the most common questions about an activity} 
 
\newduneword{spack}{Spack}{Software package manager for UNIX and mac OS systems} 

\newduneabbrev{lxplus}{LXPLUS}{Linux Public Login User Service)}{The interactive logon service to Linux for all \gls{cern} users} 

\newduneword{dunesw}{dunesw}{The base DUNE software release} 

\newduneword{posix}{POSIX}{Portable Operating System Interface - Unix standard for operating system interfaces} 

\newduneword{nutools}{NuTools}{shared code for \gls{larsoft} + NOvA + other neutrino experiments that use \gls{art}.  Includes beam and event generators and re-weighting packages}

\newduneword{singlecube}{SingleCube}{A cubical time projection chamber 30 cm on a side with a single ND-LAr pixel readout tile, used for prototyping and tests}

\newduneword{hllhc}{HL-LHC}{High-luminosity \gls{lhc}} %anne added

\hypersetup{
    pdftitle={\expshort CDR \thedocsubtitle},
    pdfauthor={\expshort Collaboration},
    final=true,
    colorlinks=false,
    linktocpage=true,
    linkbordercolor=blue,
    citebordercolor=green,
    urlbordercolor=magenta,
    filecolor=black,
    pdfpagemode=UseOutlines,
    pdfborderstyle={/S/U},  
}

\begin{document}
\pagestyle{titlepage}
\cleardoublepage

\pagestyle{titlepage}

\begin{center}
   {\Huge  DUNE Offline Computing}  %

  \vspace{5mm}

  {\Huge  Conceptual Design Report }  %

  \vspace{10mm}

\titleextra  %

\includegraphics[width=\textwidth]{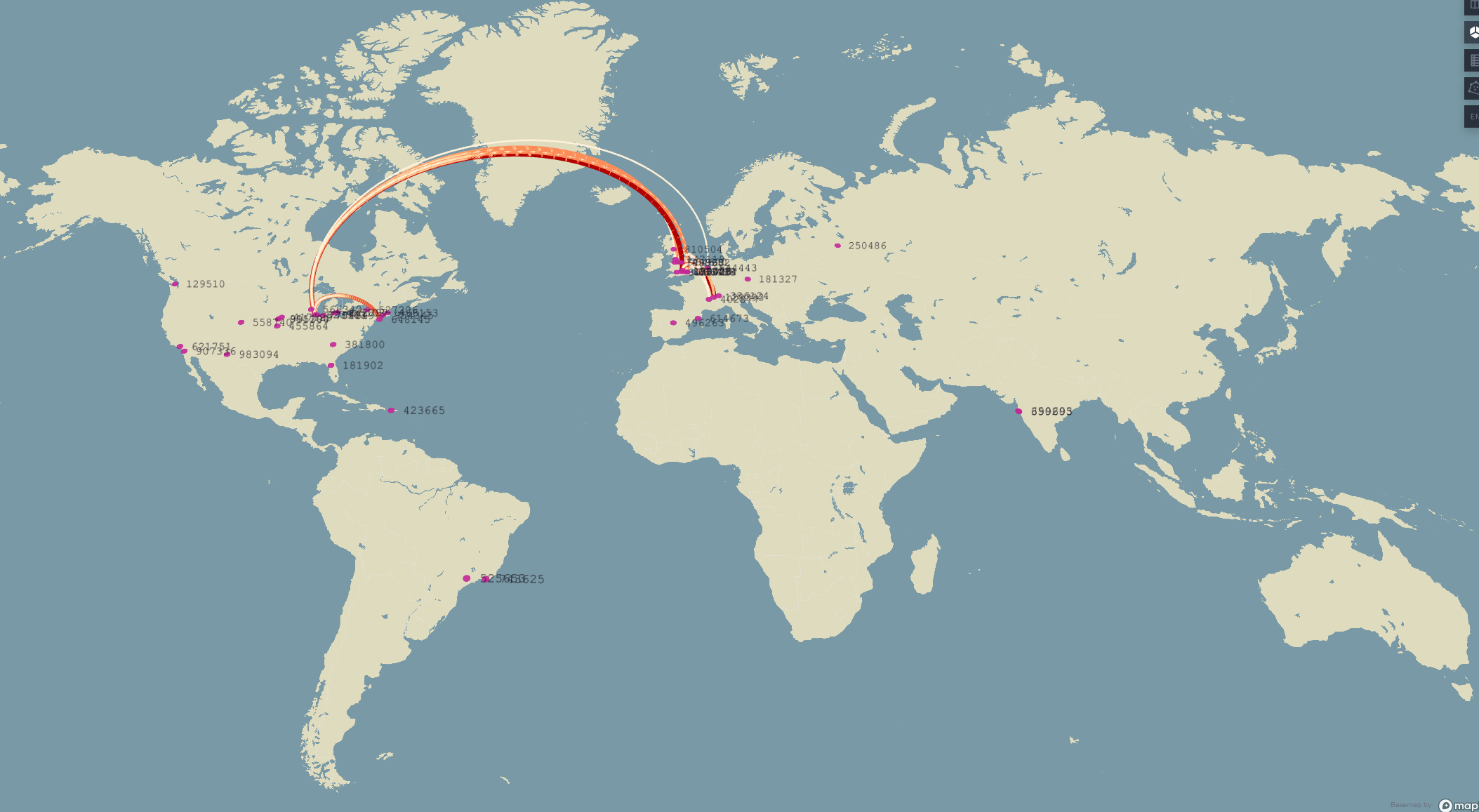}%

  \vspace{10mm}
  \today%
    \vspace{15mm}
    
    {\large{The DUNE Collaboration}}
\end{center}

\cleardoublepage
\vspace*{13cm} 
  {\small 
  
FERMILAB-DESIGN-22-01
  
This document was prepared by the DUNE collaboration using the
resources of the Fermi National Accelerator Laboratory 
(Fermilab), a U.S. Department of Energy, Office of Science, 
HEP User Facility. Fermilab is managed by Fermi Research Alliance, 
LLC (FRA), acting under Contract No. DE-AC02-07CH11359.
This work was supported by
CNPq,
FAPERJ,
FAPEG and 
FAPESP,                         Brazil;
CFI, 
IPP and 
NSERC,                          Canada;
CERN;
M\v{S}MT,                       Czech Republic;
ERDF, 
H2020-EU and 
MSCA,                           European Union;
CNRS/IN2P3 and
CEA,                            France;
INFN,                           Italy;
FCT,                            Portugal;
NRF,                            South Korea;
CAM, 
Fundaci\'{o}n ``La Caixa'',
Junta de Andaluc\'ia-FEDER,
MICINN, and
Xunta de Galicia,               Spain;
SERI and 
SNSF,                           Switzerland;
T\"UB\.ITAK,                    Turkey;
The Royal Society and 
UKRI/STFC,                      United Kingdom;
DOE and 
NSF,                            United States of America.

The ProtoDUNE-SP and ProtoDUNE-DP detectors were constructed and operated on the CERN Neutrino Platform.
We gratefully acknowledge the support of the CERN management, and the
CERN EP, BE, TE, EN and IT Departments for NP04/Proto\-DUNE-SP.

This research used resources of the 
National Energy Research Scientific Computing Center (NERSC), 
a U.S. Department of Energy Office of Science User Facility 
operated under Contract No. DE-AC02-05CH11231.
  }
\includepdf[pages={-}]{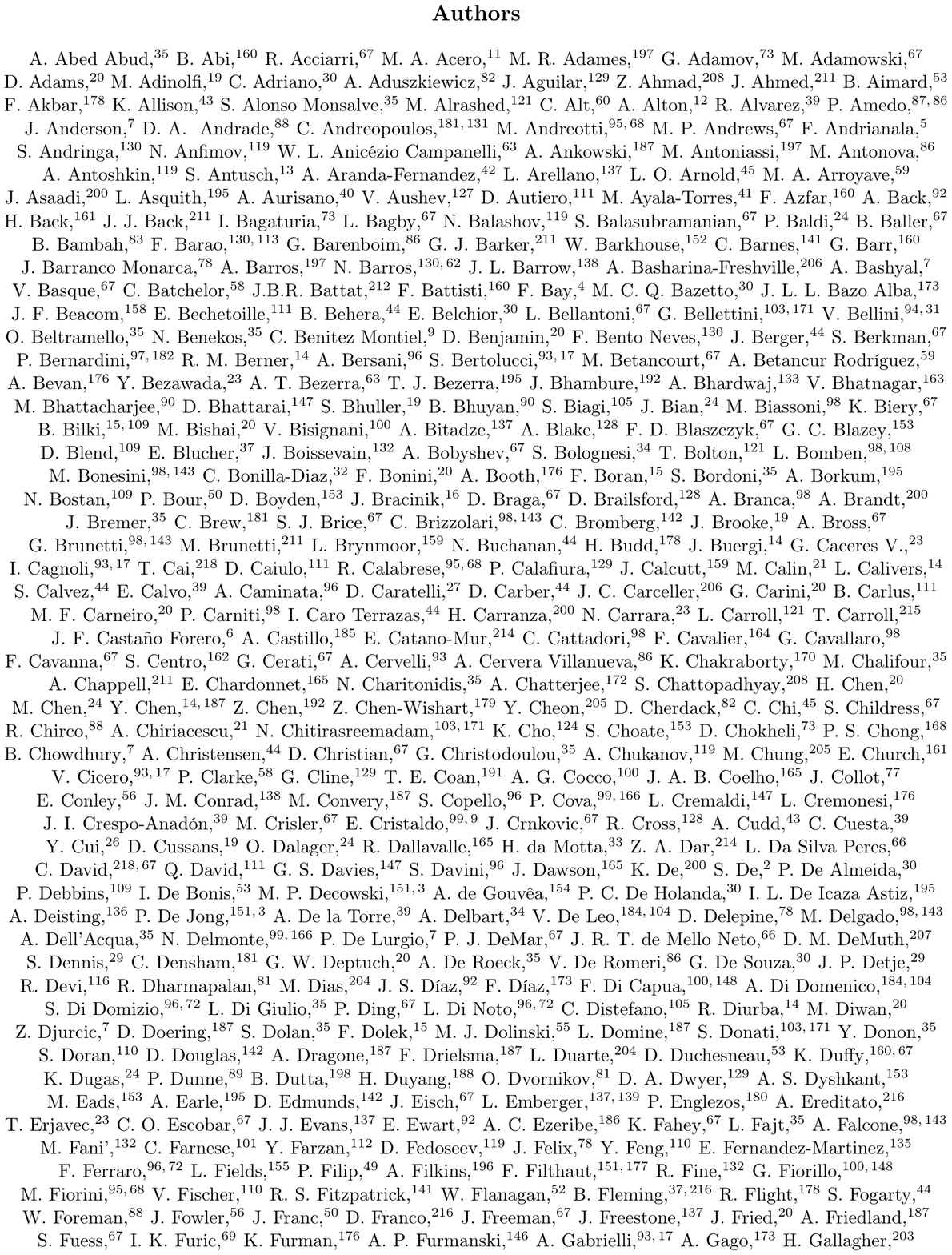}%

\renewcommand{\familydefault}{\sfdefault}
\renewcommand{\thepage}{\roman{page}}
\setcounter{page}{0}

\pagestyle{plain} 

\setcounter{tocdepth}{2}  %
\textsf{\tableofcontents}

\textsf{\listoffigures}

\textsf{\listoftables}
  \vspace{4mm}

\renewcommand{\thepage}{\arabic{page}}
\setcounter{page}{1}

\pagestyle{fancy}

\renewcommand{\chaptermark}[1]{%
\markboth{Chapter \thechapter:\ #1}{}}
\fancyhead{}
\fancyhead[RO,L]{\textsf{\footnotesize \thechapter--\thepage}}
\fancyhead[LO,R]{\textsf{\footnotesize \leftmark}}

\fancyfoot{}
\fancyfoot[RO]{\textsf{\footnotesize Conceptual Design Report --  Page \thepage}}
\fancyfoot[LO]{\textsf{\footnotesize DUNE Offline Computing}}
\fancypagestyle{plain}{}

\renewcommand{\headrule}{\vspace{-4mm}\color[gray]{0.5}{\rule{\headwidth}{0.5pt}}}

\cleardoublepage

\FPset{GlobalFDDataStorage}{30} %

\part{Overview} %
\label{part:overview}

\FPset{DatVolFDColdElecPrecision}{12} %

\chapter{Introduction \hideme{comments from AH and MK included 5/8} }
\label{ch:intro}

This document describes Offline Software and Computing for the \dword{dune} experiment, in particular, the conceptual design of the offline computing needed to accomplish its physics goals. Our emphasis in this document is the development of the computing infrastructure needed to acquire, catalog, reconstruct, simulate and analyze the data from the \dword{dune} experiment and its prototypes. In this effort, we concentrate on developing the tools and systems that facilitate the development and deployment of advanced algorithms. %
Rather than prescribing particular algorithms, our goal is to provide resources that are flexible and accessible enough to support creative software solutions as \dword{hep} computing evolves
and to provide computing that achieves the physics goals of the \dword{dune} experiment.

This chapter provides an introduction to the \dword{dune} experiment.  The last section (Section \ref{ch:intro:challenges}) summarizes the physics drivers of offline computing challenges that inform the rest of this document.

\section{Introduction \hideme{Schellman - draft based on CHEP paper}}\label{sec:intro-introduction}

\dword{dune}  will begin running in the late 2020's. The goals of the experiment include 1) studying neutrino oscillations using a beam of neutrinos sent from \dword{fnal} in Illinois to the  \dword{surf} in Lead, South Dakota, 2) studying  astrophysical neutrino sources and rare processes and 3) understanding  the physics of neutrino interactions in matter.
\dword{dune} will consist of a modular \dword{fd} located about \SI{1.5}{km} underground at \dword{surf} in South Dakota, USA, \SI{1300}{\km} from \dword{fnal}{}, and a \dword{nd} located on site at \dword{fnal} in Illinois. The \dword{dune} detectors will be exposed to the world's most intense neutrino beam originating at \dword{fnal}{}. A high-precision near detector, \SI{574}{m} from the neutrino source on the \dword{fnal} site, will be used to characterize the intensity and energy spectrum of this wide-band beam. The overriding physics goals of the \dword{dune} experiment are the search for leptonic \dword{cp} violation, the search for nucleon decay as a signature of a Grand Unified Theory underlying the Standard Model,  the observation of  \dwords{snb} from supernovae in our galaxy, and studies of solar neutrinos.  %

This document  concentrates on the neutrino oscillation and supernova capabilities of the experiment as their needs for high data volumes and high-precision measurements at low interaction energies drive the computing needs for the experiment.

When produced, the neutrino beam from \dword{fnal} will consist almost entirely of muon-type neutrinos. %
Neutrinos are known to come in (at least) three flavors that can be distinguished by their interactions -- electron %
neutrinos produce electrons when they interact via charged currents; muon neutrinos, muons; and tau neutrinos, tau particles.  But these flavors do not correspond to fixed mass states.  All three flavors of neutrinos are mixtures of mass states, much as  light polarized in the $x$ direction  can be considered a superposition of  $x^\prime$ and $y^\prime$ polarizations along  alternate axes rotated by 45 degrees.  When neutrinos propagate through space, it is the mass state that sets their wavelength and if the neutrino goes far enough, the multiple mass states  corresponding to the initial flavor state will get out of phase.  When the resulting  mixed state is later probed about its flavor, the detected flavor may be different than it was when the neutrino was produced. %
This phenomenon is known as neutrino oscillation, and has been %
observed in multiple experiments since it was first confirmed in 1998\cite{Kajita2006}.

\begin{dunefigure}
[Illustration of the neutrino flavor and mass states]
{fig:neutrinos}
{Illustration of the neutrino flavor and mass states.  The mass states are a superposition of the flavor states.  Courtesy the particlezoo.net.}
\includegraphics[height=6cm]{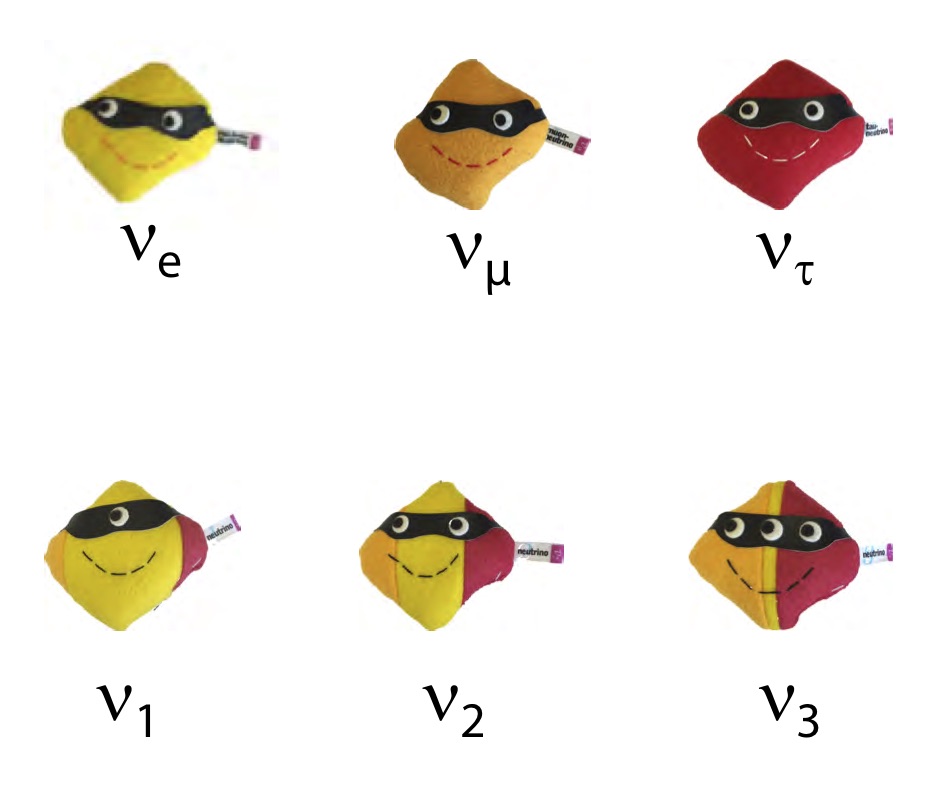}
\end{dunefigure}

\begin{dunefigure}
[Electron neutrino appearance signal and background as seen in ArgoNeut]
{fig:Argoneut}
{Electron neutrino appearance signal (top) and background (bottom) as seen in the ArgoNeut experiment\protect{\cite{Acciarri:2016sli}}.  In the true appearance signal, an electron is seen emerging from the primary vertex, then showering.  In the background interaction, a muon neutrino enters and  produces a final-state muon and photons that propagate some distance before showering.}
\includegraphics[trim={0cm 0.6cm 2.5cm 0.7cm},clip,height=6cm]{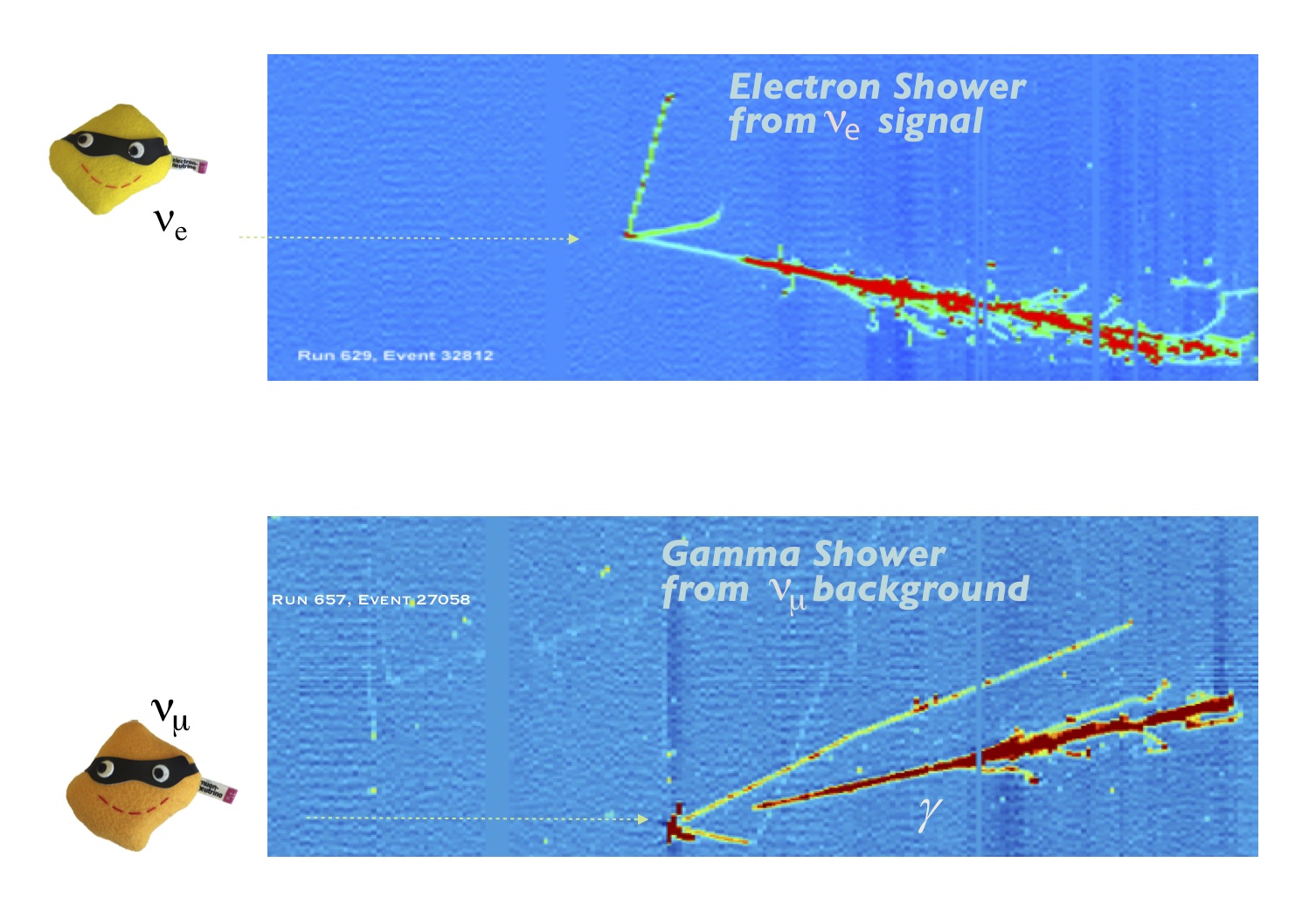} 
\end{dunefigure}

DUNE,  in particular,   wishes to understand the conversion of the muon neutrinos created in Illinois %
 into electron neutrinos at the \dword{fd} %
 in South Dakota  %
and compare that conversion rate between neutrino and antineutrino beams. The location of the \dword{fd} and energy of the neutrino beam were chosen to maximize the oscillation effect.   A difference in the conversion rate for neutrinos and antineutrinos could be evidence for matter-antimatter asymmetry in the neutrino sector, a phenomenon called \dword{cpv}.  

To make these measurements, the experiment must be able to distinguish %
electron-neutrino interactions from the dominant muon-neutrino interactions one would expect in the absence of oscillations.  %
Doing this requires a very large detector, as neutrino interactions are intrinsically rare, but also an extremely fine-grained one, as well.  Noble liquid  \dwords{tpc}, %
which read out large transparent volumes of liquid by drifting the electrons produced when charged particles from neutrino interactions ionize the liquid to charge-sensitive detectors through strong electric fields (\efield{s}), have the needed capabilities of extremely large scale and fine-grained resolution. The \dwords{tpc} are augmented by photon detection systems that provide precise timing of interactions and reconstructed particles. The proposed DUNE far detector will instrument four \SI{14x12x58}{m} %
volumes of \dword{lar} with readout granularity of $\sim$0.5\,cm.  The \dword{fd} modules will be located 4850\,ft below the earth's surface to reduce the rate of cosmic rays traversing the detector by orders of magnitude and thus allow sensitivity to very low-energy solar and astrophysical neutrinos, as well as the higher-energy neutrinos produced in the beam at Fermilab.  %
See the \dword{fd} and \dword{nd} design reports\cite{tdr-vol-1, tdr-vol-2, DUNE:2021tad} for full descriptions of the \dword{lartpc} technology. 

Additionally, physics programs focused on nucleon decay, the detection of a \dword{snb}, and other Beyond Standard Model (\dword{bsm}) signatures take advantage of the large size of the detector and flexible readout window of the \dword{daq} and \dword{lartpc}. %

The neutrino beam from \dword{fnal} will be pulsed approximately once per second, 24 hours per day during running periods, %
with 15 million pulses per year.  Because neutrinos interact  extremely rarely, we expect to detect of order 7,000 neutrino interactions/year in each of the four planned 10\,kt \dwords{detmodule} located at the \dword{fd} site in South Dakota.\footnote{This is based on the beam repetition rate of 0.83\,Hz and an estimated uptime for the accelerator complex of 56\% \cite{Abi:2020evt}.}.

A full \dword{tdr} for the first \dword{detmodule}, which uses \dword{sphd},
is available in References~\cite{Abi:2020evt, DUNE:2020lwj, Abi:2020oxb, Abi:2020loh}.  %
 A \dword{cdr} for a second detector module that implements \dword{spvd} is available in Reference~\cite{bib:spvd-cdr-edms}. This module is planned to come online soon after the \dword{sphd} module. Computing needs for both of these modules are included in the present document. 
The first two \dwords{detmodule} should go live late in this decade with commissioning of the \dword{daq} %
systems %
expected to start in 2027. %

\begin{dunefigure}
[A far detector cryostat and the horizontal drift \dshort{tpc} structure]
{fig:DUNESchematic} %
{(Left) A far detector cryostat that houses a 10\,kt \dword{fd} module with horizontal drift technology. The figure of a person at the bottom left of the image indicates the scale.  (Right) A 10\,kt  \dword{dune} \dword{fd} \dword{spmod}, showing the alternating 58\,m long (into the page), 12\,m high anode (A) and cathode (C) planes, as well as the field cage that surrounds the drift regions between the anode and cathode planes. The modular anode and cathode planes are constructed of units called \dwords{apa} and \dwords{cpa}; the blank area on the left side was added to show the profile of a single \dword{apa}.}
\includegraphics[height=0.35\textwidth]{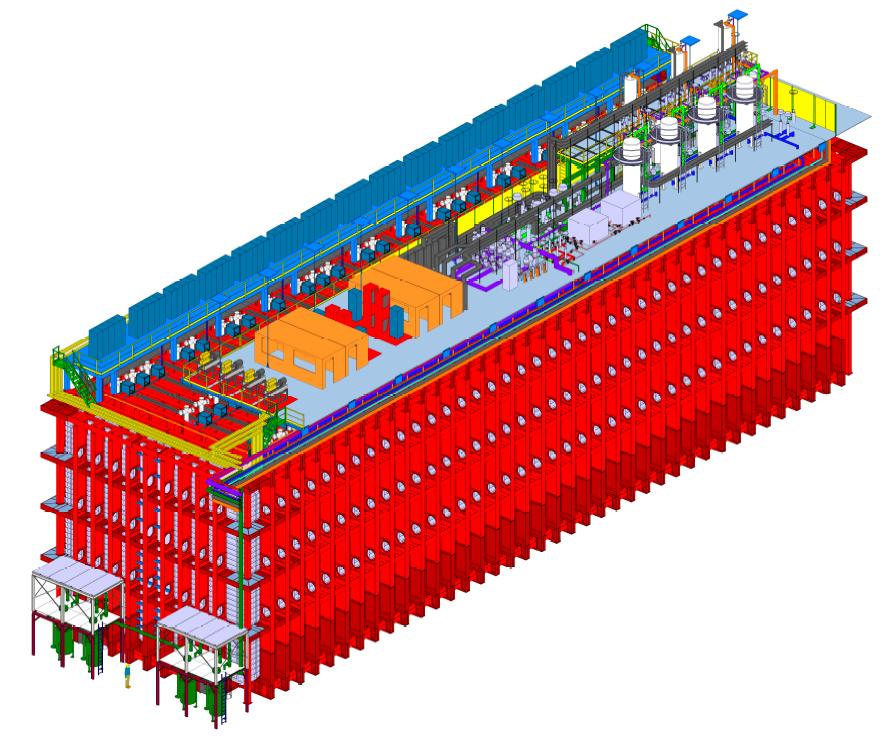}
\includegraphics[height=0.35\textwidth]{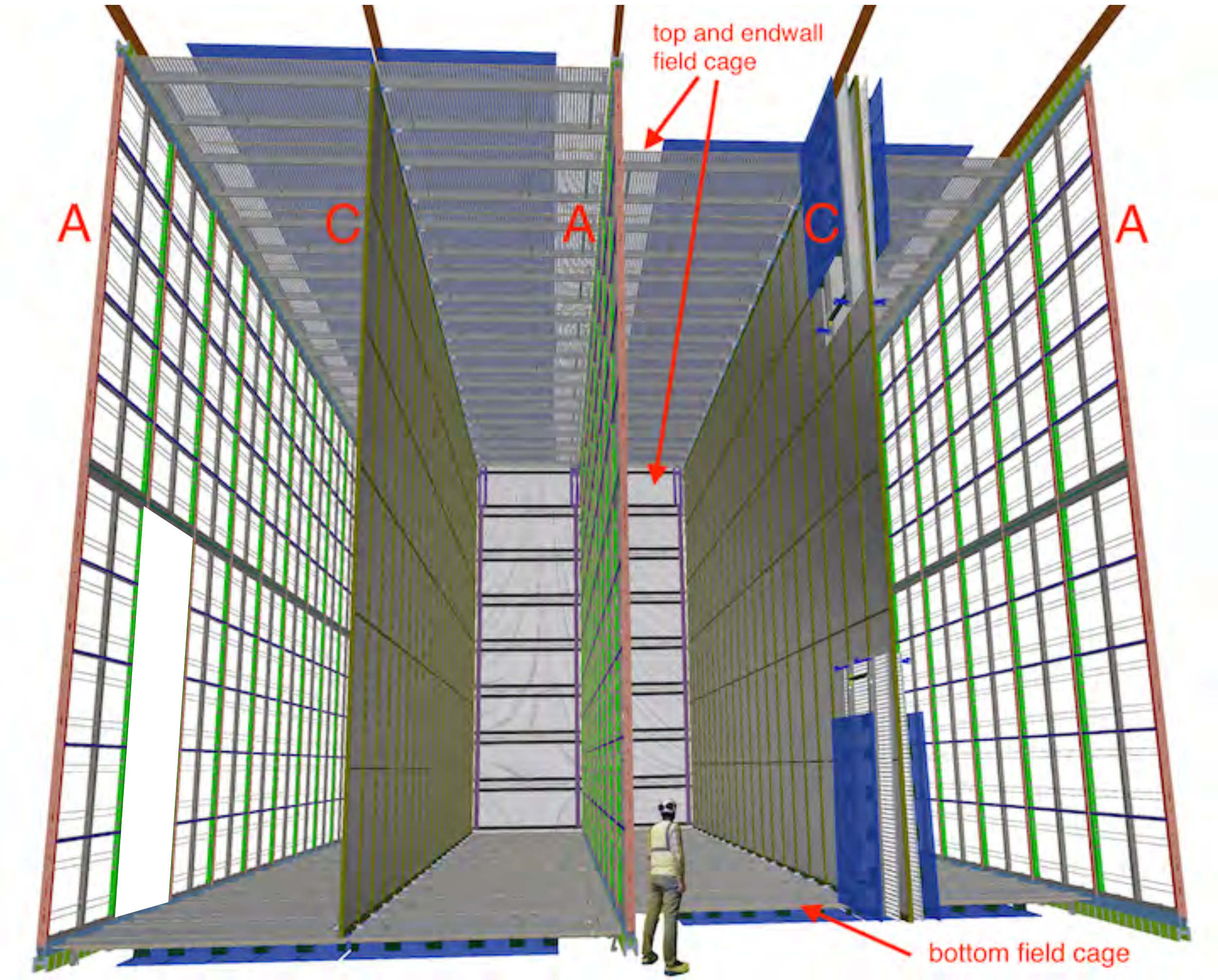}
\end{dunefigure}

\section{ProtoDUNE Tests at CERN  %
\hideme{Schellman/Pennacchio-draft}}

Building an experiment of this size requires an extensive period of prototyping.   The Argoneut\cite{Acciarri:2018myr}, MicroBooNE\cite{microboone} and ICARUS\cite{icarus} collaborations have demonstrated the capabilities of \dwords{lartpc} for neutrino detection on scales between 1 and 500 metric tons (tonnes, t) of fiducial mass.  In preparation for the \dword{dune} experiment, a campaign for testing %
full-sized \dword{fd} components in 700\,tonne  capacity cryostats %
in the \dword{ehn1} hadronic test beams at the \dword{cern} was launched in 2018.  Both \dword{sp} (horizontal drift) and \dword{dp} (liquid and gas, vertical drift) prototypes, called \dword{pdsp} and \dword{pddp} respectively, were constructed and operated. %
The complete data-taking chain from detector construction to full offline reconstruction and analysis of data was tested, and the results provided considerable insight into the computing challenges for the full \dword{dune} experiment.

\subsection{\dshort{pdsp}} %
The \dword{pdsp} experiment\cite{DUNE:2021hwx}, located at the CERN Neutrino Platform within \dword{ehn1}, began taking data %
in late 2018.  \dword{pdsp} %
collected ionization electrons and scintillation light %
directly from the \dword{lar}. The readout system consists of  %
\dwords{apa} (APAs) and a Photon Detection System (PDS).
Each \dword{apa} consists of an aluminum frame with three layers of active wires that form a grid on each side of the \dword{apa}, with each layer strung at angles chosen to reduce ambiguities in event reconstruction. Based on the drift field and the layer separation, the relative voltage between the layers is chosen to ensure transparency of the first two layers ($U$ and $V$) to the drifting electrons. These layers produce bipolar induction signals as the electrons pass through them. The final layer ($Y$) collects the drifting electrons, resulting in a unipolar signal. The drift time of the electrons provides a measurement of the x-coordinate of ionization while the z-coordinate is determined from the vertical $Y$ wires. The $U$ and $V$ layers are at $+/- 60$ degrees and the combination of $U$, $V$, and $Y$ ionization pattern collected on the grid of anode wires provides the reconstruction in the remaining coordinate perpendicular to the drift direction. The integrated area of the wire signal is proportional to the collected ionization charge.
Figure \ref{fig:tpcconcept} illustrates the principle of operation. %

\dword{sphd}, the first of the four \dword{fd} modules and described in Section \ref{sec:intro-fd}, will use horizontal-drift \dword{sp} technology with \dword{apa} readout, similar to that used in \dword{pdsp}.

\begin{dunefigure}
[Signal formation in a LArTPC with three wire planes]
{fig:tpcconcept} %
{Diagram  from  \protect{\cite{ Acciarri:2017sde}}  illustrating the signal formation in a \dword{lartpc} with three wire planes~\cite{Acciarri:2016smi}. For simplicity, the signal in the first (U) induction plane is omitted in the illustration. }
\includegraphics[trim={0cm 0.6cm 2.5cm 0.7cm},clip,height=8cm]{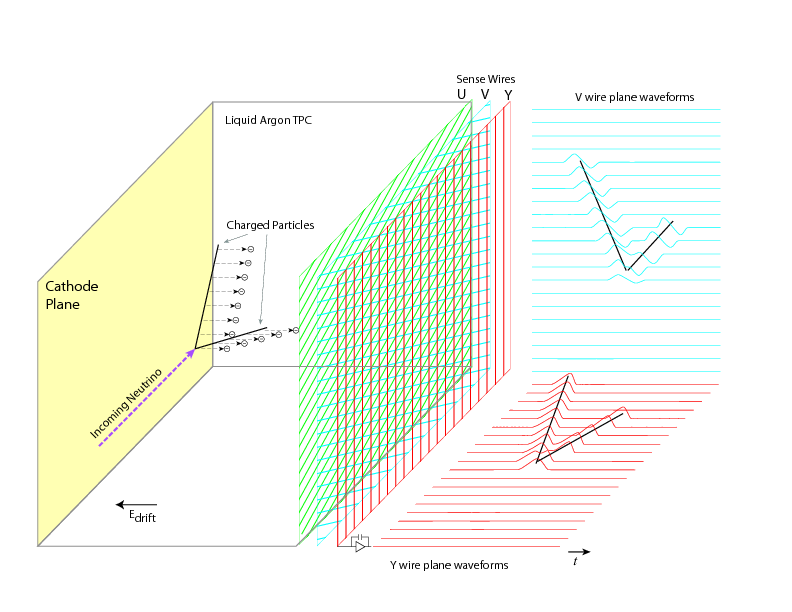}
\end{dunefigure}

The \dword{pdsp} detector~\cite{DUNE:2021hwx,DUNE:2020cqd}, immersed in a 770\,t volume of liquid argon, includes %
a cathode plane in the center and sets of three \dwords{apa} mounted on the %
sides of the liquid volume opposing the cathode, creating %
equal and opposite horizontal electric (E) fields on the two sides of the cathode plane. %
The maximum drift distance is  3\,m with a nominal voltage of 150\,kV  across that distance and a field strength of 500~V/cm.  Each \dword{apa} has 2560 channels and each channel reads out a 12\,bit \dword{adc} every 0.5\,$\mu$sec. %
For \dword{pdsp} the readout time appropriate for a 3 m drift was set to 3 msec,  resulting in 6000 12\,bit samples per channel for every readout window. 
The total data size for six \dwords{apa} is thus 140\,MB with additional header data and data from photon detectors and external tagging systems (details available in \cite{DUNE:2021hwx}), bringing the nominal \dword{tr} size to about 180\,MB.  Lossless compression of the \dword{tpc} readout data was implemented in the \dword{daq}, resulting in a final compressed \dword{tr} size of about 75\,MB.

All or part of the detector will be read out for time intervals of a few ms herein called a ``trigger record'' or ``event'' triggered by beam and/or activity in the FD.
DUNE prefers to use the term ``trigger record'' rather than ``event'' for the unit of processing, as it avoids confusion between an interaction ``event'' and a readout ``event.''

The \dword{pdsp} test beam ran at rates of up to 25\,Hz over a period of six weeks at beam momenta between 0.5 and 7\,GeV/c.  Time-of-flight and Cherenkov counters in the beamline provided beam flavor tagging.  Around 8M total ``physics'' records were written, with around 3M having beam tag information.  In total  850\,TB of raw test beam data were written, along with 1\,PB of commissioning and cosmic data. These data were successfully cataloged and written to storage at both \dword{cern} and \dword{fnal} at rates of up to 2\,GB/sec.   

Thanks to significant prior effort in the \dword{lartpc} %
computing and algorithms community, reconstruction software was quickly developed and production workflows integrated. Keep-up processing of the first reconstruction pass began soon after data taking started, and two weeks after the end of data taking was complete.  These results were extremely useful in demonstrating the capabilities of the detector; they are summarized in Volume II of the \dword{sphd} \dword{tdr}~\cite{Abi:2020evt}.  A second pass, with improved treatment of instrumental effects ranging from sticky codes, \twod deconvolution, and correction for space charge effects was completed in late 2019. Another pass with major improvements to the electrostatic modeling and reconstruction algorithms was completed in 2021. 

Below are three figures that show examples of the output of production processing algorithms. Figure~\ref{deconvolution} illustrates the signal processing stage of reconstruction, where raw \dword{adc} signals have noise and sticky codes removed and are then deconvolved to yield Gaussian hit candidates. %
This impacts the reconstruction algorithms and framework processing. Figures~\ref{wire-cell-bee} and~\ref{pandora} illustrate full pattern recognition and event reconstruction, respectively. 

Figure~\ref{deconvolution} illustrates the result of the first stage of reconstruction which has two main parts. The first applies digital noise filtering techniques to reduce the effects of thermal and coherent noise (and to counteract ADC imperfections in a small ($\sim 1$\%) of channels). The second stage performs further signal processing in two steps. First, a model of the detector response is deconvolved from the noise filtered waveforms. Importantly, this deconvolution is across both the longitudinal drift time dimension and across the `channel direction'. This is required as electrons drifting near wires will induce measurable current not just on the wire of closest approach but across multiple neighboring wires and over a drift time of about 100 $\mu$s. This deconvolution process inherently amplifies low frequency noise. To combat this, a second step locates signal regions of interest and over that region the waveform baseline is recalculated. These signal waveform regions are then input to all subsequent pattern recognition and event reconstruction as illustrated in Figures~\ref{wire-cell-bee} and~\ref{pandora}.

Compressed raw input trigger records were of order 75\,MB in size and took 500-600 seconds to reconstruct, of which around 180\,s was signal processing and the remainder high-level reconstruction dominated by 40-60 cosmic rays per readout.  Memory footprints for data processing ranged between 2.5 and 4\,GB.  Output   record sizes were reduced to 35\,MB by dropping the raw waveforms after hit finding.   Data reconstruction campaigns took of order 4-6 weeks (similar to the original data taking) and utilized up to 15,000 cores on \dword{osg} and \dword{wlcg} resources.  Job submission was done through the \dword{poms}\cite{poms} job management system developed at \dword{fnal}. \dword{poms} supports submissions to \dword{fnal}-dedicated resources and selected \dword{osg} and \dword{wlcg} sites.  Figure \ref{fig:sites} shows the distribution of wall hours used for the first pass of reconstruction in 2019.  
These metrics have influenced the ideas regarding memory utilization, event data model, and workload/workflow management in this document.

For reconstruction, data were streamed via {\tt xrootd}\cite{Behrmann:2011zz} from \dword{dcache} storage at \dword{fnal} (and in some cases \dword{cern} where a second copy was stored) to the remote sites. Despite individual processing jobs taking 15-30 hours to complete, network interruptions rarely caused job failures. These numbers and performance metrics are used later to predict future resources needs for the full DUNE experiment.

\begin{dunefigure}
[Raw and deconvolved induction U-plane signals before and after signal processing]
{deconvolution} %
{Comparison of raw (left) and deconvolved induction U-plane signals (right) before and after 
the signal processing procedure from a \dword{pdsp} \dword{tr}. The bipolar shape with red (blue) color representing
positive (negative) signals is converted to the unipolar shape after the \twod deconvolution.}
\includegraphics[width=0.49\textwidth]{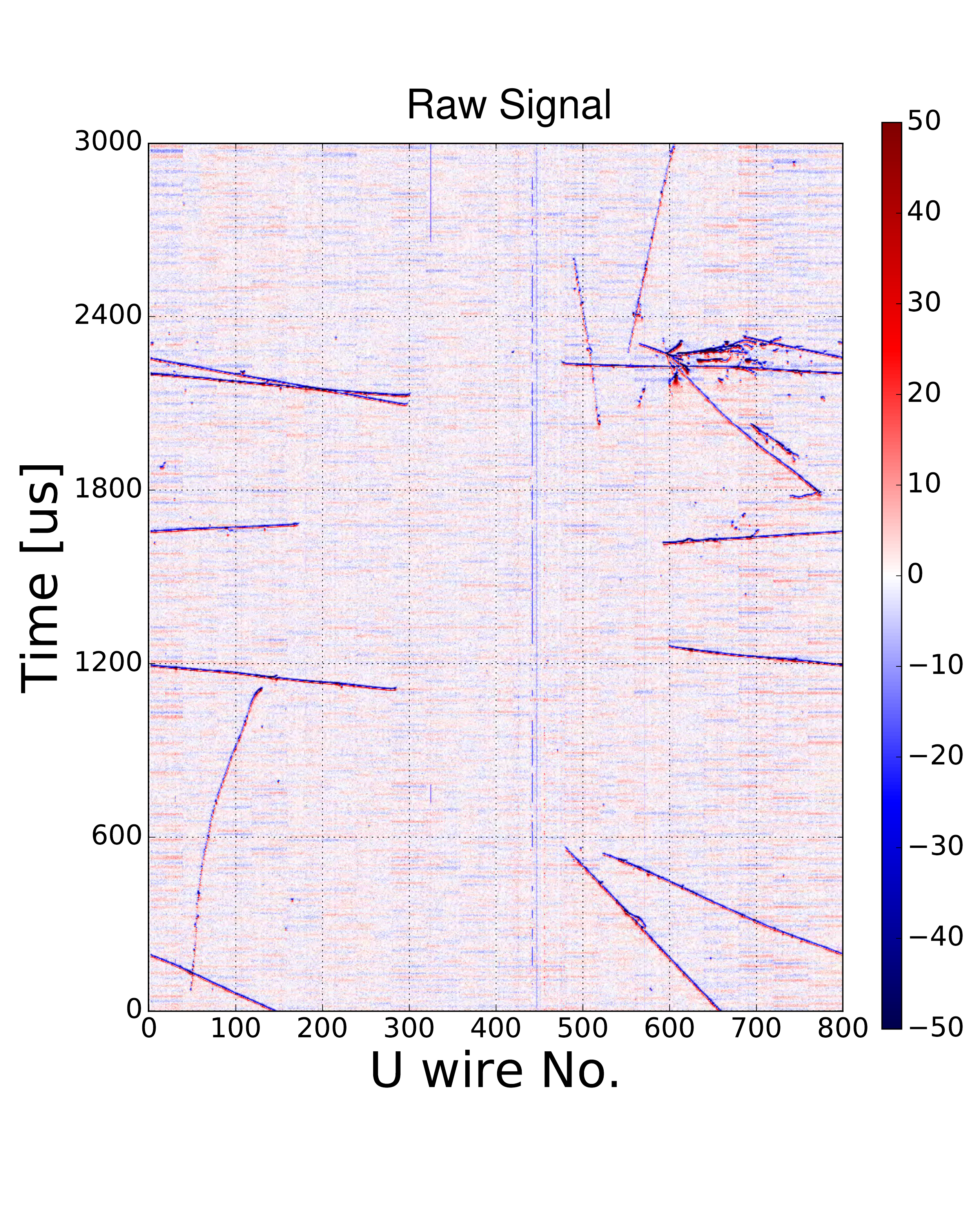}
\includegraphics[width=0.49\textwidth]{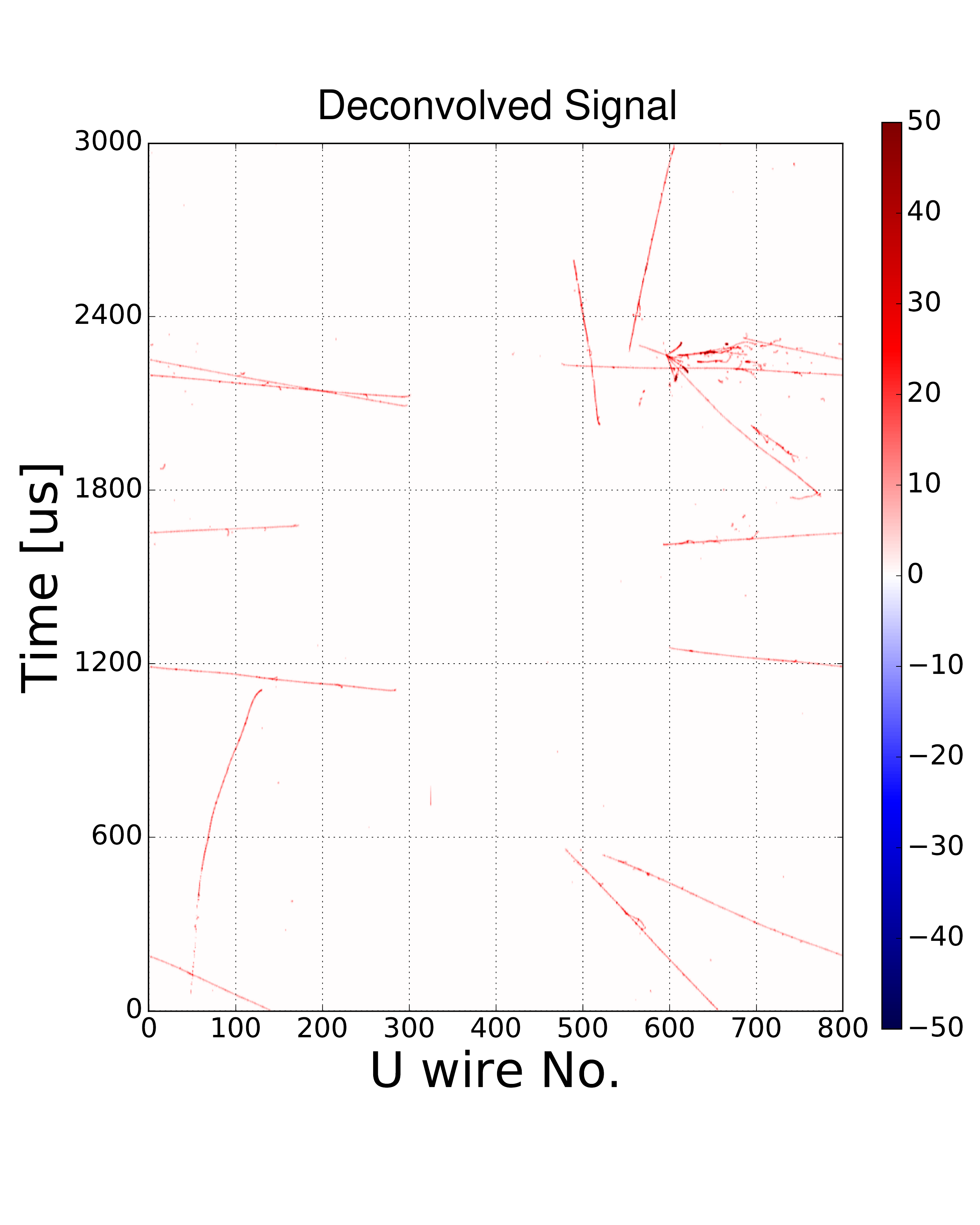}
\end{dunefigure}

\begin{dunefigure}
[Cosmic rays and beam interaction in \dshort{pdsp}]
{wire-cell-bee} %
{The \dword{pdsp} detector (gray box) showing 
the direction of the particle beam (yellow line on the very far left) and the outlines of the six \dwords{apa}. Cosmic rays
can be seen throughout the white box, while the red box highlights the beam region of interest with an interaction of the 7\,GeV beam. 
The \threed points are obtained using the Space~Point~Solver reconstruction algorithm.}
\includegraphics[width=0.9\textwidth]{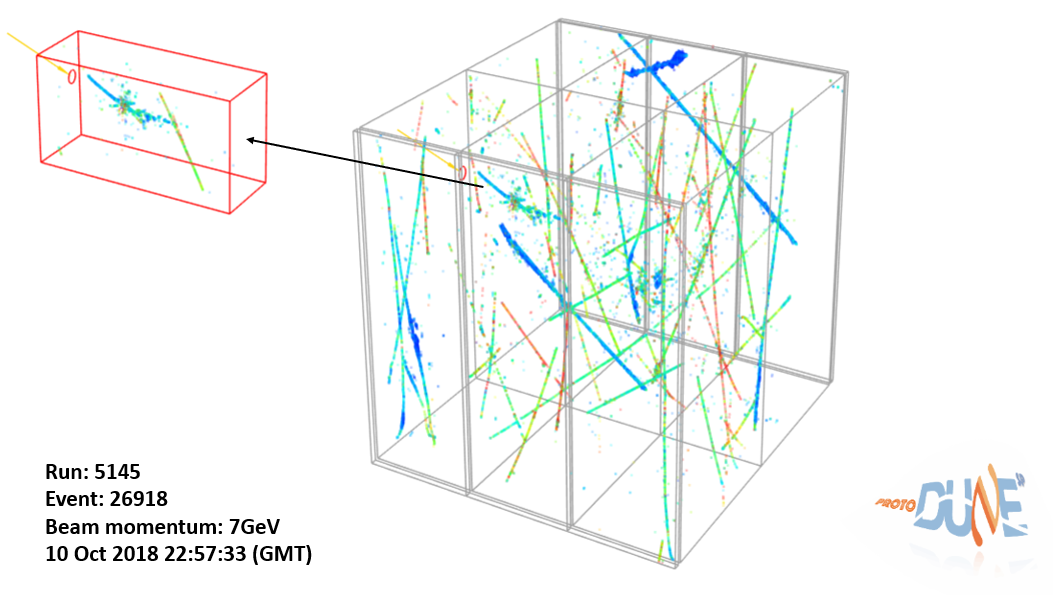}
\end{dunefigure}

\begin{dunefigure}
[Pandora reconstruction of cosmic rays and beam interaction in a \dword{pdsp} trigger record]
{pandora}
{Pandora \protect{\cite{Acciarri:2017hat}} reconstruction of cosmic rays and beam interaction in a \dword{pdsp} \dword{tr}. The left side of the figure shows the full detector volume with all interactions, including cosmic rays, and the right side shows the identified beam interaction.}
\includegraphics[width=0.8\textwidth]{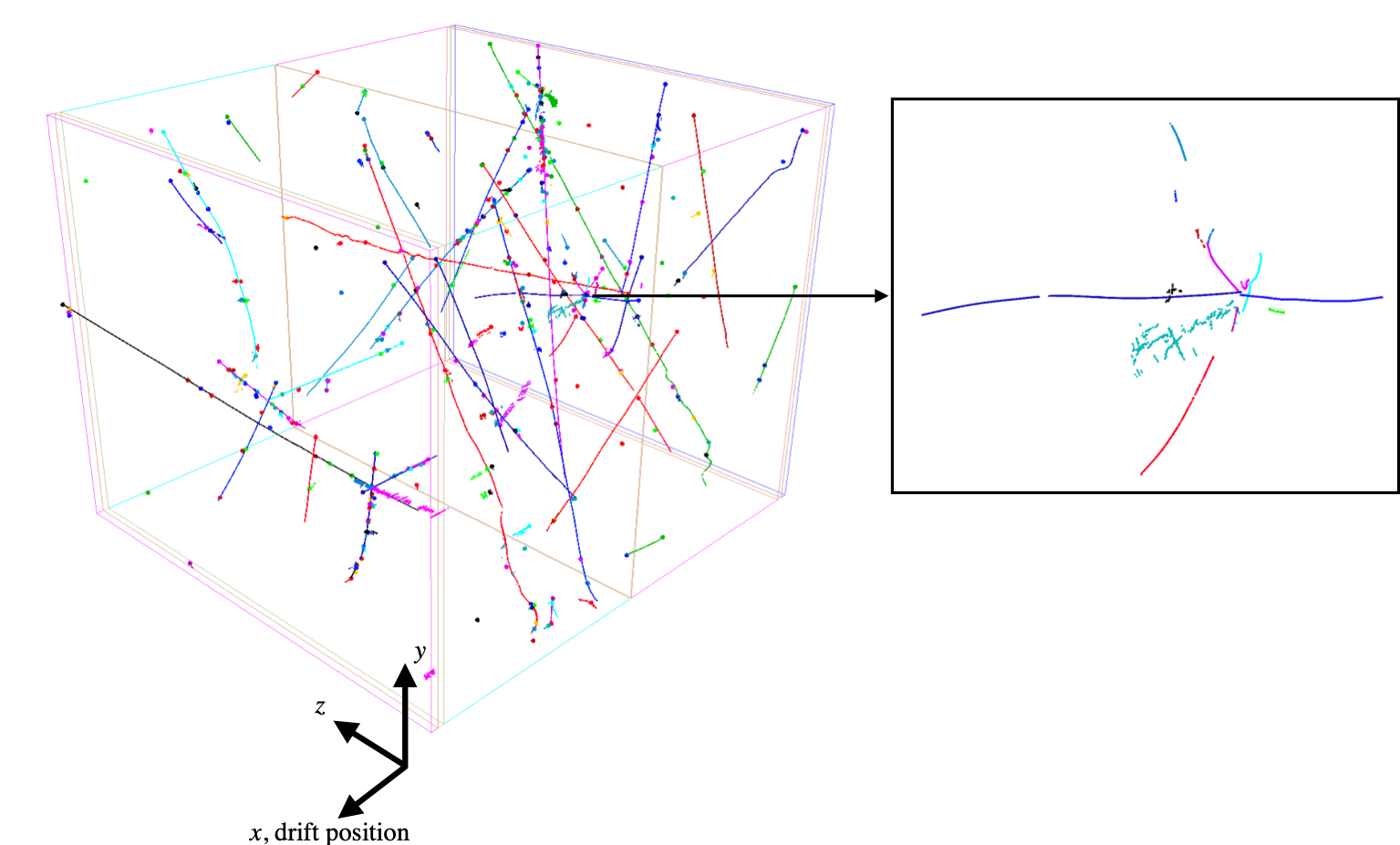}
\end{dunefigure}

\begin{dunefigure}
[Reco/sim processing distribution across sites for DUNE production, 2019]
{fig:sites} %
{Reconstruction and simulation processing distribution across sites for DUNE production in calendar 2019.  The inner circle shows national contributions while the outer circle shows individual site contributions.}
\includegraphics[height=0.65\textwidth]{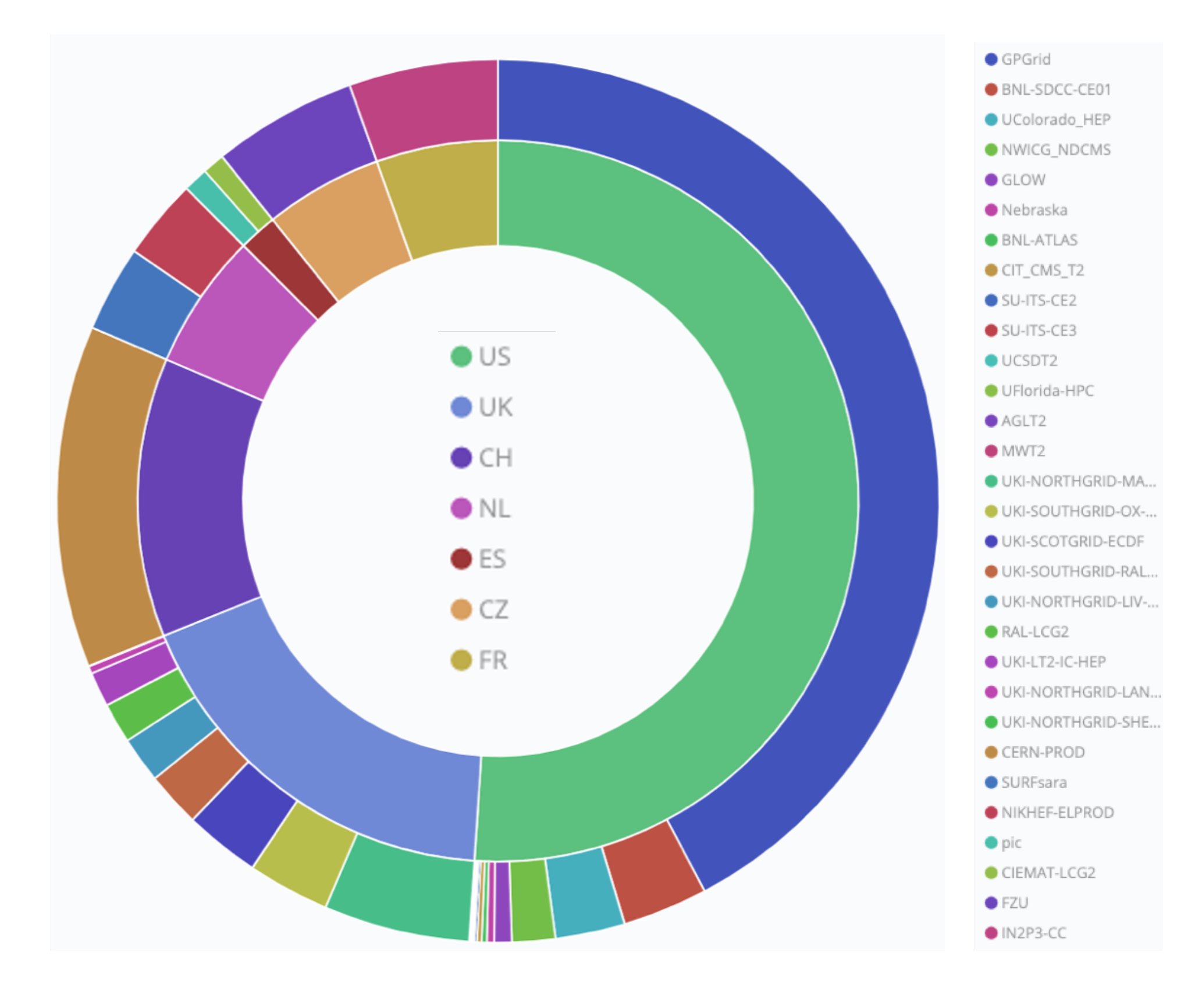}
\end{dunefigure}

\subsection{ProtoDUNE Dual-Phase and its Evolution to the Vertical Drift Design \hideme{Pennacchio - draft}}

In parallel with the single-phase horizontal-drift detector tests, a vertical drift  \dword{dp} readout prototype (\dword{pddp})
in a similar cryostat was also  tested.  \dword{pddp} was not exposed to beam but ran on cosmics in 2019 and 2020. 

The basic operating principle of \dword{pddp} is shown in Figure~\ref{dp_principle}. As in \dword{pdsp}, charged particles that traverse the active volume of the \dword{lartpc} ionize the medium while also producing scintillation
light that is detected by a Photon Detection System. The ionization electrons drift vertically upward %
toward an extraction grid just below the liquid-vapor interface. 
After reaching the grid,
an \efield stronger than the drift field extracts the electrons from the liquid up into the gas phase.
Once in the gas, electrons encounter micro-pattern gas detectors, called \dwords{lem}, with high-field regions. The \dwords{lem} amplify the electrons in avalanches that occur in these
high-field regions. The amplified charge is then collected and recorded on a \twod anode consisting
of two sets of 3.125\,mm pitch gold-plated copper strips that provide the $x$ and $y$ coordinates (and
thus two views) of an interaction.   

\begin{dunefigure}
[Principle of \dshort{dp} readout and parameters for extraction]
{dp_principle} %
{Principle of \dword{dp} readout (left), and thicknesses and HV values for electron extraction from liquid to gaseous argon, their
multiplication by \dwords{lem}, and their collection on the x and y readout anode plane (right) The HV values are
indicated for a drift field of 0.5\,kV/cm in LAr.}
\includegraphics[width=0.85\textwidth]{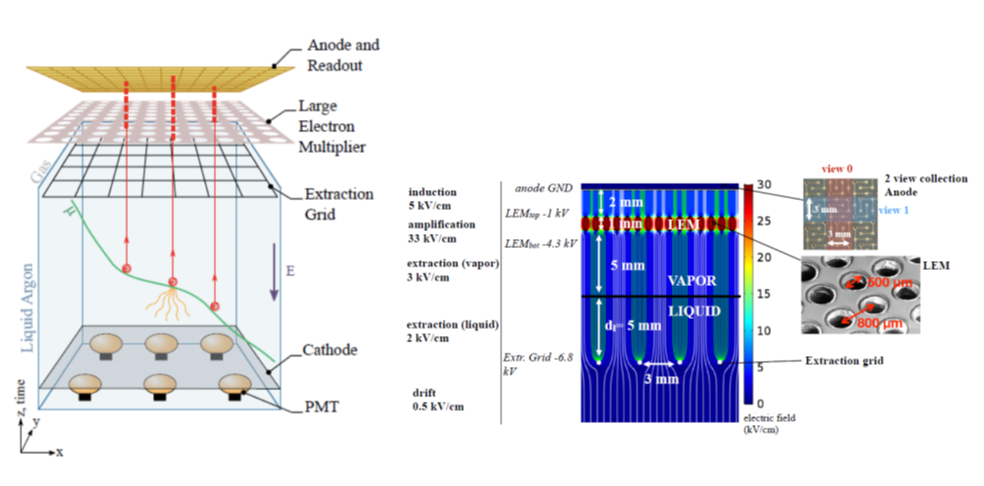}
\end{dunefigure}

The readout area surface is \SI{6x6}{m}, subdivided into four $3\times3$ m$^2$  \dwords{crp}. Each \dword{crp} is an independent detector element %
that performs electron extraction, amplification, and collection. 
The 7680 readout channels are read by   12\,bit \dwords{adc} every 0.4\,$\mu$sec. 
The \dword{pddp} detector   consists of a 700\,t volume of \dword{lar}, with  a vertical drift length of 6\,m, corresponding to a full drift window of 4\,ms (10,000 samples).

The \dword{pddp} detector began taking cosmic ray data in August 2019. Thanks to preceding data challenges, these data have been successfully integrated into the full data cataloging and reconstruction chain and were   reconstructed as they became available.
A total of 1.45M \dwords{tr} were collected; the size of the raw data files (run sequence files) was 3.2 \,GB, each file containing 30 \dwords{tr}. Cosmic ray data are displayed in Figure~\ref{dpevent}: on
the left a horizontal muon track is shown with the corresponding waveform on a channel, giving an
idea of the low noise conditions. A \dword{tr} including an electromagnetic shower and two muon decays
and a \dword{tr} with an example of multiple hadronic interactions in a shower are shown on the right.

\begin{dunefigure}
[Cosmic ray data from \dshort{pddp}]
{dpevent} %
{Cosmic ray data from the \dword{dp} prototype, \dword{pddp}.}
\includegraphics[width=0.85\textwidth]{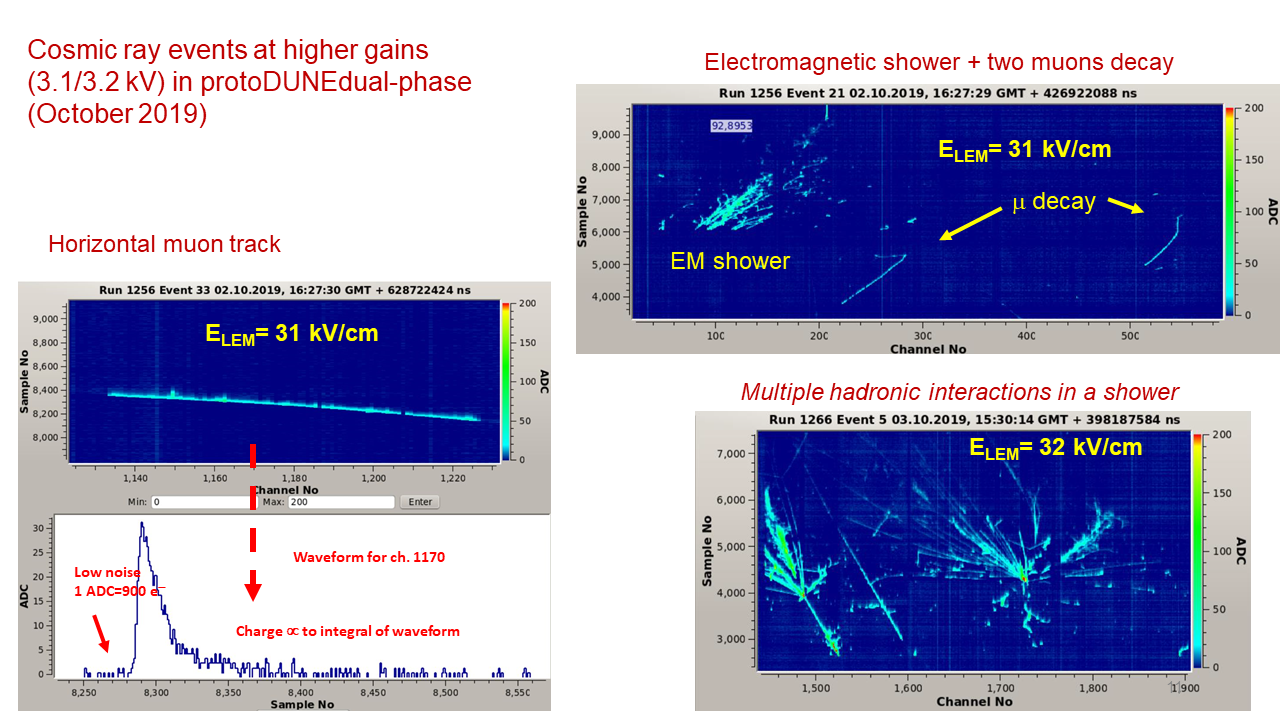}
\end{dunefigure}

All data ($\sim$\,330\,TB) taken during different campaigns   have been copied to \dword{fnal}. A subsample is composed of data sets taken during detector transient conditions, motivated by various specific testing needs;  all cosmic ray data taken in well defined and stable detector conditions in 2019 and 2020 ($\approx$\,377K events) have been processed with \dword{larsoft} by performing the reconstruction of hits and \twod tracks. A second pass, including \dword{pandora} reconstruction algorithms, started in spring 2021. 
The memory footprint is between 1.9 and 2.5\,GB.
Data management and job submission was successfully done through the same systems as \dword{pdsp}.

Experiences with the dual-phase CRP-readout and vertical-drift, and with the horizontal-drift single-phase APA-readout, motivated development of the single-phase CRP-readout, vertical-drift detector planned for
the \dword{spvd} concept. The \dword{spvd} incorporates many of the design aspects developed for the \dword{dp}, such as the \dwords{crp};   the  main  difference  with  respect  to  the  \dword{dp}  design  is   the  removal  of  the  extraction  stage  to  the  gas  phase  and  the  subsequent  charge  amplification  stage.  %
This eliminates the grid biased at high voltage (to transfer the electrons from the liquid to the gas) and the \dwords{lem} used to amplify the signal in the gas. The \dword{spvd} \dwords{crp} %
perform charge readout using %
perforated \dword{pcb} anodes with finely segmented strip electrodes %
that are immersed in the \dword{lar}. The \dwords{crp} operate in a conceptually similar manner to the HD \dwords{apa} in that there are two sets of induction strips and one set of collection strips. Each set of strips is constructed at an angle offset to the other two angles to provide disambiguation in both spatial dimensions.

\begin{dunefigure}
[Vertical drift solution with \dshort{pcb}-based charge readout]
{fig:vd_principle}
{Vertical drift solution with \dword{pcb}-based charge readout}
\includegraphics[width=0.85\textwidth]{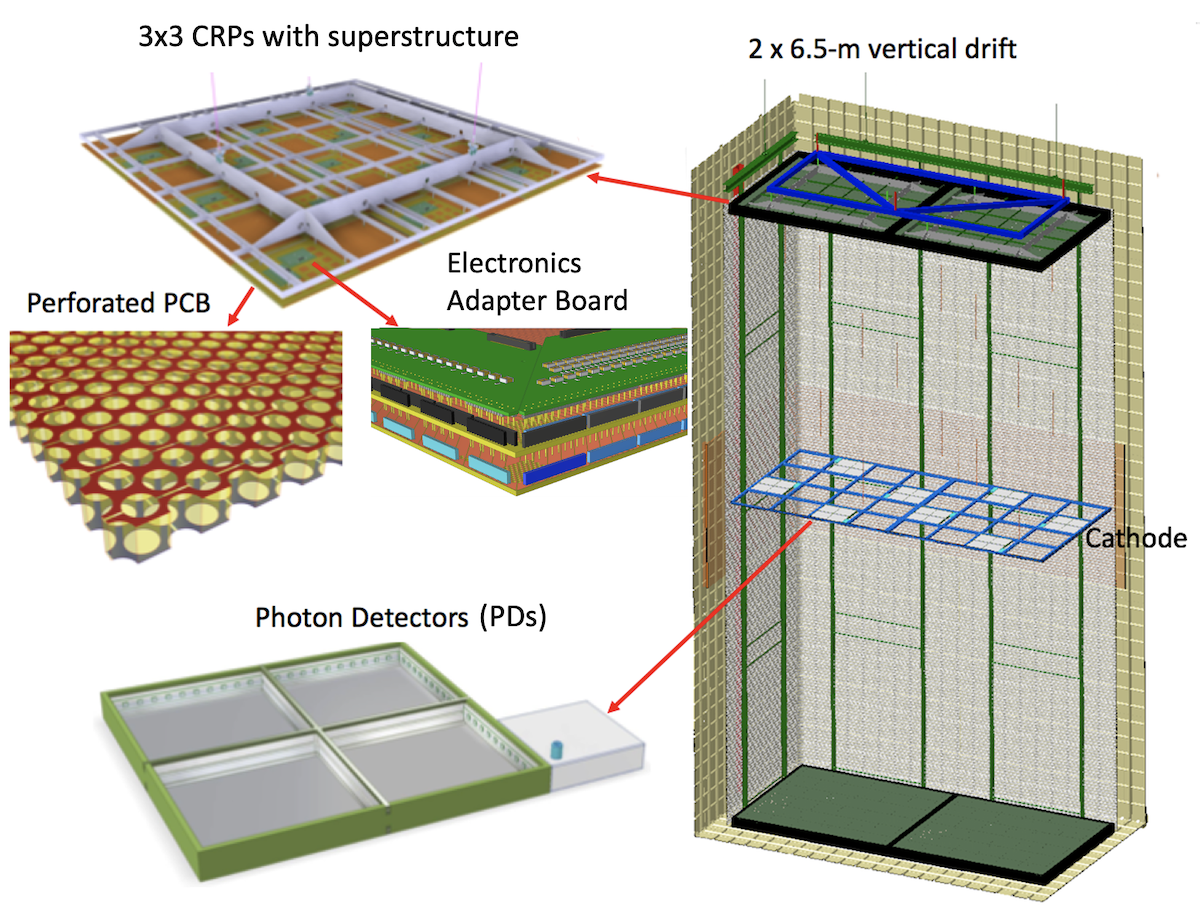}
\end{dunefigure}

 Figure \ref{fig:vd_principle} illustrates the \dword{pcb}-based charge readout concept for the \dword{vd} detector. The electron drift direction is vertical.
Two separate drift volumes of 6.5\,m are defined by a cathode plane at roughly mid-height in the
detector volume. Ionization electrons above the cathode will drift upwards; ionization electrons in the liquid below the cathode will drift downwards.
The \dword{spvd}  prototype components have undergone testing starting in 2021 with a small \coldbox; some examples of cosmics tracks are shown in figure \ref{fig:VD_coldbox_trks}. A full prototype, known as \dword{mod0}, is expected to be completed and  installed 
 inside the \dword{np02} cryostat in 2022 and 2023. Long-term operation and full characterization with a charged particle beam and cosmics will follow. 

\begin{dunefigure}
[cosmic tracks collected during Vertical drift cold box data taking]
{fig:VD_coldbox_trks}
{Examples of cosmic tracks in the Vertical Drift cold box as visualized with a raw-data event display (left) and after calibration has been applied (right).}
\includegraphics[width=0.85\textwidth]{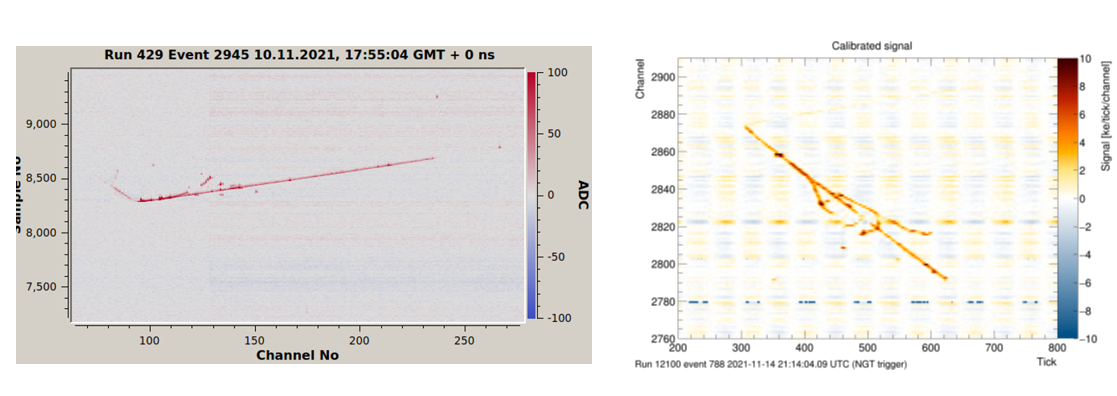}
\end{dunefigure}

\subsection{Conclusions from Prototype Tests \hideme{Schellman draft}}

\dword{protodune} %
runs are continuing and will continue through beam tests in 2022-23 at \dword{cern}.  Data cataloging, movement and storage techniques were tested before the start of the 2018-2020 \dword{pdsp} and \dword{pddp} runs and were able to handle the full rate of the experiments.   Reconstruction algorithms were also in place on time and %
produced early results that led to increased understanding of the detector and improved calibrations for a second iteration.  These tests also identified some deficiencies in our infrastructure, including incomplete schemes for the transmission of configuration and conditions information between hardware operations and  offline computing.  The first round of test beam runs have been extremely valuable; they helped identify which variables are important to transmit and they motivated the design of improved systems for gathering and storing that information. 

An additional beam and cosmic ray run of \dword{pdhd} is planned for 2022-23, with beam and cosmic tests of the \dword{spvd} design to follow in 2023-24, allowing further development and testing of our computing infrastructure before the full detector comes online in the late 2020s. Additionally, there are ongoing smaller-scale readout and noise tests of both \dword{apa} and \dword{crp} in cold boxes at \dword{cern} and small cryostats at \dword{fnal}.

\section{Far Detector %
\hideme{Schellma-draft}}
\label{sec:intro-fd}

The full DUNE \dword{fd} will begin with one \dword{sphd} module to be installed  at \dword{surf} %
starting near the end of  this decade.  A second \dword{spvd} module will be installed and commissioned in parallel.  High-intensity neutrino and antineutrino beams should arrive after a year or so of commissioning of the detector and the \dword{lbnf} beamline.  The first %
module will %
be a scaled-up version of \dword{pdsp} with 150 \dwords{apa} %
stacked two \dwords{apa} deep %
the length of the cryostat, down the center and along the long walls. %
The argon volume will be $15\times14\times62$\,m$^3$ with a total (fiducial) mass of 17\,kt (10\,kt).  Section~\ref{sec:est:FD} summarizes the expected event rates and data volumes for the first two modules.  Two additional detector modules, possibly with updated or novel technologies, %
will be added later. For now, we assume that data volumes and rates coming from other technologies, will not exceed %
the values for the \dword{sphd} or \dword{spvd}.

The detector will be sensitive to energy deposits on the order of $\sim 1$\,MeV. The decay of $^{39}$Ar,$^{42}$Ar, and other radioactive nuclei will produce a prohibitively high trigger rate if the threshold were to be set much below $\sim 5$\,MeV. On the other hand, beam neutrino interactions will typically deposit well above 100 MeV. An energy-based trigger threshold will provide near perfect detection efficiency for beam neutrino interactions.
More sophisticated triggering algorithms should also allow standalone detection of astrophysical sources, including higher-energy solar neutrinos and \dword{snb} candidates. 

The data rates will be dominated by 4,500 cosmic rays expected per module/day.  These events are vital for monitoring and aligning the detector. %
The next highest rate source of events will be calibration campaigns with radioactive and neutron sources and lasers.  In all cases, the goal is to gather data from the full volume of the detector with as fine a granularity as possible. 

Beam interactions themselves are expected to be quite rare, occurring in only 1/2000 of beam gates ($\simeq$\,2/hr)  Extraction of oscillation parameters will require both powerful background rejection, discussed in  Section 1.1, and precise calibration of the energy scale of the experiment, hence the much larger calibration samples.

Beam, cosmic ray and solar neutrino interactions are reasonably localized in time and space, involving a small fraction of a module over a few milliseconds.  The DUNE DAQ group has plans to design and implement sparse readout of Far Detectors that will allow significant reduction in data size without loss of physics information if suitable triggers are used. Details of the expected rates and data volumes are described in Section \ref{sec:est:volumes}.

\subsection{Supernova candidates \hideme{Schellma -updated early march }}\label{sec:supernova}

Supernova candidates pose a unique problem for data acquisition and reconstruction.  \dword{snb} physics in DUNE is discussed in some detail in the far detector \dword{tdr}\cite{ Abi:2020evt}, reference \cite{Cuesta:2020dyj} and a dedicated paper \cite{DUNE:2020zfm} and only summarized here. 

\dword{dune} is uniquely sensitive to $\nu_e$ through the reaction $\nu_e + \hbox{Ar} \rightarrow e^- + \hbox{K}^*$ which leaves an electron trajectory that can be used to estimate the pointing direction of the supernova explosion.  Figure \ref{fig:SNrates} shows the expected neutrino interaction rates from supernovae as a function of their distance.

\hideme{HMS 3/2 added the missing figure}

\begin{dunefigure}
[Supernova rates in DUNE as a function of distance]
{fig:SNrates}
{
Supernova rates in the DUNE detector  as a function of distance from the source, from reference \cite{DUNE:2020zfm}.}
\includegraphics[width=0.9\textwidth]{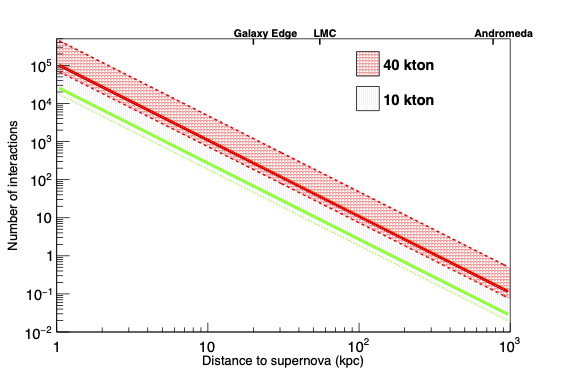}
\end{dunefigure}

A typical core-collapse supernova 10\,kpc away is expected to yield around 3,000  \dword{cc} electron neutrino interactions across four detector modules. \dword{snb} candidates will be quite different from beam interactions, having small interactions with energies in the 5-30\,MeV range spread across the full volume of the detector modules over many seconds, in contrast to the localized, coincident,  500-10,000\,MeV signature of beam neutrino interactions. These differences impose interesting requirements on the \dword{daq} and computing models for the experiment.  

\begin{dunefigure}
[Supernova interactions in the FD]
{fig:SNe} %
{
 Simulated $\nu_e$
 CC event with a 20.25 MeV neutrino, showing an electron track and ``blips" from Compton-scattered gammas. The vertical dimension indicates time and the horizontal dimension indicates wire number. Color represents charge. The top panel shows the collection plane and the bottom panels show induction planes. The boxes represent reconstructed hits. }
\includegraphics[width=\textwidth]{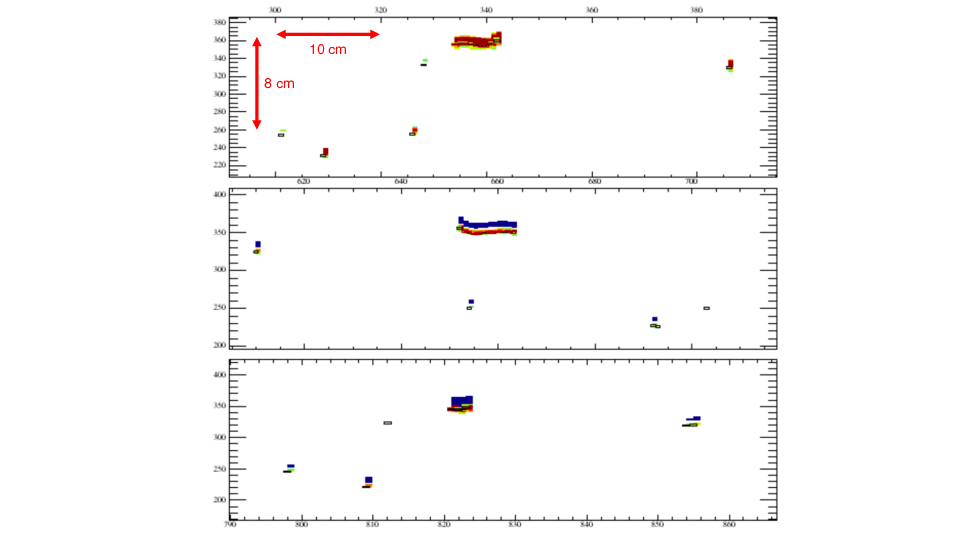} %
\end{dunefigure}

\dword{snb} physics and its influence on neutrino emission are not fully understood and will result in significant modulations of the event rates for different neutrino types  over the few tens of seconds of the burst.  DUNE's fine-grained tracking should allow significant pointing power with the most optimistic scenario of four modules and high electron neutrino fraction yielding pointing resolutions of less than 5 degrees. Other neutrino detectors worldwide will also be able to provide fast information but the \dword{dune} information will be unique in its sensitivity to electron neutrinos. Figure \ref{fig:SNe} illustrates simulated signatures of \dword{snb} neutrino interactions in the far detector. The ability to produce a reasonably fast pointing signal is extremely valuable to optical astronomers doing followup, especially if the supernova were in a region where dust masks the primary optical signal.   The need to be alert to supernovae and to quickly transfer and process the data imposes significant requirements on triggering, data transfer and reconstruction beyond those imposed by the %
beam-based oscillation physics.

The \dword{tpc} and \dword{pd} produce several signals of increasing precision.  First, scintillation light is detected by the \dword{pd}, yielding fast triggering information.  Trigger primitives based on a subset of the \dword{lartpc} information can also be searched for supernova signatures.  With the greatest resolution, a full detector readout can be mined for information across the full spatial and energy range of the detector over a period of up to 100 s.  There is also research into full-chain reconstruction on highly-condensed data. %

In reference \cite{DUNE:2020zfm} the performance and efficiency of fast triggering on scintillation light and \dword{tpc} hits are investigated.  Radiological and noise backgrounds are required to produce less than 1 false trigger/month. Both \dword{pd}- and \dword{tpc}-based triggers are sensitive to relatively low numbers of interactions ($\sim10-25$) at acceptable background rates, yielding expected sensitivities out to the Large Magellanic cloud as shown in Figure~\ref{fig:SNrates}.

Once the trigger system has identified a potential \dword{snb} signal  the \dword{daq} then records information for 100 s, yielding 140~TB of uncompressed information for a \dword{hd} module and 180 TB for a \dword{vd} module.  These data must then be transferred offsite for processing with the full algorithms as the \dword{surf} site does not currently have infrastructure designed to support reconstruction of SBN candidate trigger records underground or on the surface.

Moving  320 TB (assuming the first 2 modules) of uncompressed data   from \dword{surf} to processing centers will require, at minimum, 6-7 hours with the planned 100Gb/s network link.  This can be accelerated by implementing onboard data compression and may require upgrades to the network when the projected third and fourth modules come online.  The need to store up to 320 TB of data from a supernova candidate while also continuing normal data taking drives the size of local disk buffers at \dword{surf} and presumably requires similar reserved or rapidly-preemptable space at the centers where the data would be processed. Our \dword{protodune} experience indicates that reconstruction of \dword{lartpc} data takes between 1 and 3 sec/MB of raw data on present-day computing resources.  It will thus take of order 12,000-40,000 cores to perform a preliminary reconstruction of the data as fast as it comes in. In principle, a significant fraction of the neutrino data could be processed  in the 2-4 hours before a supernova becomes visible at optical frequencies.

\subsection{Other phenomena -- Solar Neutrinos and Beyond-the-Standard-Model Processes} %
There are multiple other measurements that can be made in  massive low background detectors such as the \dwords{fd}.  These are described in a recent White Paper \cite{Caratelli:2022llt}. These include detection of solar neutrinos and searches for \dword{bsm} processes such as neutron-anti-neutron oscillations.  Detection of these rare processes will require low radiological backgrounds and trigger systems capable of reading out the detector at rates low enough to remain well below a 30 PB/year limit on data logging.  They will require recording the full waveforms, perhaps over a limited physical region, to perform optimal signal extraction. In this document we concentrate on the oscillation, calibration and \dword{snb} scenarios as they are likely to dominate data rates in the near term.

\section{Near Detector \hideme{Junk/Muether - needs update}}
\label{sec:intro-nd}

High-precision oscillation physics requires a \dword{nd} system to allow measurement of the unoscillated neutrino flux and %
to provide improved understanding of neutrino interaction physics. 
The DUNE  collaboration is proposing a suite of near detectors optimized for these two goals. The proposed detectors are described in more detail in the Near Detector Conceptual Design Report \cite{DUNE:2021tad}.
 
 The \dword{nd} will be located in an enclosure in the path of the neutrino beam on the \dword{fnal} site 574\,meters from the target. %
 Interaction rates per beam spill are expected to be very large (at 0.83\,Hz), with 40-60 interactions per spill, including muons originating from interactions in material upstream of the fiducial volumes. Figure~\ref{fig:beamline} shows the beamline and location of the \dword{nd} on the \dword{fnal} site. There are three major subdetectors:
 
\begin{itemize}
\item A pixel readout 
 \dword{lartpc}, \dword{ndlar}, is  the most upstream of the three subdetectors shown in Figure~\ref{fig:nd}, where the beam propagates  from right to left. 
 \item Immediately downstream of \dword{ndlar} is a detector for characterizing muons exiting the \dword{ndlar}. This will be a magnetized steel range stack detector (TMS) for Phase~I of the DUNE experiment. 
 A gaseous liquid argon detector, \dword{ndgar}, which serves  as  a muon spectrometer and allows more detailed study of neutrino interactions, is planned for Phase~II operation when better control of neutrino interaction uncertainties is required.
 \item Beyond TMS/\dword{ndgar} is the \dword{sand} component of the \dword{nd} that acts as a beam monitor. \end{itemize}
 
Note that while the \dword{nd} will have the same \dword{daq} timing window flexibility as the \dword{fd}, it is not foreseen that \dword{nd} physics goals will require the use of varied time windows for trigger records, in contrast to the \dword{fd}. %

 \begin{dunefigure}
[The neutrino beamline on the Fermilab site]
{fig:beamline} %
{The neutrino beamline on the \dword{fnal} site. The near detectors will be situated 574\,m from the target and 62\,m below grade.}
\includegraphics[height=0.3\textwidth]{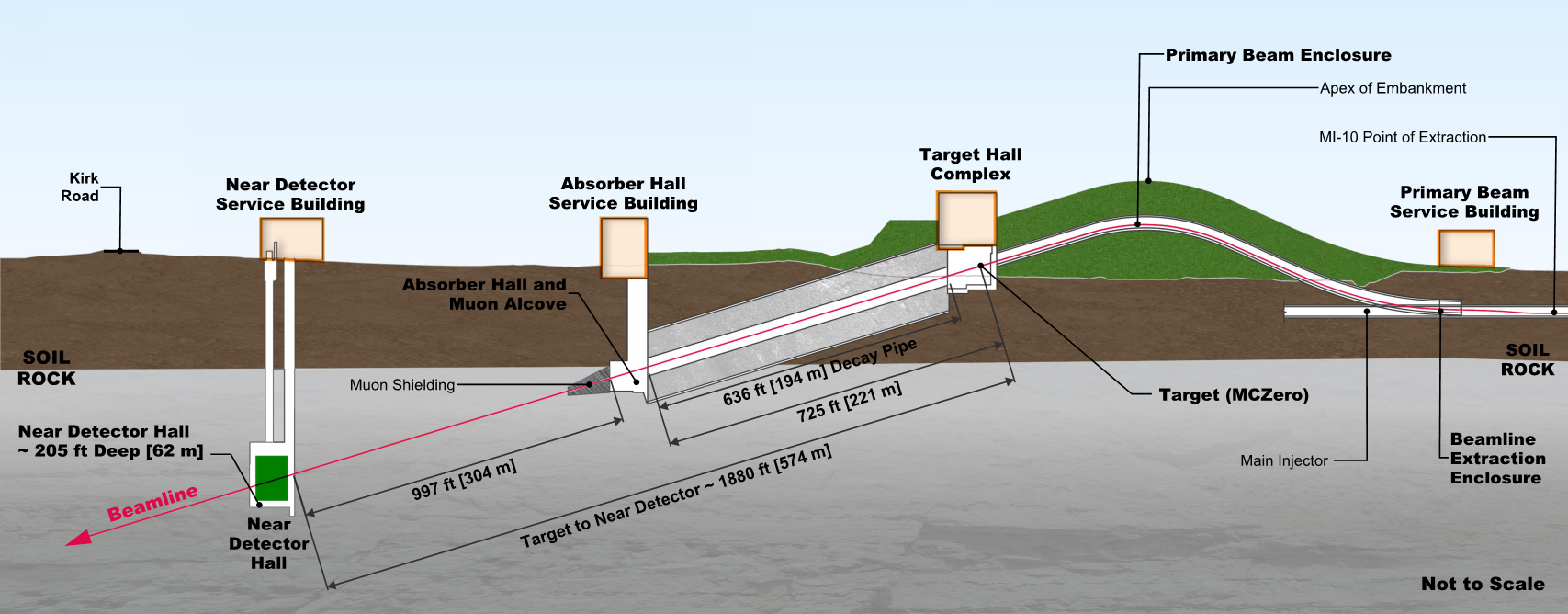}
\end{dunefigure}

 \begin{dunefigure}
[The ND systems in an on-axis configuration]
{fig:nd}
{The \dword{nd} systems in an on-axis configuration.  The beam enters from the right in this view. The \dword{sand} scintillating beam monitor remains at beam center while the pixel %
\dword{ndlar} and gaseous \dword{ndgar} \dword{tpc} detectors can be moved off-axis to make detailed studies of the neutrino flux at multiple angles.}
\includegraphics[height=0.5\textwidth]{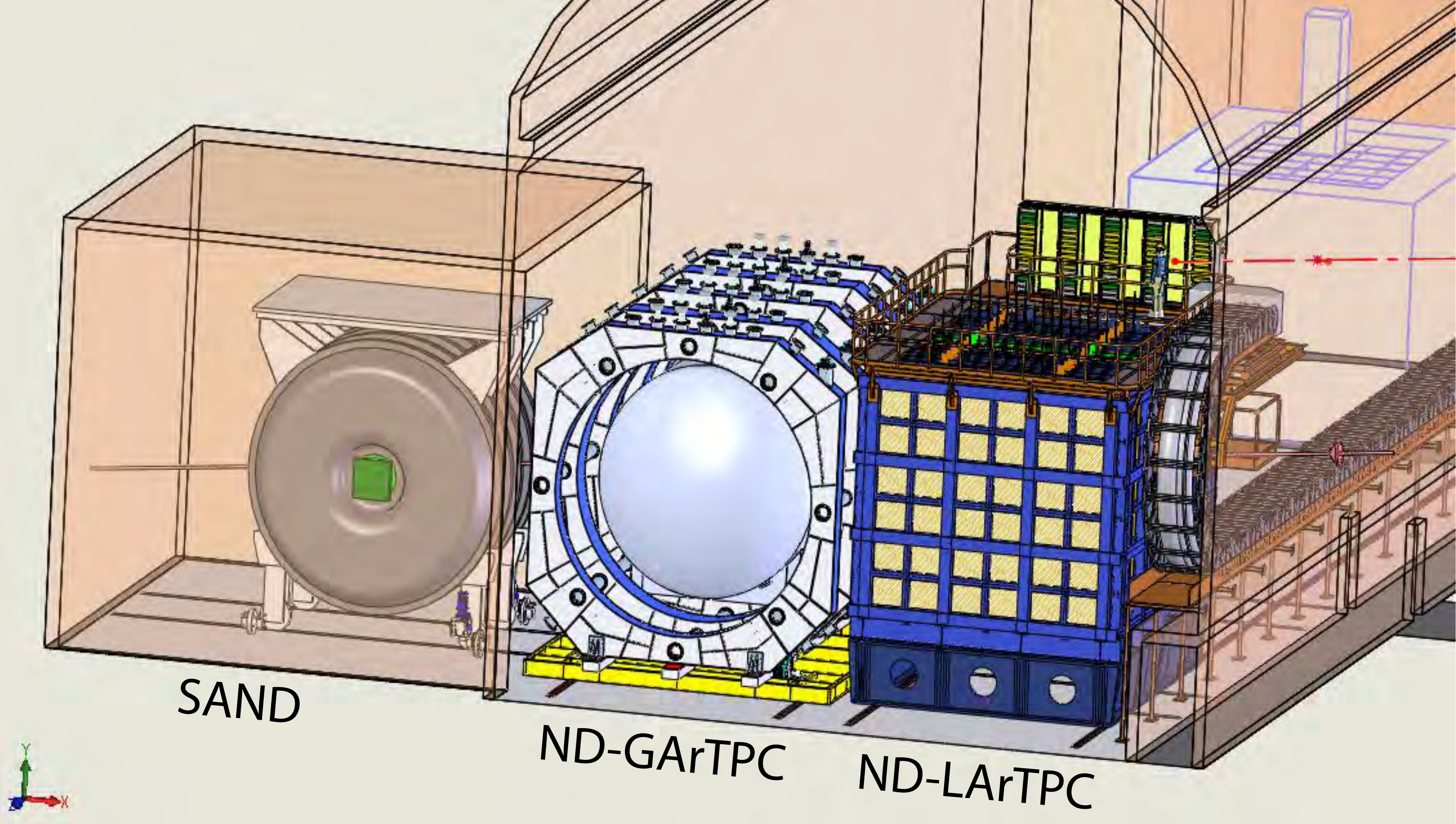}
\end{dunefigure}

 \subsection{Pixel \dshort{lartpc} - \dshort{ndlar} \hideme{Muether - draft}}
 \label{sec:intro-nd-lar}
 
 As the target material in the %
 far detector is \dword{lar}, optimal cancellation of systematic uncertainties between the near and far detectors requires that the near detector include a  \dword{lar} component to match the \dword{fd}.  However, at the intense neutrino flux and high event rate in the \dword{nd} region, occupancies will be too high to allow the \twod readout provided by conventional wire planes. A new \dword{arcube}  technology has been developed that allows pixelized charge readout and, along with modularity and highly capable light detection, provides unambiguous \threed imaging of  particle interactions.  The \dword{ndlar} component of the \dword{nd} is made up of a configuration of \dword{arcube}  \dwords{lartpc}  large enough to provide the required hadronic shower containment and statistics.  %
 
 The pixel \dword{lar} detector is designed to have 12~million $3 \times 3$\,mm$^2$ pixel channels and $\sim$4200 \dword{pd} channels.  The \dword{lartpc} will read out only pulse times and integrals, in contrast to the far detector which reads out every time slice.  The \dwords{pd} will, however, read out complete wave-forms.   A total of 3\,MB of uncompressed data is anticipated per spill from the \dword{tpc} with 5\,MB from the \dwords{pd} leading to an estimate of 144\,TB/year for uncompressed in-spill data. Calibrations and cosmic-ray data increase that data volume by around 20\%.

\subsection{Near Detector Phase I Muon Spectrometer (TMS) \hideme{Muether - draft}}
\label{sec:intro-tms}

The TMS, a low-cost, low-risk detector, will operate for the early part of the DUNE running when beam intensities are at their lowest and DUNE is not systematics limited. The TMS is a magnetized steel range stack (similar to MINOS \cite{minosNIM}) downstream of ND-LAr.  The purpose of the TMS is to measure the momentum of 1 - 5 GeV muons that exit ND-LAr by range with a momentum precision comparable to the Far Detector (taken to be 4\%). The steel will be magnetized to identify the charge of muons it detects with better than 98\% accuracy. These metrics are designed to ensure the DUNE physics program will function for the first 2-3 years of TMS operations with no degradation to physics output.

The TMS is placed 8.2 m downstream from the ND-LAr (center to center) and centered 2.51 m below the origin in $y$. The TMS consist of 200 total layers made up of alternating layers of magnetized steel plates and scintillator bars. The first 40 layers of steel are 1.5 cm thick. The rear 60 layers of steel are 4 cm thick. There is a 4 cm gap between each steel layer.  The thin (thick) central steel is given a vertically oriented downward pointing field of 1.25 T (0.9 T), while the outer steel thin (thick) steel plates have their field pointing upwards with a strength of 1.5 T (1.0 T). The scintillator is implemented as a collection of vertical bars, 3.5~cm wide by 3~m tall by 1~cm thick, arranged into modules of 1.68 m wide. Four modules are placed side by side into a layer. These scintillator layers are replicated in the gap between steel layers.  

The estimates for the TMS's data volume and raw processing needs are uncertain, but they are known to be less than the needs for \dword{ndgar}.   In the raw data, a TMS hit is 40-64 bits long.  Assuming 64 bits per hit and 550 filled buckets per spill, each with 20 hits, one arrives at a raw data volume of 700 kb/spill, or approximately 88 kB/spill. A generous factor of four for out-of-spill data is assumed.

\subsection{Near Detector Phase II \dshort{gartpc} \hideme{Muether - draft}}
\label{sec:intro-nd-gar}

The \dword{ndgar} is a magnetized detector system consisting of a high-pressure \dword{gartpc} surrounded by an \dword{ecal} and a muon system. The \dword{ndgar} measures the momentum and sign of charged particles exiting the \dword{ndlar}. In addition, for neutrino interactions occurring in the \dword{ndgar} itself, higher resolution and lower momentum thresholds can be achieved for charged particle tracks, leading to improved neutrino interaction models. This capability enables further constraints of systematic uncertainties for long-baseline neutrino  oscillation analyses.

The \dword{ndgar} is composed of 678,136 readout pads in the \dword{tpc}, and approximately 3~million channels in the \dword{ecal}.  Approximately one in five spills will generate an interaction in the \dword{gartpc}, but particles entering the gas from interactions in the \dword{ecal} will provide the bulk of the data volume.  The readout strategy will be similar to the \dword{lartpc}, with only time and integral recorded. A  data volume of 2\,MB of uncompressed data per spill is expected from the \dword{tpc}.  The calorimeter is expected to contribute approximately 1\,MB per spill of uncompressed data.

\subsection{SAND \hideme{Muether - draft}}
\label{sec:intro-sand}

\dword{sand}'s primary function is the primary beam flux monitor.  
The \dword{sand} detector comprises an active LAr target exploiting light read-out (\dword{grain}) placed upstream of a low-density tracker based on straw tube technology (STT)  The layers of straw tubes are sandwiched between radiators of configurable composition.  The vast majority of the radiator material is planned on being $\left({\rm{CH}}_2\right)^n$, though carbon radiators are also included in the design so that hydrogen measurements can be obtained by subtraction~\cite{DUNE:2021tad}. Both are surrounded by $4\pi$-hermeticity \dword{ecal}~\cite{Adinolfi:2002zx} and immersed in a 0.6~T magnetic field provided by a solenoidal magnet. The ECAL is read-out by 4880 photosensors, with an estimated 5500 total hits per spill or 33 kB of packed data per spill. The STT in the current design foresees 218,000 channels and a mean number of hits per spill of about 12,500 corresponding to 75.6 kB of packed data per spill. In the current design, \dword{grain} will be instrumented with up to 76 (32x32) \dword{sipm} matrices. The mean number of hits per spill is about 120,000 corresponding to 1.44 MB of packed data per spill. With these assumptions, the data volume from SAND comes to about 40 TB/year. The amount of data from out-of-spill cosmic rays is estimated to be 20\% of that of the in-spill data, or approximately 8~TB.

\section{Relation of Physics Goals to Offline Computing Challenges \hideme{Schellman-draft}}\label{ch:intro:challenges}

\subsection{Physics and sociological drivers}
The DUNE physics program drives several detector characteristics that pose novel computing challenges.  While the overall data volumes are smaller than those routinely handled by the large \dword{lhc} experiments, the remote detector site and unique physics goals present novel computing challenges. 

\begin{description}
\item{\bf Fine segmentation needed for electron-photon discrimination: \\}   The primary goal of the DUNE long-baseline experiment is measurement of $\nu_\mu\rightarrow\nu_e$ and $\nubar_\mu\rightarrow\nubar_e$
oscillation probabilities for GeV-scale accelerator neutrinos.   These oscillation probabilities are intrinsically low and are sensitive to backgrounds in which the neutral current process  $\nu_\mu+A\rightarrow\nu_\mu+\gamma/\pi^0+X$
produces a photon or $\pi^0$ meson that in turn produces electromagnetic showers that  
fake an oscillation signal.  Fine detector segmentation is necessary to distinguish between these scenarios. Figure~\ref{fig:Argoneut} illustrates this capability. The need for sub-cm-level segmentation drives the technology choice of \dwords{lartpc} and hence   
the number of channels.   
\item{\bf Low-energy thresholds for astrophysical neutrinos: \\}
Other important physics goals are the detection of astrophysical neutrinos from the sun, possible \dword{snb} neutrinos, atmospheric neutrinos and %
\dword{bsm} signatures in the \dword{fd}.  Astrophysical neutrinos produce lower-energy signatures, in the 1-30\,MeV range. Extracting such signals, near the noise threshold of the detector and in the presence of radiological backgrounds, requires careful attention to signal processing and zero-suppression for the \dword{fd} \dword{tpc} and \dword{pd} waveforms.  The need to optimize the low-energy threshold drives our need to carefully record waveforms with minimal processing and thus drastically increases the raw data volume. 

\item{\bf Precise energy calibrations:\\}
An additional challenge in oscillation physics is the need for accurate energy calibration in order to fully %
exploit the energy spectrum of the reconstructed neutrinos to further constrain oscillation parameters. While \dword{lar} detectors have a reputation for stability, the large volumes, complex \efield configurations, liquid motion, and potential variations in electron lifetime and drift velocity make it necessary to have large calibration data samples that span the full \dword{fd} detector volume.  Large cosmic ray and artificial calibration samples will dominate the total data volumes from the \dword{fd}. 

\item{\bf Supernovae:\\}A %
\dfirst{snb} candidate will generate 320\,TB of (uncompressed) data across the first two modules, resulting in %
thousands of data files produced over a 100\,s period. These data must be recorded at a low energy threshold due to the expected interaction energy range, but must also be analyzed quickly and coherently in order to measure the time evolution of neutrino emissions, which carries invaluable information about the supernova process itself. In addition, if \dword{dune} can quickly analyze electron-scattering interactions and separate a fraction of them from \dword{cc} %
interactions, we can provide pointing information to telescopes for optical followup~\cite{tdr-vol-2}.  %
Supernova physics drives the need for fast data transmission from the \dword{fd} to computing facilities and for robust tracking of data movement so that a full picture of the %
\dword{snb} interaction can be reassembled after signal processing.
The drastically different time scale of \dword{snb} physics also places requirements on the software framework. %

\item{\bf \dword{nd} integration: \\}
While the \dword{fd} %
modules produce large data volumes, the detectors themselves are reasonably simple, consisting of a small number of technologies and large numbers of repeating components.  The \dword{nd} is much more complex. The \dword{nd} use case is similar to other fixed target experiments such as \dword{sbnd} at \dword{fnal} and \dword{compass} at \dword{cern}.  The main computing challenge for the \dword{nd} will be integration of a large number of disparate detector technologies into a coherent whole. Here careful attention to simulation, detector geometry and configuration, and code management will be the major challenges. 

\item{\bf Analysis and parameter extraction:\\}
\dword{dune} has over one thousand collaborators spread across five continents. Those collaborators will want to analyze our data over several decades. Fortunately, once reconstruction has been done, neutrino interaction samples are generally simpler than event records at colliders and should %
allow researchers to analyze them at their local institutions. However, final parameter extraction  using large numbers of nuisance parameters remains a computationally intense problem and will require significant resources
and efficient utilization of \dword{hpc} to quickly achieve final results.

\end{description}

\subsection{Resulting offline computing challenges} \hideme{Schellman-draft}\label{intro:challenges}

DUNE offline computing faces four major challenges, some of which are unique to DUNE and others shared widely by \dword{hep} experiments.  

\begin{description}
\item{\bf Large memory footprints -}  DUNE events, with multiple data objects consisting of  thousands of channels with thousands of time samples,   present formidable challenges for reconstruction on typical \dword{hep} processing systems. Efficient processing of DUNE data will require careful attention to data formats and, likely, substantial redesign of the processing framework to allow sequential processing of chunks of data.  Chapters~\ref{ch:use} and~\ref{ch:fworks} describe the status of applications and frameworks. 

\item{\bf Storing and processing data on heterogeneous international  resources -} DUNE depends on the combined resources of the collaboration for large-scale storage and processing of data.   Tools for using shared resources ranging from small-scale clusters to dedicated \dword{hpc} systems need to be developed and maintained.   Fortunately, \dword{hep}, through the \dword{wlcg}, \dword{osg} and \dword{hsf}  has a well developed ecosystem of tools that allow reasonably transparent use of collaboration computing resources.  Chapters~\ref{ch:est}, \ref{ch:cm} and~\ref{ch:datamgmt} describe the data volumes, computing model and data management plans. 

\item{
\bf Machine learning - }  Use of machine learning techniques can greatly improve simulation, reconstruction and analysis of data. However, integration of \dword{ml} techniques into a software ecosystem of the size and complexity of a large \dword{hep} experiment requires substantial effort beyond the original demonstration.  How is the \dword{ml} trained?  What special data format or processing requirements are present? How is the algorithm versioned and preserved to ensure reproducibility?   Chapters~\ref{ch:use} and~\ref{ch:codemgmt} discuss the applications and management.

\item{\bf Efficient and sustainable use of resources -} As discussed in Chapter \ref{ch:est} \dword{dune} will use substantial computing resources.  Historically the main concern has been financial - getting the most computing possible within a given budget. However, our activities also have environmental impact, through energy consumption and the creation, use, and disposal of hardware. To make our efforts more sustainable, a driving consideration in the design of our systems is efficient use of storage and  CPU resources.  This includes not just optimization of our major workflows (Chapters \ref{ch:cm}-\ref{ch:wkflow}) but also documentation and training  (Chapter \ref{ch:train}) of end-users in efficient and error-resistant practices to avoid needless reprocessing.  

\item{\bf Keeping it all going -}  
There is a large suite of activities, that,   
while not necessarily novel, 
still needs to be done over the full lifetime of the experiment.  These activities include database design and operations, security updates, code management, documentation, training, and user support.   For example, the \dword{nd} presents few novel computing challenges in memory or CPU use but is highly complex in terms of the number of detector systems that must be integrated. Another example is the continuing evolution of operating systems and security requirements.  These require constant modifications to working systems to maintain operations.  A third activity is database design and maintenance. Here the problem is largely sociological, getting the attention of busy people to do database design and then %
populate and use the official collaboration databases. This requires continual engagement with reluctant stakeholders. These issues are discussed throughout this document with special reference to databases, Chapter~\ref{ch:db}, authentication Chapter~\ref{ch:auth}, code management Chapter~\ref{ch:codemgmt} and training and documentation Chapter~\ref{ch:train}
\end{description}
A broad suite of use cases is discussed in Chapter~\ref{ch:use}.

\cleardoublepage

\chapter{Computing Organization\hideme{Ready 3/13}}\label{ch:org}

This document chiefly describes the activities of the formal Computing Consortium that concentrates on development and operations for  the \dword{dune}. The consortium provides the hardware and software computing infrastructure that allows development and implementation of algorithms and data analysis techniques. The algorithms and techniques themselves are within the purview of the collaboration physics groups. 
The whole of \dword{dune} offline computing consists of three major focus areas, illustrated and described in Figure~\ref{fig:roles}.  The Consortium itself provides the underlying infrastructure and collaborates with the sites in delivering computing  resources. %
Chapter~\ref{ch:cm} includes a description of the global sites and their hardware contributions. The Computing Consortium and the global sites meet weekly via teleconference to coordinate operations.

\begin{dunefigure}
[DUNE Computing Subgroups]
{fig:roles}
{Offline computing roles.  The first column shows the formal DUNE Computing Consortium, mainly comprised of experts funded to perform offline computing for \dword{dune}.  The second column represents the infrastructure contributed to computing by collaborating nations and institutions and formalized through the \dword{ccb}.  The third represents the broader contributions of the collaboration that are vital to the overall effort but not managed by the Computing Consortium.}
{\includegraphics[width=0.9\textwidth]{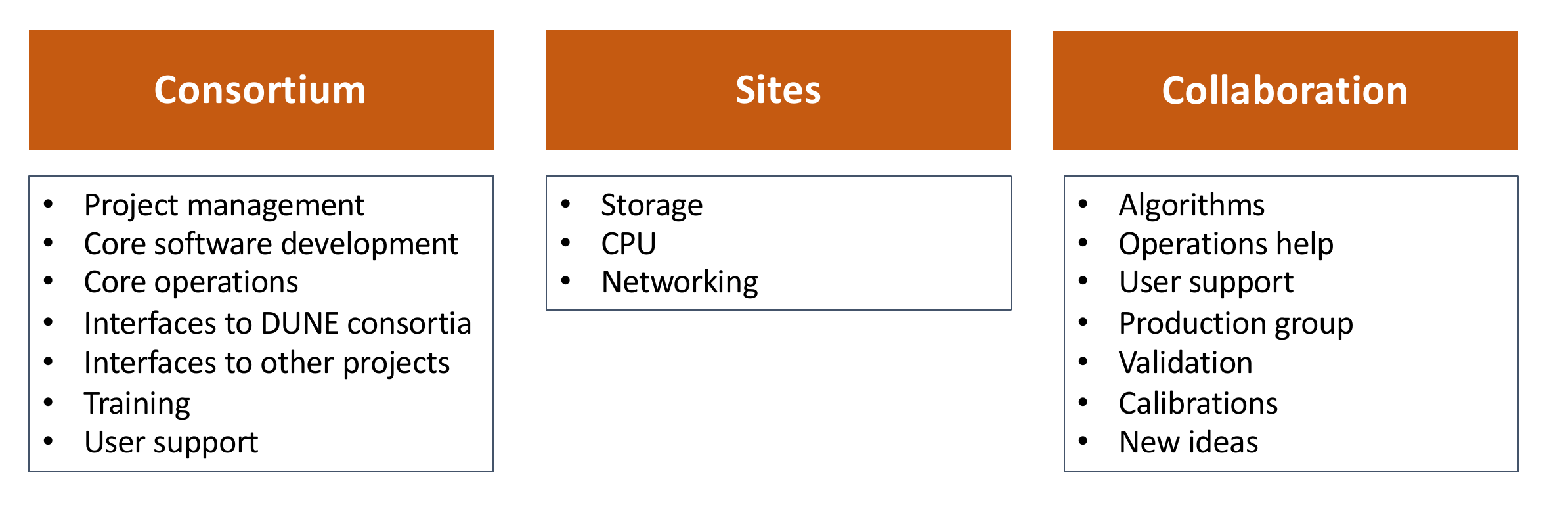}}
\end{dunefigure}

\begin{dunefigure}
[DUNE Organization Chart]
{fig:DUNEorgchart}
{DUNE Collaboration organization chart, July 2022.}
{\includegraphics[width=0.9\textwidth]{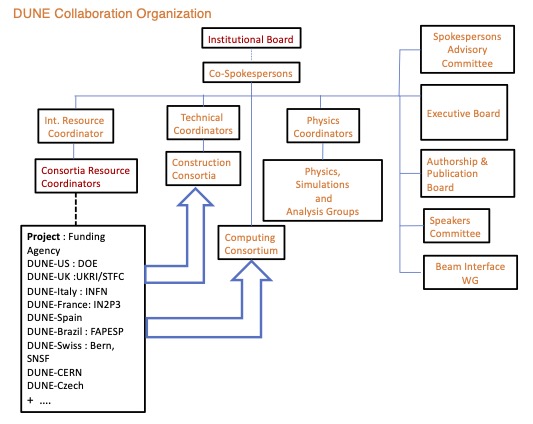}}
\end{dunefigure}

\begin{dunefigure}
[Computing Consortium Organization Chart]
{fig:orgchart}
{DUNE Computing Consortium organization chart, September 2022.  Flags for the group leaders are shown.}
{\includegraphics[width=0.9\textwidth]{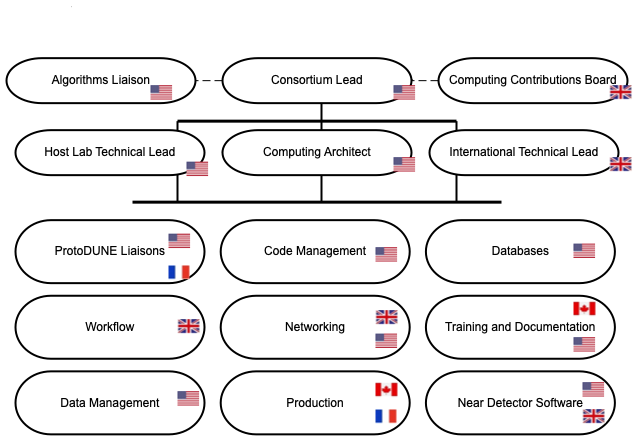}}
\end{dunefigure}

Figure~\ref{fig:orgchart} shows the Computing Consortium organization as of September 2022.  The Consortium leadership consists of a Consortium Lead, responsible for overall coordination, two technical leads, one US-based and the other from the international collaboration, and a Computing Architect.  There is an Algorithms Liaison responsible for coordinating with the Physics groups, liaisons to the \dword{protodune} experiments and to the calibration consortium. There is also an independent \dword{ccb}, described in Section~\ref{sec:ccb}, that negotiates resource contributions from international %
collaborators. 

The Computing Consortium is not formally part of the \dword{dune} construction project but has representation on the Executive Board.  The consortium is part of the interface document matrix with the construction consortia and the interface agreements %
are listed in Table \ref{tab:intro:interfacedocs}. The most important formal interfaces are (1) data-management and networking with the \dword{daq} and calibration groups and (2) databases with the \dword{daq} and hardware construction groups. The Computing Consortium also has liaisons to the physics groups and the \dword{protodune} experiments.

\begin{dunetable}
[Interface documents with hardware consortia]
{l l}
{tab:intro:interfacedocs}
{ \dword{edms} interface documents between the Joint Offline Computing (JT COM) with various DUNE hardware and physics consortia. }
Document title&EDMS ID\\
JT COM and SP APA Consortium Interface Document	&\href{https://edms.cern.ch/document/2145145}
{2145145 v.4}\\
JT COM and SP PD Consortium Interface Document	&\href{https://edms.cern.ch/document/2145146}{2145146	v.2}\\
JT COM and SP TPC Consortium Interface Document	&\href{https://edms.cern.ch/document/2145147}{2145147	v.2}\\
JT COM and DP CRP Consortium Interface Document	&\href{https://edms.cern.ch/document/2145148}{2145148	v.1}\\
JT COM and DP PDS Consortium Interface Document	&\href{https://edms.cern.ch/document/2145149}{2145149	v.1}\\
JT COM and JT HV Consortium Interface Document	&\href{https://edms.cern.ch/document/2145150}{2145150	v.2}\\
JT COM and JT DAQ Consortium Interface Document	&\href{https://edms.cern.ch/document/2145151}{2145151	v.2}\\
JT CAL/CI and JT COM Consortium Interface Document	&\href{https://edms.cern.ch/document/2145159}{2145159	v.2}\\ 
JT COM to Facility Interface	&\href{https://edms.cern.ch/document/2145167}{2145167	v.1} \\
\end{dunetable}

\section{Internal Organization}\label{org:internal}

Within the Consortium there are specific development groups responsible for particular components of the computing infrastructure.  Many of these activities are described in greater detail in separate chapters. 

\begin{itemize} 
\item ProtoDUNE Liaisons - make certain that offline computing is coordinated with ProtoDUNE data taking and analysis.

\item Data Management - responsible for storage, cataloging and delivery of data. See Chapters \ref{ch:cm}, \ref{ch:datamgmt}

\item Workflow - responsible for global coordination of computing resources. See Chapter \ref{ch:cm}, \ref{ch:wkflow}.

\item Code Management - responsible for maintaining the code infrastructure and  repositories and building common executables.  See Chapter \ref{ch:codemgmt}.

\item{Networking} -  Responsible for negotiating suitable networking capabilities from \dword{surf} to the host laboratories and from the host laboratories to the global compute resources.  See Chapter \ref{ch:netw}. 

\item{Production} - Responsible for setting up and running common  reconstruction and simulations jobs. This group includes both experts and collaboration volunteers who run the jobs. See Chapter \ref{sec:current} for a description of the current status.

\item{Databases} - Responsible for the design and implementation of databases to support offline data processing. See Chapter \ref{ch:db}.

\item{Training, Document and User support} - Responsible for development of training materials, infrastructure for and review of documentation and for directing users to experts and documentation where needed. See Chapter \ref{ch:train}.

\item Near Detector Software - responsible for coordination of the unique \dword{nd} software and integration with the main \dword{dune} infrastructure. See Section \ref{ch:use:nd}. 

\end{itemize}

\section{Funding Sources for Computing Development}
Most personnel working on \dword{dune} computing are supported by their institutions or national funding agencies as part of their base or \dword{dune} project support. 
This support is generally not  long-term %
and is subject to periodic requests %
to the relevant funding agencies.  For example, a US \dword{doe}-funded consortium of four universities and three national labs is funded specifically for \dword{dune} computing infrastructure at a level of \$1M/year through 2024, directly supporting postdocs and lab physicists, summing to $\sim 6$ FTE.  The UK,  France and \dword{cern} also make significant contributions to software development through the DUNE project while many nations contribute substantial hardware capabilities. Membership in the Computing Consortium does not have any specific implication for provision of CPU and storage resources; members may contribute personnel or hardware, or both; these contributions are handled separately. Hardware commitments are made through the \dword{ccb} %
described in the next section.

\section{Computing Contributions Board }\label{sec:ccb}

The \dword{ccb} has been set up to formalize and recognize the contribution of DUNE partners to the computing and storage capacity required for large-scale DUNE computing. 
For the present, the \dword{ccb} takes a ``nation'' as the natural unit of aggregation to set a ``size metric'' upon which to base requests for resources. The \dword{ccb} does, however, recognize that whilst within some countries centralized coordination is natural, this is not true for others. A more natural, fair-share unit of aggregation would be based on each funding agency (as used by the \dword{wlcg}), and \dword{dune} recognizes that an evolution to this model will be required in the future. Nevertheless, at present, during the integration and construction phase of \dword{dune}, the current model is providing sufficient resources and informal pledges with the assumption that institutions within each nation can (loosely) coordinate where needed.

The fair-share metric to which the \dword{ccb} will move, once \dword{dune} is taking data, %
is likely to be senior authors. %
During construction, %
the current listing of DUNE personnel in the collaboration database is a very poor proxy for the number of active contributors. Therefore, during construction all nations with %
a threshold number of DUNE participants, and that are capable of providing Tier-1 or large Tier-2 capacity to \dword{lhc} experiments, are asked to provide a ``reasonable'' share (see below). We refer to these as compute-active-nations.
This is very flexible and is up to each nation to decide if it wishes to be classified as a compute-active-nation.

At present the \dword{ccb} is composed of a Chair, one member per compute-active-nation, one member for each of \dword{fnal}, \dword{cern} and \dword{bnl}, and the Computing and Software Consortium  Management ex officio.

The Computing Consortium management produces an overall requirements document each year\cite{ccb2022} between November and January. The \dword{ccb} receives this document, and then seeks pledges to meet these requirements. As host lab, \dword{fnal} plans to provide $\sim$\,25\% of the disk and CPU capacity, as well as the primary tape service.
\dword{cern} also currently provides substantial capacity, including a second tape copy of raw data for \dword{protodune}.
The aim is for the remaining capacity to come from other contributors according to the prevailing computing model. A substantial proportion of capacity is expected from outside of the US  ($\sim$\,50\%). National contributions of at least 5-20\% are requested, depending upon the circumstance and capability of each compute-active-nation.

It is intended to use the \dword{cric} (http://wlcg-cric.cern.ch/) information system to record pledges, this work is currently in progress. In due course this process could be formalized into a (non-binding) \dword{mou}.

The \dword{ccb} may also receive requests and information from the Computing Consortium Management with respect to other non-capacity matters. The \dword{ccb} may seek to help with such requests where they pertain to national contributions. This may include promotion of requests for software engineering support to be propagated within each nation.  

Chapter~\ref{ch:cm:intro} describes the impact of  collaboration contributions in more detail. 

\section{Collaborations with other Organizations}

DUNE Computing is actively engaged with multiple other computing organizations around the globe with the intent of both drawing from and contributing to the community knowledge of computing solutions. As a member of the \dword{osg} Council, \dword{dune} is helping define the direction of High Throughput Computing within North America. DUNE Computing is also an observing member of the \dword{wlcg} and is using multiple services and solutions developed within the scope of the \dword{wlcg}. We also work closely with the \dword{hsf} and \dword{rucio} collaborations in developing modern software solutions.  DUNE intends to continue to maintain these partnerships as part of both sharing new computing developments and efficiently integrating new services into our computing model.

\cleardoublepage

\part{Single Interaction  Scale}\label{part:event} %

\chapter{Data Processing Considerations and Challenges\hideme{Anne's comments addressed 4/28}}%
\label{ch:use}
\newcommand{\ignore}[1]{{}}
Before describing the large scale data model for the \dword{dune} we begin by describing the activities and algorithms that drive that model at the \dword{tr} scale. 

\section{Introduction \hideme{Schellman - draft}} \label{ch:use:intro}

In this chapter  we describe some of the use cases for \dword{tr} level processing,   %
building on the experience gained from  \dword{protodune} and   using lessons learned from the 2018 run to understand the performance of future near and far detector modules. 

In addition to the challenges faced by \dword{hllhc} experiments, as described in the HSF White Paper\cite{HEPSoftwareFoundation:2017ggl}, DUNE (along with some dark matter and astrophysics experiments) faces considerable challenges in extracting very weak signals from large, noise-dominated data volumes.  This additional challenge places strains on memory management beyond those anticipated at collider experiments.

\begin{dunefigure}
[Processing diagram for standard ProtoDUNE (and FD) reconstruction]
{fig:ch:use:pdii}
{Processing diagram for standard \dword{protodune} and \dword{fd} data reconstruction. All processing and data transfers shown are part of offline processing. The central boxes show the processing steps, the right side shows the large-scale data flow,  and the left side shows the auxiliary information needed for processing.}
\includegraphics[width=0.9\textwidth]{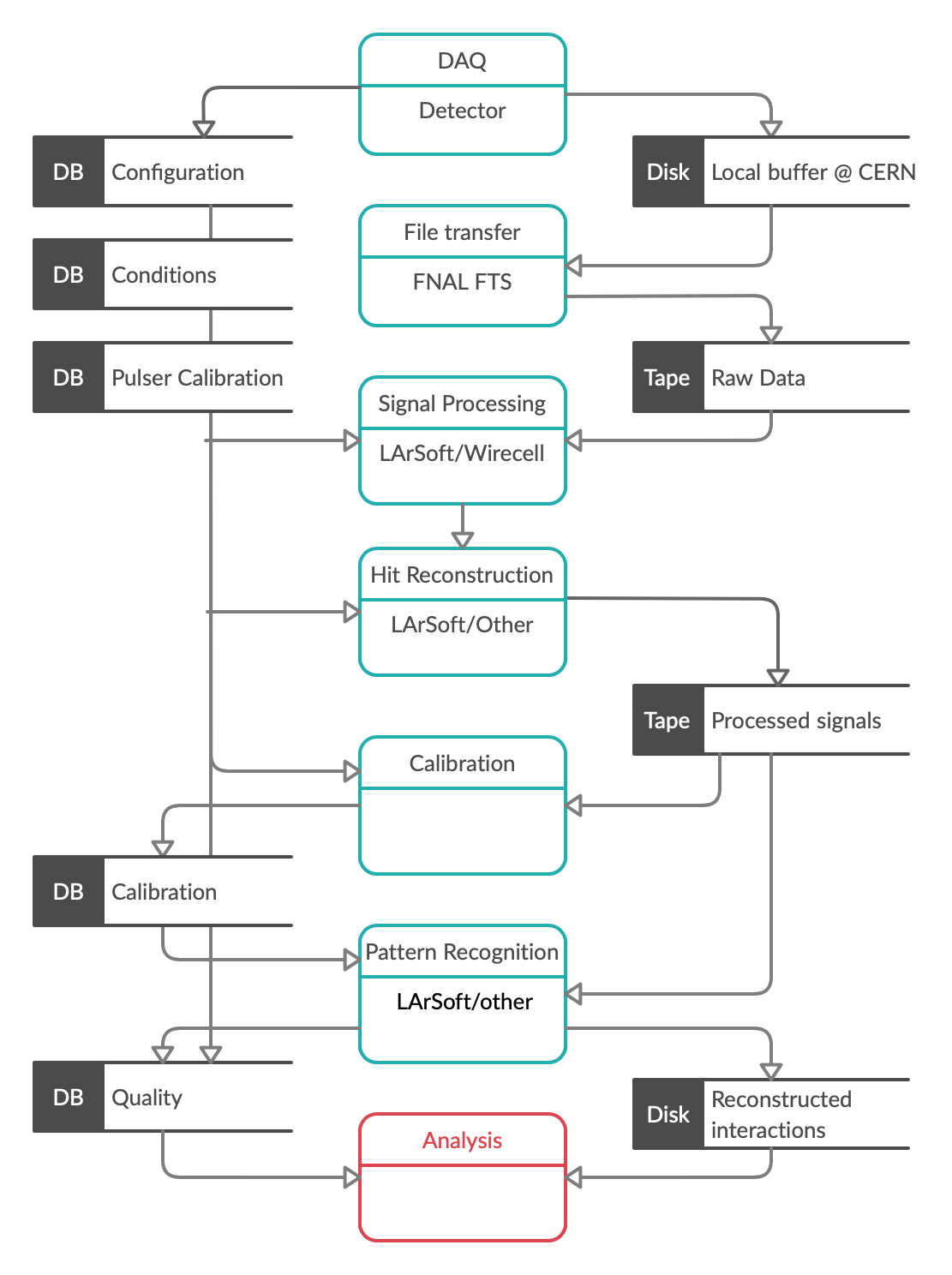}
\end{dunefigure}

\section{Data Acquisition and Storage\hideme{Schellman - draft}}

Figures \ref{fig:ch:use:pdii}, \ref{fig:ch:use:fdsimrecoflowdiagram} and \ref{fig:ch:use:ndsimrecoflowdiagram} outline the offline processing flow for raw data, protodune and \dword{fd} simulation, and \dword{nd} simulation.
Offline processing starts with the transfer of data from a disk buffer located at the experiment.

\Dword{daq} software and hardware are the responsibility of the \dword{daq} consortium, with significant interfaces to offline computing. Important overlapping factors include the management and versioning of data format specifications, communication of calibration and configuration information between online and offline computing, high-speed networking for data, high-reliability networking for configuration and monitoring, and adequate disk and CPU resources on both ends of the connection to allow both efficient data transfer and continued data taking in the event of an extended network interruption. These efforts are coordinated by dedicated liaisons, for example between the offline databases and online slow controls and formal liaisons to the \dword{pdhd} and \dword{pdvd} experimental operations groups.  Informal coordination also occurs through shared Slack channels and joint sessions at collaboration meetings. 

These overlaps already apply to \dword{protodune}, where data taking occurs at the \dword{cern} with offline databases and archival storage in the US,  and, in the future, will apply to the far and near detectors in South Dakota and Illinois, respectively. 
Current status and proposed solutions are discussed in Chapters~\ref{ch:db} (Databases) and~\ref{ch:netw} (Networking). 

\subsection{Data Transfer from the Experiment}
Data needs to be buffered locally and then transferred to permanent storage at the host lab(s).  Data transfer activities include the generation of descriptive metadata, file transfer, writing the files to permanent storage, and integrity checks.  Once data are confirmed to have successfully reached permanent storage, the local buffer may be cleared.  Current specifications call for the local DAQ buffers at the FD remote site to be able to store 3-7 days worth of all triggered DAQ data or several \dword{snb}'s worth ($\sim 640$ TB) at a minimum. 

{\it %
The remote location of the far detector, and rack-space and power limitations at both the \dword{4850l} detector level and the surface mean that many computing tasks (for example data reformatting) may need to be performed at the host labs rather than at the experimental site. The main requirement is that there be sufficient local data buffer space (of order 1 PB, or greater than one week at normal data rates) to withstand a significant network outage or staging of supernova data. 

Other challenges include 
the need for adequate lossless data compression to prevent data volume explosion. Overall our strategy is to move operations which can be done offsite, offsite, reserving infrastructure at the \dword{surf} site for activities (such as DAQ and buffering) that cannot be done elsewhere without the risk loss of data. Chapter \ref{ch:netw} describes the status and plans for networking.} 

\subsection{Control, Configuration, Conditions and Calibrations}
In addition to large-scale data transfers, online data-taking configurations and conditions need to be stored and communicated to the offline systems. High-reliability network connections are needed to ensure that control signals, and configuration and monitoring information are exchanged between the remote sites, local control facilities and the host laboratory.
This is the responsibility of the \dword{fnal} networking groups and the \dword{daq} consortium with input and assistance from offline computing.

\subsubsection{Slow Controls}
Data from slow control and monitoring systems need to be recorded for offline use.  Examples include set points and readback for \dword{hv} systems, %
pressure, temperature and the outputs of purity monitors. 
{A potential challenge:  commercial \dword{scada} systems tie DUNE to proprietary software that may have limited interfaces, require expensive licenses and need migration to new systems if the vendor stops support.  Another issue of \dword{scada} systems is the fact that they need to last for many years and are generally supported only for the version of the operating system for which they were built. It is a challenge therefore to keep those systems running for extended periods of time, despite the progress in virtualization and containers.}

\subsubsection{Online Calibrations}
Online calibration information (e.g., pedestal information for some sub-detectors and trigger time offsets relative to an absolute time standard) needs to be passed to the offline systems.

\subsubsection{Beam Monitors}
Beam monitoring information, either per spill for neutrino interactions or per beam trigger in the \dword{cern} test beams, needs to be recorded and made available for offline processing. 
The \dword{ifbeam} system used for \dword{numi} is already in use for \dword{protodune}.

\subsection{Monitoring Information} 
Local online monitoring of data and experimental conditions will be done by the \dword{daq}. %
The offline computing consortium will provide fast feedback on data movement and data quality based on the fast processing of a subset of the data transferred offsite. %

\subsection{Offsite High-level Data Filtering}

This is not yet part of the data plan but there may be a need for offsite filtering of data before %
it is written to permanent storage. 
 Initiation of a high-level trigger offsite remains a possibility if the need arises. Local space, power and cooling issues make this difficult to do at the \dword{surf}.

\subsection{Data Compression}
The data volumes and rates anticipated are significant. Storage and network resources can be optimized with zero-suppression (lossless or lossy) and data compression. Multiple avenues for data reduction are being explored, including on-board algorithms in the readout boards or downstream \dword{daq} systems, or before data are written to archival storage.  Considerations include the computational load of compression/decompression, especially of the very large data objects the far detector can produce, and irrevocable data loss in the cases of lossy methods.

In \dword{pdsp} lossless compression was performed as part of the \dword{daq}. For \dword{pdhd} this is not planned due to the need to keep processing overheads low in the \dword{daq}. Lossless compression of \dword{fd} \dword{tpc} data will need to be applied later in the processing chain, most likely using the intrinsic compression methods of the chosen data format(s).  \dword{root} allows implementation of lossless compression, as does \dword{hdf5}. The data model assumes that for \dword{pdhd} and \dword{pdvd} the same level of compression will be achievable as in \dword{pdsp}.  The \dword{nd}s currently plan to take advantage of their higher signal/noise ratio and do lossy zero-suppression.

\subsection{Outputs from DAQ and Monitoring}

The end result of data acquisition includes the transfer of the data themselves and of the %
metadata and configuration/calibration/beam information needed for offline processing. Only when the data and metadata have been successfully transferred and stored can they be deleted from the local DAQ storage.  The computing consortium is responsible for developing the software for transferring data from the DAQ disk buffer and deleting stale data, while the DAQ consortium is responsible for providing the hardware resources (CPU, memory and disk) to run the data transfer software, as well as provide alerts to the running data transfer program when new data are available~\cite{bib:docdb7123}.

\section{Simulation Chain \hideme{Junk and Schellman - draft}}

The \dword{dune} simulation chain involves a large number of steps including beam simulation, particle interaction or decay simulation, simulation of energy deposition and simulation of detector and electronics response and noise. \dword{fd} and \dword{protodune} share a common simulation framework, based on \dword{art} and \dword{larsoft}, with other Fermilab \dword{lartpc} experiments while the \dword{nd} simulation framework is still under development.
Figures \ref{fig:ch:use:fdsimrecoflowdiagram} and \ref{fig:ch:use:ndsimrecoflowdiagram} illustrate the main processing flow for the \dword{pd}\dword{pd} and \dword{nd}. 

\begin{dunefigure}
[Diagram showing simulation flow for ProtoDUNE and the Far Detector]
{fig:ch:use:fdsimrecoflowdiagram}
{This diagram shows the components of the LarSoft based Far Detector and ProtoDUNE simulation  workflow. Major external inputs are shown on the left while the main data movement is shown on the right. The reconstruction chain is illustrated in \ref{fig:ch:use:pdii}.}
\includegraphics[width=0.8\textwidth]{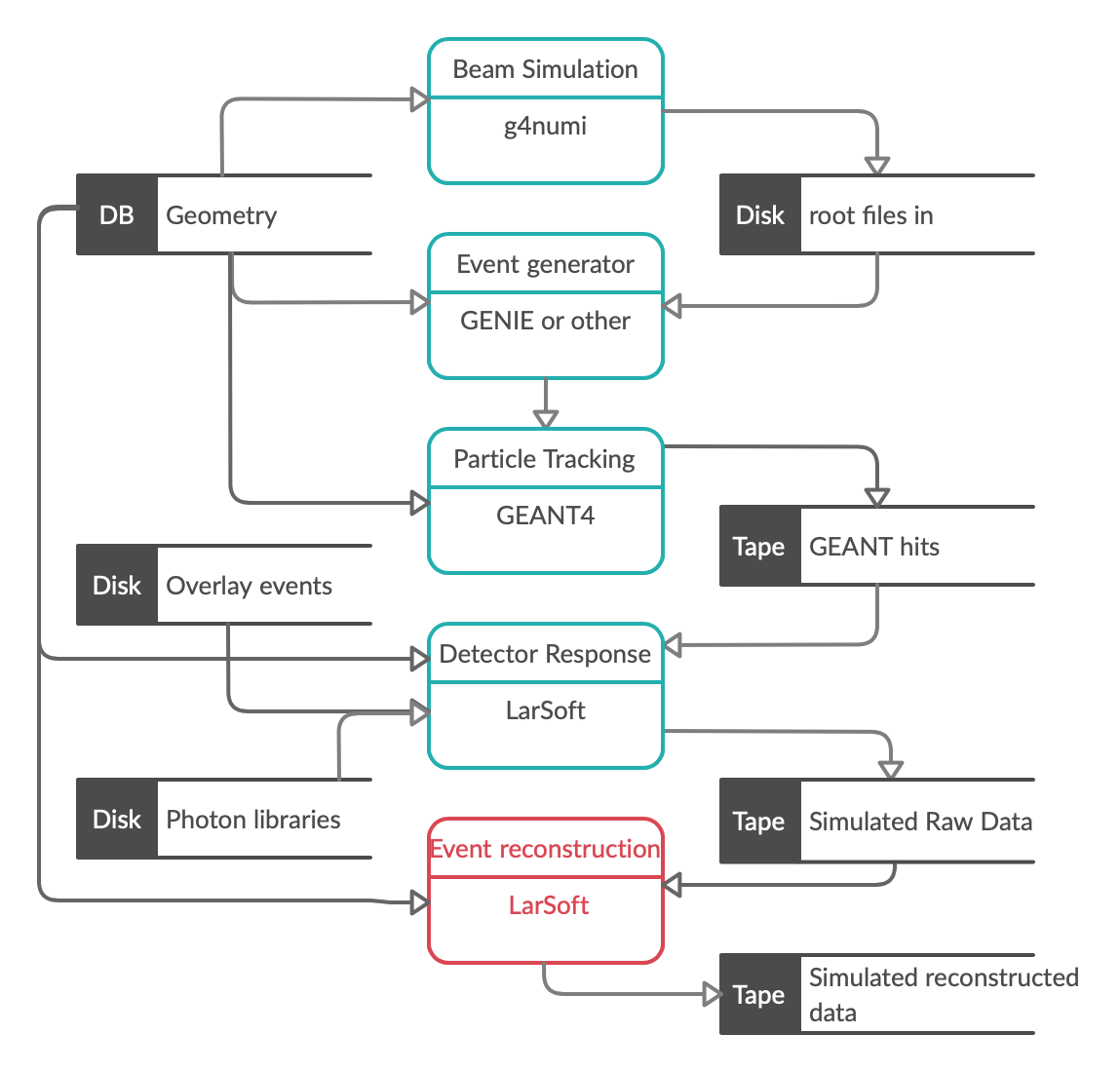}
\end{dunefigure}

In the following sections we break out in more detail the portions of this simulation chain, and enumerate the computational challenges and technical approaches that are associated with them.

\subsection{Supporting Simulation Volumes}
\label{sec:algo-use-fdsimvol}

The simulation needs and scale of Monte Carlo generation that are required for \dword{dune} are driven by uncertainty budgets of different analysis results that we are pursuing.  In the case of the long baseline oscillation analyses, the small signal rates and sources of potential backgrounds from cosmic ray induced events and from radiogenic backgrounds set the scale of the simulation that is needed to cover different systematic uncertainties.  In this respect, the amount of supporting Monte Carlo that is required for far detector portion of the long baseline oscillation measurements is estimated at approximately a factor of 10x of the experimental exposure, as measured in kT*kW*s, for the signal channel [beam] samples, and a similar 10x or greater simulation of the background channels (cosmic and radiogenic) as determined by the beam live time of detectors measured in seconds.  These estimates are based upon similar approaches taken by earlier neutrino experiments (NOvA, MINOS) which have been statistically signal limited by their far detectors and for which cosmic ray backgrounds can contribute significantly to the backgrounds.  However, it should be noted that unlike NOvA which was surface detector with only 3~m earth equivalent overburden, \dword{dune} has the advantage of a highly suppressed cosmic ray flux owing to its deep underground siting.  This may allow for background Monte Carlo samples to be smaller than our initial estimates but also may require more advanced computational techniques to generate the attenuated spectra over the large flux surfaces of the detector modules.

In contrast to the far detector simulations, the near detector complex also will require significant supporting simulation to drive the beam flux predictions for the combined near to far oscillation calculations and for the large suite of near detector based cross section physics measurements.  Similar to the far detector simulation volumes we estimate that, based on similar oscillation and cross section programs from past neutrino experiments, we will require supporting simulation at a level of 10-20x the integrated beam flux, as measured in kW*seconds for the near detector cavern complex.  In particular this supporting simulation need is driven by the beam flux uncertainties in targets and particle production models, and by neutrino induced interactions in the rock surrounding the detector cavern that result in muon fluxes through the detector volumes which overlap with signal interactions in the main detector volumes.

We also recognize that these estimates are essential minimal estimates, and that in almost all cases the \dword{dune} program benefits from additional Monte Carlo in terms of providing lower statistical uncertainties on the Monte Carlo and from the single event sensitivities that can be reached for rare event interactions and interaction topologies.  In practice, the supporting Monte Carlo sample sizes will be driven by the \dword{dune} physics groups and their specific analyses.  This is especially true if the data are categorized into many sparsely-populated categories, or if they are binned in multi-dimensional histograms as is common in differential (and doubly-differential) cross section measurements.  In these cases, even if the number of data events is very small (or even zero) in portions of the spectra, the predictions of rates in each category or bin must still be reliable in order for parameters to be extracted properly and sensitivities calculated.  In these cases, we expect the supporting Monte Carlo factor may need to be higher, in the range of  10x to 100x of that of the beam interaction sample in order to cover the low rate portions of the parameter spaces.

Similarly nucleon-decay searches and other analyses seeking rare events or limits on rare processes will require larger samples of simulated cosmic-ray events or radiogenic backgrounds than are required by the core oscillation measurements.  These samples may also differ from the core oscillation samples in the method and regions that they are generated over, both inside and near the \dword{fd}, in order to simulate the very small fractions of events that may leak into the fiducial detector volumes and be potentially selected.  Strategies to reduce the CPU and storage requirements for generating these rare search samples, such as rejecting events at the generator level if they are known not to pass exotic-search requirements no matter what happens in the detector, may be useful in optimizing the production of these samples and reducing the computing requirements and budget they represent.

As mentioned in Section~\ref{sec:algo_use_generator}, some, but not all, generator systematic uncertainties can be evaluated using reweighting techniques without having to generate, simulate, and reconstruct additional samples.  Between five and ten versions of the \dword{fd} simulation samples however will be needed to address generator systematics.  As these uncertainties are expected to factorize from detector systematics, shortcuts, such as zero-suppression and simulation in smaller geometries can be taken in order to reduce the CPU, memory and storage impact of the need to study generator systematics.

The simulated data size for the \dword{fd} is expected to consist of a different mixture of physics data types than the raw data.  The raw data volume is expected to be dominated by cosmic rays, supernova triggers, and calibrations, with a much smaller amount of data from atmospheric neutrinos and beam neutrino interactions.  These latter categories will require large multiples of the data in simulations, while the cosmic-ray sample and the supernova triggers will require a smaller simulation to raw data volume.  Various strategies can be taken especially in the supernova simulations, such as placing the stubs much more densely in the detector than in the expected raw data, to reduce the number of empty waveforms simulated.

\subsection{Beam Simulations}
The \dword{mars}\cite{abs1} simulation is used in the design of the beamline and near detector systems and  for safety and environmental calculations.  It is a significant user of CPU time. 
{The \dword{mars}\ codes are not available outside of the US due to export controls which poses some challenges in balancing compute tasks.}

Neutrino production in the \dword{lbnf} 
beamline  is simulated using the \dword{g4lbnf}\cite{g4lbnf} framework.  The ProtoDUNE beamlines were designed and simulated using the \dword{madx} framework used at \dword{cern}\cite{PhysRevAccelBeams.20.111001}.
Inputs include random number seeds and the beamline and target region geometry and materials.  Outputs are flux files containing simulated particle  trajectories.  The flux files contain sufficient information to allow reweighting based on interaction cross sections.  These flux files must be cataloged and made available to offline simulation jobs.
  The size and versioning of these files lends the distribution of flux files to remote sites to be performed using \dword{stashcache}~\cite{bib:stashcache}.  See Section~\ref{subsec:io} for more details on I/O handling for offline jobs.

\subsection{Detector Geometry Description \hideme{- NEW|| 2/19}}
\label{sec:usecases:detectorgeometry}

Offline event generation, simulation, reconstruction and visualization tasks require descriptions of the geometries of the relevant detectors.  Similar to some dark-matter and neutrino-less double-beta decay experiments, a neutrino experiment's detector is the interaction target.  Therefore, the event generator must be aware of the spatial distribution of significant amounts of material, listed separately for each element's contribution, and the local densities. Neutrino detectors contain some elements for which the neutrino-scattering cross sections are poorly known; when these do not contribute significantly to the detector volume, a simplified geometry containing only the key components is often useful.

Once the primary interaction has been generated, \dword{geant4}~\cite{geant4} requires a representation of the spatial locations of materials in order to simulate the subsequent propagation and energy depositions of all particles except neutrinos.  As was the case for generating primary interactions, a simplified geometry description is needed since a detailed geometry with a cylinder of CuBe for each wire in a horizontal-drift \dword{apa} would use a large amount of memory and CPU for a very small amount of non-fiducial interaction modeling.

The common language used in DUNE for geometry description is \dword{gdml}~\cite{gdml}.  It is stored in human-readable text files.  Versions of each detector's geometry are saved alongside software source code in the code repository.  When a release is built, these \dword{gdml} files are copied into directories that are visible to running jobs.  At the time of writing, \dword{cvmfs} is the chosen distribution method for pre-compiled code and small auxiliary files like \dword{gdml} files.   Both \dword{geant4} and \dword{root} have functionality for reading \dword{gdml} files.

One notable 
requirement %
for code versioning is that all supported versions of the \dword{gdml} files for a detector must be available in each release.  If a geometry description is updated for a release of the software, the older versions must be kept, with the old \dword{gdml} filenames, in order to ensure backwards compatibility, as data files are stored on disk and tape with the presumption of the old geometry.  DUNE thus uses a simple additional versioning system by adding a version number to each detector's \dword{gdml} filename.  \dword{larsoft} and  \dword{garsoft} jobs allow for \dword{fhicl} configuration of the detector geometry \dword{gdml} file.  A \dword{larsoft}  job, if run as input with data from an earlier LArSoft job, will check the compatibility of the requested \dword{gdml} file and that used for the input.  By default, if there is a mismatch, an exception is thrown and the job stops.  Sometimes, however, it is desired to simulate and reconstruct \dword{mc} samples with different geometries, in order to simulate misalignments, for example.  In order to accommodate these cases, the geometry consistency check can be disabled by setting an appropriate \dword{fhicl} parameter.

Some software components, such as the DUNE plug-in components to the LArSoft geometry service, have different behaviors depending on which detector is being simulated or reconstructed.  If an update to a geometry description for a detector requires a corresponding update to the software components, then a new name must be assigned to that version of the detector geometry to distinguish it at run time so the appropriate routines can be called.  Software routines supporting older geometry versions will be maintained as long as the older geometry is supported.

\dword{gdml}  files include many similar code blocks to allow simulation of detectors with repeated, similar structures.  The \dword{gdml}  files for the %
\dword{sphd} and \dword{spvd} far detector modules and the \dwords{protodune} are made with Perl scripts and a number of small shell scripts to make final adjustments.  Some of the inputs are not repetitive, such as material mixture definitions, and these fragments are hand-edited.  \dword{gdml}  files for the near detectors are made with DUNENDGGD~\cite{ref:ggd}\footnote{https://github.com/dune/duneggd}.  DUNENDGGD takes as input Python classes called ``builders'' 
and produces \dword{gdml}  files.  The near detector \dword{gdml}  file creation requires additional steps, as two of the detectors move off-axis while a third remains stationary.  Particles travel from \dword{ndlar} to  \dword{tms} and thus they must be simulated together, with a single \dword{gdml}  file.  In addition to a full hall description, detector working groups may prefer to work with a geometry that only has one detector element in it for computational convenience.

Computational convenience also drives the need for workspace geometries for the far detector modules.  A subset with only 12 \dwords{apa} instead of the full complement of 150 is a useful geometry to use for physics simulations, though care must be exercised regarding differences in event containment between the workspace geometry and the full geometry.

\subsection{Neutrino Event Generation}
\label{sec:algo_use_generator}

Extracting physics results from the DUNE experiment requires comparing the observed data with simulation, 
which includes detailed simulations 
of the physics processes under study as well as of the response of the detectors.  These physics processes can be simulated by any of several 'neutrino event generators', including \glsunset{genie}\dword{genie}~\cite{Andreopoulos:2009rq}, \glsunset{nuwro}\dword{nuwro}~\cite{NuWro2012}, \glsunset{gibuu}\dword{gibuu}~\cite{Gallmeister:2016dnq}, \glsunset{neut}\dword{neut}~\cite{Hayato:2009zz}, and others. 

Neutrino event generators are used to simulate neutrino interactions within a detector volume.  Input requirements are the neutrino beam flux, random number seeds and the detailed detector geometry. The output consists of four-vectors for final-state particles, where the final state of the interaction includes the decay of any short-lived particles and the subsequent particles produced. DUNE currently uses \dword{genie}~\cite{GENIE} v2.12.10 as the default, with the \texttt{DefaultPlusValenciaMEC} physics list.   An upgrade to GENIE~v3 is in progress.  Other event generators may also be used.

A large fraction of DUNE's analysis effort will go into studying the effects of the neutrino-nucleus interaction models encoded in the generators~\cite{tdr-vol-2}.  Central values and systematic uncertainties for predictions of cross sections as functions of neutrino flavor and energy for each analyzed final state are needed for physics analyses.  Some systematic uncertainties can be approximately evaluated by reweighting a central default sample generated with one generator with a specific choice of steering parameter values to mimic the output of the generator with a different choice of steering parameter values, or to that of a different generator entirely.  This is not possible in all cases, due to underpopulation of some regions of phase space in the default generator.  As has been the case with other high-statistics neutrino experiments, DUNE's physics working groups will participate actively in generator validation and tuning.  It is not yet known how many full-statistics sets of simulated data will be needed in order to cover the non-reweightable generator systematic uncertainties, but an estimate is that between five and ten such sets will be needed.

\subsection{Non-Beam Interaction Simulations}
Simulations of non-standard model physics (e.g., neutron decay) and of cosmic ray and rock muon backgrounds are also used.  Cosmic-ray simulations are performed with \glsunset{corsika}\dword{corsika}~\cite{Wentz:2003bp,Dembinski:2020wrp} for detectors on the surface, and MUSUN/MUSIC for detectors deep underground~\cite{Kudryavtsev:2008qh,LBNEDOCDB9673}.  Radiological decays are modeled with BXDECAY0~\cite{Ponkratenko:2000um} and \dword{larsoft}'s  radiological generator.

Factorization of the simulation into a generation stage and a detector simulation stage is a common practice 
in collider experiments, such as ATLAS and CMS.  The fact that the interactions simulated by generators for collider physics happen inside an evacuated beampipe means that the details of the detector geometry and materials are not relevant for most event generation, and lists of four-vectors of particles emerging from a primary vertex will suffice.  In a neutrino experiment, however, the detector material is the target material, and hence the generators must be aware of the detector geometry and materials, which affects the structure and performance of the generator code.  Currently \dword{genie}, \dword{corsika}, MUSUN/MUSIC, BXDECAY0 are integrated with \dword{larsoft}.  \dword{larsoft} also has custom generators for radiological decays and a text-file interface providing arbitrary four-vectors as inputs to the simulation. \dword{genie} is also integrated with \dword{garsoft}.

{ DUNE can rely on strong international efforts to produce standard generators, but it should be noted that these efforts are in general external to the \dword{dune} Collaboration. One challenge is the complexity of integrating multiple simulation code versions into our framework to ensure that there is timely evolution of the generators used as the physics understanding evolves.}

\section{Far Detector and ProtoDUNE Detector Simulations}

A large amount of code re-use and sharing via the design of \dword{larsoft} has allowed the shared development of simulation algorithms for \dword{pdsp}, \dword{pddp}, \dword{pdhd}
and the \dword{spvd}  detector proposals.  Only the geometry, the field description,  and the anode plane channel models need to be updated; the rest of the simulation chain is re-used.  It should be noted that modules and workflows may need to be developed in order to apply detector-specific calibrations or physical effects, but that the general framework appears to be flexible enough to handle all such needs in simulation. The \dword{nd} simulation differs substantially and is discussed below in 
Section~\ref{ch:use:nd}.

Two classes of detector simulation for DUNE exist at the time of writing: parameterized fast simulations and full detector simulations that combine particle propagation codes such as \dword{geant4} that record energy deposits in active elements with detector response simulations of electron drift trajectories, photon paths and electronics performance. The output of both the full simulation and the fast simulation can be used as input to make Common Analysis Format (CAF) files. The Common Analysis Format consists of a simplified ROOT-based ntuple and corresponding library of analysis tools, CAFAna, that was developed by the NOvA collaboration and is currently used by NOvA, DUNE, and SBND for the final stage of physics analysis.

\subsection{Fast simulations}
Parameterized, or ``fast'' detector simulations involve smearing truth-level physics quantities based on expected detector performance metrics, such as acceptance and energy resolution.  These fast simulations are useful when optimizing detector designs, and for engaging physicists outside of the DUNE collaboration.  

\subsection{Particle Propagation Simulation}
Full simulations are based on detailed geometry models and \dword{geant4}~\cite{Agostinelli:2002hh,Allison:2016lfl}, and are needed for computation of precise physics sensitivities.

A detector simulation such as \dword{geant4}~\cite{Allison:2016lfl} or FLUKA~\cite{Bohlen:2014buj} is used to simulate the interaction with the detector material and signals produce by the particles described by the four-vectors produced by the event generator. Simulation input requires the geometry and 4-vectors, while the outputs consists of the true interactions and energy deposits in the active detector materials. 
Energy thresholds need to be quite low O(100s keV), as the detector systems are sensitive to physics processes at MeV-scale and above. The large uniform detector volumes in \dword{lar} detectors compensate for the low thresholds by requiring fewer \dword{geant4}-volume boundary crossings in a typical particle trajectory.  
Hadronic cross sections on \dword{lar} are not yet fully understood.  DUNE has developed a reweighting framework 
\dword{Geant4Reweight}~\cite{Calcutt:2021zck} to allow reuse of existing simulation as interaction rates are refined.{
Due to the high granularity and low energy thresholds, interaction records can become very large. Simulated \dword{pdsp} single drift-time trigger records for 6 APAs are currently 200-300\,MB, substantially larger than the raw waveforms coming from the actual detector.}

\subsection{Detector Response Simulation}

The propagation of low-energy drifting electrons and scintillation photons, signal induction  and electronics response are then simulated in a separate step.

\dwords{lartpc} require detailed models of electron trajectories through a potentially charged fluid.  Photon detection requires ray tracing over long distances.  Once electrons and photons have been propagated to the detection systems, simulation of the electrical response must also be done. Inputs to the detector response simulation are the output of the detector simulation, charge distribution in the liquid, absorption parameters and the detailed readout system geometry. Outputs are simulated streams of bits in the electronics.

Drifting electrons are simulated parametrically using a model based on the measured drift velocity, longitudinal and transverse diffusion coefficients, and a parameterized model of space charge.  This last effect is particularly pronounced at \dword{pdsp} and \dword{pddp} due to the large number of cosmic rays crossing the detector volume, giving rise to distortions in the apparent positions of particles of up to 30\,cm.  %
Based on the experience gained from  unintentionally grounded electron diverters in \dword{pdsp}, external imperfections need to be simulated as they can be the cause of field distortions.

Once the electrons drift to the anode plane in wire-based \dwords{lartpc} in the simulation, a detailed \twod model of the wire responses is applied using \dword{wct}~\cite{wirecell,ref:wire_cell_toolkit,Abi:2020mwi}.  The two dimensions are wire number and time, and the effects of induced currents on neighboring wires are included in the simulation.  The electronics response function is folded in to a final model of the observed waveforms.  Simulated waveforms have been compared with real ones  in \dword{pdsp} and are found to be very similar.
In the pixel-based \dword{ndlar} and \dword{ndgar}, the electronics simulation is at a simpler level, as the electronics have not been fully demonstrated.

\dword{protodune} experience indicates that, although simulation is reasonably fast relative to reconstruction (3 to 1), memory utilization, even for \dword{protodune}, %
is very high in simulation jobs.  Process sizes of 5-6\,GB in memory are typical for trigger records in which six \dwords{apa} are read out.  High energy \dword{fd} interactions are expected to span more detector volume (although not the full size of the detector) and thus require even more memory.   Multi-threaded solutions are being investigated, but a large fraction of a job's memory footprint is occupied by trigger record data and not shareable memory such as code segments and geometry description.  Efforts are focused on reducing the per-thread memory requirements.  Multiple %
simulation and reconstruction passes may be needed since calibration of the electron drift model incorporating space-charge effects requires a reconstructed dataset, and the output of this calibration is used for subsequent simulations and reconstruction.  A similar calibration is required for the recombination model. 

\subsection{Photon Detector Simulation \hideme{Paulucci - draft}}\label{ch:use:pd}

Photon simulation in large detectors is known to be highly computationally intensive due to the need to trace large numbers of photons over large distances.  \dword{dune} has several approaches to this problem. 

The sequence for photon simulation is:
\begin{enumerate}
\item perform particle track simulation in \dword{geant4} to produce the energy deposits along the track;
\item  calculate the number of photon/electron emissions at each vertex where energy is deposited; \item 
simulate the photon transport in the detector (either full simulation or fast simulation); and
\item  reconstruct the photons into offline objects such as individual PDS hits and collective flashes of hits.
\end{enumerate}
 
In \dword{larsoft}, after particles are propagated using \dword{geant4}, energy deposits along tracks are recorded so that the number of ionization electrons and scintillation photons generated at each step can be estimated using the available ionization and scintillation methods. Once the number of photons is determined, the fraction of those photons that will actually reach a given photon detector is usually estimated using one of several fast light-simulation methods. This procedure can be followed for both 128\,nm photons (Ar scintillation) and 176\,nm photons (Xe scintillation) to account for the wavelength-dependent Rayleigh scattering in the simulation to cover the possibility of Xe-doped \dword{lar}. 
Since a copious number of scintillation photons (25000\,$\gamma$/MeV at 500\,V/cm) is produced in \dword{lar}, it is very demanding computationally to propagate all photons individually using \dword{geant4}. Instead,  fast simulations for photons are implemented.
The full \dword{geant4}  photon simulation CPU time per event depends on the energy deposited. 
In the fast simulation methods, using a library,  semi-analytic methods or machine-learning, the CPU time depends on the granularity of the detector. Generally speaking, the performance of the three fast simulations methods is at the same level and the usage of any particular choice will depend on the physics requirements of the sample. 
 
\subsubsection{Optical Library Method}

This method consists of dividing the cryostat volume into smaller parallelepiped-shaped regions called voxels and creating a lookup table 
(the optical library) 
that  stores the visibility of each photon detector to photons  generated within a given voxel.
This optical library is created using the full \dword{geant4} simulation to generate photons anywhere inside a given voxel, with random direction and polarization, and then store the fraction of those photons that land on the optically sensitive region (visibility) of a given photon detector, identified within \dword{larsoft}  by its optical channel. When using the fast simulation, \dword{larsoft}  will retrieve the Optical Library and store its information to directly transform the number
of photons generated in a given step along a particle's track into the number of photons landing on each optical channel.
This method can satisfactorily 
be used as a fast simulation method, but nevertheless, 
its performance greatly depends on the size of the voxel and the number of photons being generated per voxel. Increasing the number of voxels in a library will improve the description and reduce the bias at the cost of a large increase in memory consumption. Increasing the number of photons per voxel will provide much better statistics and also largely increase the amount of time dedicated to generating the optical library. %

\subsubsection{Fast Simulation with Generative Neural Networks}

This method relies on a generative neural network  trained on the photon detection system. 
The input to the network is the vertex where the photons are emitted, and the output is the mutual visibility of each photon detector/emitter pair. %
The generative model can be trained ahead of time using a full  \dword{geant4} optical photon simulation with photons emitted from random vertices in the detector, after which it can be frozen to a computable graph and deployed to the production environment (\dword{larsoft}  framework).

When the computable graph is loaded in \dword{larsoft}, it quickly emulates photon transport by  computing the visibility %
 of each photon detector calculated from  the photon emission vertex along the particle's track.
This method is 20 to 50 times faster than the \dword{geant4} simulation while keeping the same level of detail for particle tracks, such as the number of energy depositions and the precision.
The model inference also requires a relatively small amount of memory. The samples for candidate  %
far detector-like geometries show the required memory for the model inference is around 15\% of the \dword{geant4} simulation. Further, this memory use is not directly correlated to the size of the detectors. %

\subsubsection{Semi-Analytical Photon Detection}
In these methods, large numbers of photons are generated at different points within the cryostat volume and propagated using \dword{geant4}. Once those propagation distributions have been simulated, Gaisser-Hillas functions are fitted to the  number of 
photons reaching the photon detectors as a function of source-detector distance and relative angle. The resulting parameters are used during event simulation  in order 
to extract the fraction of photons produced at a certain point that arrives at a given sensor.
 
\subsubsection{Comparison Between Fast Simulation Methods}
A semi-analytic model and the use of  optical library models for fast simulation have been compared to the full light simulation in \dword{geant4} by the \dword{sbnd} experiment. Their studies used a highly segmented and large-photon-count optical library, created with $\sim$1.6M voxels, with 0.5M photons being 
generated in each voxel (total of $7.9 \times  10^{11}$ photons), resulting in a  1.2\,GB file size. A similar optical library for the DUNE $1\times2\times6$ geometry (volume of $7 \times 12 \times 13.9$ m$^3$) would be prohibitively large.

All DUNE optical libraries produced so far have larger voxels and fewer total photons simulated per voxel 
in comparison to  \dword{sbnd}, one optical library of which 
was used as reference for the comparison of the two modes.
It has been reported that the optical library struggles to properly describe light signals generated closer to the detectors and more on-axis (up to $\sim50\deg$). This is a known issue caused by the intrinsic discontinuity of the voxelization schemes. In the \dword{protodune} optical library, the visibilities are smoothed using neighboring voxels  to minimize this effect.

However, 
\dword{sbnd}, observes better performance from the
semi-analytic model, with improved  resolution close-on axis
(3.6\% vs 5.6\%), and less than 1\% bias. In contrast, the
optical library is systematically biased (2.5-4.9\%), in particular for
the larger/closer signals. These differences, together with the very high memory
consumption (of several extra GB) during simulations when using an
optical library, motivate the choice of the semi-analytic model as the
default for fast simulations in the DUNE \dword{fd}.

\section{Near Detector Simulations\hideme{Junk and Muether - draft}}\label{ch:use:nd}

\subsection{Common Simulation Tools}

The near detector simulation chain is diagrammed in Figure \ref{fig:ch:use:ndsimrecoflowdiagram}. The flux, geometry and generator stages are common across the ND detectors. 
The model used to predict the neutrino flux at the near detector is a \dword{geant4}-based simulation of the incident proton beam, the target, the focusing horns, decay pipe and the hadron absorber, called \dword{g4lbnf}~\cite{DUNE:2020ypp}, and the dk2nu package.  The same simulation is used to predict the flux at the far detector, which differs from that at the near detector due to the angular and position acceptance of the near detector, as well as neutrino oscillations.  Uncertainties on the flux are estimated using the \dword{ppfx} framework developed by the \dword{minerva} collaboration~\cite{Aliaga:2016oaz,AliagaSoplin:2016shs}.
The detector geometries are generated with the DUNENDGGD package which produces a detector suite in the hall in GDML format, as described in Section~\ref{sec:usecases:detectorgeometry}. Neutrino interactions are simulated with \dword{genie} version 2.12.10, with  the \texttt{DefaultPlusValenciaMEC} physics list. \cite{Andreopoulos:2009rq}. An upgrade to GENIE~3 is in progress. Particles exiting the struck nucleus are propagated through the detectors using a \dword{geant4}-based model called edep-sim~\cite{ref:edep-sim}, which produces \dword{root} trees containing the simulated energy deposit information and associated truth labels. Detector-response simulations are handled independently for each subdetector as described below. 

\begin{dunefigure}
[Diagram showing the components of the Near Detector simulation and reconstruction workflow]
{fig:ch:use:ndsimrecoflowdiagram}
{This diagram shows the components of the Near Detector simulation and reconstruction workflow.  Common tools include flux models, geometry specification tools, event generators, and GEANT simulation.  The detector response and reconstruction is special to the subdetectors, and the reconstructed data are collected in common analysis format (CAF) files.  Data are to be reconstructed in the same way as simulated data.}
\includegraphics[width=0.8\textwidth]{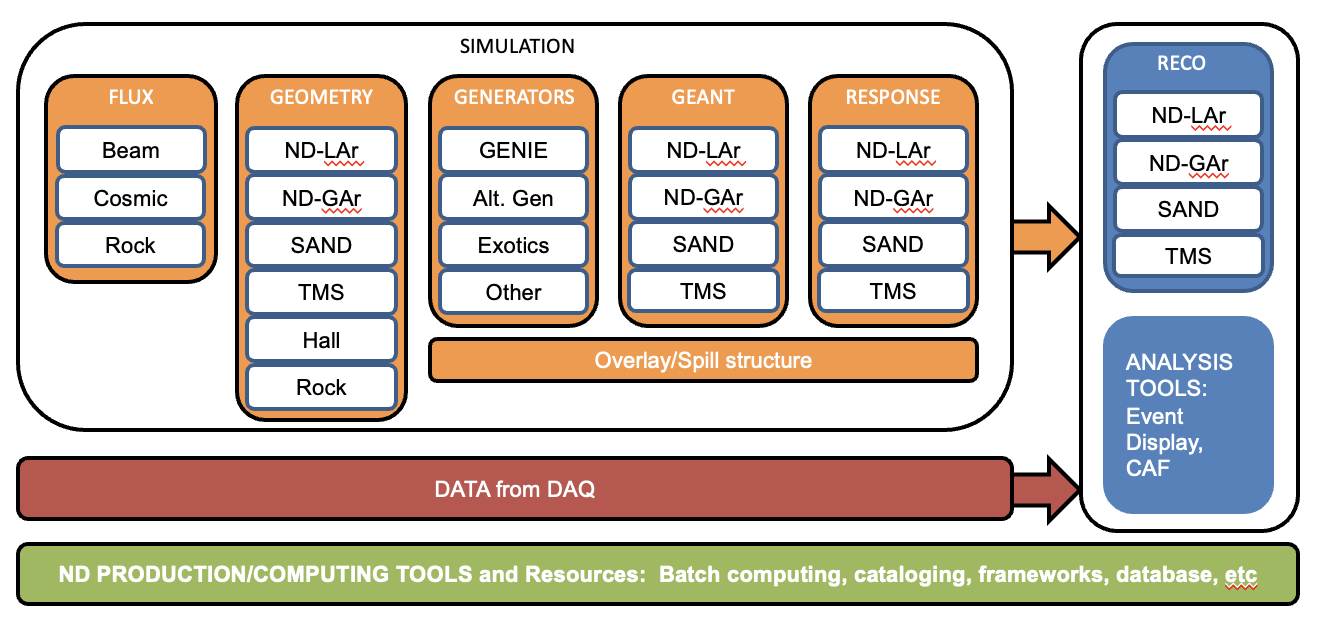}
\end{dunefigure}

\subsection{ND-LAr Detector Simulation}
The ND-LAr detector simulation is performed through a set of fully GPU-optimized algorithms. The software, called \texttt{larnd-sim} \cite{larndsim}, is written entirely in Python and compiled on the GPU using the Numba wrapper for CUDA. This highly-parallelized implementation provides orders of magnitude processing speed improvements compared to a classic CPU one.

The software takes as input energy deposits in the active volume as simulated by {\texttt{edep-sim}}. The {\texttt{edep-sim}} output ROOT files are converted into the HDF5 format by a dedicated script. The simulation takes into account quenching and drifting effects for the ionization electrons in the liquid argon and calculates the current induced on the pixels placed at the anode of each module. Subsequently, the electronics simulation produces the analog-to-digital counts and timestamps in the same format as the LArPix ASICs \cite{larpix}. The number of photons reaching the light readout system is calculated using a GPU-optimized lookup table. The output is saved in the HDF5 format. 

The high flexibility of the software allows easy implementation of different detector geometries. The software has already been employed to perform comparisons between the simulation and the data coming from the ArgonCube Module-0 and Module-1 runs \cite{module0} and it has been successfully used to produce full-spill ND-LAr simulations.

\subsection{TMS Detector Simulation}
Currently detector response is handled with resolutions incorporated at the hit level by smearing reconstructed quantities. A full detector response simulation is in progress. 

\subsection{ND-GAr Detector Simulation}
\label{sec:usecases_ndgardetsim}

The \dword{garsoft} suite of \dword{art}-based modules is used to simulate the \dword{ndgar} detector.  It performs similar functions to \dword{larsoft}'s \dword{geant4} and detector simulation modules.  \dword{garsoft} includes interfaces to the following event generators:  \dword{genie}, CRY~\cite{bib:cry2007,Cosmic-CRY}, and a text-file generator.  It also contains a \dword{larsoft}-like interface to \dword{geant4}, called {\tt garg4}.  In order to be used in conjunction with the other near detectors in a full hall simulation, \dword{garsoft} also includes a module to import energy deposits from {\tt edep-sim}, and format them internally as if they had been made by {\tt garg4}.

Once the energy deposits are available, a module that simulates the division between ionization and scintillation quanta is invoked.  Electron drift through the gaseous medium is parameterized, and longitudinal and transverse diffusion are modeled by sampling from Gaussian distributions of widths that grow with the square root of the drift time.  The effect of attenuation due to attachment of drifting electrons on electronegative impurities in the gas is also simulated at this time. 

The anode planes are modeled as re-purposed ALICE %
\dwords{iroc} and \dwords{oroc} ~\cite{Dellacasa:2000bm}.  Charge is amplified by avalanches on the anode wires, and induced charge on the nearby readout pads provides the charge detection.  The pad response functions are taken from Ref.~\cite{Dellacasa:2000bm}.

The response of the calorimeter and the muon systems are parameterized versions of the energy deposit information, localized to the detector cells in which the energy is deposited.

\subsection{SAND Detector Simulation}
\label{sec:usecases_sanddetesim}

\subsubsection{SAND Simulation Components}
The SAND detector response simulation is performed by several tools developed in C++ and Python. The digitized detector response waveforms are built starting from the energy deposits provided by edep-sim, after the simulation of the detector response.
 
 \begin{itemize}
\item{ECAL:} The energy deposits in the scintillating fibers are converted into the number of generated photons randomly extracted by a Poisson distribution with mean of 25 photons per MeV of energy deposited. The number of photo-electrons is obtained taking into account the attenuation according to the fiber length traveled by the photons. The arrival time to the photo-sensors is assigned to each photo-electron taking into account the scintillation decay time, the propagation time along the fiber and a random jitter. The amplitude of the signal of the photo-sensor is proportional to the number of photo-electrons collected in a certain time window and the time of the signal is obtained simulating a 15\% constant fraction discriminator.
 
\item{STT:} For each energy deposit in a straw tube, the electronic signal arrival time at the read-out is evaluated considering the electron drift time to the wire, the electric signal propagation along the wire and a random jitter. The timing of the signal is given by the earliest electric signal reaching the tube read-out. The amplitude of the signal is proportional to the sum of the energy deposit in a certain time window.
 
\item{GRAIN:} Argon scintillation in GRAIN is simulated using \dword{geant4}. For each energy deposit, a random number is extracted from a Gaussian distribution, with the mean value determined by the argon light yield and the fraction of the energy deposit that ends up producing photons (already calculated by edep-sim), and a dispersion given by the square root of this number. If the mean is less than 20, a Poisson distribution is used instead. For each photon, a random momentum, polarization, position along the step, time of emission, and energy are determined. The properties of argon are included: the singlet to triplet ratio~\cite{PhysRevB.27.5279}, the fast and slow scintillation time constants~\cite{PhysRevB.27.5279}, Rayleigh scattering~\cite{Seidel:2001vf,ISHIDA1997380,SneepMaartenUbachsWim2005}, and the absorption length~\cite{Jones:2013bca}.
The photons are then collected by cameras. A camera is composed of an optical system (either mask or lens) and a (32x32) \dword{sipm} matrix as photo-sensor. The simulation includes the geometry of the cameras, whether being lens-based or mask-based. The number of photons collected in each channel in a given time window and the time of the first photon impinging in each channel is stored for producing a two-dimensional image for each camera. 
The photo-sensor response is simulated with custom software written in Python and OpenCL for GPU-accelerated computing. The \dword{sipm} matrix response waveform is computed with a heuristic function from the timing of impinging photons on the matrix surface. The \dword{sipm} characteristics such as PDE, crosstalk, afterpulses, and dark count rate are included in the simulation. A small quantity of white and rf noise is added as well. The waveform is then quantized and integrated over a fixed time window to simulate a DAQ composed by an ADC and a TDC for timing over a fixed threshold.
\end{itemize}

\subsection{Overlays}
In order to build full beam spills simulated events on rock, detector fiducial volumes and the corresponding anti-fiducial sample are overlaid to reproduce expected beam timing and intensity. The OverlayGenie package is the current overlay tool in use but alternatives which are more integrated are being developed. More details are listed in \ref{sec:usecases_overlays}.

\section{ProtoDUNE Simulation Experience \hideme{Yang - draft}}

The \dword{fd}/\dword{protodune} \dword{larsoft}-based  detector simulation framework has already been tested successfully in \dword{protodune} and is described more fully in references \cite{DUNE:2021hwx,DUNE:2020cqd}.

The \dword{pdsp} simulation includes beam particles, cosmic ray interactions and radiological backgrounds. The beam particle species and momentum distributions are from the \dword{geant4} simulation of the H4-VLE beam line at CERN, which consists of $e^{+}$, $\pi^{+}$, $p$, and $K^{+}$ particles at 0.3, 0.5, 1, 2, 3, 6, and 7\,GeV/$c$. Cosmic ray interactions are produced with the \dword{corsika} generator. Radiological backgrounds, including $^{39}$Ar, $^{42}$Ar, $^{222}$Rn, and $^{85}$Kr, are also simulated using the RadioGen module in \dword{larsoft}. The primary particles are tracked in the \dword{lar} using the \dword{geant4} package. 
The ionization electrons are drifted towards the wire planes %
and the effects of recombination, attenuation and diffusion are simulated. The accumulation of positive ions in the detector modifies the trajectory of ionization electrons and the strength of the \efield, which is 
known as ``space charge effects''; this is simulated using the measured distortion and \efield maps. The electronic signal is simulated by convolving the number of electrons reaching each wire plane with the field response and electronics response. The field response is modeled using the \dword{garfield}~\cite{Veenhof:1998tt} package. 
The electronics response uses the parameterization from a \dword{spice} simulation with the average gain and shaping time measured in the \dword{protodune} charge injection system. 
\section{General Simulation Considerations}
\label{sec:use-sim-considerations}

In addition to the specific needs for particular detectors there are additional common considerations for simulation. 

\subsection{Overlays or Mixing}
\label{sec:usecases_overlays}
An efficient method for accurately simulating backgrounds in high-rate or high-noise environments is to overlay the simulated interaction information on real data.  For example, the environment in \dword{protodune} or the near detector includes multiple beam and rock muon interactions in the same readout as %
the interactions of interest.  Libraries of properly-formatted raw data can be combined with simulated hits from a simulated event to yield a realistic simulation of an interaction in the real detector. This requires a library of such interactions and careful matching of the running conditions of the external raw events to the desired simulated events.  It is also important to include the same electron lifetime and space-charge effects in the signal simulation as are present in the overlaid data, in order for the combined data to be consistent.

DUNE has not yet integrated overlays into the \dword{protodune} or \dword{fd}  frameworks. %
To do this we will need to distribute large volumes of overlay data to remote sites while maintaining randomness. Other experiments such as MicroBooNE and MINERvA have done this successfully. 

\subsection{Reweighting}
If intermediate simulation steps  are stored they can be used to reweight interactions to reflect new knowledge about the underlying cross sections, neutrino flux, detector materials or detector properties. Examples include the \dword{ppfx} beamline simulation reweighting system and the implementation of different cross section and final-state interaction models in event generators.   In the case that a particular portion of phase space is under-represented by a generator, or not represented at all, but nonetheless is populated by a desired alternative generator, additional samples must be generated with the alternative generator.

\subsection{Outputs from Simulation}
At this point we have simulated data that mimic the raw data coming from the detector, and %
the next steps are calibration and reconstruction. In all cases, descriptive metadata need to be produced that facilitate locating the simulation samples on persistent media, understanding their contents, and reproducing them if need be.

\section{ProtoDUNE and Far Detector Reconstruction \hideme{Junk, Muether and Schellman-draft}}

A detailed description of the \dword{pdsp} and Far Detector reconstruction algorithms is given in Ref.~\cite{DUNE:2020ypp}.  This section outlines those aspects of the reconstruction processing that are directly related to software and computing issues.
Figure~\ref{fig:ch:use:pdii} shows the data flow for regular reconstruction in \dword{protodune}.  
There are, in principle,  multiple stages in reconstruction that are each well suited to different computer architectures.  We expect the full \dword{fd} data processing to follow a similar path to that used by \dword{protodune}.  

Figure \ref{fig:ch:use:fdagg} illustrates  readout structures for full DUNE \dword{fd} modules.  A readout consists of a large number (up to 150) of  30\,MB \dword{apa} readout fragments and a number of smaller readout fragments that are much smaller than a 30\,MB APA fragment.  Details of trigger record volume can be found in Tables  \ref{tab:est:usefulfdhd} and \ref{tab:est:usefulfdvd}. These need to be processed and  recombined into a single readout record. 

\begin{dunefigure}
[Data aggregation diagram for FD]
{fig:ch:use:fdagg}
{Data aggregation cases for the far detector. The top case shows information for normal beam or calibration readouts. A single file of $\sim 10$\,GB size contains several complete \dwords{tr} with their boundaries designated by the dashed lines.  \dword{tpc} \dword{apa}'s, \dwords{pd},  trigger primitives and a trigger  are recorded for each trigger readout.  In addition, a manifest, which describes the relations between the data, is stored either in the file or in external metadata.  The bottom case is a \dword{snb} readout, in which thousands of 5-10\,ms time slices must be read out over 100\,s.  The solid lines denote file boundaries. How data are ordered, by geographical position or by time,    is not  yet specified.}
\includegraphics[width=0.8\textwidth]{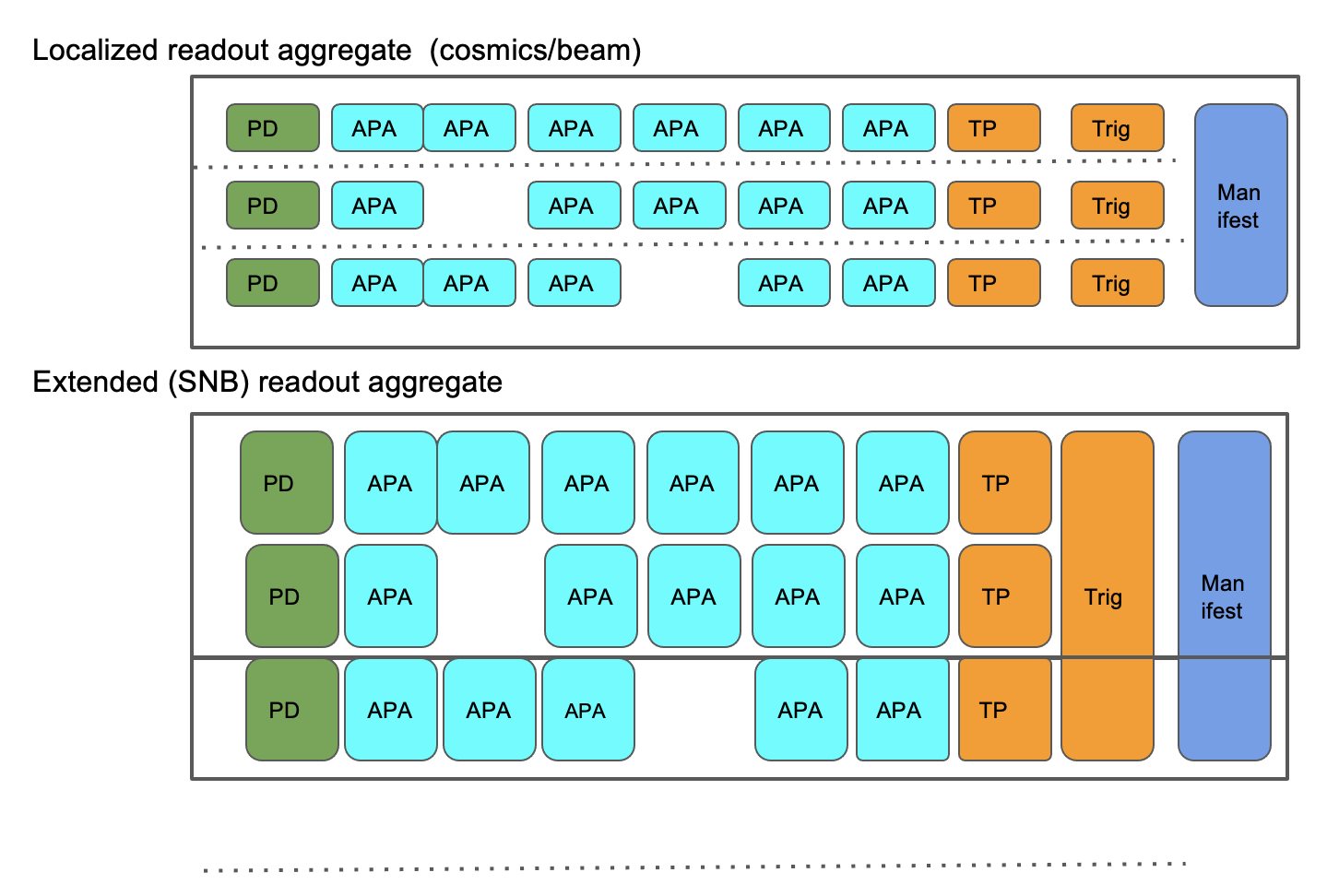}
\end{dunefigure}

Each trigger record contains a large number of waveforms, one for each readout channel.  In \dword{pdsp}, the waveforms typically were 6000 time ticks long, with one \dword{adc} sample per time tick.  A detailed description of the reconstruction procedures, starting with these waveforms, is given in~\cite{Abi:2020mwi}, and summarized briefly here.

\subsection{Signal Processing \hideme{Schellman-draft}}
\label{subsec:signal_processing}

The initial offline processing stage, labeled ``signal processing'' in Figure~\ref{fig:ch:use:pdii}, holds particular significance and challenges.
Such a stage must be applied to data from both the \dword{pd} and \dword{tpc} readouts.
It is the latter which poses the bulk of the challenges and here it is the focus.
The source of the challenges is simple.
The processing is relatively intensive in terms of CPU and memory as evaluated on a per-trigger basis and it is data intensive as it must be applied to the entirety of the TPC readouts.
Unlike subsequent stages, there is no reasonable pre-selection criteria to reduce the input to this stage which has not already been exploited by the DAQ.
The initial stage consists of a sequence of two primary data transformations.

The first is known as data preparation or noise filtering.
The raw data from the DAQ is input after being decoded and loaded into dense per-channel ADC waveform arrays.
The transformation then attempts to remove features unrelated to those caused by ionization electron signal while at the same time producing no significant alteration to those signal components.
One important goal of this processing is to attenuate non-thermal noise such as that due to RF emissions in the cryostat and which induce waveforms coherently across many channels.
Also mitigated are unwanted waveform features such as ADC bits with error codes ($\sim$1\%) and transient artifacts.
Some general-purpose algorithms have been developed by the LArTPC community but substantial effort is required to understand and invent solutions to problems specific to each given detector. 
The output of this first step is effectively still in the form of full, dense ADC waveforms.

The second major transform is known as signal processing. 
It produces ``signal regions-of-interest'' (signal-ROI).
It does so in two tightly coupled steps. 
First, a \twod model (in the space of channel vs sample time) of the per-plane detector response is deconvolved from the prepared, filtered ADC waveforms. 
This produces waveforms which are in units of number of ionized electrons in the region near a wire and per time sample.
However, due to the bipolar nature of the detector response on the induction planes, this deconvolution necessarily amplifies low frequency noise.
Left alone, this noise would overwhelm the resulting signal waveforms.
To combat this, another algorithm operating in the time domain identifies small, regions-of-interest where signal is found to be above the low-frequency noise.
Within each signal region of interest, the baseline of the waveform fragment is recalculated, thus effectively operating as an adaptive high-pass filter.
The resulting data consists of very sparse, unipolar waveform fragments and substantial data reduction is achieved while retaining almost all signal information.
This is a key operation for any induction-based LArTPC detector and the current best incarnation was developed and vetted in MicroBooNE~\cite{Adams:2018dra,Adams:2018gbi} and applied to ProtoDUNE~\cite{ref:pdune_signal_processing} and has since been applied to many other detectors including the prototypes for DUNE \dword{spvd}. 
These latter, strip-based detectors pose new challenges related to the required precision of the detector response model. 
This is an area of ongoing study but initial results show good performance.

This initial stage (both major transformations) can run in parallel, multi-thread form at the level of data from one \dword{apa} (or equivalent detector module unit).
In particular, the \dword{wct}~\cite{wirecell,ref:wire_cell_toolkit} in which the signal-processing and some forms of noise filtering are developed, implements an execution model that supports this level of parallelism.
Furthermore, the \dword{fft} and other algorithms used heavily in this stage benefit from SIMD acceleration~\cite{Dong:2022wxg} such as provided by GPU devices. 
These forms are also supported by and being further developed in \dword{wct} implementations.
This execution model provides flexibility to scale this key stage from \dword{protodune} to DUNE by mapping these two very different data volumes to different classes of computing facilities.

\subsection{Reconstruction Strategies}

Two primary reconstruction strategies have been developed in the LArTPC community, have been applied to \dword{protodune} and are expected to be applied to DUNE FD.
They may be classified as ``view-first'' vs ``slice-first''. Note that representations of \dword{pdsp} trigger records of different approaches can be seen in Figures \ref{fig:vis:larsoft3dreco} and \ref{fig:vis:bee}.

In ``view-first'', data from each wire plane view is initially processed. 
Peaks are found in the waveform fragments output by the signal-processing and each is fit to a Gaussian model in a process called ``hit-finding''. The Gaussian model is convenient as a baseline for hits, but it should be mentioned that not all waveforms will be Gaussian in shape and that there are limitations to this algorithm.
These hits are then used as inputs to reconstruction algorithms, such as the SpacePointSolver~\cite{DUNE:2020ypp}, \dword{pandora}~\cite{Marshall:2015rfa}, TrajCluster~\cite{ref:trajcluster}, and \dword{pma}~\cite{ref:PMA}, which identify clusters, tracks and showers in \threed by associating objects in the three \twod views. 
The calorimetry modules sum up and calibrate the charge deposits for use in energy reconstruction and \dword{pid}. 
The parameters of the clusters, tracks and showers are stored in \dword{root} trees that end users can analyze rapidly and repeatedly.

In ``slice-first'', data from one time slice across all wire plane views is initially processed. 
The slice is chosen to span four to six samples in time which exploits the fact that signal processing naturally produces over-sampled waveforms and thus the data may be further reduced with no information loss.
Next, a tomographic inversion is performed for each slice with an algorithm called ``wire-cell''~\cite{Qian:2018qbv}, which is provided by the Wire Cell Toolkit (\dword{wct}).
Stacking the per-slice inversions over the full readout results in sparse 3D regions of the detector volume that localize ionization electrons in a manner consistent with the original signal waveforms.
The ambiguities inherent in the tomographic nature of the \dword{tpc} also leads to portions of the solution not truly containing any ionization. 
Further constraints, such as consistent charge in all three views and simple connectivity between regions of neighboring slices, are applied to remove or reduce these ambiguities.
This is done in a model-free manner in order to not bias the results.
When bias-free constraints are exhausted, a local track-like model is used to allow collapsing regions of ambiguity, making the results yet more sparse and more tightly surrounding regions containing ionization.
These results finally enable very accurate reconstruction of quantities such locations of interaction vertices, track branch points, and $dQ/dx$ and from there $dE/dx$ along tracks.
These algorithms were initially developed for MicroBooNE~\cite{MicroBooNE:2021ojx} where they have been found to outperform others~\cite{MicroBooNE:2021rmx, MicroBooNE:2021nxr}. 
There is an ongoing effort to generalize and ``port'' these algorithms into \dword{wct} for use by DUNE and other LArTPC detectors.

Either strategy results in a 3D model of the ionization distribution which is composed of a number of spatial clusters connected in some way through the 3D space of the detector volume.
With these clusters as input, a process called ``flash matching'' attempts to associate processed \dword{pd} signals with the TPC clusters in order to determine when in time or equivalently where along the drift axis the cluster of ionization occurred.
The offline software must arrange for these two data flows to merge on a per-trigger-record basis.

From this relatively low-level reconstruction, various high-level reconstruction techniques follow.
For example, separating track-like energy deposits from shower-like deposits is a key part of many DUNE analyses. 
In one case, this is accomplished with a \dword{cvn}, {\tt EmTrackMichelID}, listed in Table~\ref{tab:protodune_cpu_reco_by_module}.
The signal processing is run in two modules: ``caldata'' includes all but 2D deconvolution and signal-finding which are handled in ``wclsdatasp''.
It is one of the most CPU-intensive operations in the \dword{pdsp} reconstruction chain, when the algorithm is run on a grid node lacking a GPU. 
Recently, however, a \dword{gpuaas} technique has been developed~\cite{Wang:2020fjr}, enabling a speed-up on the order of a factor of ten, though it depends on the ratio of CPU-only nodes to GPU resources.

\pagebreak

\begin{longtable}
{l r}
\caption[Processing time for reconstruction modules for a \dword{pdsp} event]{Wall-clock module execution times for the reconstruction of a typical \dword{pdsp} event, in seconds on a 2020 vintage processor with an HS06 rating of 11~\cite{bib:HS06}.  Processes taking less than 1 second are not shown. The event is a data event from Run 5809, a 1\,GeV beam run.
Note that the electromagnetic shower reconstruction module EmTrackMichelId  dominates. } \\ \toprowrule
  \rowcolor{dunesky}
Module Label & time/event (sec)\\ \toprowrule
RootInput(read)                          &     0.14\\%7283          \\
beamevent:BeamEvent                      &      1.6\\%54           \\
caldata:DataPrepByApaModule              &      83.0\\%24          \\
wclsdatasp:WireCellToolkit               &      79.9\\%16          \\
gaushit:GausHitFinder                    &      1.6\\%342          \\
reco3d:SpacePointSolver                  &      9.4\\%157          \\
hitpdune:DisambigFromSpacePoints         &      1.5\\%541          \\
pandora:StandardPandora                  &      39.9\\%87          \\
pandoraTrack:LArPandoraTrackCreation     &      4.6\\%221          \\
pandoraShower:LArPandoraShowerCreation   &      3.7\\%432          \\
pandoracalo:Calorimetry                  &      2.1\\%152          \\
pandoracalonosce:Calorimetry             &      1.9\\%852          \\
pandoraShowercalo:ShowerCalorimetry      &      3.3\\%953          \\
pandoraShowercalonosce:ShowerCalorimetry &      3.2\\%698          \\
emtrkmichelid:EmTrackMichelId            &      233.8\\%4          \\
c%
anodepiercerst0:T0RecoAnodePiercers      &      1.1\\%673          \\
pandora2Track:LArPandoraTrackCreation    &      11.6\\%25          \\
pandora2calo:Calorimetry                 &      4.9\\%932          \\
pandora2calonosce:Calorimetry            &      4.5\\%118          \\
pandora2Shower:LArPandoraShowerCreation  &      4.2\\%815          \\
pandora2Showercalo:ShowerCalorimetry     &      4.2\\%29           \\
pandora2Showercalonosce:ShowerCalorimetry&      3.8\\%562          \\
RootOutput(write)                        &     2.8\\%458               \\
{\bf Total:}                             &     {\bf 507.9}      \\ \colhline
\label{tab:protodune_cpu_reco_by_module}
\end{longtable}

\section{Near Detector Reconstruction\hideme{Junk/Muether - needs revision}}
\label{sec:nd-reco}

The \dword{nd} Reconstruction primary exists as separate reconstructions for each individual ND sub-detector (described below). The ND software group are developing a common data model which will allow inter-detector software connection. For instance, an event matching algorithm between the \dword{ndlar} and TMS has been developed. Output from the Reconstruction will be written to common analysis files (CAFs) that will enable physics analysis (Figure \ref{fig:nd:reco}). The current ND data model is outlined in the DUNE ND Data Model document~\cite{bib:docdb24735}.

\begin{dunefigure}
[ND Reconstruction Flow Chart Example]
{fig:nd:reco}
{An example ND reconstruction workflow is shown for simulation with TMS and LAr (MLReco) data producing track matching information and stored in CAF files.}
{\includegraphics[width=0.8\textwidth]{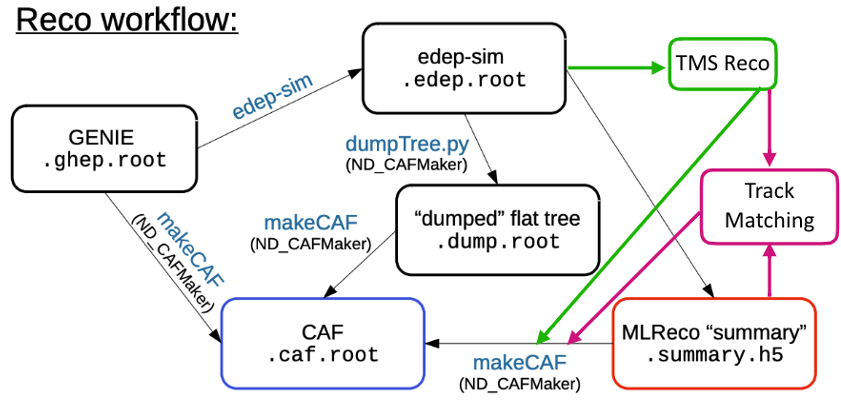}}
\end{dunefigure}

\subsection{Near Detector Liquid Ar (NDLAr)}
\label{sec:nd-lar-reco}

 Reconstruction algorithms for the \dword{ndlar} include both traditional and machine-learning based chains. The primary reconstruction is an automated machine-learning based package which has been demonstrated on full-spill events simulated in the native 3D detector readout. A parallel Pandora based reconstruction is also being prepared. These reconstruction algorithms are currently being tested on both simulated and prototype data. Rough estimates for CPU processing time are listed in Table \ref{tab:NDCPUPerEvent}.
 
\subsubsection{ND LArTPC ML Reconstruction }

A full data reconstruction chain for the LArTPC at DUNE ND consists of multiple machine learning algorithms optimized for multiple reconstruction tasks. An overall software design can be found in~\cite{Drielsma:2021jdv} and details of individual algorithm can be found in~\cite{Domine:2019zhm,Domine:2020tlx,DeepLearnPhysics:2020hut,Koh:2020snv,NeutrinoUQ}, including evaluations performed on a public LArTPC simulation dataset~\cite{Adams:2020vlj}. This reconstruction software is based on earlier development of applying powerful deep learning methods to LArTPC experiments~\cite{Acciarri:2016ryt,Radovic:2018dip,Adams:2018bvi}. In particular, these algorithms include deep convolutional and graph neural networks as well as analytical physics models (e.g. Multiple Coulomb Scattering for estimating momentum). The reconstruction chain takes voxelized 3D images (i.e. data recorded by the detector) as an input, and produces output that organizes the following information:
\begin{itemize}
    \item A list of neutrino interactions
    \item A list of particles in each neutrino interaction
    \item Attributes of individual particles
    \begin{itemize}
        \item Particle type and kinematic information
        \item A cluster of pixels that resemble the particle's trajectory
        \item Particle hierarchy (e.g. a parentage information for a particle produced through a decay process)
        \item Calibrated uncertainty quantification for (most of) the attributes listed above
    \end{itemize}
\end{itemize}

This reconstruction chain is end-to-end optimizable: all algorithms can be optimized by a gradient-based technique~\cite{Rumelhart1986} simultaneously in a completely automated fashion, unlike traditional software, without requiring any human interventions. This saves weeks and months of manual tuning process required for a traditional counterpart. The software is developed on top of the CUDA programming ecosystem~\cite{10.1145/1365490.1365500} and the PyTorch machine learning library~\cite{NEURIPS2019_9015},
which allow the use of NVIDIA Graphics Processing Units (GPUs) to accelerate computation. Furthermore, the implementation allows multi-GPU parallelization and is also compatible to distributed computing (i.e. inter-node parallelization exploiting fast network interconnects such as infiniband to expand parallelization beyond a single server) to unlock the power of modern infrastructures including High Performance Computing (HPC) clusters. While many deep neural networks require excess amount of computing power and are thus often impractical to execute on CPUs, this software can be run on CPUs within a reasonable time thanks to an implementation of innovative sparse matrix multiplication that exploits the sparsity of LArTPC image data. In terms of computing needs, the process of optimization is typically performed on a single NVIDIA A100 GPU and takes about a week starting from scratch. It takes a day or two for a typical fine-tuning of the chain when the underlying detector physics modeling changes.  Discussion of access to and resource estimates for GPU-based algorithms is discussed in more detail in Section \ref{sec:model:hpc}. %
 
 \subsubsection{Pandora based Reconstruction}
 Pandora is a multi-algorithm approach to particle reconstruction that has been used effectively for other LArTPC detectors, including the MicroBooNE detector and the ProtoDUNE detectors. As it is currently a leading reconstruction option for the DUNE Far Detector, it is thus ideal that it can also be adapted to work with the DUNE Near Detectors. Currently the effectiveness of the pre-existing 3x2D reconstruction chain when used with ND-LAr is being assessed, with the plan being to develop a native 3D approach and tailored algorithms as the next stage of work.

\subsection{TMS}
\label{sec:tms-reco}

The output of the \dword{tms} reconstruction will inform the \dword{ndlar} reconstruction as to which tracks exited the \dword{ndlar} and were reconstructed in the \dword{tms}. A stand-alone \dword{tms}-reconstruction also enables monitoring of the inclusive neutrino event rate, using neutrinos interacting on the \dword{tms} fiducial volume.

The current \dword{tms} reconstruction approach is running the track finding and reconstruction simultaneously through a Kalman filter, or having a separate track finding algorithm identify track candidates which it feeds to the Kalman track reconstruction.

\subsubsection{Track finding}
For the track finding stage, we are currently investigating three approaches. They all run in each view separately, and the 2D to 3D track matching and merging is done at a later stage.

\begin{itemize}
    \item Perform a Hough transform \cite{Hough:1959qva} in each view separately, and find which hits intersect the primary Hough line. Traverse the hits and group neighboring hits into a track candidate. Take remaining hits (not belonging to a candidate) and repeat process to see if there is a $N^{th}$ track candidate. Due to the bending in the $x-z$ view, the Hough transform is often more successful in the $y-z$ view, and the neighbor clustering in $x-z$ is critically important.

    \item Sort the hits in $z$, and run a custom A* \footnote{https://en.wikipedia.org/wiki/A*\_search\_algorithm} path finding algorithm from first to last hit in $z$. The intent is to find the shortest path between the first and last hit, only traversing cells that have been hit. If no path is found from start to finish, repeat A* path finding by running the algorithm until the hit with second largest z, and continue.
    
    \item Sort the hits in $z$, and run a Kalman filter from first to last hit in $z$. For this to succeed, the Kalman filter needs to account for the energy-loss, bending and multiple-scattering happening in the iron and scintillator layers. 
\end{itemize}

\subsubsection{Track reconstruction}
Currently, the lepton candidate is assumed to be the longest track in the track finding. After the reconstruction stage, it is envisioned the lepton candidate is assumed to be the most energetic track that fits a muon hypothesis. Depending on the final design of the \dword{tms}, the track reconstruction will operate on 2D or be combined into 3D views. 

Preliminary studies show track length is by far the most important feature for a good muon KE measurement, with the energy deposits and track bending being secondary contributions. The current Kalman implementation in the \dword{tms} software includes effects from energy loss and multiple scattering, inspired by similar implementations in \dword{minos}, \dword{minerva}, and \dword{t2k}. Efforts towards integration of the \dword{ndlar} and \dword{tms} reconstructions are also ongoing.

\subsection{Pixel-Based Gaseous Argon TPC Reconstruction}
\label{sec:algo:reco:gartpc:pixels}

The \dword{ndgar} consists of two primary detectors -- a copy of the ALICE pixel-based \dword{tpc} in a 10-bar gas consisting predominantly of argon, surrounded by a calorimeter and a superconducting magnetic coil.  A muon system is envisaged outside of the calorimeter but is not included in the simulation or the reconstruction at the time of writing.  Data are unpacked and hits are found on the per-readout-pad waveforms, similarly to how the initial stages of reconstruction for a \dword{lartpc} are followed.  Hits are then clustered into TPC clusters, which reduces the memory and CPU usage of subsequent steps, and also increases the spatial resolution of individual clusters.  Vector hits are found by grouping TPC clusters into short line segments, and the vector hits themselves are grouped into track candidates by a pattern recognition module.  A track fit based on a Kalman filter then finds the best estimates of the track parameters.  Vertices are then found using tracks with nearby endpoints.  Because there is a cathode in the middle of the drift volume in the nominal \dword{ndgar} design, 
a cathode stitch module is then run to associate track segments on either side of the cathode, bring them together by solving for the best interaction time that lines the segments up, and also moves the associated tracks and vertices. 
The calorimeter consists of a mixture of strips and pads.  Calorimeter hits are found in the \dword{sipm} waveforms provided by the calorimeter \dword{daq}, and they are clustered together to form reconstructed three-dimensional energy deposit objects with positions, directions, and energies.  Calorimeter clusters are associated with tracks, and they must match both in position and in time.  Before association with a calorimeter cluster, a track's position along the drift direction is uncertain due to the unknown interaction time.
The nanosecond-scale timing resolution of the calorimeter allows the updating of track positions along the drift direction.

\subsection{SAND}
\label{sec:sand-reco}

For neutrino interactions in the Straw Tube Tracker (STT), a first rough estimation of the vertex position is based on the STT digit spread in both the XZ and YZ views, where Y is the vertical axis, Z is the projection of the beam direction on the horizontal plane and X is perpendicular to the first two in a right-handed coordinate system. The vertex position is then used to apply a conformal transformation to the coordinates of the YZ digits $(y,z \rightarrow u,v)$. The YZ tracks are then identified as clusters of digits in the v/u distribution. The particle momentum in the bending plane is estimated with a circular fit of the YZ digits. To measure the dip angle, a transformation is first applied to the coordinates of the XZ digits $(x,z \rightarrow x,\rho)$ to linearize the trajectory, followed by a linear fit. A Kalman Filter based track reconstruction for the STT is under development.
The longitudinal position and timing of the energy deposit in an ECAL cell is obtained by comparing the measured timing in the photo-sensors on both ends of the cell. Adjacent cells are grouped into a cluster and the position, timing and energy deposit of each cell is used to obtain the position, timing, total energy deposit and direction of the cluster.
The reconstruction for GRAIN lens-based light readout takes as input the two-dimensional images produced by the sub-set of the cameras which have observed each event. The first step of the reconstruction is aimed at identifying the track projections in the images and fitting them. The algorithm is based on the Hough transform~\cite{Hough:1959qva}, multi-Otsu thresholding~\cite{1979:ots} and DBSCAN~\cite{Ester96adensity-based} clustering to isolate, select and fit the tracks in each image. The second step combines the two-dimensional information to obtain a three-dimensional reconstruction of the event. Assuming the pinhole approximation, pixels associated to a reconstructed track are projected back into 3D space to find a volume compatible with their 2D projections. The final outputs are then the reconstructed tracks, an estimate of the energy deposited from each track based on the amount of collected light, and the 3D vertex candidate.
The direct 3D reconstruction for GRAIN with Hadamard masks consists of a custom reconstruction algorithm that obtains a 3D map of the deposited energy. It is based on a probabilistic and combinatorial approach: each hit on the camera pixel is propagated in the detector volume through each mask hole, with an appropriate weight assigned to the segmented volume units (voxels), which measure 1 cm$^3$.  The algorithm is designed to be run on \glsunset{gpu}\dwords{gpu} to provide the required performance.
The event analysis in case of mask-based light readout is not so different from the lens-based case. Clustering algorithms and fit procedures are used for the 2D reconstruction of the tracks. The single-pinhole approximation is assumed to combine the signals from different masks and to get the 3D view of the event. A new algorithm to avoid wrong associations of light pixels on different masks is under study. Also, for the masks, the final results are the vertex identification looking at the most probable 3D voxels and the calorimetric measurement.

\subsection{Common ND analysis files}
\label{sec:nd-caf}

The final output format used to conduct physics analysis will be a simplified
{\it analysis tree} file whose format has not yet been fully defined, which contains a reduced subset of the full simulation
and reconstruction output. The first version of the analysis tree format is
based on the ND CAFs, which are produced by applying a parametric response
model to the \dword{geant4}-level ND simulation for use in long-baseline and other analyses. It is recognized that a more detailed definition of the simplified analysis will be needed in the future, but given the preliminary nature of the ND design greater definition is not possible right now.

\begin{dunetable}
[Average \dshort{ndgar} event wall-clock module execution times]
{l r}
{tab:garsoft_cpu_reco_by_module}
{Average wall-clock processing time in seconds for reconstruction modules for \dword{garsoft} events consisting of just one interaction in the gas.  The processor had an HS06 rating of $\sim$11. An actual spill will contain of order 60 interactions, mostly  in the calorimeter, though many tracks will pass through the gas.  Reconstruction of more complex events with future software is expected to take more CPU per event than shown below.}
Module Label & time/event (sec)\\ \toprowrule
RootInput(read)                             &    0.00\\%0353765       \\
init:EventInit                         &    1.31275e-05       \\
hit:CompressedHitFinder                &    0.00488308        \\
tpcclusterpass1:TPCHitCluster          &     0.0091922        \\
vechit:tpcvechitfinder2                &     0.0103787        \\
patrec:tpcpatrec2                      &     0.0130245        \\
trackpass1:tpctrackfit2                &     0.014211         \\
vertexpass1:vertexfinder1              &    0.000851081       \\
tpccluster:tpccathodestitch            &     0.0269436        \\
track:tpctrackfit2                     &     0.0135842        \\
vertex:vertexfinder1                   &    6.19847e-05       \\
veefinder1:veefinder1                  &    9.96417e-05       \\
sipmhit:SiPMHitFinder                  &    0.00153375        \\
sscalohit:CaloStripSplitter            &     0.0547754        \\
calocluster:CaloClustering             &    0.00492599        \\
trkecalassn:TPCECALAssociation         &    0.000237503       \\
TriggerResults:TriggerResultInserter        &    2.36397e-05       \\
RootOutput                                  &    3.6783e-06        \\
RootOutput(write)                           &    0.214077         \\
{\bf Total}                                  &     {\bf 0.369802}       \\ 
\end{dunetable}

\section{Calibration \hideme{Schellman-draft}}

Before final pattern recognition can be applied to physics events,  the processed hits from calibration samples (subsets of the full data, sometimes with special conditions) are run through specialized pattern recognition and used to derive high-quality calibration constants which are stored in the conditions database for future use.  Inputs include processed hit data but also detailed information about the configuration of the calibration system.  This step will likely be done many times, especially at the start of the experiment. Calibration samples may be taken quickly and may be very large, for example 500\,TB for a full laser calibration of the far detector. They will require occasional fast processing of PB-scale data samples. 

 The large size of some calibration samples means that fast processing for monitoring and fast application will require peaks in data storage and processing at rates considerably higher than normal data taking.

\section{Visualization}
\label{sec:visualization}
Both raw and reconstructed data from each of DUNE's detectors must be visualized graphically in multiple ways in order to successfully design, commission and operate the detectors, and to extract meaningful physics results.  While reconstruction is automated and hand-scanning is no longer used to extract physics results, event displays are critical for a large number of necessary steps:
\begin{enumerate}
    \item {\bf Detector Design Development}   While designing detectors, it is important to visualize interactions in the design in order to ascertain which interactions are easy to identify and measure and which ones are difficult, in order to refine the design.
    \item{\bf Optimizing reconstruction algorithms}  Comparisons of true particle trajectories and energies with reconstructed versions thereof in simulated interactions provides guidance for improving reconstruction algorithms.  Quantitative metrics such as completeness and purity can be optimized, but it is easier to do so when viewing an event display that shows which pieces of which interactions have failed to reconstruct well.
    \item{\bf Operating prototypes and learning from them}  Data from prototypes, such as the \coldbox{}es, the \dwords{protodune}, and \dword{iceberg} must be analyzed quickly to ascertain what the noise level is and to search for the presence of signals.  This is part of an iterative experimental process to commission the prototypes by searching for noise sources, checking \dword{hv} and cryogenic performance, drift medium purity, etc.  Dead channels or other artifacts such as long-range induction effects are plainly visible on raw data displays.  Various artifacts may be addressable with software, such as correcting for front-end amplifier artifacts, correlated noise, and \dword{adc} issues.  These are all discovered first with event visualization.
    \item{\bf Interpreting calibration data}  Laser calibrations, neutron source data and radioactive source data must be inspected first before use, as unanticipated artifacts may confound the intended calibration use.
    \item{\bf Providing educational and public outreach materials}  Introducing new students to DUNE's physics program involves showing them what interactions look like in each of our detectors.  Materials for DUNE's public-facing web sites and press releases also require high-quality data visualizations.
\end{enumerate}
Each of these uses places different requirements on the visualization software.  Furthermore, the event display programs must be responsive and easy to use, though specialist functions may require additional configuration.

\subsection{Two-Dimensional Event Displays}
\label{sec:visualization:2d}

The data from the \dword{sphd} and \dword{spvd} modules are inherently two dimensional, with \dword{adc} values read out per wire (or strip) for each sample time.  Inspecting the raw waveforms side by side in a grayscale or colored map for each of the three views is critical for understanding the features of the raw data.  \dword{larsoft} provides such a display,
optimized for X-Window connections.  Users can zoom in on rectangular subsets of the data and clicking on the image brings up a one-dimensional waveform plot for the corresponding channel.  An example of such a display for a \dword{pdsp} event with cosmic rays is given in Figure~\ref{fig:vis:larsoftraw}.  An example of a two-dimensional display of reconstructed tracks, vertices and showers is given in Figure~\ref{fig:vis:larsoftreco}.  

\begin{dunefigure}
[LArSoft Raw Data Display]
{fig:vis:larsoftraw} 
{Example raw data display produced with \dword{larsoft}.  One \dword{apa}'s worth of data are shown for a \dword{pdsp} trigger record collected in October 2018. The top panel shows the collection-plane data, the next panel V-plane data, and the bottom colored panel shows U-plane data.  Below those three is a single channel's waveform.  The colored panels show the pedestal-subtracted \dword{adc} values for each channel on the horizontal axes with sample time in 500~ns ticks on the vertical axes.  The time between samples is 500\,ns.  White vertical lines in the collection plane correspond to channels that have been flagged as noisy or dead.}
\includegraphics[width=0.9 \textwidth]{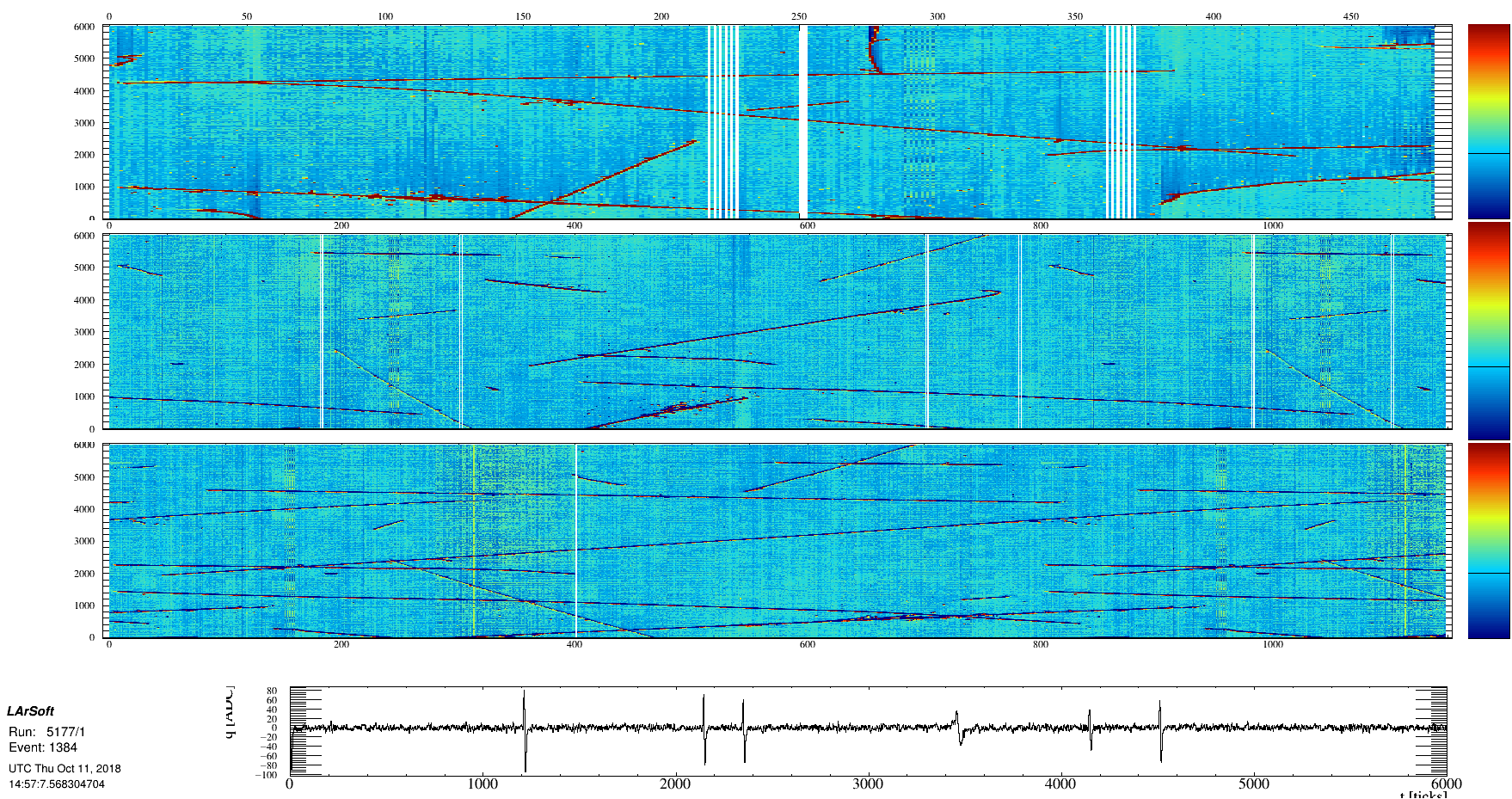}
\end{dunefigure}

The raw data display provided by \dword{larsoft} is best run interactively.  The user can step through the events in a file, forwards or backwards, or skip to a desired event number.  This event display is not built for non-interactive use.  There are use cases in which large numbers of raw event displays need to be produced in a batch.  For this purpose, dedicated modules have been written in DUNE's \dword{larsoft}-based software that provide two-dimensional displays of raw data and data that have been processed through each step of data preparation. Examples are shown in Figure~\ref{fig:vis:dataprep} after four stages of data processing. While there is definitely room for development of features and improved ease-of-use, this is currently the best resource for raw data viewing.

\begin{dunefigure}[Two-Dimensional Displays of ProtoDUNE-SP Data in Stages of Data Preparation]{fig:vis:dataprep}{Example batch-mode event displays for a collection plane showing background reduction in successive stages of data processing.
For each plot, the horizontal axis is the readout sample time in 500~ns ticks, and the vertical axis is the channel number.
The color scale represents the charge for each channel averaged over five ticks
with the range chosen to make the noise visible.
Signals from charged tracks appear mostly in black and are off scale, well
above the noise level.  From Ref.~\cite{DUNE:2020cqd}.}
\includegraphics[width=0.49\textwidth]{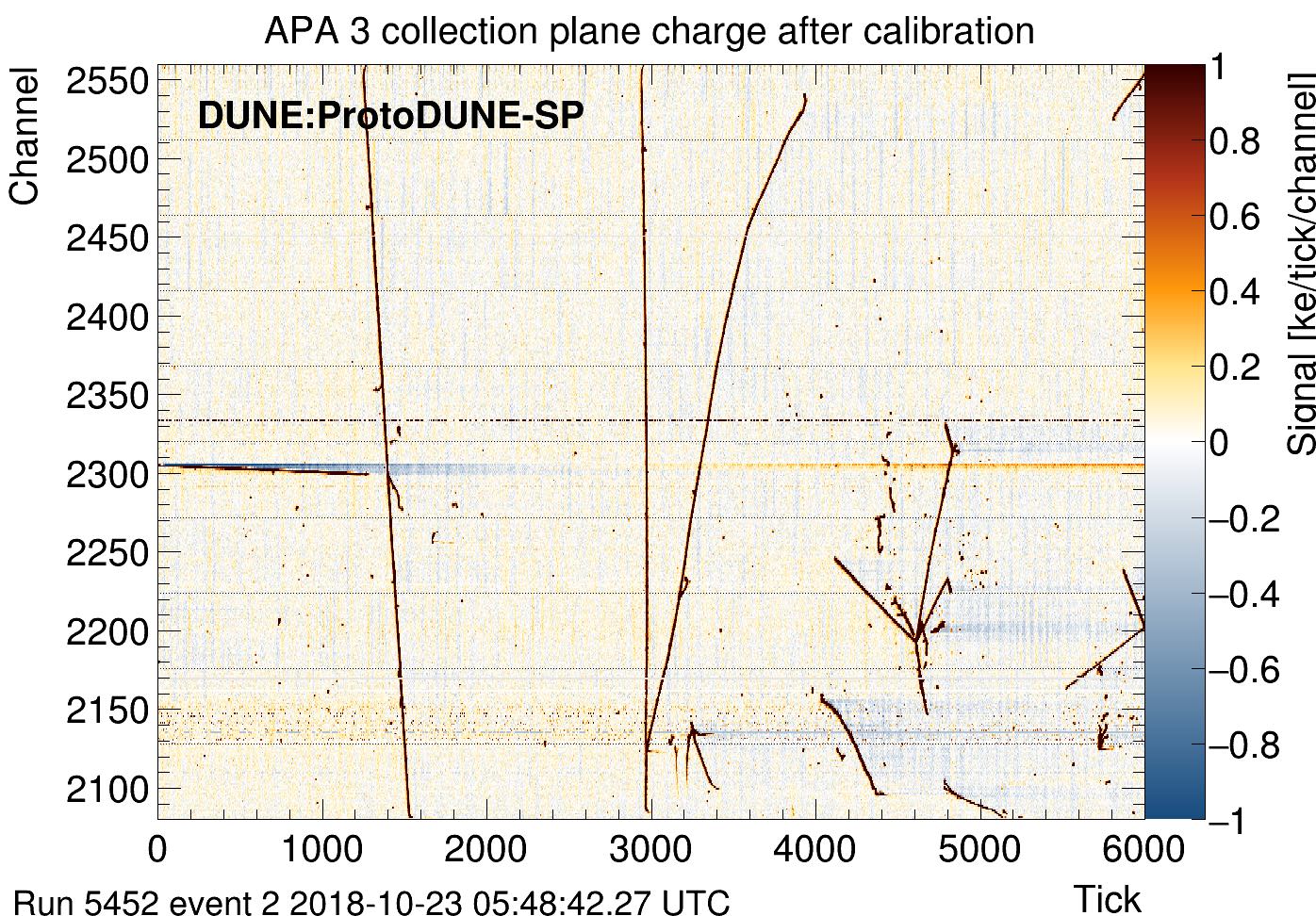}
\includegraphics[width=0.49\textwidth]{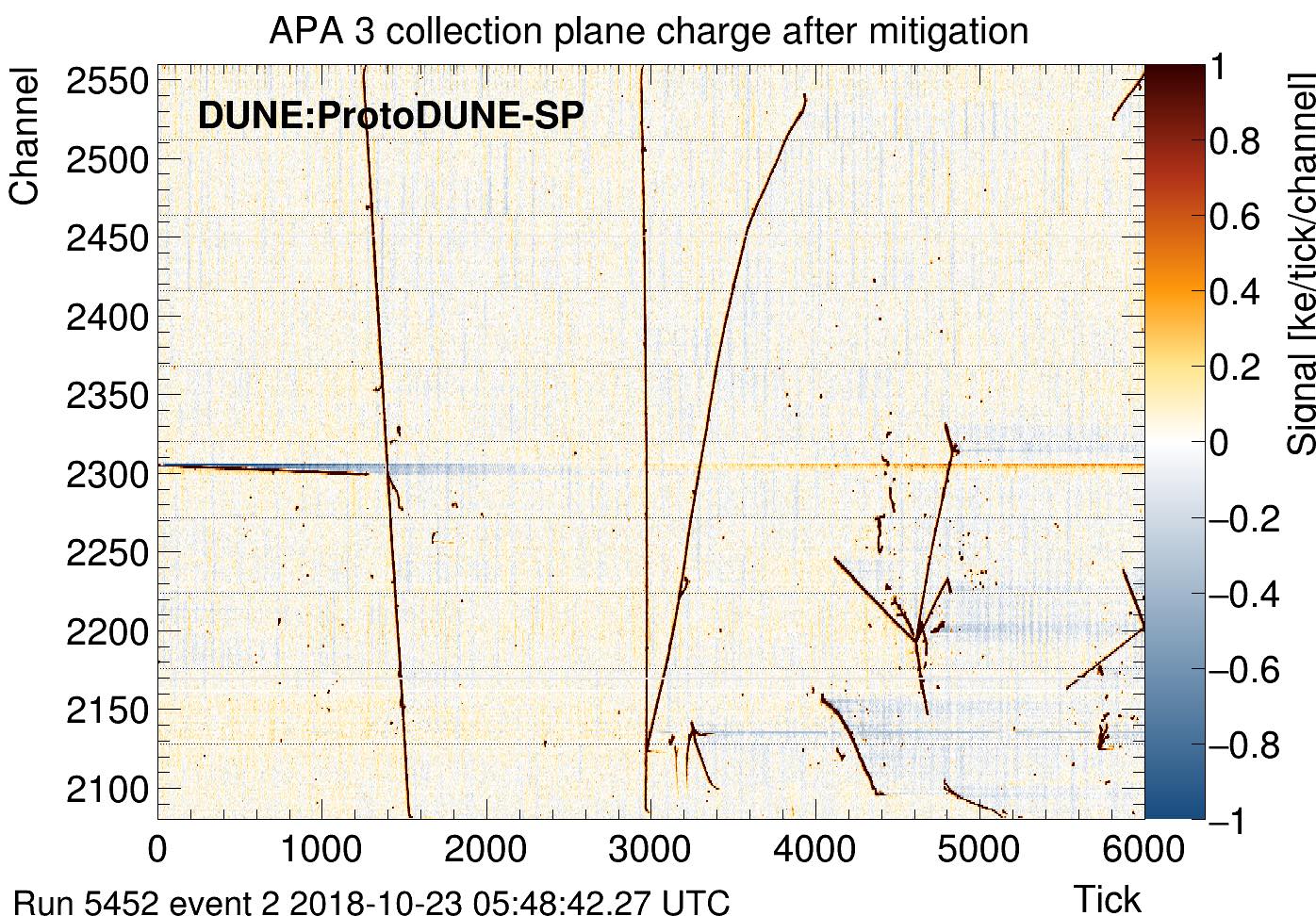}
\includegraphics[width=0.49\textwidth]{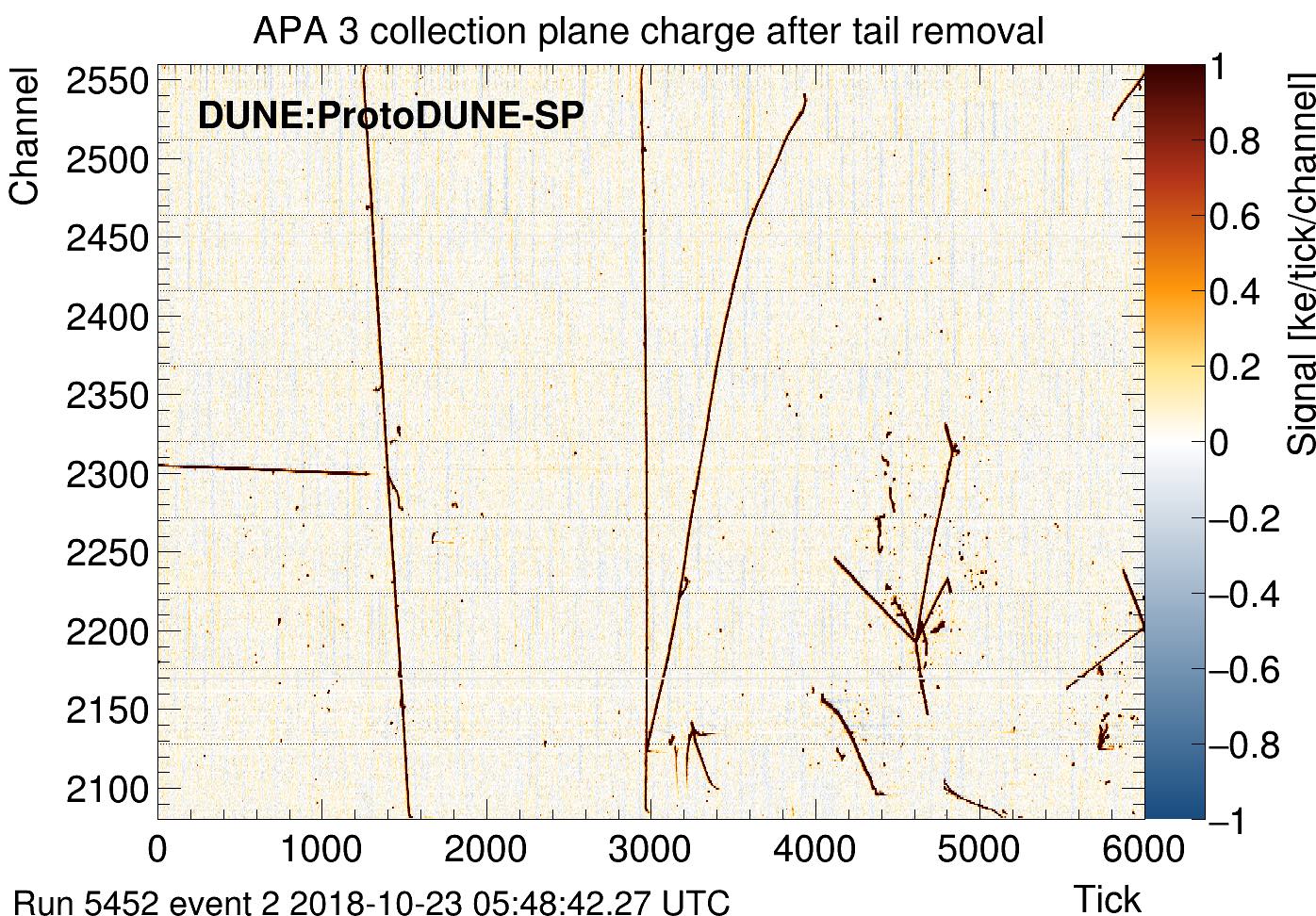}
\includegraphics[width=0.49\textwidth]{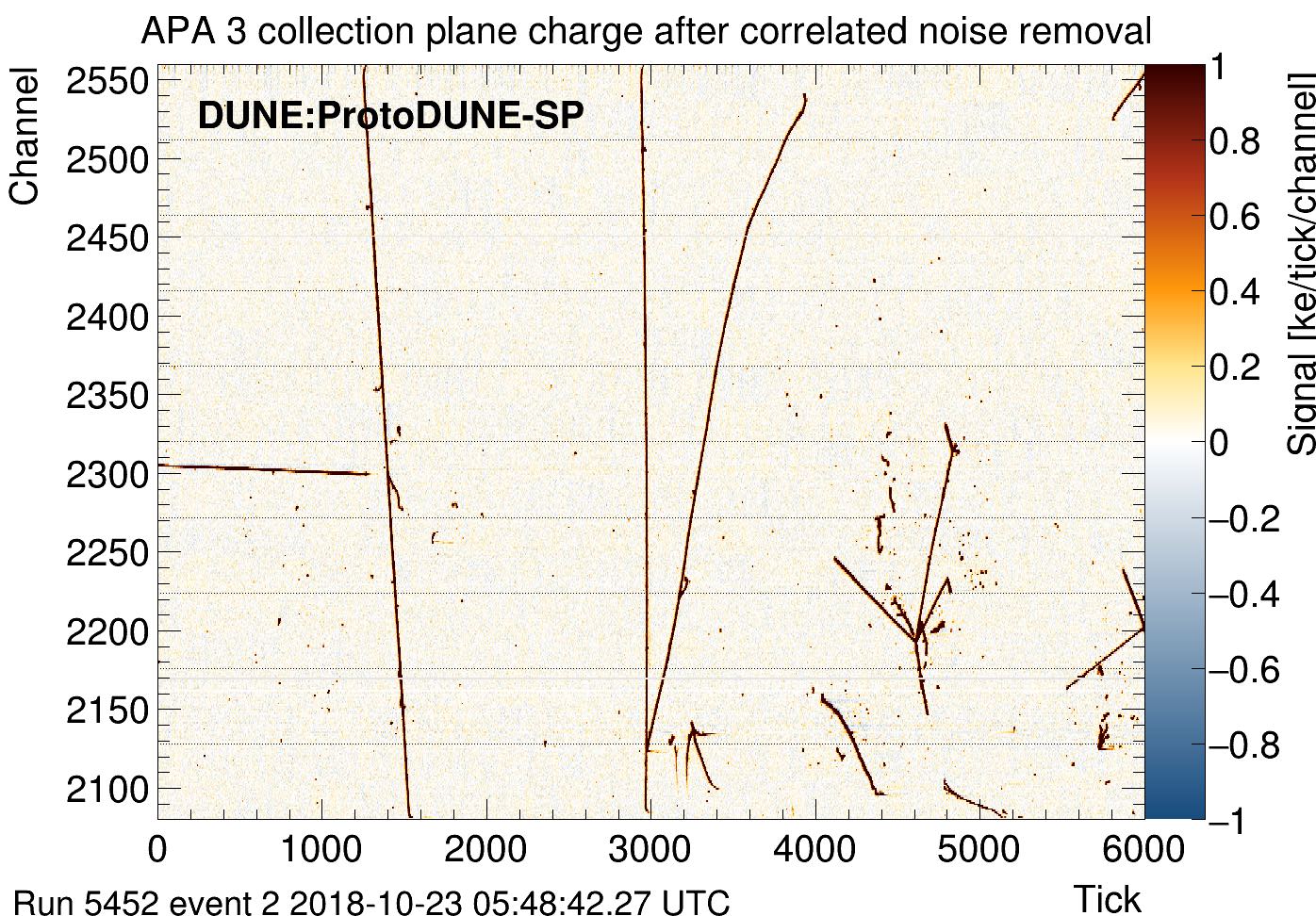}
\end{dunefigure}

\begin{dunefigure}
[LArSoft Reconstructed Data Display]
{fig:vis:larsoftreco} 
{Example reconstructed data display produced with \dword{larsoft}.  One \dword{apa}'s worth of data are shown for a simulated \dword{pdsp} trigger record. The top panel shows the collection-plane data, the next panel V-plane data, and the bottom colored panel shows U-plane data.  Below those three is a panel showing a single channel's deconvolved waveform and fitted hits.  Reconstructed hits, space points and tracks are displayed in the top three panels.}
\includegraphics[width=0.9 \textwidth]{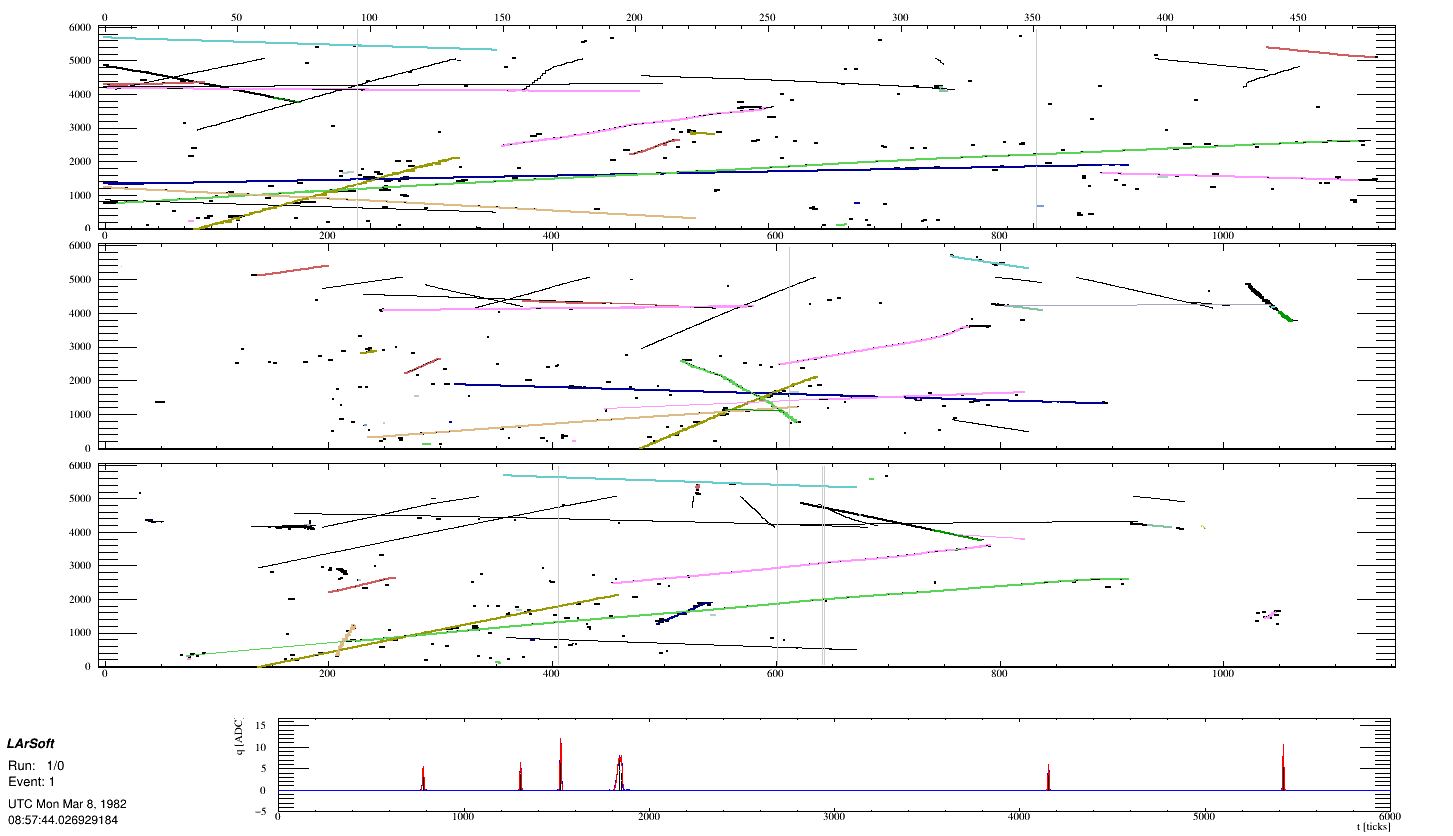}
\end{dunefigure}

\subsection{Three-Dimensional Event Displays}
\label{sec:visualization:3d}

 LArSoft also provides a three-dimensional event display, showing the same reconstructed objects.  An example is given in Figure~\ref{fig:vis:larsoft3dreco}.  Only reconstructed objects are available in three dimensions as the data in the ProtoDUNEs and the Far Detector modules are two-dimensional.
 
\begin{dunefigure}
[Three-dimensional LArSoft Reconstructed Data Display]
{fig:vis:larsoft3dreco} 
{Example three-dimensional reconstructed data display produced with \dword{larsoft}, showing data from a \dword{pdsp} trigger record, for all \dwords{apa}.  Reconstructed space points are shown in gray while shower cones are drawn in red.  The display can pan, zoom, and rotate under user control.  The \dword{larsoft} \threed event viewer can also produce animated gifs.}
\includegraphics[width=0.8 \textwidth]{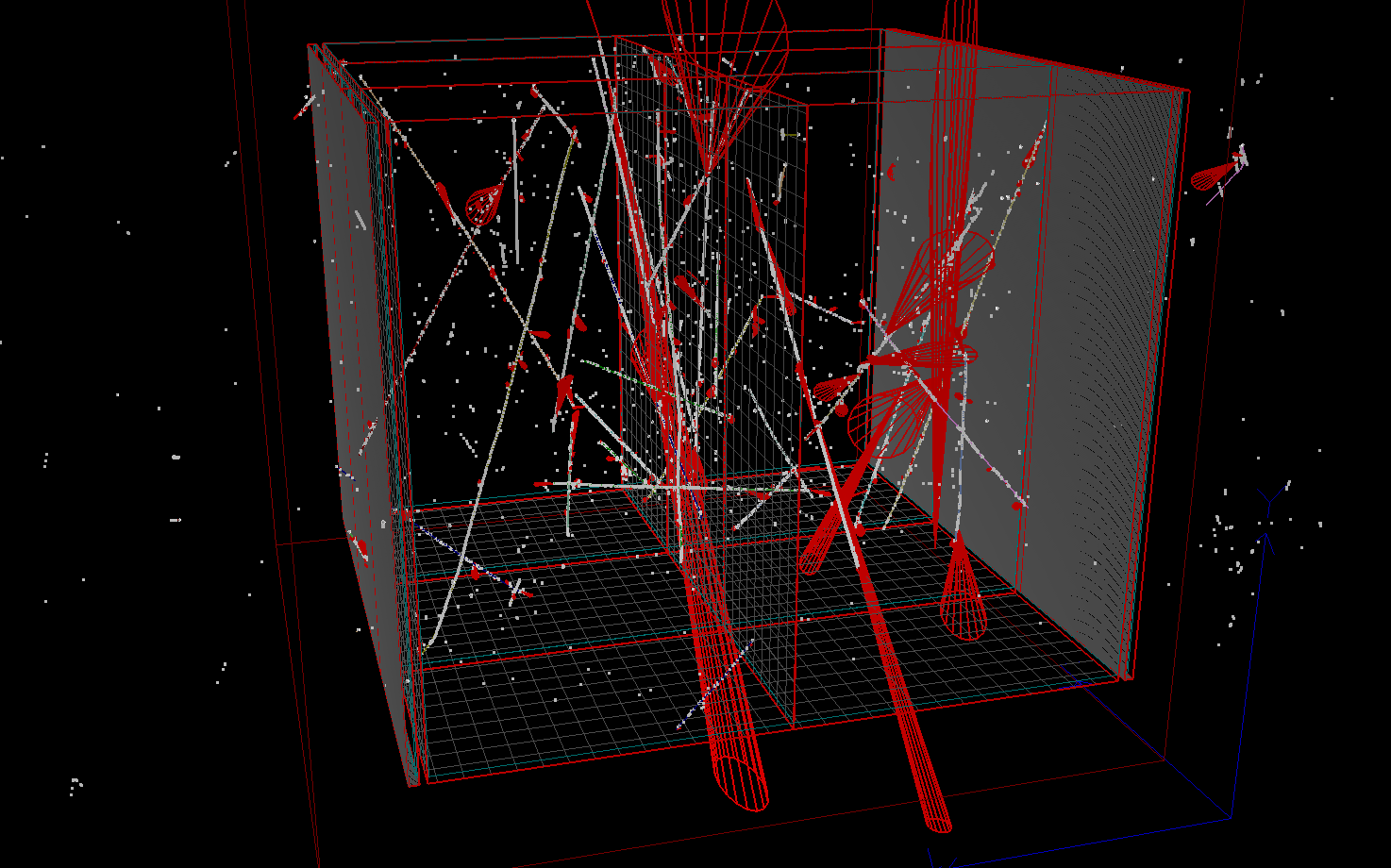}
\end{dunefigure}

\subsection{Web-Based Event Displays}
\label{sec:visualization:web}

In addition to the \dword{larsoft}-based event displays, DUNE has multiple web-based event displays.  One, called the Bee event display, is made for displaying \dword{pdsp} events in a web browser.  The server is hosted at Brookhaven National Laboratory and contains a short list of typical \dword{pdsp} events, one of which is shown in Figure~\ref{fig:vis:bee}. No software needs to be installed on the user's computer, and no data files need to be copied.  The user only needs a modern web browser and the experience is enhanced with a capable graphics card.

\begin{dunefigure}
[Three-dimensional ProtoDUNE-SP Bee Data Display]
{fig:vis:bee} 
{Example three-dimensional reconstructed data display for a \dword{pdsp} data trigger record produced with the web-based Bee event display.  Users interact with the data needing only a modern web browser on their computer.  They can pan, zoom, rotate, and change display colors, point sizes, and transparency.}
\includegraphics[width=0.9 \textwidth]{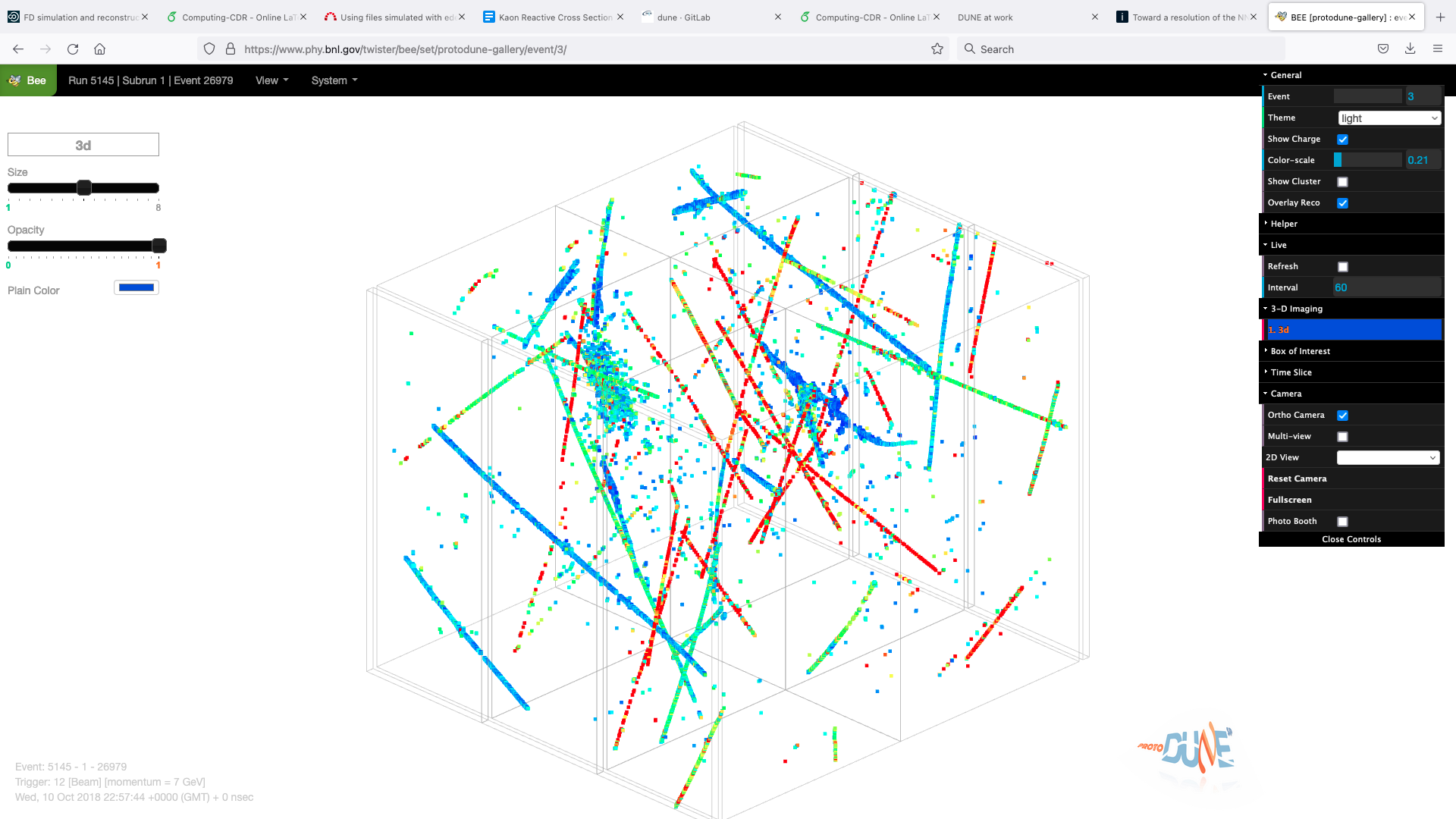}
\end{dunefigure}

Another web-based event display is called WebEVD.  It runs as a \dword{larsoft} module, and it sets up a local web server owned by the user process on an interactive computer.  The user then must ssh in to the interactive computer running the web server, with a specified port forwarded.  The user then connects to that port with a local web browser and interacts with the event display with it.  This user-owned web server which is only visible via authenticated ssh has the advantage that it provides the user with full control over which data are to be displayed, as the web server is not centrally managed.  An example display of two-dimensional deconvolved waveforms for a simulated $\nu_e$CC event in a \dword{fd} workspace geometry run is shown in Figure~\ref{fig:vis:webevd2d}.  An example three-dimensional event display from a \dword{pdsp} data run produce using WebEVD is shown in Figure~\ref{fig:vis:webevd3d}.

\begin{dunefigure}
[Two-dimensional WebEVD of a simulated $\nu_e$CC interaction.  ]
{fig:vis:webevd2d} 
{A simulated $\nu_e$CC interaction in a far detector workspace geometry run. The display is of charge reconstructed on collection wires (recob::Wire). The horizontal axis is the $z$-axis (beam direction) in the detector. The vertical axis corresponds to drift time ($x$ position in the detector). Both axes are labeled in cm. Rendered with WebEVD. }
\includegraphics[width=0.9 \textwidth]{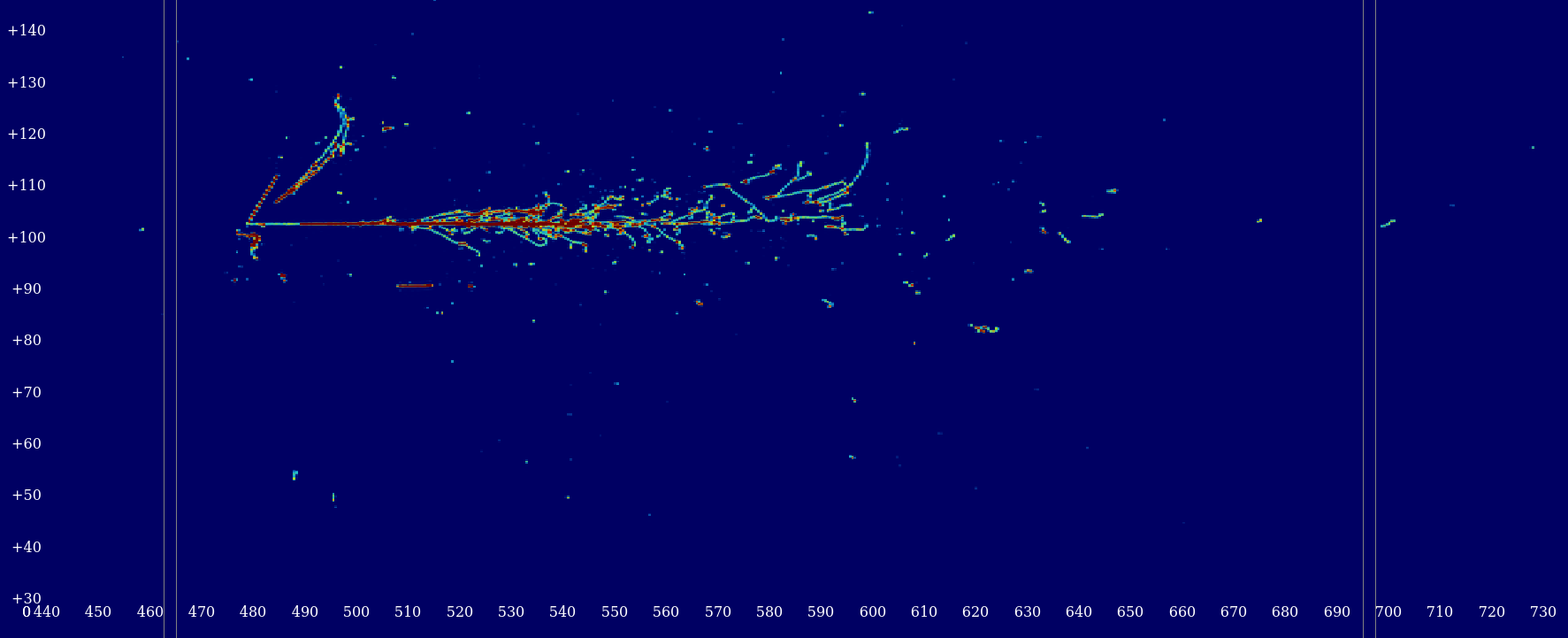}
\end{dunefigure}

\begin{dunefigure}
[Three-dimensional WebEVD of a ProtoDUNE-SP trigger record. ]
{fig:vis:webevd3d} 
{ Example three-dimensional reconstructed data display for a \dword{pdsp} data trigger record, rendered with WebEVD. Reconstructed tracks and space points are shown in blue, and detector elements are shown in red.}
\includegraphics[width=0.9 \textwidth]{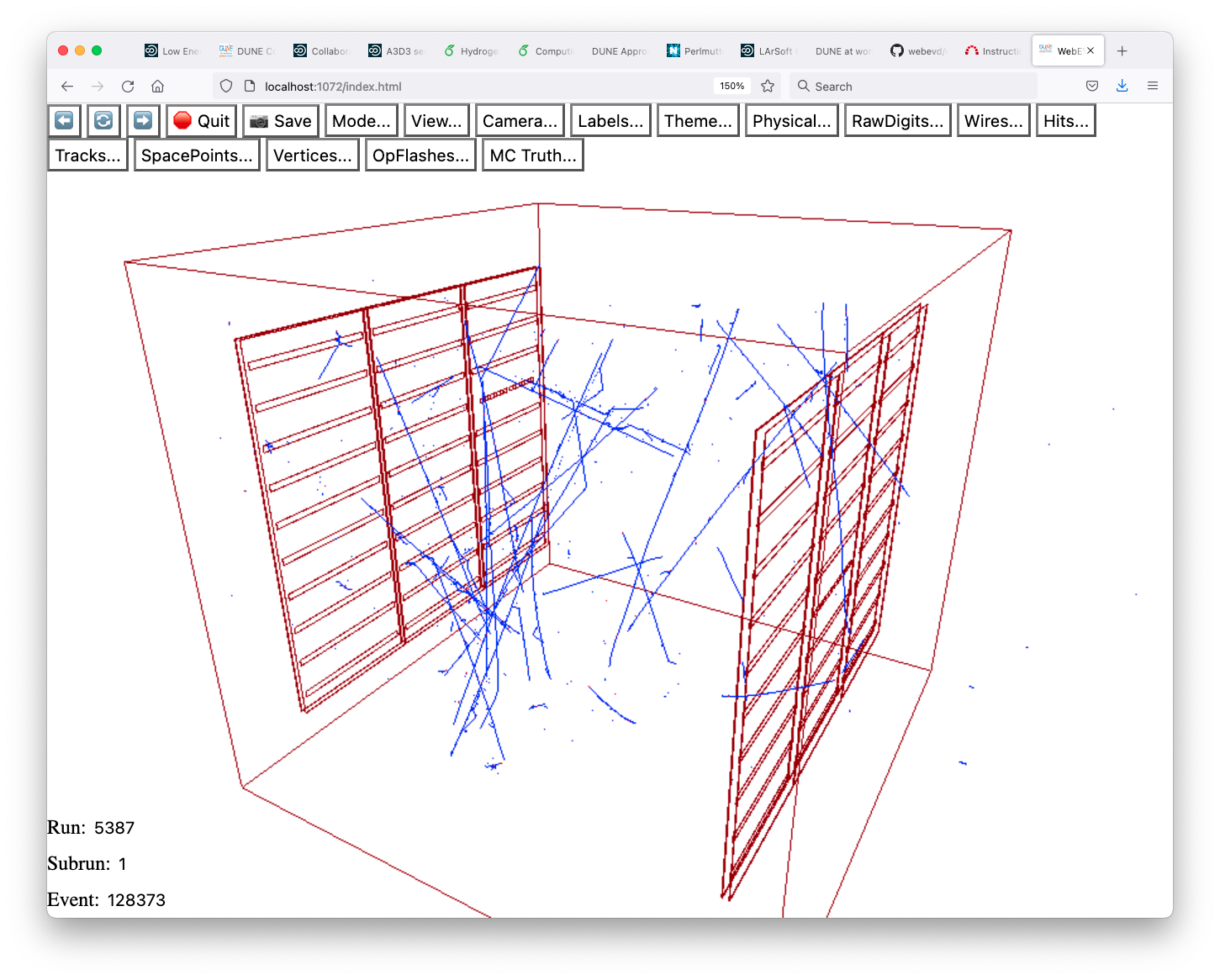}
\end{dunefigure}

\subsection{Near Detector Event Displays}
\label{sec:visualization:neardetector}

An example of a reconstructed event in \dword{ndgar} is shown in Figure~\ref{fig:vis:garsoftreco} rendered with the \dword{root}-based event display in \dword{garsoft}.  The same event is shown in Figure~\ref{fig:vis:garsoft2}, rendered with the TEve-based event display in \dword{garsoft}.

\begin{dunefigure}
[GArSoft reconstructed data display, ROOT-based]
{fig:vis:garsoftreco} 
{Example reconstructed $\nu_\mu$CC event produced with \dword{garsoft}, using an ALICE-inspired detector geometry.  The reconstructed vertex is indicated with a yellow dot, and tracks are shown in green, blue, and red.  Reconstructed calorimeter clusters are shown with green tetrahedra.  The annotations giving the \dword{mc} particle identities were added by hand.}
\includegraphics[width=0.9 \textwidth]{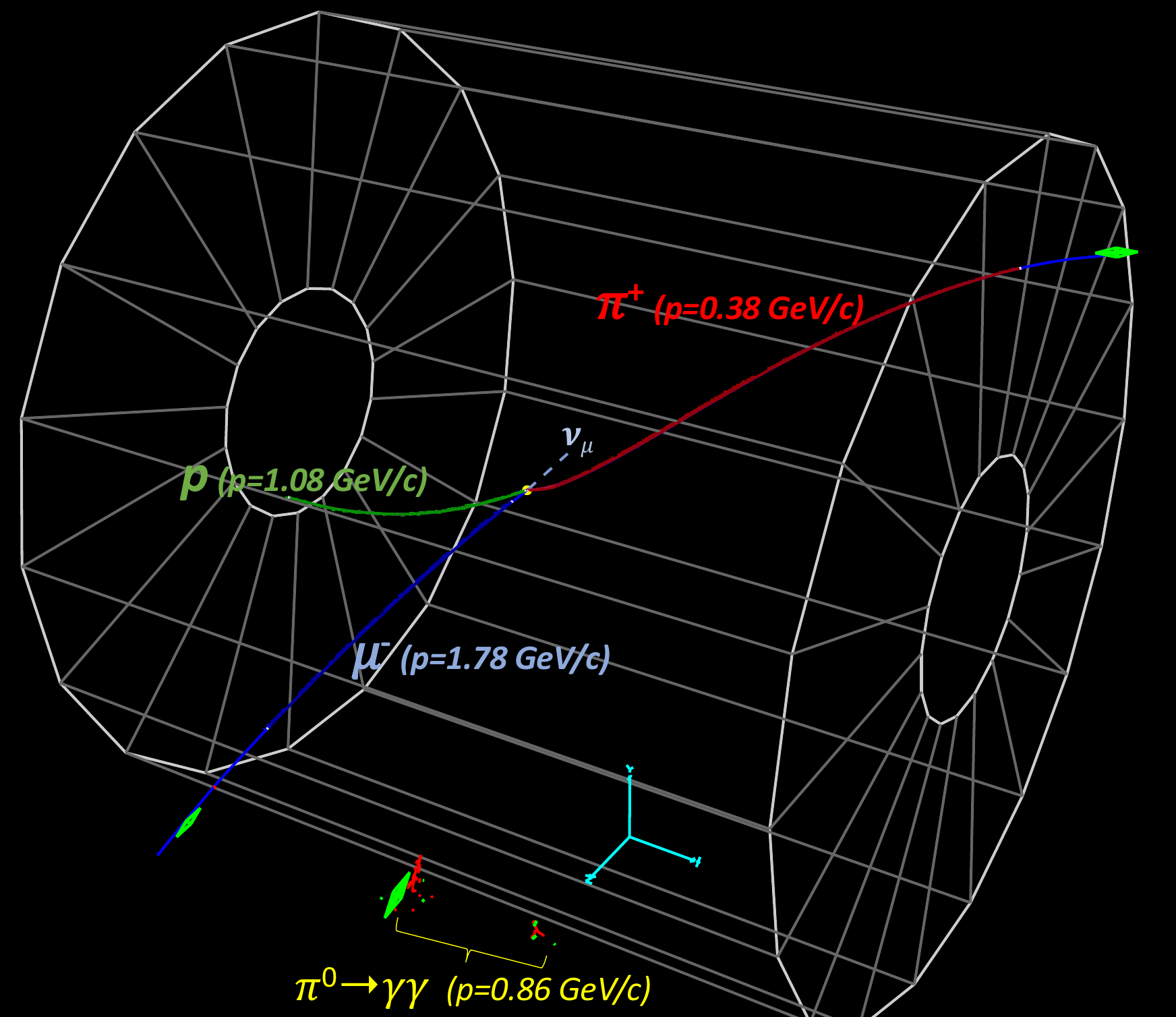}
\end{dunefigure}

\begin{dunefigure}
[GArSoft Reconstructed data display, TEve-based]
{fig:vis:garsoft2} 
{The same \dword{ndgar} event as shown in Figure~\ref{fig:vis:garsoftreco}, rendered here with  \dword{garsoft}'s TEve-based event display.}
\includegraphics[width=0.9 \textwidth]{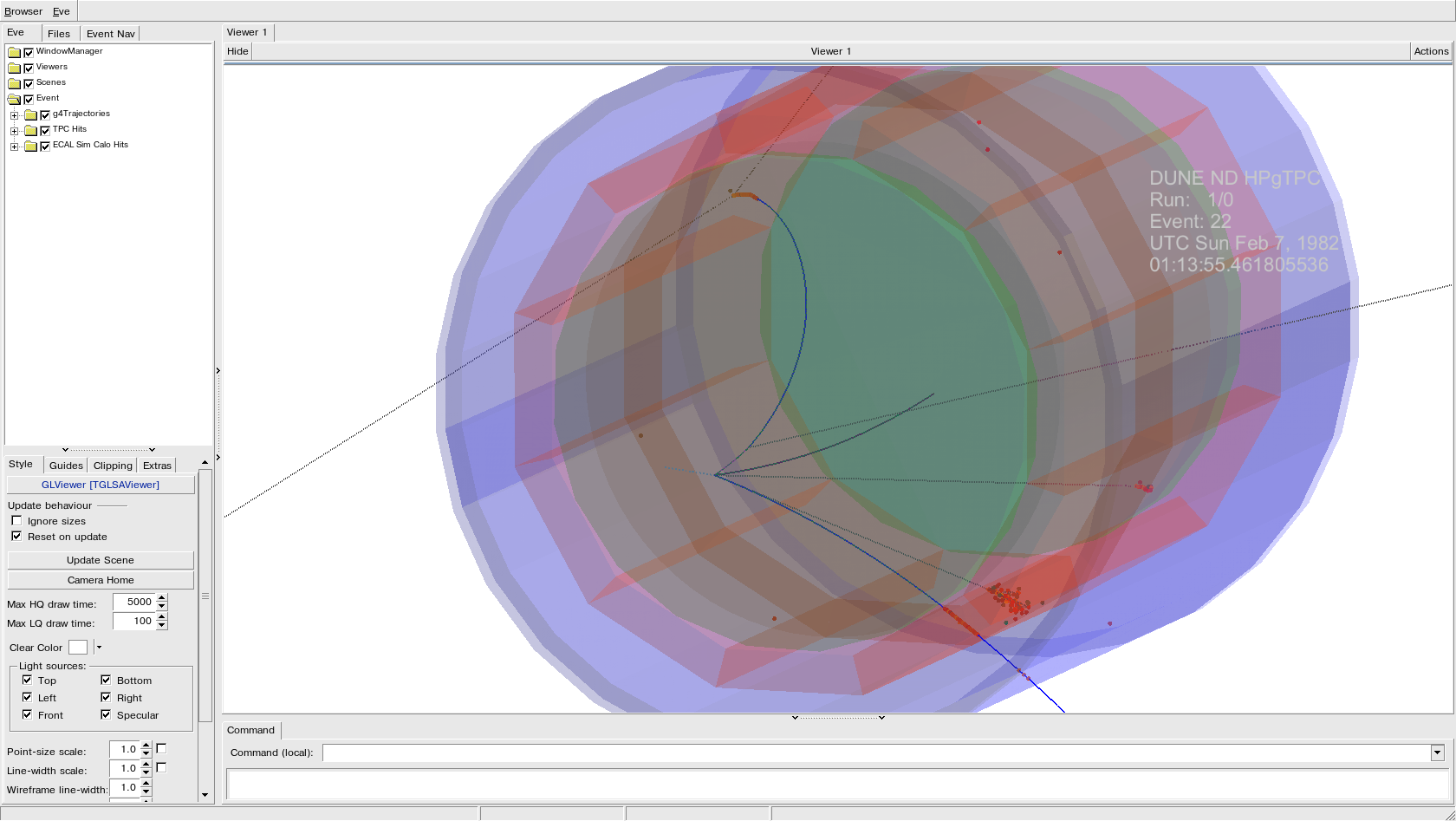}
\end{dunefigure}

The third component of the near detector, \dword{sand}, consists of a straw-tube tracker and a calorimeter inside a solenoidal magnet.  A small cryostat containing liquid argon, called \dword{grain}, %
is to be installed upstream of the straw-tube tracker inside the \dword{ecal}.  Design and performance studies are underway.  Basic event visualization tools have been developed to aid in the development of the reconstruction algorithms.  An example display of simulated and reconstructed particles in \dword{sand} is shown in Figure~\ref{fig:vis:sandevd}.

\begin{dunefigure}
[SAND Event Display]
{fig:vis:sandevd} 
{A rendering of a simulated full spill of neutrino interactions in the \dword{sand} detector. Magenta stars indicate true neutrino interaction vertices, blue curves indicate true simulated muon tracks, and thick green curves show proton tracks  Thin red lines indicate electrons and photons.  The red circles show the outline of the \dword{ecal}.  Blue dots indicate calorimeter hits.}
\includegraphics[width=0.9 \textwidth]{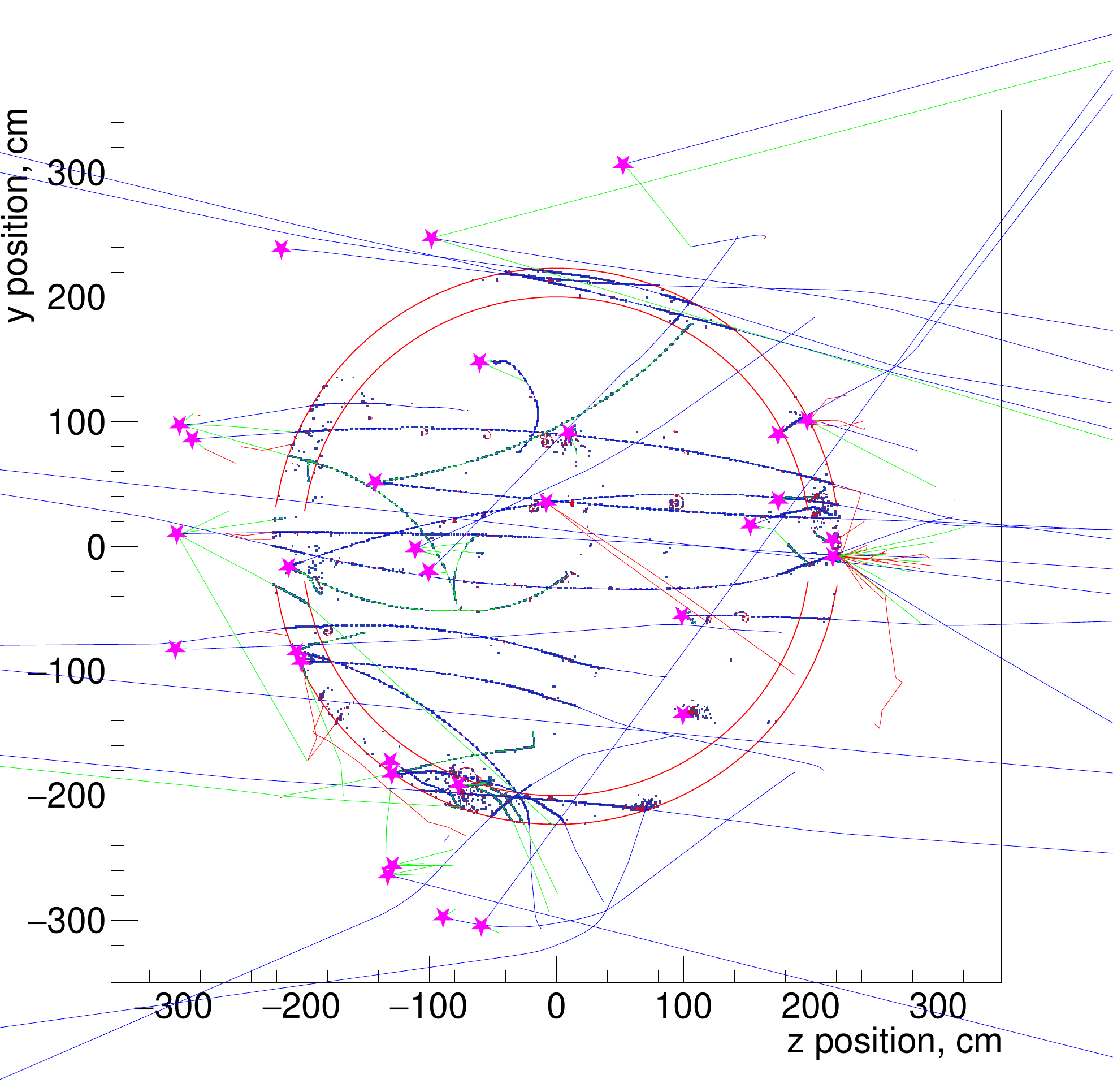}
\end{dunefigure}

\section{Analysis of Reduced Data Samples}

\subsection{Analysis Sample Production}
At some point in the processing, specialized datasets and streams based on trigger type, reconstructed event information, and intended use need to  be defined and produced.

The interaction data, which are in the output format supplied by the full reconstruction are then reduced and reconfigured into analysis formats for use by users. 
In the long run, this processing will be done as coherent production steps but may also be done by small groups while the data formats and procedures are being developed.

CAF (Common Analysis Files), the standard analysis-level file, is an example of a reduced data sample. DUNE's primary analysis framework for oscillation physics is currently the \dword{cafana} framework, originally developed for the NOvA experiment \cite{Backhouse:2015wlj,  bib:cafana}. The \dword{highland} framework used by T2K is also used for some analyses.

  This phase of processing is I/O limited and can put considerable strain on storage and network resources.  The CAF framework and calibration drive the need to put most of the reconstructed data and simulation on disk at distributed sites as described in Part~\ref{part:model}.
A typical \dword{pdsp} data or simulation pass produces  10,000--100,000 2-8\,GB reconstructed files and  then produces much smaller tuple outputs for analysis.  These reduction jobs stream using \dword{xrootd} and are I/O bound. Preliminary monitoring studies indicate that average input rates of 5-30\,MB/sec per process can be achieved when the source \dword{rse} and processing are co-located (as is possible at the largest facilities) and the data can be streamed via the \dword{lan} within the site, with aggregate rates of up to several GB/sec observed when many batch processes run at a given site.  We are currently mining data access records to measure rates and reliability as a function of source and sink.

\subsection{Reduced Analysis Samples}

Reduced data analysis samples  are the samples users see and analyze when they are not developing new reconstruction  or calibration algorithms. Analysis samples should be useful and as small as possible.  Analysis codes should not need to read from the central databases but may need to access small local replicas. 

Data analysis may be done on local clusters, on collaboration grid facilities using shared data samples, or in dedicated analysis facilities that offer advanced access methods such as the Columnar Analysis framework \dword{coffea}~\cite{Smith:2020pxs}.  %
This phase of processing is typically very I/O bound and requires fast access to smaller data samples.

An important question is what auxiliary information is needed and how it will be delivered. In particular, geometry and run-quality information are often needed in final analysis stages while direct access to detailed electronics calibrations is less often needed. 

\subsection{Current Practices}

A survey of data analysis users was done in February 2022.  Analyzers were contacted through the physics groups and analysis Slack channels.  We received 29 responses with users from institutions in eight countries dominated by the US(14) and UK(8). 
Responses were spread across multiple physics efforts: \dword{protodune}	(13),
Near Detector	(5),
Low Energy FD	(9),
High Energy FD	(7) and 
Calibration	(5).

Figure \ref{fig:ch:use:survey} summarizes the results.  \dword{larsoft} is used by most users (presumably for small scale tests or tuple creation).  Users then rely on their own ROOT-based C++ or Python code for the final analysis steps.  A small number of users use shared analysis frameworks such as \dword{cafana}, \dword{nuisance}~\cite{Stowell:2016jfr} and \dword{highland}.  Jupyter notebooks are becoming popular. %

Users analyze their reduced samples (which range in size between 5 GB and 10 TB) on the \dshort{fnal} interactive machines, via the grid batch systems and on local clusters and desktops.

Although the physics groups tend to use the same technologies to analyze their data, the particular data content of their tuples differs.  Many users generate their own formats (at considerable expense in time and effort). Physics groups are beginning to move to common shared production as the content choices stabilize.  A single common format with common content is unlikely to emerge given the wide range of content different use cases (calibration, far detector low-energy signal, near detector, \dword{protodune} studies) that are already present.

The current goal for computing infrastructure is to support a wide range of formats and analysis strategies while encouraging shared, well-documented, cataloged samples. 

\begin{dunefigure}
[DUNE analysis code survey responses]
{fig:ch:use:survey}
 {User responses to the analysis survey. Panels show the data formats used, the analysis frameworks used, the locations where the jobs are run, and the locations where code is kept.}
\includegraphics[width=6 in]{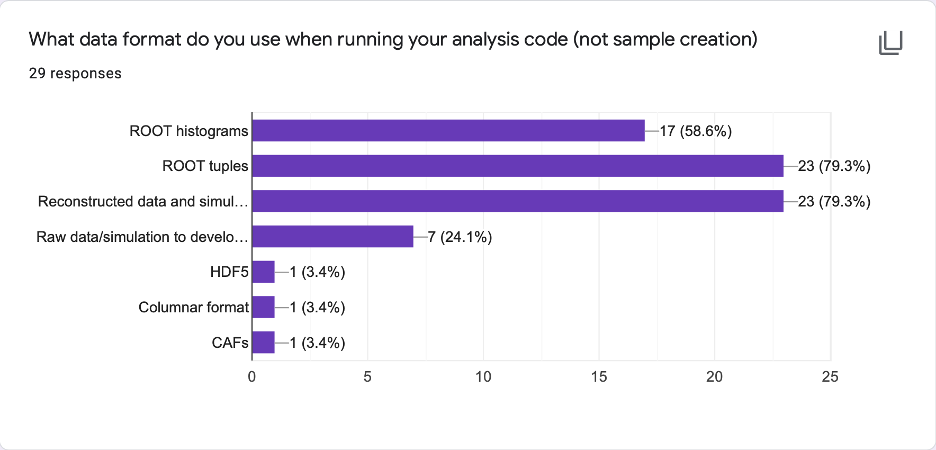}
\includegraphics[width=6 in]{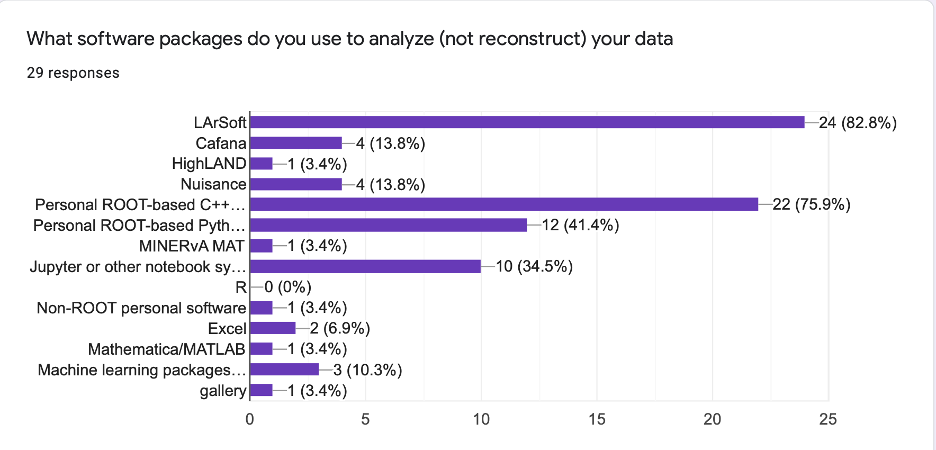}
\end{dunefigure}
\begin{dunefigure}
[DUNE analysis code survey responses 2]
{fig:ch:use:survey2}
 {User responses to the analysis survey (continued). Panels show the data formats used, the analysis frameworks used, the locations where the jobs are run, and the locations where code is kept.}
\includegraphics[width=6 in]{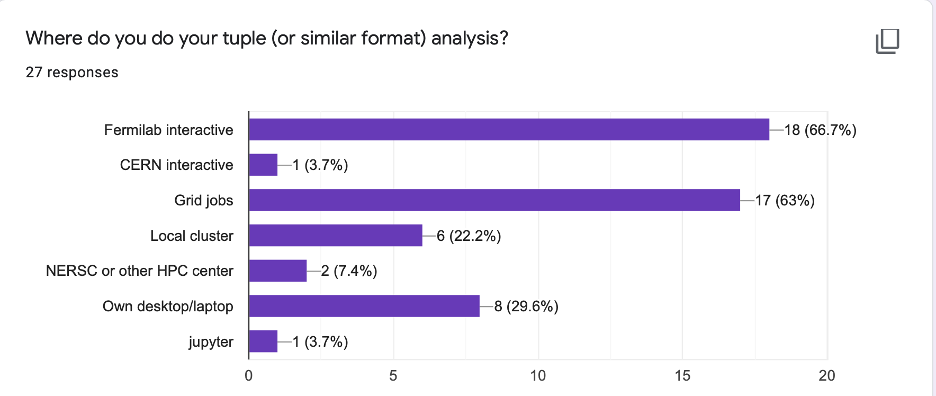}
\includegraphics[width=6 in]{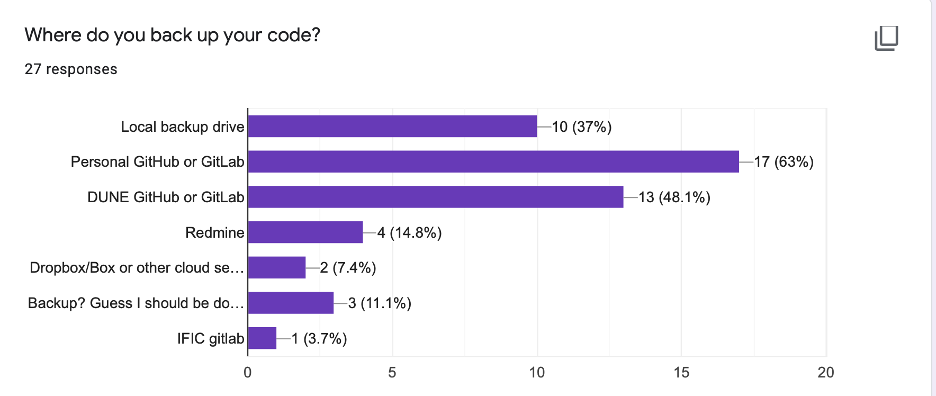}
\end{dunefigure}

\ignore{\subsection{Example of \dshort{protodune} Analysis \hideme{draft}}

\fixme{(anne) is this section really needed for a CDR? Lots of undefined terms and details, and the reader might get bogged down - I didn't review}

One of the \dword{protodune} analyzers %
provided information on workflow for a typical ongoing analysis.

\subsubsection{Ntuple Production}

Standardized analysis ntuples are produced on the grid from the reconstructed PDSPProd4a simulation samples (133 TB, 691,200 events)  and  PDSPProd4 data (48 TB, 1,295,813 events )  for an example 1 GeV/c beam momentum setting. The tuple production code pdspana  partially reruns partial reconstruction to include updated calibrations and then extracts physics information and stores it in root ntuple format. The full MC ntuple took 2,332 CPU hours to create. The data ntuple took 335 CPU hours to create. This processing is found to be IO bound even when reading from local dCache storage at $\sim 20-30 $MB/s. The outputs are 14 and 13 GB in size respectively and cataloged in the \dword{sam} system.  

\subsubsection{Preliminary Event Selection Studies}

A set of python scripts are then used to produce distributions of important reconstructed observables. These are used to refine event selections, including determining cut values to use and estimating efficiencies and purities. These scripts use pyroot to interface with the data/mc files and to create/display the distributions. These studies  are done interactively on a dunegpvm virtual machine at Fermilab.
Refined ntuples are then produced with bad events (mainly those missing beam information) filtered out and labels applied to simulated information. The event selection code uses ROOT’s `RDataFrame' class to perform filtering and labeling. This takes about 20 minutes to perform for the full ntuples.  These final samples contain 159,083 data and 170,238 simulated events. 

\subsubsection{Cross Section Analysis Extraction}

The filtered/labeled ntuples are then passed to custom  ROOT based C++ fitting code for cross section extraction. Multiple types of extraction fits are performed with varying input parameters and samples, including data, full simulation and toy samples.  
Individual extractions are run interactively or on the grid for fake data tests and the ultimate real data fit. They can take O(5-45) minutes depending on how the fit is configured (i.e. what percentage of the input MC is used or how many parameter throws are performed post-fit to propagate error to measured/extracted cross sections). Sets of 1000 toy fits – where the nominal MC is systematically and statistically varied and used as fake data – are performed to study the overall behavior of a specific configuration of the fit.

}

\section{Machine Learning Training and Implementation} \label{sec:google}
Neutrino science, and \dword{hep} more generally, has been increasingly relying on \dword{ml} methods for track and shower reconstruction, particle ID, event selection and classification, and more.  
Many simulation and reconstruction algorithms in the DUNE software stack can profitably use \dword{ml} algorithms, and some already do.  While the inference stages of these algorithms can run quickly on each event, the training stages require significant resources. In some cases, the inference stages run much more efficiently on specialized hardware.  Some non-\dword{ml} algorithms also benefit from specialized hardware. Multiple groups are working on implementing some of \dword{dune}'s algorithms on \dwords{gpu}.  A notable example is reference~\cite{Wang:2020fjr} in which \dword{protodune} reconstruction workflows on simulated far detector neutrino interactions are implemented via the \dword{sonic} framework developed at Fermilab. This hybrid framework enables the use of remote \dwords{gpu} as call-outs within the normal {\it art}/LArSoft workflow.  In particular, the electromagnetic shower algorithm, {\tt EmTrackMichelId}, runs a \dword{cnn} inference step. When run on a CPU, it is the algorithm that takes the most CPU time, as seen in  Table \ref{tab:protodune_cpu_reco_by_module}.  To speed it up, this algorithm was run on commercial (Google Cloud) remote \dwords{gpu}. Patches of detector data were sent to the remote processors over TCP/IP and inference results were returned.  In the initial test, event processing, which is dominated by that step, was sped up by a factor of 17. Similar approaches are very promising for the future.  

Other \dword{ml} algorithms export data from {\it art}/LArSoft jobs in formats that can be read by external \dword{ml} packages, and the training is done by a separate program.  Algorithms for reconstructing vertices, $\nu_\mu$CC and $\nu_e$CC events are written as \dwords{cnn} and trained in this way.  The \dword{cnn} parameters from the training step are stored in a file that is then loaded by an {\it art}/LArSoft job that then calls TensorFlow to run the \dword{cnn} inference during the reconstruction phase.  An experimental infill algorithm that interpolates data for dead or broken channels also operates in this way.

\section{Parameter Estimation}

Once the experimental observations are made (i.e. the number of and energies of different charged current interactions in the near and far detectors), these observations must be linked back to the underlying physics models explored through a formal process of parameter estimation.  In its most basic form, parameter estimation takes the form of a ``fitting'' procedure where by {\em observed} spectra are fit to {\em prediction} spectra based upon PMNS or other oscillation models.  However, the statistical nature of the observations combined with the experimental uncertainties intrinsic to the measurements, and the natural degeneracies that are present in different neutrino interaction and oscillation models, leads to a statistically ill defined backwards transformation problem (i.e. the mapping from the observation space to the model space is not a single valued transform).  In particular the formal process of parameter estimation relies on a number of statistical techniques that determine the most likely regions of parameter space and their neighborhoods which would result in the observation that was made by the detectors.

In the case of the neutrino oscillation measurements, the combination of the low statistics of many of the observations (i.e. the $\nu_e$-CC and $\bar{\nu}_e$-CC events in the far detector) along with the periodic form of the transition probabilities of the oscillation models, lead to a violation of Wick's theorem and prevents the use of Gaussian statistical approximations for the interpretation of $\chi^2$ distributions and other frequentist statistical techniques.  Instead, neutrino oscillation parameter estimations from experiments such as NOvA \cite{NOvA:2021nfi} have relied on Feldman-Cousins based techniques which profile over nuisance parameters, and are able to obtain proper statistical coverage of multi-dimensional parameter spaces.  Parameter estimates from other experiments such as T2K have used Bayesian techniques combined with Hamiltonian Monte Carlo integration techniques \cite{T2K_2021} to probe similar parameter spaces.

The issues with these techniques, which will become even more challenging for DUNE, is that their dimensional scaling runs with both the dimensionality of the fundamental parameters of interest being estimated, and also with the dimensionality of the auxiliary or nuisance parameters that are included in the computations.  In the case of the modern results from NOvA, pairwise estimation of PMNS parameters (i.e. $\sin^2\theta_{23}$ and $\Delta m^2_{32}$ or $\sin^2\theta_{23}$ and $\delta_{CP}$) are extracted against a field of $\mathcal{O}(50)$ nuisance parameters, which are profiled over.  These computations require repeated fitting/minimization computations against pseudo-experiment distributions that are ``thrown'' using Monte Carlo techniques to provide adequate coverage of the underlying model and observation spaces (i.e. both the neutrino interaction model spaces and the smearing and intrinsic resolutions of the detector apparatus spaces need to be varied to provided realistic coverage of the actual observation.)  In the case of experiments like NOvA, which are exploring the parameter spaces at $2-3~\sigma$ significance levels at $< 5\%$ induced computational uncertainty this can result in the need to generate and fit in excess of 9k pseudo-experiments per point in the primary parameter space near the $3\sigma$ confidence boundaries.

Taken as a whole, parameter estimation has proven to be a significant portion of the total computational budget for modern neutrino experiments.  In this respect the parameter estimation portion of NOvA results required 73 million core hours of computation in 2018, 78.5M in 2019, 127M in 2020 and most recently the 2022 neutral current results required 55 million core hours (although this result was only report to the $2\sigma$ significance level).  In all of these cases, the experiment was able to obtain these resources from the HPC facilities at NERSC and were able to transition their codes to run first on Edison, then Cori-Haswell, Cori-KNL, and most recently the Perlmutter system.

Techniques and infrastructure for executing these computations using high performance computing platforms (HPCs) have been developed by the ASCR-supported SciDAC program, in conjunction with NOvA and SBN.  These codes have been used by NOvA and SBN to obtain their published results starting in 2018 and have become the {\it de facto} standard for parameter estimation in these experiments.  It is expected that DUNE will adopt this approach, and build upon the performance improvements that have been so far pioneered by these projects.

\section{Summary: Characteristics of Large-Scale Processing Tasks}

A large number of complicated tasks have been listed above.  Table \ref{tab:tasks} summarizes the computing characteristics of the most prominent ones per MB of data. In combination with data volumes for any given task, these estimates can be used to predict processing needs as described in Chapter~\ref{ch:est} and to optimize data and job placement as described in Chapters~\ref{ch:datamgmt} and~\ref{ch:wkflow}. 

\begin{itemize}
    \item Simulation requires flux files (and possibly overlay libraries) as input but produces much larger outputs. The impact on networks is negligible but the memory needs are substantial.
    \item Reconstruction of data is assumed to be done by streaming the input raw data to most likely a computing site that is separate from the raw data storage. The processing time/MB is large enough that the impact of streaming raw data on networks is minimal. 
    \item Ntuple creation and calibration require multiple passes over the reconstructed samples. Either local data stores or streaming from ``nearby'' sources is optimal as network speeds become important. Section~\ref{ch:model:perf} describes studies of job throughput as a function of disk and CPU locality. 
    \item Parameter estimation has been done using the \dword{hpc} resources at \dword{nersc}, and it is anticipated that additional HPC sites will be available in the future.  In this case, a small amount of input data is processed simultaneously with systematic variations across a large number of cores. 

\end{itemize}

\begin{dunetable}
[Summary of computing resources needed per file]
{l r r r r r r }
{tab:tasks}
{Summary of resources needed per file for compute intensive tasks. The total CPU needed for a task can be determined from the total size of the sample and these numbers.  Here the time is in seconds on a processor with rating around 11 HS06.}
Use case	&memory &	input file size &	output file size 	&	CPU time 	&	input  	& cores/job		\\
units	&GB	& MB	&MB	&	sec	& MB/s	&		\\

Simulation+reco	&		6	&	100	&	2000	&	27000	&	0.00	&	1	\\
data reco	&	4	&	8000	&	4000	&	60000	&	0.13	&	1		\\
tuple creation	&	3	&	4000	&	1	&	100	&	40.00	& 1		\\
calibration	&	3	&	4000	&	1	&	100	&	40.00	& 1		\\
Parameter estimation &	1 &		400 &		1 &		600000&		0.00&		68000 \\
\end{dunetable}

\cleardoublepage

\chapter{Frameworks\hideme{Laycock and Norman - needs review by Anne  4/28}}\label{ch:fworks}

The \dword{dune} presents a unique challenge for data analysis and data processing software frameworks. 
\dword{dune} has an ambitious physics program that spans numerous physics topics including precision neutrino oscillation measurements, searches for proton decay, sensitivity to nearby supernova explosions, and more. Moreover, these physics interactions occur at vastly different timescales, from nanoseconds to 100s of seconds, and software frameworks that are fully capable of adapting to these varied timescales efficiently will be very important. Furthermore, accomplishing timely results will require that data from the \dword{dune} far and near detectors  be efficiently processed using modern computing techniques. Given evolving computing architectures, traditional \dword{hep} software must be adapted to run on new generations of advanced and accelerated computing resources such as \dword{hpc}, \dwords{gpu} and other novel computing architectures.  

Development of this type of modern computing pipeline and analysis structure will require reoptimization and extensions of the existing software frameworks used by \dword{dune}, and shared to a large extent by the greater neutrino community to which \dword{dune} and its computing ecosystem belong.
In this Chapter, we provide a brief overview of our existing code base followed by a description of our process for designing the optimal frameworks for data processing and analysis that will be needed by the time \dword{dune} begins operations.  

\section{Defining a Framework}\label{sec:framework:def}

\subsection{The Data Atom}

Modern \dword{hep} experiments generate large volumes of primary detector data, as well as auxiliary and ancillary data from secondary systems. Along with detector data, vast amounts of simulation which mimic these detectors and their responses provide estimates and predictions for the different physics processes that may be observed in the experiments.  The management, flow, transformations, analysis and interaction of these data and their algorithms require a well-defined framework for governing and sequencing these tasks.  In particular, traditional \dword{hep} analyses data processing and analysis frameworks are designed to provide "event loops" which allow experimenters to step through the data and simulation in a manner that can break up the information into discrete chunks which represent physical processes.  In this chapter, these discrete chunks are referred to as "data atoms" and contain a fundamental assumption that they can be treated as independent units of work. In a neutrino experiment, such as \dword{pd} or the DUNE near detector, this ``atom" of data may well contain several neutrino or hadron interactions of direct interest in addition to 50-100 cosmic rays or background muons.  

Historically in \dword{hep} experiments, the data atoms have been associated with well defined external time structures which govern the data production mechanism and/or collection process.  For collider experiments these are often associated with a particle accelerator beam or bunch crossing, while for fixed target experiments they have often been associated with pulsed beam extraction to target stations (often referred to as a "spill"), or substructure within the beam spills defined by the data acquisition and triggering systems (i.e. the data atoms are the individual triggers).  

In the case of the \dword{dune} physics mission and the resultant system designed to accomplish that mission, the definition of these data atoms is not as simple or as static. %
In particular in the \dword{dune} vernacular, the data atom does not map in a one-to-one manner with traditional ``event'' terminology that is used in collider physics.  Rather the data atom represents a spatial and temporal extent of interest and can vary depending on the contextual nature of where within the processing chain we are focused.  This is not a new concept within the neutrino community, and predecessors of DUNE have used varying terminology to refer to the subsetting of triggered and free running readouts, whether the NOvA ``slice'', the MINOS ``snarl'' or other such ill-named terms for the simple, contextual subsetting of time windows into variable collections of data.  In particular, this approach is needed because 
while neutrino beam interactions have natural timescales of nanoseconds, the timescale of supernova neutrino bursts, proton decay searches, and long-lived, BSM particle searches span up to 100s of seconds. The nature of these varying times scales within the \dword{dune} physics program results in data atoms which vary in size and structure by orders of magnitude, and may be spread across numerous organizational units, data structures and even permanent storage records such as files. 
Furthermore, apart from the neutrino beam interaction case, which should have a limited number of predetermined readout times, the data production mechanism is not provided by a deterministic external source and is random in nature.
\dword{dune} data atoms could represent a) an arbitrary time window representing a time period of interest b) a time window representing the drift of ionization across the detector volume c) a spatially and temporally isolated region of the detector readout d) a sub-region of interest within a spatially/temporally distinct portion of the detector or e) other segmentations of the \dword{dune} detector readouts corresponding to potential physical signatures with time structures defined by the physical process. Having a framework that can adapt to varying time structures, multiple input files, and data structures is a very important feature for \dword{dune} to accomplish its vast physics goals.

\subsection{Key Software Framework Concepts}

A software framework is a software structure and engine that can ingest and apply transformations and filters to the individual data atoms regardless of their underlying nature and related time structure.  The framework has a responsibility for controlling and managing the I/O associated with ingesting the data atoms, for sequencing and scheduling the algorithms and transforms that run on the data, and for providing logging, provenance and other bookkeeping tasks associated with describing and enumerating how the data was handled and new information derived.  

To these ends, the framework provides mechanisms (such as "plug-in" structures) for experimenters to develop modular algorithms which can be included or excluded from a given analysis chain and be run either serially or in parallel depending on the data context.  The framework also provides mechanisms for data to be passed between different algorithmic portions of the data processing and analysis chain while maintaining the data's coherence and integrity from the standpoint of the computing platform.  In some framework implementations this data passing mechanism is provided by a {\it data store} which includes controlled forms of data access and locking.  
Frameworks which use these types of methods for data passing need to then additionally manage memory footprints of the stores to ensure that they can operate within the technological memory capacities provided by the platforms on which the codes run.  Other frameworks can provide less rigid data pipelines for information passing which can be advantageous when dealing with large data streams and which can better accommodate ingress and egress data paths with respect to the bulk memory footprints, but which may not be compatible with certain types of parallel processing and parallel algorithms, or with algorithm scheduling systems.

The other key feature that modern frameworks provide, is the ability to configure and re-configure the content and flow of the execution modules without the need to rewrite the underlying algorithms.  This aspect of a framework as a configuration driven entity is key to their flexibility and their ability to satisfy the needs of the many different and varied analyses that are proposed for \dword{dune}.  Without a robust configurable framework, different data processing tasks and analysis chains could be spread across a wide variety of codes, likely with significant duplication, which could severely hamper the collaboration's ability to understand the derivation of results and ensure their integrity during scientific reviews.  A robust configuration-driven framework system allows users to make changes to the values of individual parameters in order to investigate the data or simulation's response to those changes, and it allows for reordering, restructuring and inclusion/removal of algorithmic blocks without the need to edit or recompile the core physics codes.  This in turn promotes code stability, reuse, and facilitates modern code design and debugging principles.  

Another key concept that defines modern data processing frameworks is their construction and organization as advanced state machines.  In contrast to older, linear execution techniques, this design allows for the execution of more complicated workflow topologies, parallel analysis chain topologies, and dynamic execution models.  This state machine organization and behavior also allows the frameworks to emulate better the behavior of other systems that are used throughout  \dword{hep}, such as those encountered in the data acquisition systems realm, and to adapt more readily to parallel and asynchronous data access systems where the coherence of the system's ``state'' can be monitored and maintained between framework controlled state transitions (i.e. between configuration, initialization, algorithm execution, data serialization, and finalization stages).  

It should be noted that in modern framework definitions and designs, there is typically an interplay and hand off between the framework which is executing the algorithm code, and the higher level \dword{wfms} layers which are responsible for the batch scheduling and delivery of the framework ``jobs'' to compute resources at different computing sites or on different types of computing hardware.  Modern frameworks are typically designed to be aware of this macro level scheduling layer, and will often provide hooks to allow for the management layers to either transmit or receive information from the framework jobs. Information regarding framework configuration and various types of diagnostics can help both the framework and the \dword{wfms} layers more efficiently execute their missions, e.g. a framework will often provide monitoring diagnostics regarding the overall progress of the job which then allows for the management layer to stage queued work or initiate data transfers between sites.  Equally, the \dword{wfms} may provide ``late binding'' configuration information to the framework which allows it to determine at runtime sources of input data, or locations for output data which may be site specific or coordinated by the higher level management tools.

Lastly, modern frameworks as used by \dword{hep} today are also designed and responsible for adapting to  heterogeneous computing resources.  Modern frameworks must, due to the push towards exascale computing platforms, be able to execute their codes or bind/link their codes across not only different operating systems, but across different hardware architectures.   In the case of \dword{dune} this will be an enabling technology that allows \dword{dune} algorithms to run on platforms ranging from commodity desktop/laptop systems, to large scale grid computing centers, and across the exascale era leadership computing facilities being built by the \dword{doe} and other major national and international computing centers. These facilities' architectures will span from the classic x86 CPU architectures to advanced, many-core GPU and AI/ML tuned accelerator systems.  The framework's ability to deal with this diversity is a requirement from \dword{dune} and is discussed in detail in the following sections.

\section{Current status}\label{sec:framework:status}

Currently a set of frameworks has been used for the simulation of the \dword{dune} detectors, and processing of data from the \dword{protodune} detectors.  These frameworks have been developed or adopted based on the immediate needs of the sensitivity studies, simulation studies and \dword{protodune} running.  These frameworks are based on earlier frameworks that have been used in the neutrino community and reflect in part the features that are needed for long and short baseline neutrino analysis.

For the \dword{fd} simulation and reconstruction efforts, the \dword{protodune} detectors and \dword{ndgar} studies, the \dword{art} framework which was developed at \dword{fnal} is the primary framework being used throughout the collaboration.  The \dword{art} framework, was originally developed based on the CMSSW framework used by the CMS experiment, but was designed to meet the specific needs of the neutrino and muon science communities.  The \dword{art} framework was designed as a multi-experiment framework starting in 2011~\cite{Green:2012gv}, and is currently used by 11 different major experiments in the \dword{hep} community including \dword{dune}, NOvA, MicroBooNE, ICARUS, SBND, Muon g-2 and Mu2e.  It is developed, maintained and supported by \dword{fnal}'s \dword{scd} and through a stakeholders committee consisting of the experiments using the framework.

The framework provides the data processing loop, manages memory, interfaces to I/O tools, defines uniform mechanisms for defining, associating, and persisting data products, provides a uniform mechanism for job configuration, stores job configuration information in its output files, and manages messages, random numbers and exceptions.  The framework is highly modular and common modules which are shared across experiments have been developed for common infrastructure components such as the accelerator beam information systems at \dword{fnal}, various data management systems and data cataloging systems, and database systems for calibration and conditions data access.  Use of the \dword{art} framework has allowed  \dword{dune} collaborators and other experimenters to quickly interface with \dword{fnal} facilities, and with other facilities hosted by \dword{cern} and other sites, in a common and consistent manner.  The framework also allows for dynamic (runtime) configuration of the code modules, which has allowed  additional common code libraries to be developed and run on top of the \dword{art} base. 

In particular, the \dword{larsoft} toolkit is a collection of \dword{art} plug-ins and associated algorithm code, configuration files, static data such as geometry specification files and photon visibility maps which has been developed for the \dword{lartpc} detector community.  \dword{larsoft} provides the interface to neutrino event generators such as \dword{genie} and cosmic ray generators and simulations such as \dword{corsika} and CRY,
detector simulation via \dword{geant4}, custom simulation and reconstruction software, event displays and tutorials.  Experiment-specific metadata and configuration database plug-ins assist in batch workflow organization.  Like \dword{art}, \dword{larsoft} is supported by \dword{fnal}'s \dword{scd}, and through collaboration and contributions from participating experiments.

It is worth noting that while \dword{art} has some multithreading capabilities, these features are not extensively utilized by either \dword{dune} or \dword{larsoft}. A significant amount of work is needed to transition \dword{dune} modules, \dword{larsoft} modules, and most implementations of \dword{art} services into thread-safe software so that the multithreading capabilities of \dword{art} would become the standard workflow within \dword{dune}.

For the gaseous detector simulations, the \dword{garsoft} software package is used.  \dword{garsoft} is patterned on \dword{larsoft} as a layer of common algorithms which runs on top of the \dword{art} framework.  It provides a robust toolkit for simulating and reconstructing data from the \dword{ndgar} concept detector or for other gaseous argon detectors.  Like \dword{larsoft}, it provides interfaces to event generators and \dword{geant4}, custom simulation and reconstruction software, and event displays.  It also provides simulation and reconstruction for \dword{ndgarlite}.  Unlike \dword{larsoft}, \dword{garsoft} is written, maintained, and supported only by the \dword{dune} collaboration and is not shared with other experiments in the neutrino or \dword{hep} community.

In addition to the \dword{ndgar} software, the \dword{dune} collaboration's near detector groups have developed different tool suites for simulation and analysis of the proposed near detector designs.  The \dword{ndlar} software, which represents a group of standalone tools for simulating and reconstructing pixel-based \dword{lartpc} data.  These tools and toolchains have been developed for simulating and analyzing \dword{singlecube} prototype data.  The \dword{sand} software efforts are based on collaborators' experience with the \dword{kloe} detector and its software.  Portions of the \dword{sand} software are being used to provide simulation for the magnet and calorimeter systems, while new software is currently being developed for the \dword{3dst} and other components of SAND.  These software stacks are specifically being developed with maximum flexibility due to the design stage of the SAND concept.  This software flexibility is important at this stage in the design in order to allow studies of different detector designs which can be used between the \dword{3dst} and the calorimeter.  These tools are currently only used by the near detector group and have an independent software structure from the long baseline groups.

For higher level analysis tasks, sensitivity studies, and event selection, \dword{dune} is currently leveraging the 
\dword{cafana}
framework.  \dword{cafana} is a high level framework based on the \dword{root} analysis software stack, which is designed to work with \dword{root} TTree's (ntuples). As mentioned in the previous chapter, \dword{root} will be transitioning to \dword{rntuple}~\cite{Blomer:2020usr, ROOTTeam:2020jal} as a replacement for TTree ntuples on the DUNE timescale, and so \dword{cafana} will also need to transition.  \dword{cafana} is designed to provide bookkeeping facilities for neutrino flux information and exposures, which are needed for long baseline oscillation measurements and for short baseline cross section measurements.  \dword{cafana} is designed to work with output from data that has been processed with \dword{art} and \dword{larsoft} and provides a more interactive environment for experimenters to explore the data, by using the \dword{root} scripting and interpreter interfaces, with specific \dword{cafana} libraries and functions.  \dword{cafana} provides facilities for high level event selection, normalization of distributions, re-weighting distributions,  and more.

The distinguishing features of the two frameworks, \dword{art} and \dword{cafana}, are that \dword{art} works at the individual event record level, while the \dword{cafana} framework allows for high level event selection but operates primarily on the resulting ensemble level distributions.  

\section{Framework Requirements \hideme{Laycock/Norman - needs update}}

\subsection{Requirements Process} \label{sec:req:proc}

Given the unique challenges that \dword{dune} data pose, in 2020, the collaboration assembled a task force to examine and define needs and requirements for the \dword{dune} software frameworks that are needed to accomplish the core physics missions of \dword{dune}.  The task force was charged with identifying ``physics use cases''  which could then be translated into software requirements that could be imposed on the current framework system, and applied to any new software systems\cite{bib:docdb21934}.  The collaboration then approached the \dword{hsf} who assembled a panel of software framework experts from various experimental backgrounds to review these requirements and assess their compatibility with existing \dword{hep} software. The report \cite{bib:docdb24423} produced by this panel was discussed in a workshop involving the panelists and task force members, which resulted in a followup workshop to address the outstanding issues, mostly around the topic of concurrency and mapping \dword{dune} software and data processing to \dwords{hpc} and leadership scale computing facilities.  The findings of the panelists, and the summary of the concurrency workshop \cite{bib:docdb24426}, have been incorporated into the \dword{dune} software framework requirements documented here.  In particular a number of issues unique to \dword{dune} were identified which must be addressed as \dword{dune} and its software moves forward.  We present these in the following sections with details regarding their need and their impact to the \dword{dune} computing model.

\subsection{Summary of Software Framework Use Cases} \label{sec:fworks:use_cases}
\hideme{HMS 3/13 I added a table - probably need a row for a single ND event}

The complete set of use cases that was identified for \dword{dune} has been enumerated, annotated and reviewed in detailed in \cite{bib:docdb21934} as well as the \dword{hsf} report of section~\ref{sec:req:proc}.  As a result we present and discuss here the major requirements form those reviews that impact the computing R\&D that DUNE will need to address and the use cases which have driven (listed in Table \ref{tab:fworks:use_cases}) the collaboration to place them upon the framework develop plans.

\begin{dunetable}
[Summary of computing resources needed per trigger record]
{l r r r r r r }
{tab:fworks:use_cases}
{Summary of resources needed per trigger readout for compute intensive tasks in \dword{protodune} and the \dword{fd}. These numbers assume that reconstruction will load individual subunits such as \dwords{apa} into memory one at a time, possibly in parallel across multiple cores. Note that the output size of SNB reconstruction is not listed because a the time of writing this CDR there was no consensus estimate for algorithm output.}
Use case	&memory &	input size &	output  size 	&	CPU time 	&	input/core  	& cores/readout		\\
units	&GB	& MB	&MB	&	sec	& MB/s	&		\\

Simulation+reco	&		6	&	100	&	2000	&	2700	&	0.00	&	1 to 150	\\
FD reco	&	4	&	80	&	4000	&	600	&	0.13	&	up to 150		\\
SNB reco	&	2	&	3.2$\times10^8$	&	** 	&3-10$\times10^8$		&	0.13	&	10-30,000		\\
\end{dunetable}

One striking feature of the use cases that \dword{dune} needs to support is the diversity in scale of the data atom.  Somewhat counter-intuitively, although the raw trigger records are themselves very large, the data atom (a single \dword{apa} or \dword{crp}) is relatively small [O(10 MB) per ROI], and compatible with the same high-throughput computing workflows adopted by most \dword{hep} experiments.  Meanwhile at the analysis level although the quantity of data related to a single trigger record is small, nuisance parameter extraction requires correlations across all trigger records in an analysis dataset, which is the effective data atom.  This is a very good fit to high performance computing, particularly if the analysis framework can take advantage of many \dword{hpc} nodes at the same time. Experience from the \dword{hsf} panelists encouraged \dword{dune} to separate the analysis and production use cases when considering the software framework design.

\subsection{Unique Software Framework Requirements for DUNE}
\label{sec:fworks:urgent}

While section \ref{sec:fworks:use_cases} provides an overview of use cases, certain unique use cases drive the needs of the \dword{dune} software framework to be fully compatible with the physics missions.   We summarize these driving requirements here as a summary of the major work that needs to be executed for \dword{dune} to have a software framework capable of supporting its physics program.

Driving use cases and consequences on framework development:

\begin{itemize}
\item Far detector partial region simulations and subsetting of temporal and spatial data.
\end{itemize}

While the actual visible energy signatures from the simulation of physics processes in the \dword{fd} are likely to span only a subset of a detector module, as individual interactions are confined to a reasonably small volume and a prompt temporal footprint, the generation of these simulated interactions is not.  This is due to the manner by which the current generation of detector geometries and simulations treat the relevant volumes, the overlaying of noise, cosmic ray induced activity, and other background processes.  Without substantial effort to restructure how these processes are accurately simulated across the detector volume, while preserving the detector response, all of the simulated hits regardless of locality (within the sub-volume, typically 10-20\% of the total  module, spanned by the interaction) have to remain in memory at once during the initial simulation stages (i.e. we have to generate all activity across the detector as a whole and then after the full event topology along with its background overlays are determine, we can subset the event the discard information that falls outside of the primary regions of interest).  This memory footprint for the full detector is substantial and presents difficulties for scaling to current and future computing platforms.  The framework will need to be able to handle the management of these simulations and in particular will need to be able to subset them efficiently and effectively ``page'' them in and out of the active memory in order to propagate information between stages.  Simulations of a 2x6 \dword{apa} region in the \dword{hd} detector have already been demonstrated, but this needs to be extended and made a native function of the framework's memory management and handling systems.  This is also an area where auxiliary detectors, and in particular the photon detectors may add substantially to the memory footprint as discussed in section \ref{ch:use:pd}.

\begin{itemize}
\item Far detector partial region reconstruction and subsetting trigger records for data processing
\end{itemize}

Reconstruction algorithms running within the \dword{art} framework, are already running into memory limitations when processing the 6 \dwords{apa} \dword{protodune} data trigger records.  While most \dword{fd} Trigger Records, whether initiated by cosmic rays, beam neutrino interactions, or  atmospheric neutrino interactions, will result in activity in only one \dword{fd} module, there are still extensive air shower events and other physics events, even at the 4850' level of \dword{surf}, which results in correlated activity spread across multiple modules.  Processing data from a full \dword{fd} module, with hundreds of thousands of channels and thousands of samples/channel will require sophisticated memory management to handle not only ingesting the data, but working with it through the different algorithmic transforms that are required for the digital signal processing, the reconstruction tasks, and for writing out the resulting information.  There is however natural parallelism in the \dword{dune} detector corresponding to the APAs.  A framework designed to operate with knowledge of this parallelism, would be able to work with the individual readout subunits such as the \dwords{apa} or \dwords{crp} and process them either individually, sequentially, or in parallel in such a manner as to manage the total memory footprint while maximizing processing throughput.  In contrast current frameworks, such as \dword{art}, are not designed in this manner and can not natively handle this type of subsetting and scheduling.

\begin{itemize}
\item Temporal and spatial ``stitching'' of readout trigger records into new extended windows
\end{itemize}

  In the event of a supernova burst, or other multi-messenger or long time duration events, the data from multiple files (or storage objects) will need to be combined across file boundaries to form complete views of the interactions
  and, perhaps more crucially, to avoid edge effects in the \dword{fft} (see section \ref{subsec:signal_processing}).  In the case of a SNB Trigger Record, the data from ten's of seconds of readout will be 100's of TB in size and may have thousands of physical interactions spread across the active volume.  Proper analysis of extended events like this require that data from different portions of the underlying DAQ readout be ``stitched'' across boundaries in both the detector spatial dimensions and also in time.  However, current \dword{hep} frameworks make the implicit assumption that data atoms are independent of all other data atoms and that the temporal ordering of data atoms can be ignored when scheduling data processing tasks.  To overcome this, the framework that \dword{dune} uses will need to handle data that has a well-defined temporal ordering, and combine data from adjacent readouts, taking into account overlap regions, effective sidebands, and duplication of data products.  This is a highly non-trivial task and one for which R\&D is required for developing the methods for scheduling these types of merger operations within the context of a parallel or distributed computing environment.

\begin{itemize}
\item Contextual switching of primary data atom types for driving event loops
\end{itemize}
  
  In contrast to the far detector supernova readout, the \dword{dune} near detector due to its proximity to the target station, high fluxes, and cavern siting, will have many neutrino induced interactions within or crossing the detector volume during each LBNF spill window.  The Gaseous Argon Near Detector Component (ND-GAr) expects approximately 60 overlaid interactions per spill.  Most of these interactions will originate in its calorimeter, which has fast timing capabilities that can be used to disambiguate the interactions.  Disambiguating these interactions will allow for a single trigger window to be effectively unfolded into the particle content of each interaction.  Similarly, the \dword{lartpc} near detector is expected to have on the order of 20 interactions per spill which can also be disambiguated and separated out.  From the framework perspective this subsetting of a data window into what has been colloquially referred to as ``slices'' changes the primary context of items that need to be looped over and analyzed.  The \dword{dune} framework will need to be able to handle this type of a contextual switch, so that it can run appropriate reconstruction and identification algorithms over each of the interactions, instead of over the original readout window.  In much the same way that current frameworks are not designed to handle ``stitching'' of data atoms, current frameworks similarly are not designed to break data atoms down into smaller atoms while retaining the required bookkeeping information, and then looping and scheduling algorithm modules over these.

Given the potential for pressure on memory resources, a requirement on the framework is:

\begin{itemize}
    \item The framework must provide fine-grained control of memory, managing the memory needed by data products in the {\it data store} or its equivalent.
\end{itemize}

\subsection{Modular Framework Design} \label{subsec:fworks:modular}

For developers, highly modular code must be encouraged, allowing evolution or replacement of sub-algorithms that lend themselves to particular approaches.
The codebase will therefore likely contain several alternatives for (sub)algorithms, the choice of which would depend on the available hardware.  The framework will need to run in heterogeneous and potentially dynamic environments where the availability and type of co-processors may be known late.  While the design of this technically challenging aspect is left to framework developers, it is worth pointing out the added challenge of developing algorithms for diverse hardware if the framework does not easily allow it.  Therefore:

\begin{itemize}

\item The framework must separate the persistent data representation from the in-memory representation seen by algorithms

\item The framework must separate data and algorithms 
\end{itemize}

For developers, fast turnaround time is crucial:
\begin{itemize}
 \item The framework should allow a rapid development setup with minimal overhead both in terms of writing code, compilation and startup time (including configuration)

\end{itemize}

It is noted that a sufficiently advanced Configuration system (see section \ref{subsec:fworks:config}) would be able to achieve this.

\subsection{I/O Requirements for the Simulation/Reconstruction Framework}

Given the uncertainty in the choice of data format:

\begin{itemize}
\item  the framework must support reading and writing different persistent data formats

\item The framework I/O read functionality must be backward compatible across versions.

\item The framework I/O layer needs to provide a mechanism for user-defined schema evolution of data products.
\end{itemize}
The \dword{dune} data model allows for Trigger Record data to be stored in persistable representations which are generated by customized hardware or which are optimized for specific acceleration hardware or computing systems.  As a result, the data model expects that data will have custom “packed” representations that do not conform to 32-bit or 64-bit little-endian words. Furthermore, compression of the raw waveform data will be performed in the DAQ, though some data may arrive uncompressed.  Some highly compressible data products may benefit from dedicated compression algorithms scheduled to run before output. Therefore, 

\begin{itemize}
\item the framework I/O layer must provide a mechanism to register/associate and to run/apply custom serialization/deserialization and compression/decompression algorithms to data on write/read of that data from a persistable form.

\item The framework I/O layer should support compression on output data in a manner that is transparent to users and is configurable.  It must be possible to disable the automatic compression of output data.
\end{itemize}

Experience shows that it is highly desirable to be able to configure a maximum file size such that output files are the correct size for efficient storage and for units of data processing; currently a file size of several GBs is considered optimal.  As this requires the closure of an existing output file and creation and opening of a new file (with sensible filename) then this needs to be addressed at the framework level.

\begin{itemize}
\item The framework I/O layer should allow a configurable maximum output file size and provide appropriate file-handling functionality.
\end{itemize}

Frameworks need to provide configurable, flexible I/O access so that experiments can control the output of their jobs in fine-grained detail.  This is needed to save on storage and also experimenter time, as smaller datasets take less time to analyze than larger ones.
 
\begin{itemize}
\item The framework needs to support skimming/slimming/thinning for data reduction.
\end{itemize}

Similarly, processing and analysis is made much more convenient (or even possible) if the skimmed/ slimmed/ thinned output streams can be associated with information in other streams that may be stored separately.  Some analyzers may need the auxiliary data streams while others may not, and so a framework job that produces outputs for collaboration use would need to read all of the necessary streams.  This also allows efficient use of storage, as data does not need to be co-located in the same file to be available for processing.  In the case of writing, I/O cost can be very efficiently amortized if several output streams can be written based on one input file.

\begin{itemize}
\item The framework needs to be able to read and write several parallel data streams.  Labeling of data objects across streams should be intuitive and not error prone.  Provenance information should support correlating related data objects across streams.

\end{itemize}

A common offline job need that goes beyond the $1 \rightarrow 1$ input to output data model is mixing real Trigger Record data with Monte Carlo simulation.  Monte Carlo simulation is often not sufficiently realistic, perhaps it fails to capture the time dependence of detector conditions or the generator or detector simulation simply lacks sufficient accuracy for the physics use case.  In this case, Monte Carlo simulation can be augmented by adding actual detector data, e.g. to embed single tracks or entire Trigger Records into simulated data and reconstruct them as if they were actual Trigger Records.  This can lead to a $2 \rightarrow$ many input output situation with asynchrony in both input and output. 

\begin{itemize}
\item The framework needs to allow experiment code to mix simulation and (overlay) data.

\end{itemize}

\subsection{Reproducibility and Configuration}

The reproducibility of physics results and the knowledge of how physics results were obtained is essential to \dword{dune} and to the neutrino community as a whole.  It must be possible both to replicate physics results using identical input data, or to repeat an analysis using a different set of input data with an identical sequence of identically configured algorithms.

For the purpose of these requirements, we consider data to be reproducible if for integer fields 
the products in a given event record are bitwise identical. 
Floating point numbers in a given event record are the same to within the floating point precision of the machine they were generated on.
The ordering of data products within an event record are the same.
These criteria for reproducibility are designed to allow \dword{dune} to simulate “rare” events with well defined single event sensitivities, and to perform numeric transform operations on data values across different hardware platforms and runtime concurrency topologies.
\begin{itemize}

\item It is highly desirable that the framework broker access to random number generators and seeds in order to guarantee reproducibility.

\item The framework must provide a full provenance chain for any and all data products which must include enough information to reproduce identically every persistent data product.  By definition, the chain will also need to include sufficient information to reproduce the transient data passed between algorithm modules, even though the product is not persisted in the final output.  
\end{itemize}

This will include the computing architecture, including any specialized hardware used, on which the application is run as well as the runtime environment, execution model and concurrency level that the application used.  It is likely that a full picture of the necessary metadata will also require information only known to the \dword{wfms}.  In such a complex environment, it is highly desirable that the framework provide support to allow the effect of configuration changes and computing environments to be easily understood.

\subsubsection{Configuration} \label{subsec:fworks:config}

Configuration is distinct from initialization of the framework objects; configuration happens first
and it is particularly important for multi-threaded frameworks to have a fully configured framework instance in the initialization phase.
%.  Given that \dword{raii} is an important concept for multi-threaded design, it is best to have a fully configured framework instance in the initialization phase.  
Given the abundance of variations that make up \dword{hep} workflows a robust and easily programmable configuration system is a foundational component of all modern frameworks.  Some use a strictly declarative language and some use a Turing complete language.  The former must be augmented with scripts that write the declarations in a Turing complete language (usually it is part of the \dword{wfms}) because it turns out that control flow is a requirement.  Given this, there is a requirement that:

\begin{itemize}
\item The framework should provide a configuration language as a foundational component so that it can ensure coherence of its configuration.
\end{itemize}

As discussed in section \ref{sec:fworks:urgent}, the \dword{wfms} needs to be able to supply framework configuration parameters such as input file(s) or random number seeds to each framework application instance, which it should do using the framework’s configuration language.  To minimize errors, these parameters should be self-describing and validatable, and so:

\begin{itemize}
\item the framework should provide a suitable API so that algorithm writers can ensure their required parameters are self-describing and validatable.

\item The framework should provide the concept of parameter sets that are nestable.  The set of all parameter sets that define a framework application instance should be identifiable, referred to here as a FrameworkConfigID.  Tracking that identity is one of the ingredients necessary to ensure scientific reproducibility.  However some parameters do not (and should not) change the algorithmic results, such as a debug print flag.  Independent of the state of such a flag it should be possible to define equivalence between FrameworkConfigIDs.

\item The framework configuration system needs to have a robust persistency and versioning system that makes it easy to document and reproduce previous results.  It must be possible to create, tag, check-sum, store and compare configurations.  This configuration management system should be external to the framework or data files so that configurations can reliably be reused and audited.

\item The resulting state of the configured framework and its components should be deterministic and reproducible given a set of environmental conditions which include the available hardware, operating system, input data, etc.  
\end{itemize}

Related to the discussion on rapid turn-around in section \ref{subsec:fworks:modular}, but also important central features of the configuration system:

\begin{itemize}
\item it is desirable that it should be possible to only configure those framework components required for a particular data processing use case.

\item it must be possible to derive the input data requirements for any algorithm, in order to define the sequence of all algorithms needed for a particular use case.
\end{itemize}
 
\subsubsection{Conditions Configuration}

In addition to the Trigger Record data, data processing requires access to other data from various sources, for example slow controls, detector status, and beam component status.  Such data is referred to generically as \dword{condmeta}, introduced in section \ref{subsec:db:conditions_metadata}, and also includes e.g. detector calibrations and any data external to the Trigger Record.  The time granularity or “interval of validity” of this data varies by source and is typically of much coarser granularity than individual Trigger Records, e.g. calibrations may be valid for months of data taking.  Meanwhile there are often several versions of \dword{condmeta} and correlations between \dword{condmeta} is not uncommon, making the coherent management of \dword{condmeta} a challenge in itself.  For this reason, \dword{condmeta} management should be external to the framework.

Access to the external conditions data should preferably proceed via REST interfaces that support loose coupling of the framework and the external \dword{condmeta} management system.  As \dword{condmeta} may need to be transformed from its persistent format into a format required by an algorithm, and as multi-threading makes the cache validity of \dword{condmeta} complicated: 

\begin{itemize}
\item the framework must provide a thread-safe conditions service that is a single point of access to \dword{condmeta}.  

\item The configuration of the framework conditions service should ideally be via one configuration parameter (a global tag). 
\end{itemize}

Developers must not hard-code conditions data in their algorithms, although 

\begin{itemize}
\item it must be possible to override a subset of global tag configured conditions for testing purposes.  
\end{itemize}

Developers usually find it convenient if such alternative conditions payloads can be provided outside of the main managed conditions system, e.g. via a local file.

\subsubsection{Concurrency} %

The intrinsically parallel structure of the DUNE detector data, and the types of digital signal processing and machine learning algorithms that are being used to analyze it, benefit highly from both the multi-core CPU architectures and new hardware based accelerators such as GPUs, Tensor processor units and other many-core architectures.  However, while there is significant benefit to using these heterogeneous architectures, there is also a cost in terms of the complexity of needing to manage the concurrent operations and ensure that they are computationally safe and reproducible.

In the context of DUNE, the framework, in being able to schedule module work, will need to be both thread aware and accelerator aware.  By this we mean that the framework will need to ensure that the modules it schedules and executes in parallel threads, or through offload to accelerators, preserve the coherent state of the memory, data, I/O and other resources that the framework controls and serves as a gateway to.  

While the goal of the DUNE collaboration will be to write modern, high parallelized, thread- and accelerator-safe algorithm codes, we recognize that there will be code within the DUNE code base that will rely on serial programming techniques.  We also acknowledge that portions of the DUNE code base will leverage external libraries which may or may not be thread safe, or which may need exclusive access to accelerators.  For these reasons we impose upon our framework the following requirement: 

\begin{itemize}

\item The framework must be able to schedule thread-safe and non-thread-safe task modules, while ensuring coherence of the data, memory and I/O systems.  
\end{itemize}

\begin{itemize}

\item The framework must be able to schedule tasks modules which require access to hardware accelerators, while ensuring coherence of the data, memory and I/O systems.  
\end{itemize}

At the same time the framework must be able to perform this scheduling in the context of different thread pools or memory maps which are commonly controlled  through different parallelism toolkits (i.e. OpenMP, Intel TBB, Intel OneAPI, OpenACC, etc...).  This needs to be done due to the way in which different toolkits can interact and result in oversubscription of resources.   For DUNE, where researchers may be leveraging major data science, machine learning or other commercial libraries, this is particularly important because it may imply that the task modules need to provide ``hints'' which will help the framework schedule the modules or change the sharing model that is used during portions of the code's execution.

\begin{itemize}
\item The framework should be compatible with the use of [external] multi-threaded data processing techniques and access to co-processors resources.  The framework should be capable of scheduling these resources in an efficient manner with respect to a given runtime environment. 
\end{itemize}

This type of parallel scheduling and access problem is an active topic across more domains than just HEP.  Developing a framework system that meets DUNE's needs will require both significant development work and not-insignificant R\&D with regards to the computer science and data science techniques which will ultimately be used.  This investment is however has the potential to reap massive benefits to DUNE in terms of the execution times of the algorithm codes, and the ability of the collaboration to use the exa-scale computing platforms that will be dominate and supported across the DoE leadership computing facilities.

\subsubsection{Externals including Machine Learning}
Machine learning is already heavily used in analysis of \dword{protodune} data and has become an integral part of scientific analysis across HEP.  The developed by \dword{dune} framework should give special attention to machine learning inference in the design, both to allow simple exchanges of inference backends and to record the provenance of those backends and all necessary versioning information.  In particular the framework needs to support the injection of AI/ML model configuration and be able to describe not only the foundational model but the training state of the models or inference systems.  It is tempting to require that the framework natively support a foundational model description language, such as the Open Neural Network Exchange (ONYX) standard and leverage its ``model zoo'' as a path towards \dword{dune}'s use of ML.  However while this is a current push within the ML community in 2022 for the description of natural language processing, facial recognition and other major AI/ML enabled technologies, the scope and timing of the \dword{dune} framework development would advise against a requirement that effectively names an implementation specification.   

Instead, we require that the framework support configuration of AI/ML systems through deterministically defined configuration systems, and that this system provide the framework with provenance information, or provide a derivable method for obtaining provenance information, sufficient to fully describe the state of any external ML system and allow for reproducibility of that state at a later time with the same fidelity as originally described or configured.

In addition, the framework should be able to work with both the \dword{nd} and \dword{fd} data on an equal footing and within the same job, which may require the simultaneous configuration, instantiation, management, and scheduling of multiple models or model instances based on the detector context.

In this respect, we require that the use of external libraries relating to machine learning, machine learning inference or other functions which are not part of the core framework or algorithms code stacks of the experiment be registered and treated by the framework proper in such a manner as to preserve the integrity and reproducibility of the data analysis.

\begin{itemize}
\item The framework should give special attention to machine learning, machine learning inference, and the use of external libraries and codebases in its [the framework's] design, both to allow simple exchanges of ML/external backends and to record the provenance of those backends, their states, and all necessary versioning information to permit full reproducibility of the scientific results.

\end{itemize}

\subsection{Analysis\hideme{Norman/Laycock - still needs more}}

Based on the experience and recommendations of the \dword{hsf} panel experts, the software framework used for large scale production may be different to that preferred for late stage data analysis. This difference may not necessarily be due to technological challenges or missing functionality in the primary production framework, but may arise due to sociological factors that can, and have in the context of other HEP experiments, played a large role.  It was noted by the \dword{hsf} panel that physicists may be willing to sacrifice performance, capability, and flexibility in favor of simplicity.  In these cases, it has been observed that lighter-weight, less capable, less performant, less stable and less supported frameworks may grow organically from members of the collaboration, and even become preferred methods for addressing niche topics.

That being said, there is no clear dividing line separating the higher level analysis from the use cases served by the data processing framework, and the choice may vary depending on the use case or analysis group in question.  In the following discussion, requirements more often associated with analysis use cases are discussed, but data processing framework developers should also consider them as many can, and should, be common to the analysis chains.

Analysis tasks include the extraction of oscillation parameters, comparisons between data and Monte Carlo, extraction of calibration and detector performance parameters, cross-section measurements, measurements of atmospheric, solar, and supernova neutrinos, and searches for non-standard phenomena.  For these tasks, the event-by-event paradigm is not a good match in all cases.  In contrast to the event-by-event model, ensemble distributions and representative spectra are the quantities of interest and while they can be generated by iteration or looping over event records, neutrino candidates, or data-atom style data organization, they often may  undergo substantial processing in their own right, whether that be in the form of renormalizations and re-weighting, or different type of peak finding, filtering, de-convolutions and other similar ensemble operations. For example, the main work of an oscillation fit is the evaluation of many different combinations of oscillation and nuisance parameters against a given distribution of observed energies. Yet, for particle level analysis, while time or spatial slices provide sub-Trigger Record control flow, particle candidate control flow ignores the Trigger Record structure entirely.  Indeed it is this contextual switching between the fundamental quantity of interest, and the ability to perform and propagate operations upon these contextually defined quantities that defines our analysis environment. 

These 
analysis problems also tend to lend themselves to ``columnar'' analysis techniques and organizations of the data.  While it is true that the manner by which we acquire the data is inherently ``row-wise'' given the way we generate our readout records in a temporal sequence, there is nothing that prevents the data from being effectively transposed into a columnar or table based structure at later points in the processing chain.  These tabular organizations are almost universally more efficient in terms of ensemble operations, memory efficiency, data parallel operations, and other factors which influence the late state analysis process.  As mentioned in previous sections, \dword{dune} aims to support both paradigms for data organization as it has been shown to have benefits not only in our custom analysis codes, but in leveraging machine learning and other external products which expect these types of data frame oriented views of the information.

It should also be noted that analysis work will be undertaken by a large number of collaborators with varying levels of experience, in comparison to the mainline or centralized data processing efforts which will be engaged in by smaller teams of individuals with more expertise in the data processing workflows.  In these use cases, simplicity, rapid execution, feedback, and iteration are of paramount importance:

\begin{itemize}
\item The analysis framework should have a low entry-level in terms of software expertise.
\end{itemize}

Analysis files, of course, are derived from the data-processing framework files, and it must be possible to reconstruct this history. Due to the very large number of Trigger Records expected to be summarized in a single analysis file, the size requirements, and the fact that per-Trigger Record information remains available in the parent files, we require:

\begin{itemize}

\item Analysis files must record their parent framework files, but no Trigger Record provenance is required.  The full provenance information need not be retained in analysis files as this could easily become larger than the data itself.
\end{itemize}

One common and insidious class of mistakes is errors in exposure accounting and normalization. This is also a problem that is entirely solvable at the technical level.  Each individual Trigger Record (beam spill or other trigger) has exposure associated with it, whether POT or livetime or both. When filling a summary histogram from processing Trigger Records, the exposure should be calculated and stored as an integral part of the histogram, and operations between histograms should take correct notice of the exposure, e.g. ratio of one large exposure sample to a smaller exposure sample should produce a dimensionless ratio that has allowed for the differing exposures.

\begin{itemize}

\item The framework must have native support for exposure accounting (POT and livetime), so as to make errors of this sort difficult.
\end{itemize}

All but the simplest analyses require a treatment of systematic uncertainties. There are three main technical means by which these systematics can be introduced. The most common, and most convenient, is reweighting. For example, the effect of various cross-section and flux uncertainties may be encapsulated by applying weights to Trigger Records of certain categories, to increase or decrease their representation in the final spectra. Secondly, Trigger Record data may be shifted. For example, an energy scale uncertainty may be most conveniently represented by rewriting of Trigger Records to increase or decrease reconstructed energies by a certain amount. Finally, the least convenient method is alternate simulation samples. The profusion of files requiring processing and bookkeeping makes this a heavyweight option, but in the case of uncertainties early in the analysis chain with complex effects, it may be the only way to handle them accurately. The treatment of systematics is cross-cutting across all analyses, it is important it is handled correctly, and the framework is able to offer substantive technical assistance. In addition to being able to handle multiple input data streams:

\begin{itemize}
\item The analysis framework should provide some means of cross-referencing (labelling) multiple input streams to correlate them in order to facilitate evaluation of systematic uncertainties.
\end{itemize}

For oscillation analysis, it will be important to work with both \dword{nd} and \dword{fd} data. Whether in an explicit joint fit, or where extracting constraints from the \dword{nd} to apply to the \dword{fd} analysis, there must be a uniformity in the treatment of various systematics. In general, experience gained with the \dword{nd} (where the majority of analysis work is likely to happen) should be transferable to the \dword{fd}.  This re-emphasizes the importance of the framework making minimal assumptions about the Data Model.

\begin{itemize}
\item The analysis framework should be able to work with both ND and FD data on an equal footing, and within the same job.
\end{itemize}

Experience shows that oscillation fits accounting for large numbers of systematic uncertainties are resource intensive, while analysts will likely only have access to modest local resources for prototyping and development.  Therefore the framework should make scaling and concurrency transparent both to the analyst and the developer as far as possible.  The use of declarative analysis techniques should be strongly encouraged to support this even when co-processors (and low-level implementation) changes.

\begin{itemize}
\item The analysis framework should easily scale from local resources such as a laptop, up to multi-node compute at an \dword{hpc}.
\end{itemize}

It is noted that MPI-enabled frameworks written in python already exist and would be a good match to the above requirements.

\section{Development Plan and Effort Profiles}

The production use cases for the second run of the \dword{protodune} detectors can be handled by the existing \dword{art}/LArSoft based software framework and toolkits.  We therefore base our development timelines on the need to develop our next generation framework for \dword{dune},
assuming that the framework will need to be commissioned and ready for production prior to far detector installation.  
It will be validated against the late stage \dword{protodune} analyses, and also used for the first detector data for the far site systems, and for pre-cursor systems which will be need to be commissioned prior to the far detector installation. Proto-DAQ systems will produce readout data prior to the formal start of detector operations that must be processed through the offline computing system.

Currently the compute architecture landscape is evolving rapidly, a challenge being experienced by running experiments with large code bases.  \dword{dune} is engaging with the \dword{hsf} to ensure that progress made elsewhere can be effectively utilized in its own software frameworks.  The general strategy is to tackle \dword{dune}-specific challenges, discussed in Section \ref{sec:fworks:urgent}, as soon as possible, and then work on consolidating best practice on heterogeneity from the wider community into a core framework.  Following the delivery of a core framework, algorithmic code will be migrated by the wider collaboration, similar to the development pattern used by previous \dword{hep} collaborations that was reported by the \dword{hsf} panel \cite{bib:docdb24423} (see appendix).

The development plan starts with dedicated effort to accommodate the dynamic time window needs of the full \dword{dune} physics program, detailed in Section \ref{sec:fworks:urgent}.  Given that this as-yet-unknown solution is both central -- and critical -- to the success of the framework, \dword{fnal} has funded LDRD investigations in 2022/2023 to provide proof of principle and prototype data handling systems.  Successful completion of these prototypes will significantly reduce risk associated with the extended readout events and allow us to reduce our contingencies relating to later phases of the development, and in particular to the supernova burst science case. The work will also investigate techniques for efficient minimization of runtime memory footprints and restructuring of simulation chains to accommodate the far detector simulations.

The next major development work concerns the rapidly changing landscape of compute architectures.  The shift towards heterogeneous and accelerated architectures by major platform manufacturers such as Intel, AMD and NVidia, are very relevant to \dword{dune} due to the numerous compute problems that are well-matched to accelerators.  These include a mixture of both highly accelerate-able signal processing and signal transform work that dominates the early stages of the data processing chain, and emphasis on AI/ML techniques which are more prominent in the later stages of the pattern recognition and tracking, event identification and higher level analysis computing chain.  
After the LDRD work, \dword{dune} will incorporate best practice from \dword{hsf}, following a clearer \dword{doe} direction for future facilities expected on a timescale of around 2024.

An estimate of the effort needed for core software framework development and to migrate algorithmic code uses the \dword{hsf} panel report appendix, which provides approximate numbers for several experiments with diverse frameworks. In that report the HSF collected information from major HEP experiments which had either developed a new ``green-field'' framework to meet their experimental needs, or had migrated from a legacy framework to a newly developed framework which was capable of satisfying their scientific needs.  The report included information from LHCb, CMS, Atlas, ALICE and NOvA along with details of the development and migrations which were undertaken.  When the information that was provided was normalized by the size of the code stacks that were migrated or developed, a fairly consistent estimate was obtained across the experiments with some deviation based on simultaneous DAQ software development that was undertaken by ALICE.  Our resulting estimate of 6 core FTE years was then obtained by scaling the current DUNE codebase and including both a moderate allowance for the expanded feature/requirement set that has been outlined here and modest contingency.  Based on this estimation methodology, 6 FTE years is estimated for core framework developer effort from 2024-2027, and 10 FTE years is estimated to be needed for algorithmic code migration to deliver a fully functional framework in 2027.  It should be noted that of this effort the initial core framework development requires highly specialized domain expertise in frameworks, workflows, and parallel computing techniques.  This domain expertise profile is consistent with the highly technical computational science staffs that are available from the collaborating laboratories, and which have significant experience in similar projects within the HEP communities.  The majority of the algorithm and physics code migration, in contrast, requires domain expertise aligned with the neutrino sciences and is well covered by the wider scientific expertise of the \dword{dune} collaboration.

With these effort profiles we believe that a realistic development pathway to a full production framework for DUNE could be completed by mid to late 2027.  The development effort timeline can be summarized as commencing with high risk R\&D in 22/23 (covered by \dword{fnal} LDRD), Core Framework effort from 24-27 (requiring funding for new expert effort), and algorithm migration in 26/27 (combining effort from the collaboration at large with support from Core Framework experts).
Increased expert effort on core software framework development beyond the foundational LDRD-funded work will therefore be needed to deliver the core software framework for data processing, and this will form part of the next round of multi-institutional/national funding requests for the period 2024-2027.  A core framework delivered by 2026 would allow two years for algorithmic code migration with full support from core framework experts.

\cleardoublepage

\chapter{Databases \hideme{Buchanan and Laycock -Anne's comments accepted}}
\label{ch:db}

\section{Introduction  \hideme{Buchanan and Laycock - draft} }
\label{sec:db:intro} 

In order to accommodate the large range of metadata that will be tracked by \dword{dune}, the \dword{dune} \dword{db} structure comprises several databases specific to the information, or metadata, that they contain. 
The subset of all \dword{dune} metadata that is required, in the strict sense of being absolutely necessary, for data processing and analysis needs to be carefully identified and assessed.  We refer to this subset of metadata as "\dword{condmeta}", see section \ref{subsec:db:conditions_metadata}.
It is critical that users be able to access \dword{condmeta} throughout the full data processing and analysis chain with as little burden as possible. To achieve this, users will interact with a centralized high-level interface   described in Section~\ref{sec:db:conditions}.

The \dword{dune} experiment is expected to operate for 20-30 years and the \dword{dune} databases need to be reliably maintained and operated for that entire period. 
In order to accommodate this requirement, the database system should not rely on implementation solutions that possess the risk of becoming unavailable during the operation period. 
Although no solution is without risk, the general \dword{doe} lab strategy has been to adopt open-source, non-proprietary solutions, and this is the strategy that \dword{dune} will follow. Currently the databases housed at \dword{fnal} use the open-source PostgreSQL (\dword{postgres}) relational database management system. Postgres is supported by the \dword{scd}. 
\dword{dune} would also like to benefit from the close collaboration between \dword{doe} laboratories to improve service availability and mitigate single-point-of-failure risk by having secondary databases at other \dword{doe} labs.  \dword{bnl} is a good example of a lab that already provides database services to international experiments using the same \dword{postgres} technology.

It is expected that there will be reconstruction and analysis jobs distributed across large numbers of traditional grid-based and \dword{hpc} systems and that database access will need to be able to scale appropriately. Additionally, it is important to ensure that users are able to work on analysis tasks when unable to access the database directly through a network connection. 

Some of the database solutions outlined in this document have been deployed and tested during Run I of the \dword{protodune} experiment. Experience coming from %
this will be briefly described in the sections below, when relevant. %
The \dword{protodune2} experiment will provide a further testbed for the database systems proposed for \dword{dune}.  

\subsection{Conditions Metadata}
\label{subsec:db:conditions_metadata} 

\dword{condmeta}
is defined as the information necessary to understand the context of physics data, e.g., beam data or calibrations.
Metadata can be indexed by either %
time, run, or fraction of a run (subrun). 
An \dword{iov}
defines the period, indexed by any of the three options, for which a given metadata is valid.
Alignment constants are an example of \dword{condmeta} that will likely be valid for several runs. 
The \dword{dune} \dword{cdb} will have APIs that provide easy and transparent access to all \dword{condmeta} independently of technical details.

Time-indexed metadata will, in some cases, be sampled at a rate higher than typically needed by offline users. In these instances, the metadata will be filtered down, or interpolated, to a lower rate for inclusion in the \dword{cdb}. In general, metadata falls into two categories, interpolated and non-interpolated, where an example of the latter would be run-indexed values pertaining to run configuration. Interpolated metadata, e.g., values read back from the slow control system, can be interpolated through a rolling average or updated on changes to the values. The method used will depend on the natural variation of the value being recorded and the physics use case. 
The database group will provide interpolated values to match the physics requirements of the experiment.  

Conditions metadata will in general be stored in appropriate databases but there will be some cases where it is more reasonable to include the metadata with the raw trigger records instead. An obvious example of this is metadata that changes at the individual trigger-record level, i.e., trigger-bit information. 

The following table contains classes of \dword{condmeta}:

\begin{dunetable}
[\dword{condmeta}]
{l l l} 
 {table:metadata}
 {Example \dword{condmeta} values and types, and the databases in which they are stored. (I) indicates interpolated metadata.}
 Conditions Metadata Type  & Example(s) & Database  \\ \toprowrule

Run Configuration   &  Start Time, config file & Run Configuration \\ \colhline  
Detector Conditions (I)  & TPC high voltages & Slow Control   \\ \colhline
Beam Conditions (I)  &  Horn polarities, beam current & IFBeam  \\ \colhline  
Hardware Information & Component history &  Hardware/QC   \\ \colhline  
Calibration Constants & Channel gains  & Calibration  \\ \colhline 
Physics/Hardware Locations & Channel maps & Geometry  \\ \colhline  
Data Quality & Good runs list & \Dword{dqm}  \\  
\end{dunetable}

A \dword{dune} metadata task force was assembled in 2020 and a resulting report discussing the interfaces between online and offline systems can be found in reference~\cite{bib:docdb22983}.

\section{Conditions Database}
\label{sec:db:conditions} 

The \dword{cdb} is a high-level centralized database that provides an easy-to-use interface for users and reduces the number of database connections required by offline processes.
This will ensure that jobs will be ``lightweight'' and processing time will not be extended due to database accesses.  The granularity and content of the \dword{cdb} will be well-matched to offline needs by design, which will allow the heavy use of caching technologies to optimize resource usage. The design of the interface between the data processing software framework (see chapter \ref{ch:fworks}) and the \dword{cdb} will play an important role if all of the access patterns needed by \dword{dune} are to be well supported. Experience from other \dword{hep} experiments with distributed computing resources show that careful design of this interface is crucial to the success of the computing model\footnote{Overlays (see section \ref{sec:usecases_overlays}) are an example of a workflow that can cause issues for \dword{cdb} access if this interface is not well designed.}.
Figure~\ref{fig:dbmap} shows the relationship between the \dword{cdb} and the other \dword{dune} databases. %
The "Master Metadata Store" is an intermediate database that allows for potentially heavy data interpolation tasks to run without creating extra load on the other, often critical, database services.  It provides separation between the online and offline worlds at the cost of partial data duplication. A design utilizing the Master Metadata Store also allows for maximum flexibility without the need for a predetermined schema. The final database system design for \dword{dune} will benefit from experience on \dword{protodune} with this design.

\begin{dunefigure}
[Map of \dword{dune} databases]
{fig:dbmap} 
{Map of \dword{dune} databases showing the \dword{cdb}, the Master Metadata Store and the lower-level \dwords{db}. The arrows illustrate the flow of metadata. Offline access to \dword{condmeta} is made through the \dword{cdb}, which contains a subset of information from the Master Metadata Store. }
\includegraphics[width=.9\columnwidth]{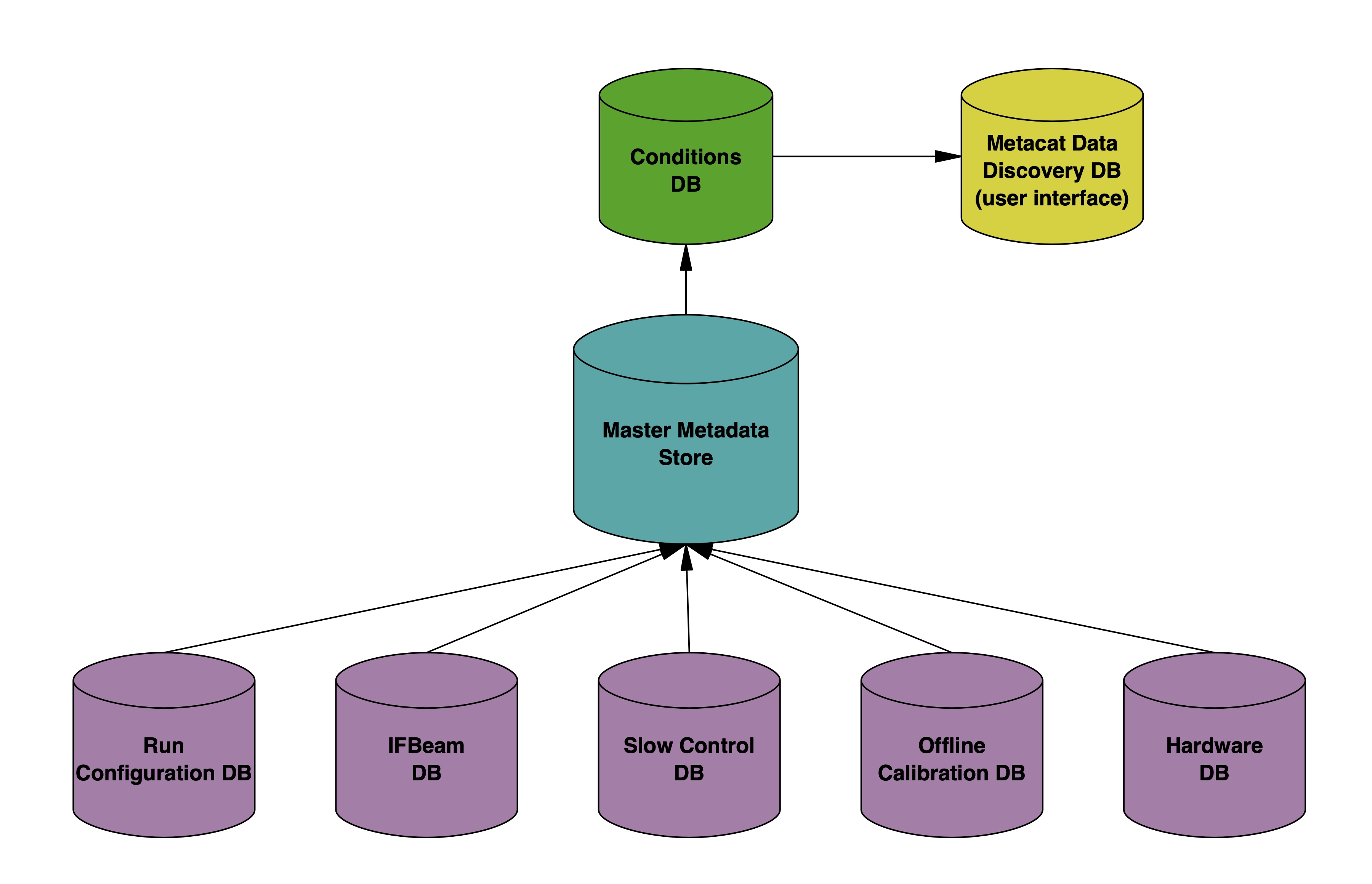}
\end{dunefigure}

\subsection{Conditions Database for ProtoDUNE II}

In order to provide a balance between the availability of the largest set of metadata possible  and allowing schema evolution,  the \dword{protodune} II \dword{cdb} will employ an unstructured approach utilizing an unstructured database \dword{ucondb}~\cite{bib:ucondb}. It is expected that the schema for the conditions DB will evolve during the lifetime of the experiment and therefore it is advantageous to be able to avoid the constraint of committing to a schema in advance. The \dword{ucondb} stores metadata from several specific databases and sources, like the \dword{daq} system, in ``blobs'' corresponding to temporal periods (run, time blocks, or \dwords{iov}). Each blob will contain a \dword{json}-formatted record of metadata. Folders will be used to hold metadata corresponding to time and run keys. Tools will be provided to correlate between the two.

The following is an example fragment of \dword{daq} metadata from the first run of  \dword{protodune}. A typical file from this run was on the order of 10~MB.  

\begin{verbatim}
{
Start of Record
Run Number: 5185
Packed on Oct 11 22:19 UTC
#####
boot.fcl:
#####
DAQ_setup_script: \
"/nfs/sw/work_dirs/dune-artdaq_artdaq_v3_03_00_beta/setupDUNEARTDAQ" 
PMT_host: "localhost" 
PMT_port: 5400 
debug_level: 1 
partition_number: 0 
tcp_base_port: 15000 
request_port: 3000 
request_address: "227.128.12.25" 
table_update_address: "227.129.1.128" 
routing_base_port: 10010 
zmq_fragment_connection_out: 17437 
...
}
\end{verbatim}

\subsection{Conditions Database for DUNE}

For \dword{dune}, the interface between the \dword{cdb} and the software framework will become particularly important.  The consensus among several \dword{hep} experiments was reported in an \dword{hsf} whitepaper \cite{Laycock:2019ynk}. The use of a lightweight relational database schema allows a single point of entry, a "global tag", to configure \dword{condmeta} access for the framework.  Evolving the \dword{protodune} solution to benefit from that experience is foreseen by the database group. Further details on the development plan can be found in section \ref{sec:db:conddbdev}.

\section{Run Configuration Database  \hideme{Buchanan and Laycock - draft}}
\label{sec:db:config}  

The run configuration database contains the intended configuration of the detectors and \dword{daq} during data collection -- physics or otherwise. 

Metadata contained in the run configuration database includes hardware settings, run type, and run start and end times. Table~\ref{table:runconfig} contains some examples of typical metadata that will be contained in the run configuration database. 

\begin{dunetable}[Run configuration database example]
{l  l } 
{table:runconfig}
{Example metadata values and types stored in the run configuration database.}
 Metadata Value & Type  \\ [0.5ex] 
 
Start of run   &  Time \\ \toprowrule
Readout window size  & Integer  \\ \colhline
Readout trigger type  &  Integer \\  \colhline
Readout firmware version &  Integer \\  \colhline
Baseline start &  Integer \\  \colhline
Shifter comments &  Text \\  \colhline
Run end status & Integer \\  
\end{dunetable}

The majority of run configuration metadata comes from the configuration files used by the \dword{daq} system during run execution. Some additional metadata collected at the end of the run, or shortly thereafter, may also be included. Examples are run completion status and comments made by the shifter during the run or in run-related checklists.

Parameters used to configure the run will be collected and packed into \dword{json}-formatted blocks in a single blob corresponding to a \dword{daq} run.   

\subsection{Run Configuration Database for ProtoDUNE II}
\label{sec:runconfigPD}

 Following the completion of a run, the run configuration parameters corresponding to the run are read from a MongoDB~\footnote{MongoDB\textcopyright, \url{https://www.mongodb.com}} set up by the \dword{daq} group for immediate archiving of run configurations. The \dwords{daq} metadata is then packed into a single "blob" of key-value pairs in \dword{json} format. Any additional information, such as end-of-run time, are added to the blob, which is then transferred to the \dword{ucondb} at \dword{fnal}. A typical metadata blob is on order 10\,MB in and contains more information than most users will want to use. An additional step of reducing the metadata is performed to produce a subset of metadata needed by offline users. The reduced set of metadata is stored in a single table in a relational database referred to as the ``run history database.'' An interface is provided to users, enabling them to retrieve run numbers and file locations based on queries of the history database. Figure \ref{fig:protoconditions} shows a diagram of the flow of metadata from \dword{protodune} DAQ to users.

\begin{dunefigure}
[Flow of metadata from \dshort{protodune} DAQ to user interface]
{fig:protoconditions} 
{Flow of metadata from \dword{protodune} \dword{daq} to user interface.}
\includegraphics[width=.9\columnwidth]{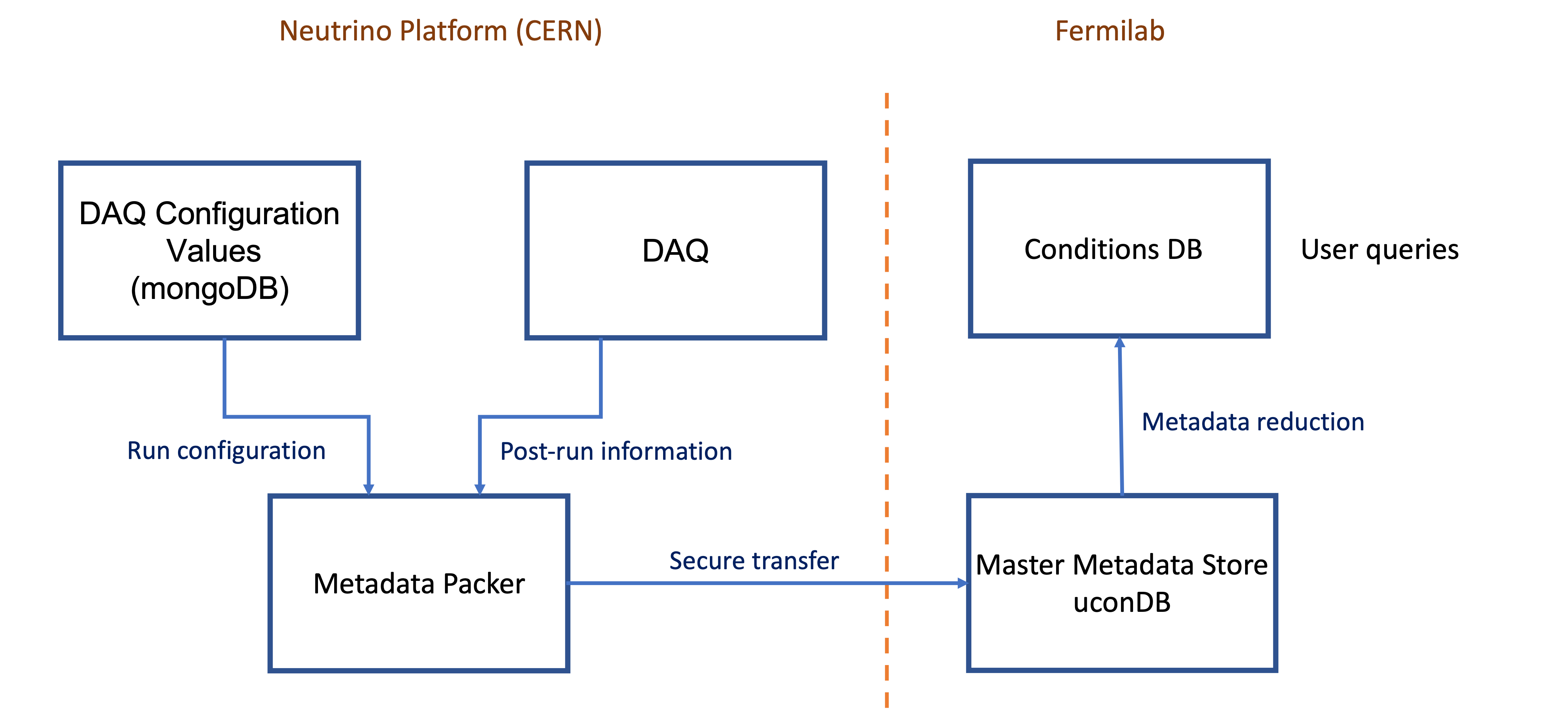}
\end{dunefigure}

\section{Data Quality and Monitoring Database  \hideme{Buchanan and Laycock - draft}}
\label{sec:db:dqm}  

The \dword{dqmdb} contains monitoring histograms and metadata derived from data collected during operation of the \dword{dune} detectors. The \dword{dqmdb} is an online database and the histograms it stores to assess and monitor data quality are critical for the operations of \dword{dune}, but not directly relevant for offline data processing and analysis. The derived data quality metadata, which include boolean flags indicating the results of the online data quality assessment algorithms, are relevant for offline analysis and this small subset of \dword{dqmdb} data will require an interface with the \dword{cdb} either directly or via an offline replica of the data quality database.

\section{Offline Calibration Database  \hideme{Buchanan and Laycock - draft}}
\label{sec:db:calib} 

The calibration database contains calibration constants determined from collected data corresponding to  \dwords{iov}. The \dword{condmeta} from the calibration system will result from offline calculations using data collected from the \dword{dune} detectors.
There will generally be multiple versions of calibration constants corresponding to the same \dword{iov}, and the \dword{cdb} will provide coherent access to the appropriate version of these calibrations via e.g., the global tag mechanism.

\section{Slow Control Database \hideme{Buchanan/Laycock - draft}}
\label{sec:db:slowcontrol}  

The slow control \dword{db} contains metadata specific to the state of detectors at the time those data were collected as well as before and after. Examples of slow control metadata are measurements of power supply voltages and currents, and temperatures. Each slow control quantity corresponds to a particular device. The slow control \dword{db} metadata is time-indexed and hence must be interpolated. Additionally, different devices will be sampled at different rates.

The slow control metadata is captured via a Supervisory Control and Data Acquisition (\dword{scada}) system that is the responsibility of the Slow Control and Monitoring group. The \dword{scada}  system pushes values to a back-end database, where the \dword{db} flavor is tied to the \dword{scada} solution. 
The \dword{scada}  system can provide data reduction through filtering prior to insertion of metadata into the back-end \dword{db}, which reduces the workload on any API used to move the metadata to the \dword{cdb}. 

\subsection{Slow Control Database for ProtoDUNE II}
\label{sec:slowcontrolPD}

The \dword{protodune} experiment has been using an Oracle\footnote{Oracle\textcopyright, \url{https://www.oracle.com}} back-end \dword{db} for the slow control system. As 
\dword{scd}
does not support Oracle, the information from the Oracle database must be extracted and moved into a Postgres \dword{db} at Fermilab. Any filtering of the metadata not handled by the \dword{scada}  system when populating the Oracle database can be handled by the API that transfers the Oracle records to Postgres.

No data filtering was provided by the \dword{scada} system for \dword{protodune} Run I but for Run II it is expected that the slow controls \dword{condmeta} will be filtered based on the physics needs.

\section{Beam Conditions Database - IFBeam  \hideme{Buchanan and Laycock - draft}}
\label{sec:db:ifbeam}  

The beam conditions database, \dword{ifbeam}\cite{ifbeam}, will contain metadata related to the condition extracted beam and corresponding diagnostics.  The functional form of this database is essentially the same as that of the slow control database. A large number of devices are sampled into the \dword{ifbeam} \dword{db}. The \dword{ifbeam} metadata transferred to the conditions \dword{db} will be a coarser subset of the original set.

Quantities contained in the \dword{ifbeam} \dword{db} include beam currents, horn currents and polarities, and beam monitoring instrument metadata.

\begin{dunetable}
[Example IFBeam metadata]
{l  l } 
{table:ifbeam}
{Example metadata values and types stored in the \dword{ifbeam} database.}
 Metadata Value & Type  \\ \toprowrule [0.5ex] 
Horn 1 Polarity &  Integer \\ \colhline
Horn 2 Polarity  & Integer  \\ \colhline
Beam current & Float \\  
\end{dunetable}

\section{Hardware Database  \hideme{Buchanan and Laycock - draft}}
\label{sec:db:hwdb}  

The principle purpose of the hardware database (\dword{hwdb}) is to track the lineage of hardware components and record the results of their \dword{qc} tests. In this context a component can be a sub-detector module or any of the individual parts comprising it. For example, a readout board is a component as is a mezzanine daughter board or programmable logic chip mounted on the readout board. The lowest level component tracked within the \dword{hwdb} will be unique to the corresponding hardware system. 

A requirement of the \dword{hwdb} is that any component, or part, stored in the database must have a unique identification number assigned to it, coordinated by the \dword{dune} Integration group. %
The part numbers will be designated as shown in Table~\ref{table:partsid}. The project field corresponds to \dword{dune} detectors (D), integration (I), LBNF (L), and future project (P). The project identifier is allocated by the project management team while the other identifiers are left for the various hardware consortia to assign. There are additional fields not listed as they are not relevant to the \dword{hwdb}.  More details of the parts identification number can be found in~\cite{bib:cernedms2505353}.

\begin{dunetable}
[Hardware database component IDs]
{l l l l l l} 
{table:partsid}
{Unique parts identification number assigned to each component stored in the hardware database.}
Project & System ID & Subsystem ID & Item Type ID & Dash & Item Number  \\ \toprowrule 
D/I/L/P & 01-99 & 001-999 & 00001-99999 & - & 00001-99999 \\  
\end{dunetable}

Hardware \dword{db} metadata will reflect the complete lifetime of the detector component, including the following:

\begin{itemize}
\item Procurement
\item Fabrication
\item Quality control testing
\item Shipping and storage
\item Installation
\item Maintenance 
\end{itemize}

The relationships between components will be reflected in the \dword{hwdb}. Metadata corresponding to multiple instances of events such as \dword{qc} tests will be handled using time series within the database. 

The database group will provide an interface to the \dword{hwdb} and each hardware consortium will be required to ensure that their metadata is inserted into the database. Given the wide range of hardware consortia, the international nature of the experiment and the fact that individual consortia must manage their own construction projects, it is 
beyond the scope of the database group to dictate how the consortia will handle data entry workflows for the \dword{hwdb}.  Instead, the database group will consult with the consortia, and in particular their database group liaisons, developing simple APIs, providing documentation for the \dword{hwdb} and advising on efficient methods for working with it.  Sharing of tools will be very strongly encouraged to reduce the duplication of effort as far as possible.

\section{Service and Maintenance  \hideme{Buchanan and Laycock - draft}}
\label{sec:db:service}  

Most, if not all, of the \dword{dune} databases will operate in advance of the full \dword{dune} experiment coming online and these databases will need to be maintained and serviced once they are operational. 

The second run of the \dword{protodune} experiment (\dword{protodune2}) will employ a suite of databases that will be the precursors to the full database system that will be in place for \dword{dune}. Each of these databases (run configuration, beam instrumentation, conditions, slow controls, and hardware) will require stable monitoring, maintenance, and service to address operational issues that will arise in the lead up to and during the running of the \dword{protodune} II experiment. 

Monitoring will be achieved using automated web-based tools and responses to offline database issues will be made within an 8-hour period corresponding to a typical operation or production ``shift'', i.e. 5 days a week\footnote{The assumption is that offline database services will not require 7 days a week coverage.}. For \dword{protodune2} databases will be located at both \dword{fnal} and \dword{cern}, both of which have a long history of database support.  Moving on to \dword{dune}, the expectation is that another \dword{doe} lab like \dword{bnl} would share database operations with \dword{fnal}.

\section{Development Plans  \hideme{Buchanan and Laycock - draft}}

There are a number of database-related projects where R\&D is needed or underway, utilizing effort and expertise from both \dword{doe} labs and universities. Recent effort (about 3 postdoc FTEs for 3 years) has been added through \dword{dune} computing-focused \dword{doe} support. Coordination is provided by the Database group and attempts to balance the short-term needs of \dword{protodune} with the longer term needs of \dword{dune}.

\subsection{Conditions Database Development}
\label{sec:db:conddbdev}

The \dword{cdb} is the primary interface to \dword{condmeta} for all \dword{dune} distributed computing resources and this task includes designing and developing caching strategies.  The \dword{dune} Database group will collaborate with the \dword{hsf} to benefit from the experience of other \dword{hep} experiments, as well as the experience gained from \dword{protodune}, to develop a robust system while optimizing development effort.
The development for the \dword{cdb} is estimated to be 6 FTE years with that effort largely flat as it is spread over the following sub-tasks.

\subsubsection{Conditions Database Core Design}

The \dword{hsf} \dword{cdb}s group identifies the following points as key features of a good \dword{cdb} design:

\begin{itemize}
    \item    Loose coupling between client and server using RESTful interfaces
    \item    The ability to cache queries as well as payloads
    \item    Separation of payload queries from metadata queries
\end{itemize}

These guiding design principles are likely to remain valid throughout the lifetime of \dword{dune}, while implementation will need to evolve with technology.

\subsubsection{Conditions Database and Software Framework}

The interface between the \dword{cdb} and the Software Framework is defined by the \dword{api} %
of the \dword{cdb}.  The \dword{hsf} Conditions Database group is in the process of defining a generic \dword{api}, again using the experience of existing \dword{hep} experiments to understand best practice for supporting all read and write use cases.  \dword{dune} will collaborate with the \dword{hsf} on this API definition and share experience on implementations.  It is noted that, in line with the \dword{hsf}, given the heterogeneity of compute hardware that \dword{dune} needs to use, a common implementation is considered to be less important than a common API and sharing experience.

\subsubsection{CDB Service Robustness and Distributed Computing}

In addition to the cache-friendly design of the central \dword{cdb} service itself, a major issue for a critical (for offline) service like the \dword{cdb} is resilience such that a robust and reliable service can be provided.  \dword{dune} plans to take advantage of synergies with other \dword{doe} labs, particularly at \dword{bnl} which hosts \dword{cdb} services for Belle II and other experiments.  The backend database technology (\dword{postgres}) is common between \dword{fnal} and \dword{bnl}, and generally \dword{cdb} services, under the umbrella of the \dword{hsf}, are expected to evolve to look more and more similar.  \dword{dune} will 
therefore be well-positioned to benefit from progress in understanding multi-site database deployments assuming
that these favorable circumstances encourage the necessary investigations.
This could
potentially include having a first class backup service that would support writing as well as reading functionality, in addition to the existing redundancy of read functionality via intelligent use of caches at sites.

\subsubsection{Database Access for High-Performance Computing Facilities}

As \dword{dune} will utilize \dword{hpc} facilities for some analysis tasks it is important that \dword{db} access is manageable when tens, or hundreds, of thousands of processes are distributed across an \dword{hpc} cluster. Studies on how scaling on such systems can be handled without overwhelming the \dword{cdb} when an enormous number of simultaneous queries are required. Again, \dword{dune} plans to leverage the experience of other \dword{hep} experiments and contribute back to the common tools for using \dwords{hpc} at scale.

\subsection{Conditions Metadata Format and Serialization}

Similarly to the discussion in chapter \ref{ch:format}, the data format and serialization of \dword{condmeta} deserves careful attention to avoid problems including technology lock-in.  The need to interact with many different groups, essentially all providers of conditions (and other persistent) metadata, is the main driver of the amount of effort required for this task, estimated to be 4 FTE years with that effort concentrated in FY2024-FY2026.

\subsection{Slow Control Database Development}

The \dword{scada}  system chosen for Run I and II of the \dword{protodune} experiment was WinCC with Oracle as the back-end \dword{db}. This system is well tested at scale and it is relatively trivial to transfer information from the Oracle \dword{db} to a \dword{postgres} \dword{db} located at \dword{fnal}. As discussed earlier, for \dword{dune} it would be beneficial to use a solution that enables a \dword{postgres} back-end 
and recent developments allow WinCC to use \dword{postgres} (and other) backends.  The \dword{scada} system for \dword{dune}, which at the time of writing has not been decided, may also bring new challenges
Direct collaboration is envisaged here in order to prevent a disconnect between the \dword{scada} system, which will generate massive amounts of data, and the subset needed for offline data processing.  The development effort from the Database group is estimated to be 2 FTE years with that effort front-loaded to work with \dword{dune} \dword{scada} experts.

\subsection{Hardware Database Development}

The core functionality of the Hardware Database is described in section \ref{sec:db:hwdb} and the first version is expected to be complete in FY22.  Experience gained as the consortia use the database are expected to motivate feature requests, many of which will require urgent attention to support the construction phase of \dword{dune}.  Looking further ahead, some attention will be needed to ensure the maintenance of this crucial database for the long lifetime of the collaboration.  Some information in the Hardware Database will correspond to Conditions Metadata (calibrations), thus requiring the transfer of data while retaining provenance. The development for the Hardware \dword{db} is estimated to be 2 FTE years with that effort concentrated in FY2022-FY-2023.

\subsection{Run Configuration Database Development}

Transfer of Run Configuration information for offline use was a major problem for \dword{protodune} Run I, and significant effort will be used to improve the situation for Run II.  The evolution of both the \dword{dune} \dword{cdb} and \dword{daq} system could imply a similar amount of effort moving to \dword{dune}.
The development for the Run Configuration \dword{db} is estimated to be 1 FTE year with that effort concentrated in FY2022 and again in FY2026.

\subsection{Other Database Development}

The \dword{dune} Calibration system is expected to generate very large datasets that will be processed to create \dword{condmeta} that needs to be transferred to the \dword{cdb}.  Meanwhile the Beam conditions database stores far more data than needed for the data processing Conditions Metadata, and effort will be needed to find efficient solutions for data reduction.  Finally, the Data Quality Monitoring Database is under the control of the \dword{dqm} group, with some amount of data needing to be transferred to the \dword{cdb}.  All of these connections could imply additional development work which, if arising from offline-only constraints, should be supported by Database group effort.
The development effort here is estimated to be 1 FTE year with that effort responding to needs as they appear and hence having a flat profile.

\subsection{Database Access Tool Development and Documentation}

Guided by use cases, tools will be developed for all of the \dword{dune} Databases to enable 
direct access for studies.  Most offline DB access will be made to the Conditions DB via the \dword{dune} software framework, transparently to users, but there will be cases where expert users will need direct access to the various databases.

Documentation is another critical aspect of the database system. This includes descriptions of the DB system components and training materials, that will need to be updated as development work proceeds. The number of use cases, databases and groups involved drive the significant amount of effort required here.
The effort for tool development and documentation is estimated to be 2 FTE years with that effort spread over time as needs arise.

\subsection{Person Power Estimates}

The personnel needs will be largely front-loaded as the database systems are researched, implemented, and tested. The database requiring the most effort will be the \dword{cdb}, with significant effort required for the data format, other databases, and tools.

\cleardoublepage

\part{Global Computing Model  }\label{part:model} %

\chapter{Data and Processing Volume Estimates \hideme{4/22  HMS changes from Anne and JZ}}

\label{ch:est}

\section{Introduction \hideme{Schellman}}%
To understand the resources needed for \dword{dune} Computing and the development needed to utilize those resources, we start with an estimate of the data volumes and CPU needs from bottom-up estimates. 
In this chapter, we describe %
the assumptions that go into the estimates of data volumes and describe possible methods of reducing the total volumes while retaining physics capabilities. 
These assumptions have been coded into a python-based model and are updated frequently based upon changes to the \dword{fd} and \dword{nd} designs, and the physics requirements.

\dword{dune}'s detectors will produce information from a variety of technologies.  We anticipate that raw data volumes will be dominated by the digitized waveforms from the \dword{lar} detectors, and to a lesser extent from the \dwords{pd}. \Dwords{lartpc} read out over long time windows, while the \dword{pd} detectors read out only above threshold. We find that the \dword{lartpc} information dominates at the raw data level and drives the total data volume. %

The \dword{daq} system can reduce these raw data volumes by several means, including:
\begin{itemize} 
\item short readout windows tailored to one drift time;
\item triggered readout of particular time slices;
\item triggered readout of specific detector sub-regions; %
\item lossless compression; %
\item lossy zero suppression; and/or
\item hardware pattern recognition.
\end{itemize}

Overall, we assume that the above methods can reduce data volumes from the hundreds of exabytes that would be produced by continuous readout to a manageable 30\,PB/year. For beam and calibration events our assumption is that readouts of \dword{lartpc} detectors will  generally be confined to a single drift time window.  \Dword{snb} readouts can last up to 100~s and would consist of up to 40,000 drift time units. The exact time segmentation for a \dword{snb} readout has not yet been determined but is limited by reasonable file sizes to much less than the full readout.

\section{ProtoDUNE Experience\hideme{Schellman - draft}}
\label{sec:est:ProtoDUNE}  

Our estimates  are largely based on our experience with the  %
\dwords{protodune} that ran at the \dword{cern} in 2018 (\dword{pdsp}) and 2019 (\dword{pddp}). 

The \dword{pdsp} detector, which used \dword{sp} \dword{sphd}  technology, was read out by six \dwords{apa} and a mix of \dwords{pd}. The corresponding first \dword{fd} module will have 150 \dwords{apa} and  \dwords{pd} based on the \dword{arapuca} %
technology. The second \dword{fd} module will use \dword{sp} \dword{spvd} and  will share some technology with  \dword{pddp}.
Table~\ref{tab:est:usefulpd} summarizes the parameters from the \dword{pdsp} experience that have been used to predict data volumes for the \dword{fd}.   The memory footprint per \dword{apa} (\dword{crp} for \dword{spvd}) was estimated from the algorithms and frameworks available for \dword{protodune}.  Work is ongoing to process the raw data separately for each \dword{apa} (\dword{crp}) so that the memory usage is not needed for all \dword{apa} (\dwords{crp}) simultaneously, but instead the work can be serialized or parallelized.

\subsection{ProtoDUNE Single Phase Experience}

The \dword{pdsp} data have been processed through three reconstruction campaigns with the first operating just-in-time as data arrived, in  ``keep-up'' mode.
  There have also been several \dword{pdsp} simulation campaigns.  This work has resulted both in publications \cite{DUNE:2021hwx,DUNE:2020fgq,DUNE:2020cqd,DUNE:2020vmp} and  in robust estimates of the computational characteristics of the data and processing. The characteristics of the \dword{pdsp} data are summarized in Table~\ref{tab:est:usefulpd}.
For example, the uncompressed \dword{sp} raw data for a single \dword{pdsp} trigger record were observed to be around 178\,MB in size, which is the amount expected for the number  of \dword{tpc} channels read + a 20\% overhead for other detectors and headers.  Compressed \dword{sp} raw data averages 70\,MB, consistent with compression by a factor of 2.5.  

 \begin{dunetable}[Useful quantities from the ProtoDUNE experience]{lrr}{tab:est:usefulpd}
{Useful quantities for computing estimates for \dword{hd}
readout based on \dword{pdsp} experience. These numbers assume 12-bit readout. Hit reconstruction combines signal processing and hit finding. }%
Quantity&Value&Explanation\\
\toprowrule
Number of APAs&6\\
Number of channels/APA&2,560&\\
Readout time & 3 ms&\\
\# of time slices & 6000&\\
Single APA readout &23 MB& Uncompressed  estimate\\ \colhline
Full detector readout &178 MB& Uncompressed real \\ \colhline
Full detector readout &70 MB& Compressed real \\ \colhline
Effective compression factor &2.5&\\ \colhline
Beam rep. rate&4.5 Hz&Average\\ \colhline
Hit reconstruction CPU time/APA& 30 sec&from MC/ProtoDUNE\\ \colhline
Pattern recognition CPU time/event & 400 sec&from MC/ProtoDUNE\\ \colhline
Simulation CPU time event & 2,700 sec&from MC/ProtoDUNE\\ \colhline
Memory footprint/APA&0.5-1GB&ProtoDUNE experience\\ 
\end{dunetable}

\subsection{ProtoDUNE Dual Phase Data}

\dword{pddp} recorded signals using two \dwords{crp} during the 2019 run.  Observed data size without compression  was 110\,MB.  %
In fall 2021, \dword{sp} \dword{vd} %
readout was tested in a smaller \coldbox and successfully reconstructed. Full \dword{protodune} tests of both \dword{hd} and \dword{vd} technologies with beam and cosmic rays are slated for 2022-24. The data volume estimates in Section~\ref{sec:est:volumes} include data and simulation for this second \dword{protodune} test campaign.

\section{Far Detector Data Volume Estimates }%
\label{sec:est:FD}  

The raw data volume estimates presented here are based on the %
Spring 2022 
\dword{fd}  and \dword{nd} designs. %
This assumes that the first module will be \dword{sphd}, the second module \dfirst{spvd}, and a third and fourth module equivalent in terms of data volume to \dword{sphd} will be added in the mid 2030s.  Estimates are provided for the Phase~I and Phase~II near detector (\dword{nd}) designs. A separate discussion is provided for each early \dword{fd} module type and the \dword{nd}. The summary plots and table at the end include data and simulation for all detectors.

\subsection{Horizontal Drift}
For the initial \dword{hd} \dword{fd} data volumes, we use our \dword{pdsp} experience and assume that raw data sizes and hit-finding CPU times scale with the number of \dword{apa}s, while pattern recognition and simulation times scale with the number of interactions. 

 \begin{dunetable}[Useful quantities for computing \dshort{hd} data volume
estimates]{lrr}{tab:est:usefulfdhd}
{Useful quantities for computing estimates for \dword{hd}
readout based on the \dword{daq} requirements document of January 2022.  CPU times are scaled from \dword{pdsp} assuming all detectors are used in data unpacking, signal processing and hit finding, but interactions are confined to a subsection of the detector not much larger than \dword{pdsp}.}%
Quantity&Value&Explanation\\
\toprowrule
{\bf Far Detector Horizontal Drift}\\ \colhline
APAs per module& 150&DAQ spec.\\
TPC channels&	384,000&DAQ spec.\\
TPC channel count per APA&	2560&DAQ spec.\\
TPC ADC sampling time& 512 ns&DAQ spec.\\
TPC ADC dynamic range&	14 bits&DAQ spec.\\
FD module trigger record window &	2.6 ms&DAQ spec.\\
Extended FD module trigger record window&	100 s&DAQ spec.\\
Size of uncompressed trigger record&	3.8 GB&DAQ spec.\\
Size of uncompressed extended trigger record &	140 TB&DAQ spec.\\
Compression factor &TBD&\\
Beam rep. rate&\beamreprate&Untriggered\\ \colhline
Hit finding CPU time&4500 sec&from MC/ProtoDUNE\\ %
Pattern recognition CPU time &1500 sec&from MC/ProtoDUNE\\ %
Simulation time CPU time event & 2700 sec&from MC/ProtoDUNE\\ %
Memory footprint/APA&0.5-1GB&ProtoDUNE experience\\ 
\end{dunetable}

The DUNE Data Volume document \cite{bib:docdb14983} %
describes the expected event rates for various signatures in a \dword{fd} module.  These can be combined with the above numbers to provide  the integrated data estimates shown in Table~\ref{tab:est:hdfdrates}. 

 \begin{dunetable}
 [Horizontal drift data volumes] {|l |r r r |}{tab:est:hdfdrates}
{Data sizes and rates for different processes in each horizontal drift detector module.  Uncompressed data sizes are given. As readouts will be self-triggering, a 2.6\,ms drift-window readout time is used instead of the 3\,ms for the triggered \dword{pdsp} runs.  We assume beam uptime of 50\% and 100\% uptime for non-beam science. These numbers are derived from References~\cite{bib:docdb24732} and \cite{bib:docdb14983}.}
Process & Rate/module & \qquad size/instance &\qquad  size/module/year\\
\toprowrule
Beam event & 41/day & 3.8 GB&30 TB/year\\
Cosmic rays &4,500/day&  3.8 GB& 6.2 PB/year\\
Supernova trigger& 1/month& 140 TB& 1.7 PB/year\\
Solar neutrinos&10,000/year&$\le$3.8 GB&35 TB/year\\
Calibrations&2/year&750 TB& 1.5 PB/year\\
\colhline 
Total& & &9.4 PB/year\\
\end{dunetable}%

\subsection{Far Detector Module with Vertical Drift Readout}

 Tables~\ref{tab:est:usefulfdvd} and~\ref{tab:est:vdfdrates} summarize the expected data rates and volumes from physics signals of interest in  a \dword{spvd} \dword{detmodule}. %
 The data volume  corresponding  to calibration events in the vertical drift module will likely be similar to the numbers shown for the horizontal drift in Table~\ref{tab:est:hdfdrates}; a more detailed estimation is ongoing. 

\begin{dunetable}[Useful quantities for computing \dshort{spvd} data volume
estimates]{lrr}{tab:est:usefulfdvd}
{Useful quantities for computing estimates for \dword{spvd}
readout based on the \dword{daq} requirements document of January 2022.  CPU times are scaled from \dword{pdsp} assuming all detectors are used in data unpacking, signal processing and hit finding, but interactions are confined to a subsection of the detector not much larger than \dword{pdsp}.}%
Quantity&Value&Explanation\\
\toprowrule
{\bf Far Detector Vertical Drift}\\ \colhline
CRPs per module& 160&DAQ spec.\\
TPC channels&491,520&DAQ spec.\\
TPC channel count per CRP&	3,072&DAQ spec.\\
TPC ADC sampling time& 512 ns&DAQ spec.\\
TPC ADC dynamic range&	14 bits&DAQ spec.\\
\dword{vd} module trigger record window &	4.25 ms&DAQ spec.\\
Extended FD module trigger record window&	100 s&DAQ spec.\\
Size of uncompressed trigger record&	8 GB&DAQ spec.\\
Size of uncompressed extended trigger record &	180 TB&DAQ spec.\\
Compression factor &TBD&\\
Beam rep. rate&\beamreprate&Untriggered\\ \colhline
Hit finding CPU time&6,000 sec&from MC/ProtoDUNE\\ %
Pattern recognition CPU time per event &1,500 sec&from MC/ProtoDUNE\\ %
Simulation CPU time per event & 2,700 sec&from MC/ProtoDUNE\\ %
Memory footprint/CRP&0.5-1GB&ProtoDUNE experience\\ %
\end{dunetable}

 \begin{dunetable}
 [%
 Vertical Drift Far Detector data volumes] {|l |r r r |}{tab:est:vdfdrates}
{Data sizes and rates for different processes in a far detector module based on vertical drift technology. 
Uncompressed data sizes are given. As readouts will be self-triggering, an extended 4.25\,ms readout window is used.  We assume beam uptime of 50\% and 100\% uptime for non-beam science.  The interaction/readout rates are derived from references~\cite{bib:docdb16028} and~\cite{bib:docdb14983}.
} 
Process & Rate/module & \qquad event size  &\qquad  size/module/year\\
\hline
Beam event & 41/day & 8 GB& 63 TB/year\\
Cosmic rays &4,500/day&  8 GB& 12.5 PB/year\\
Supernova trigger& 1/month& 180 TB& 2 PB/year\\
Solar neutrinos&10,000/year&46 TB/year\\
Calibrations&2/year& & 1.5 PB/year\\
\hline 
Total& & &16 PB/year\\
\end{dunetable}%

\subsection{Far Detector Summary}

Overall, bottom-up estimates yield data volumes of around 9.4\,PB/year and 16\,PB/year for each \dword{sphd} and \dword{spvd} module respectively.  Lossless compression and restriction of the readout to geographical regions of interest should reduce this volume substantially. However, additional modules will  increase these rates.  A maximum rate of 30\,PB/year of raw data across all modules and modes of operation has been specified and it is assumed that \dword{daq} and detector configurations will change to meet this specification.  We  note that 30\,PB/year is  an average of 1.3\,GB/sec, less than the rates already demonstrated for \dword{protodune} %
\dword{daq} and storage.  In principle, at 1 CPU-sec/MB of uncompressed input (from \dword{pdsp} experience), a few thousand cores (current FNAL - $\sim$11 HS06~\cite{bib:HS06} each) could keep up with these data rates,  but this throughput must be maintained over many years.   In addition, \dword{snb} candidates may %
produce bursts %
requiring much higher \dword{daq} and processing rates. %

\section{Near Detector Data Volumes }%
\label{sec:est:ND}  
This section is based on the estimates provided in the near detector (\dword{nd}) \dword{cdr}~\cite{DUNE:2021tad}. Abbreviated descriptions of the \dword{nd} components are given in this document in Section~\ref{sec:intro-nd}, along with the detector parameters that drive the annual data estimates. Summaries of the simulation and reconstruction workflows for the \dword{nd} components are given in Sections~\ref{ch:use:nd} and~\ref{sec:nd-reco}, respectively.

The expected annual data sizes from the \dword{nd} are summarized in Table~\ref{tab:nd_data_volume_estimates}. Due to the much higher data density in the \dword{nd} as compared with the \dword{fd}, CPU times/beam spill are expected to be much higher and are estimated to be 300\,CPU/sec/spill using current processors and the 1.2\,MW beam; i.e., a total of $1.5\times 10^7$ spills/year. Simulated data samples will need to be an order of magnitude larger and thus require at least 10 times the CPU power.  This leads to a rough estimate of CPU needed for \dword{nd} reconstruction and simulation of approximately 3,000 core-years/year.  Replacement of  \dword{tms} with \dword{ndgar} in Phase~II, along with an increase in the detector occupancy for all three detectors, is expected to increase the required CPU per spill.  Table~\ref{tab:NDCPUPerEvent} accounts for the change in the detector configuration from Phase~I to Phase~II but not the effect of occupancy, which can be mitigated by improvements to reconstruction algorithms.

\begin{dunetable}[Near Detector Data Estimates]
{l r}
{tab:nd_data_volume_estimates}
{Annual DUNE \dword{nd} raw data volume estimates.  No compression is assumed.  \dword{tms} is assumed to be installed in DUNE Phase I while \dword{ndgar} is assumed to be installed in DUNE Phase II.}
Type & Volume/year\\ \toprowrule
    {\bf \dword{ndlar}}     &  \\
    \quad\quad In-spill data & 144 TB \\
    \quad\quad Out-of-spill cosmics & 16 TB\\
    \quad\quad Calibration & 16 TB\\
    \quad\quad Total & 176 TB \\\toprowrule
    {\bf \glsunset{tms}\dword{tms} (DUNE Phase I)}           & \\
    \quad\quad In-spill data & 2 TB \\
    \quad\quad Out-of-spill cosmics & 8 TB \\
    \quad\quad Calibration & 1 TB\\
    \quad\quad Total & 11 TB \\\toprowrule
    {\bf \dword{ndgar} (DUNE Phase II)}           & \\
    \quad\quad In-spill data & 52 TB \\
    \quad\quad Out-of-spill cosmics & 10 TB \\
    \quad\quad Calibration & 6 TB\\
    \quad\quad Total & 68 TB \\\toprowrule
    {\bf \dword{sand}}        & \\
        \quad\quad In-spill data & 40 TB\\
    \quad\quad Out-of-spill cosmics & 8 TB\\
    \quad\quad Calibration & 1 TB \\
    \quad\quad Total & 49 TB \\\toprowrule
    {\bf Total ND (Phase I)} & {\bf 236 TB}\\
    {\bf Total ND (Phase II)} & {\bf 293 TB}\\
\end{dunetable}

\begin{dunetable}
[CPU estimates for Near Detector]
{l r}
{tab:NDCPUPerEvent}
{Preliminary CPU estimates per event for the DUNE near detector components, in seconds.}
Type&time/event\\ \toprowrule
    {\bf LArTPC} &  \\
    \quad\quad Monte Carlo gen+sim & 100 s \\
    \quad\quad Reconstruction & 60 s\\\toprowrule
    {\bf TMS (DUNE Phase I)} &  \\
    \quad\quad Monte Carlo gen+sim & 50 s\\
    \quad\quad Reconstruction & 5 s\\\toprowrule
  {\bf ND-GAr (DUNE Phase II)} &  \\
    \quad\quad Monte Carlo gen+sim & 100 s\\
    \quad\quad Reconstruction & 10 s\\\toprowrule
    {\bf SAND} & \\
    \quad\quad Monte Carlo gen+sim & 100 s\\
    \quad\quad Reconstruction & 10 s\\
\end{dunetable}

\section{Data Retention Assumptions}
\hideme{Schellman - draft}
\label{ch:est:retention}

The Computing Consortium has made the decision to require all data located on DUNE storage elements to be registered within the Data Management system and have a  data retention policy associated with each file. 
The goal is to assure that there is no ``dark'' data occupying storage capacity and that these data retention policies will help minimize total resource needs and our ability to recycle/reuse storage media. A short summary of the currently planned retention policy is listed in Table~\ref{tab:est:retention}.

 \begin{dunetable}[Data Retention Policies]{llrrrr}{tab:est:retention}
{Retention policies by data tier}
Tier&Description&Tape copies& Lifetime &Disk Copies& Disk Lifetime\\ \toprowrule
Raw & Physics data& 2 & indefinitely & 1 & 1 year\\ \colhline
Test & test and commissioning & 1 &6 months &1 & 6 months \\ \colhline
Waveforms & processed waveforms & 1 & 10 years & 1 & 1 month \\ \colhline
Reco & pattern recognition &1 & 10 years & 2 & 2 years\\
\end{dunetable}

\section{Data Tiers and Flow}

\subsection{Data Tiers \hideme{Schellman/Norman }}\label{sec:dataflow}

As the data from the \dword{dune} detectors and simulation systems are processed, they evolve through a series of \dwords{datatier}.  The \dword{datatier} concept was first adopted by the D0 and CDF collaborations two decades ago and is an important component in the definition of datasets in the data catalog.
The algorithms which are run against data, transform it from one \dword{datatier} to another in what is effectively a directed graph.  This allows deterministic workflows to be constructed in a modular fashion and for dependencies in the data representation to be identified.  This also allows for different branches and paths to be taken logically in the data lifecycle and for the overall computing model to manage these transforms that advance the state of the data in much the same manner as one would expect from a simple state machine.

Most importantly, this concept of data tiers allows for different information and informational representations to be present in each tier.  This allows easier management and storage of data, (as down selections can be applied at different tiers to reduce overall storage footprints) while still permitting specialized applications to be developed that require specific data products or representations be present in the data. %
This approach also allows for parallel treatments of simulation and real detector data, which is key to the analysis process. 

\begin{dunetable}
[Data Tier Summary]
{l l  c l l} 
 {table:format1}
 {Example data tiers. We show the number of copies for each version (1 version/year) and the proposed lifetimes for each data type on tape and disk.  The total volumes of data/year are discussed in Section \ref{sec:est:volumes}.}
Data Tier & Produced by  & Copies & Disk Lifetime  & Tape Lifetime \\ [0.5ex] 
PDSP raw data&DAQ&2&few months&$>20$years\\
PDSP full-reconstructed (data)&pattern recognition&1&2 years&5-10 years\\
PDSP detector-simulated (MC)&geant4+detsim&1&transient&5-10 years\\
PDSP hit-reconstructed (MC)&hit finding&1&transient&5-10 years\\
PDSP full-reconstructed (MC)&pattern recognition&1&2 years&5-10 years\\
PDSP root-tuple&data reduction&1
&2-3 years&5-10 years\\
\hline
FD raw data&DAQ&2&few months&$>20$years\\
FD hit-signalproc (data)&signalproc&1&--transient&--\\
FD full-reconstructed (data)&pattern recognition&1&2 years&5-10 years\\
FD hit-reconstructed (MC)&pattern-recognition&2&few months&$>20$years\\
FD full-reconstructed (MC)&pattern recognition&1&2 years&5-10 years\\
\end{dunetable}

This diversity in tiers can be seen by examining the projected estimates for the \dword{dune} data and retention policies for each tier.  Tables \ref{table:format1} and \ref{table:format2} summarize some of the more common formats for \dword{dune}, along with their typical trigger sizes and expected retention schedules. Full volume estimates over the course of the experiment are given in Section  \ref{sec:est:volumes}. For more details on the processing flow for data tiers, see Figure \ref{fig:ch:use:pdii}. Examples of common \dwords{datatier} are:

\begin{description}
\item{\bf raw} for raw detector data.  
\item{\bf simulated} simulated data from event generators with \dword{geant4} simulation only.  The geant4 step is computationally expensive so these data may be kept long-term.
\item{\bf detector-simulated} geant4 simulation with additional detector response simulation.   Different variants of, for example space charge effects, overlays and photon libraries are applied here. 
\item{\bf hit-signalproc} future tier for data that has had basic data preparation signal processing up through the deconvolution step performed.  This stage of processing is well suited to HPC's so this data-tier may be produces as an intermediate step. 
\item{\bf hit-reconstructed} for data that has had initial reconstruction of regions of interest.  This step is currently not stored but may play in important role in future if hit finding and reconstruction are done on a different architecture than the full reconstruction algorithms.
\item{\bf full-reconstructed} for both data and simulation.  This tier has advanced reconstruction to identify higher-level physics objects and interaction regions. %
\item{\bf root-tuple} Small common root-tuples summarizing the outputs of the full-reconstructed tier. These would be inputs for final analysis, including columnar analysis.
\item{\bf reco-dst}  An intermediate stage between full-reconstructed and analysis tuples. As we are still in the algorithm development phase, we have not yet converged on this format which likely will be <10\% of the reconstructed stage. 
\end{description}

Table \ref{table:format2} shows the estimated event sizes and rates for the different data tiers.  These are used to generate the much more detailed data volume predictions below.

\begin{dunetable}
[Event sizes and number by tier]
{l | r r r} 
 {table:format2}
 {Summary of data sizes (in MB) and numbers, when running, for the 3 majors DUNE efforts.  The Near Detector (ND) and ProtoDUNE need full detector simulation  while the Far Detectors (FD) only need simulation in the region ($\sim 15$\%) around the interaction. } 
 Data Tier & ProtoDUNE (MB)& DUNE FD (MB)& DUNE ND(MB)\\
Raw & 70--140&3750--8000& 10\\
Simulated Tiers & 200-300&&20-50\\
Reconstructed Data & 35 & 175 &20\\
Number of raw M-Events/year& 10& 2.2&25\\
Number of sim M-Events/year& 10& 10&100\\ 
Peak dates&2018-2025&2028-2040&2028-2040\\
 \end{dunetable}
 
 \subsection{Far Detector data flow}
 
 A possible \dword{fd} workflow motivated by the \dword{pdsp} experience is described below.   Data may be stored temporarily at intermediate steps, for example if the optimal site for processing moves from an HPC to a more typical batch processing center.
 
 \begin{description}
 \item{\bf Raw data} 3750--8000 MB for a single readout of a full module, is processed to produce:
 \item{\bf hit-signalproc} waveforms that may be stored temporarily and  then be processed further to produce:
 \item{\bf hit-reconstructed} hit regions of interest. These are likely 100-200 MB/readout, both because of data reduction within a single waveform and the ability to exclude large regions of the detector far away from the interaction. Those hits are then used for calibration and higher level pattern recognition producing:
 \item{\bf full-reconstructed} which contains the hits and higher level tracking information. Those data are then used to produced much smaller data summaries for later analysis. 
 \end{description}
 
 Figure \ref{fig:ch:use:pdii} in Chapter \ref{ch:use:intro} illustrates this process. 
 
 As the bulk of the far detector data will be calibration samples, we can anticipate a need to process and temporarily store the early steps in reconstruction multiple times soon after it is taken but the permanent store of reconstructed data for physics analysis will be orders of magnitude smaller. However, at this point we make a very conservative estimate that all raw data will be reconstructed and the results stored.

We assume that we will be performing a full reprocessing of all raw data once/year with limited new simulation campaigns on a similar cadence. Generally the 2-year retention policy for reconstructed samples assures that the most recent two versions are readily available while older versions would need to be staged from tape and only available for special use-cases.

The processed waveforms are of considerable interest for machine learning algorithms but constitute a very large data volume.  Given limited disk resources, we will likely need to prioritize disk access to the full reconstructed data by physics analyzers, possibly with a subset of the waveform information. Full access to the waveforms will likely require reprocessing of samples from archival storage. 

\section{Model Studies for Data and CPU Needs \hideme{Schellman 3/4 updated - draft}}
\label{sec:est:volumes}

Given the above estimates we can  estimate total disk and CPU needs every year.  The July 2022 version of these numbers is documented in~\cite{bib:docdb24732}.  Parameters are entered via a \dword{json} file and results generated using python scripts. Figure~\ref{fig:est:events} and~\ref{fig:est:sim-events} shows the assumed number of events/year for each detector type.  

Planned simulation event volumes (shown in Figure~\ref{fig:est:sim-events} are a factor of 2-4 times the raw data rate. We assume that the rate of triggers that are not useful is at least 50\% even for \dword{protodune} and the \dword{nd}, and, in the case of the \dword{fd}, far larger.  This implies a ratio of 4 to 8 between simulation and data.  Discussions of strategies to optimally use the simulation data volume are given in Sections~\ref{sec:algo-use-fdsimvol} and~\ref{sec:use-sim-considerations}. 

For downstream CPU and disk calculations, data and simulation are shown together and simulation tends to equal or dominate the resource needs due to a combination of event size, CPU time and number of events.

\begin{dunefigure}
[Event estimates]
{fig:est:events}
{Numbers of raw data events from the detectors per year used in the model data volume estimates. Left is through 2025, right is the same through 2040.  }
\includegraphics[width=0.49\textwidth]{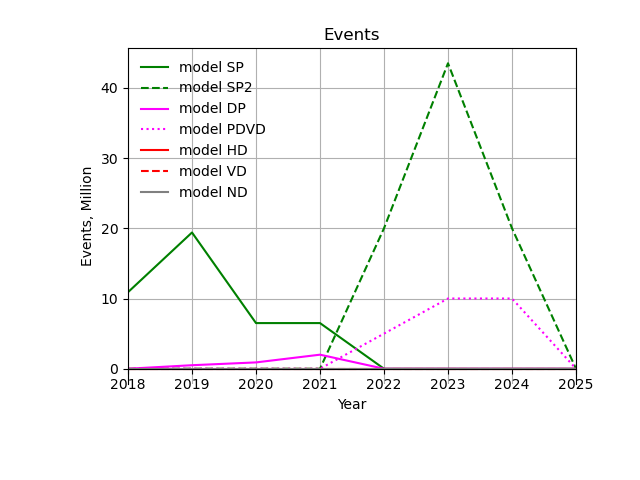}
\includegraphics[width=0.49\textwidth]
{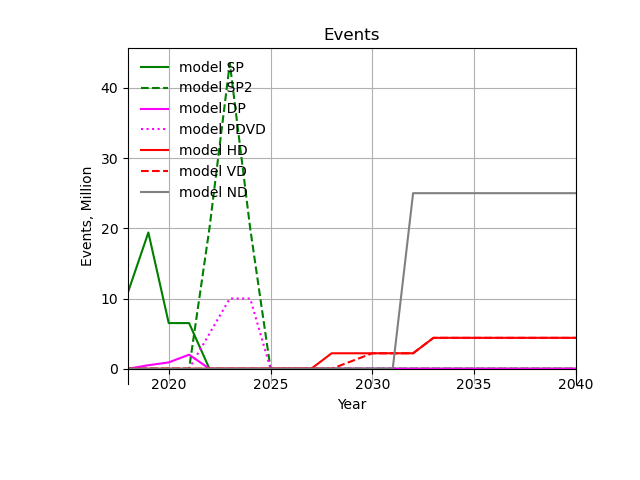}
\end{dunefigure}
\begin{dunefigure}
[Simulated Event estimates]
{fig:est:sim-events}
{Numbers of simulated data events produced  per year used in the model data volume estimates. Left is through 2025, right is the same through 2040.  }
\includegraphics[width=0.49\textwidth]{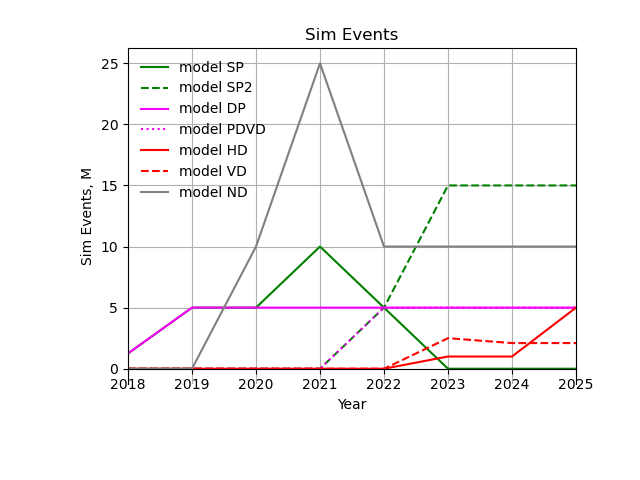}
\includegraphics[width=0.49\textwidth]
{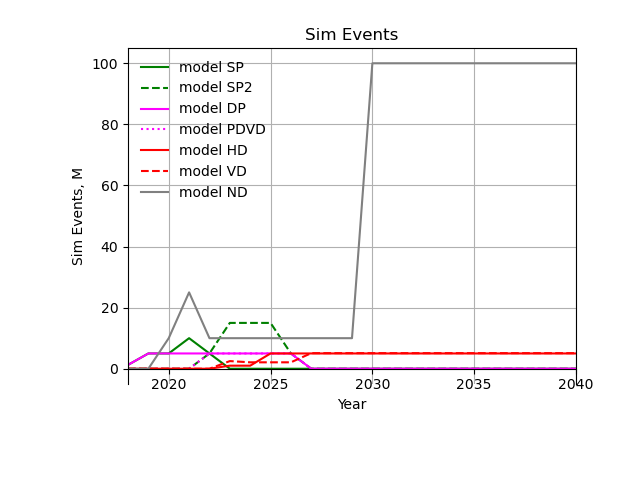}
\end{dunefigure}

CPU and size/readout are drawn from the above estimates. 

Accessing data from tape imposes large lead times due to competition for resources within and outside of \dword{dune}.  Our strategy is to archive all raw and reconstructed data and simulation on tape but to also retain raw data on disk for short periods (six months) for calibrations and reconstruction and to place several recent versions of reconstructed data and simulation on disk in both the US and Europe to optimize access times.  This motivates the following retention strategy and the data placement strategy discussed in~\ref{ch:place}. 

\begin{itemize}
\item Two copies of raw data are retained indefinitely.
\item DAQ commissioning data is marked ``test'' and one copy is retained on disk (and tape) for six months. 
\item Detector commissioning data is marked ``study'' and one copy is retained on disk  for six months and on tape for up to five years, depending on physics needs. 
\item Reprocessing (likely just pattern recognition on processed hits) is performed on the full data sample once/year   and two copies are retained on disk for two years.  This ensures that all data have been processed with the most recent version of the reconstruction and that the two most recent versions are readily available on disk.
\item Analysis-related CPU estimates include calibration.  Current experience indicates that analysis efforts are  equivalent in CPU utilization to reconstruction and simulation but produces smaller outputs. 
\end{itemize}

Data lifetimes on tape are likely to be longer than the retention policy states but expired data may be overwritten or will not be migrated to new tape technologies.  

Figures~\ref{fig:est:disk}, \ref{fig:est:disk-cms},\ref{fig:est:tape},\ref{fig:est:tape-cms}, \ref{fig:est:cores} and~\ref{fig:est:cores-cms} illustrate the estimated storage and CPU needs through 2025 and 2040.  In the early years, \dword{pd} %
and \dword{nd} prototype tests dominate while commissioning and operation of the first (and second) \dwords{detmodule} and the \dword{nd} become important after 2025. We include actual numbers for 2021. CPU and disk utilization were lower in 2021 due to not performing a fifth reconstruction pass that year %
and delays in distributing a second copy of reconstruction output to remote sites. 

The estimates to 2040 are also shown compared to estimates from the \dword{cms} experiment\cite{CMSComputingNeeds} through the HL-LHC era.  DUNE is estimated to use a few \% of CMS's CPU needs, around 10\% of CMS's disk and up to 20\% of CMS's tape by the end of the 2030's.

\begin{dunefigure}
[Disk estimates]
{fig:est:disk}
{Estimated size of various disk samples in PB. This estimate includes retention policies and multiple copies. Left is through 2025, right is the same through 2040. The points show actual use in 2021 which was lower than planned due to delays in distributing second copies of samples to remote sites.}
\includegraphics[width=0.49\textwidth]{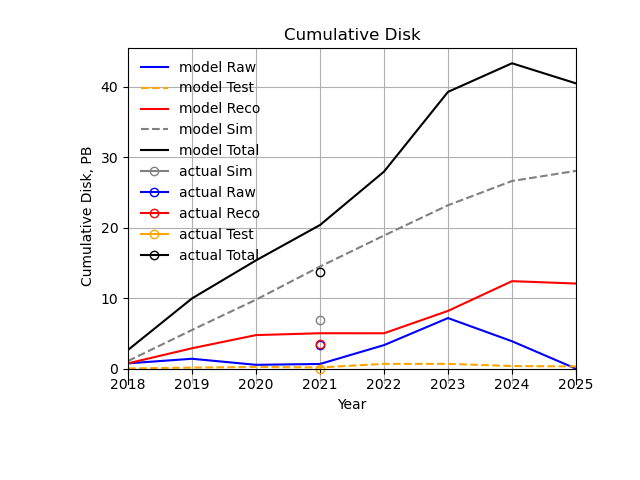}
\includegraphics[width=0.49\textwidth]{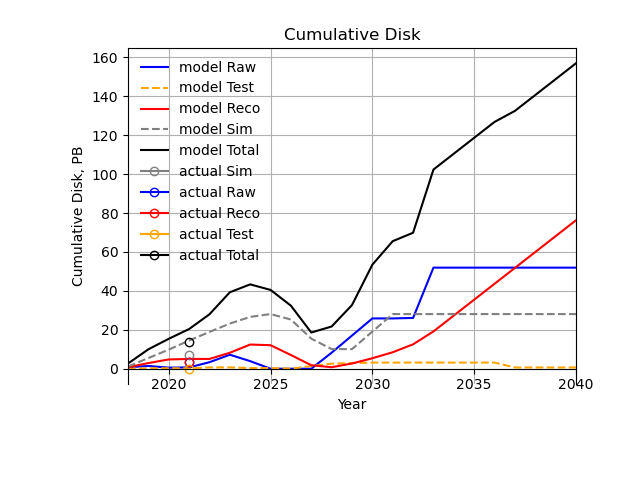}
\end{dunefigure}

\begin{dunefigure}
[Disk estimates]
{fig:est:disk-cms}
{Estimated size of various disk samples in PB for DUNE and CMS at the HL-LHC for comparison. This estimate includes retention policies and multiple copies. The points show actual use in 2021 which was lower than planned due to delays in distributing second copies of samples to remote sites.}
\includegraphics[width=0.49\textwidth, trim=0.0in 0.30in 0.0in 0.0in]{graphics/IntroFigures/compare2LHC/Parameters_2022-07-10-2040-Cumulative-Disk.png}
\includegraphics[width=0.49\textwidth]{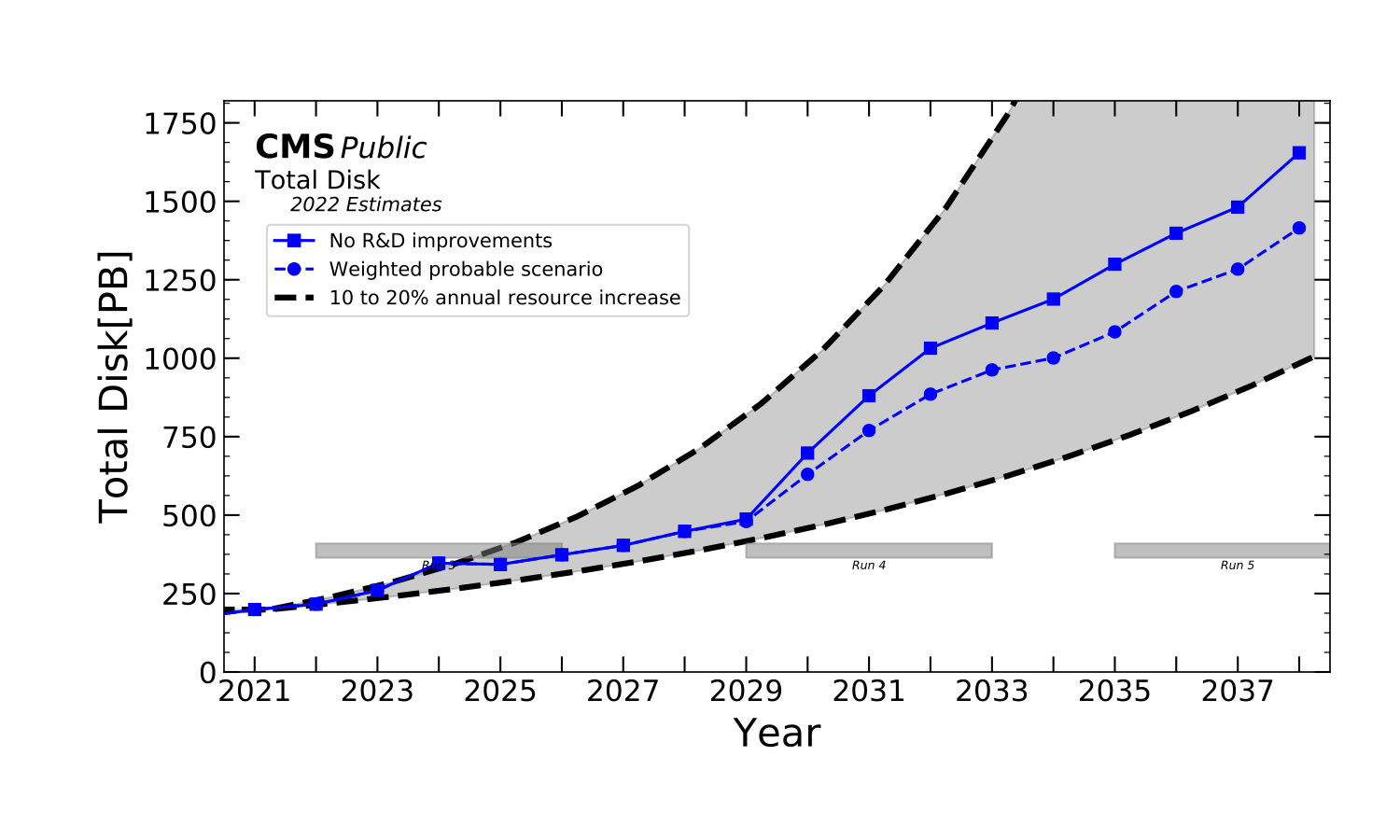}
\end{dunefigure}

\begin{dunefigure}
[Tape estimates]
{fig:est:tape}
{Estimated size of various DUNE tape samples in PB. This estimate includes retention policies and multiple copies. Left is through 2025, right is the same through 2040.}
\includegraphics[width=0.49\textwidth]{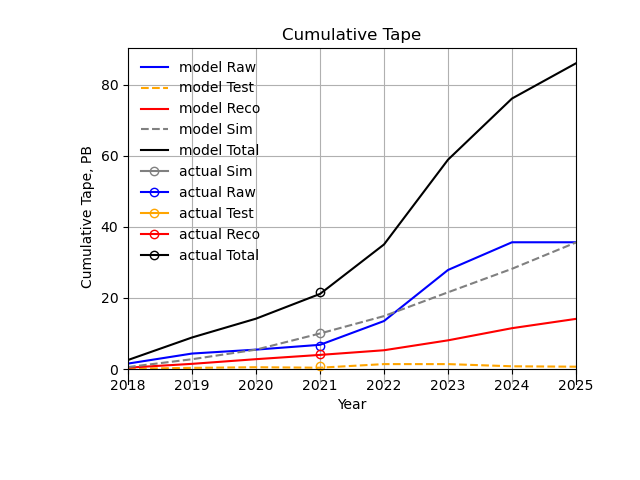}
\includegraphics[width=0.49\textwidth]{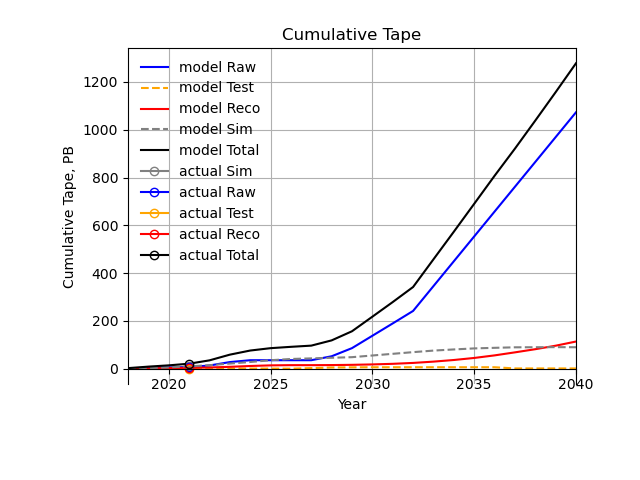}

\end{dunefigure}

\begin{dunefigure}
[Tape estimates compared to CMS]
{fig:est:tape-cms}
{Estimated size of DUNE tape volumes compared to CMS HL-LHC estimates. This estimate includes retention policies and multiple copies. }
\includegraphics[width=0.49\textwidth, trim=0.0in 0.30in 0.0in 0.0in]{graphics/IntroFigures/compare2LHC/Parameters_2022-07-10-2040-Cumulative-Tape.png}
\includegraphics[width=0.49\textwidth]{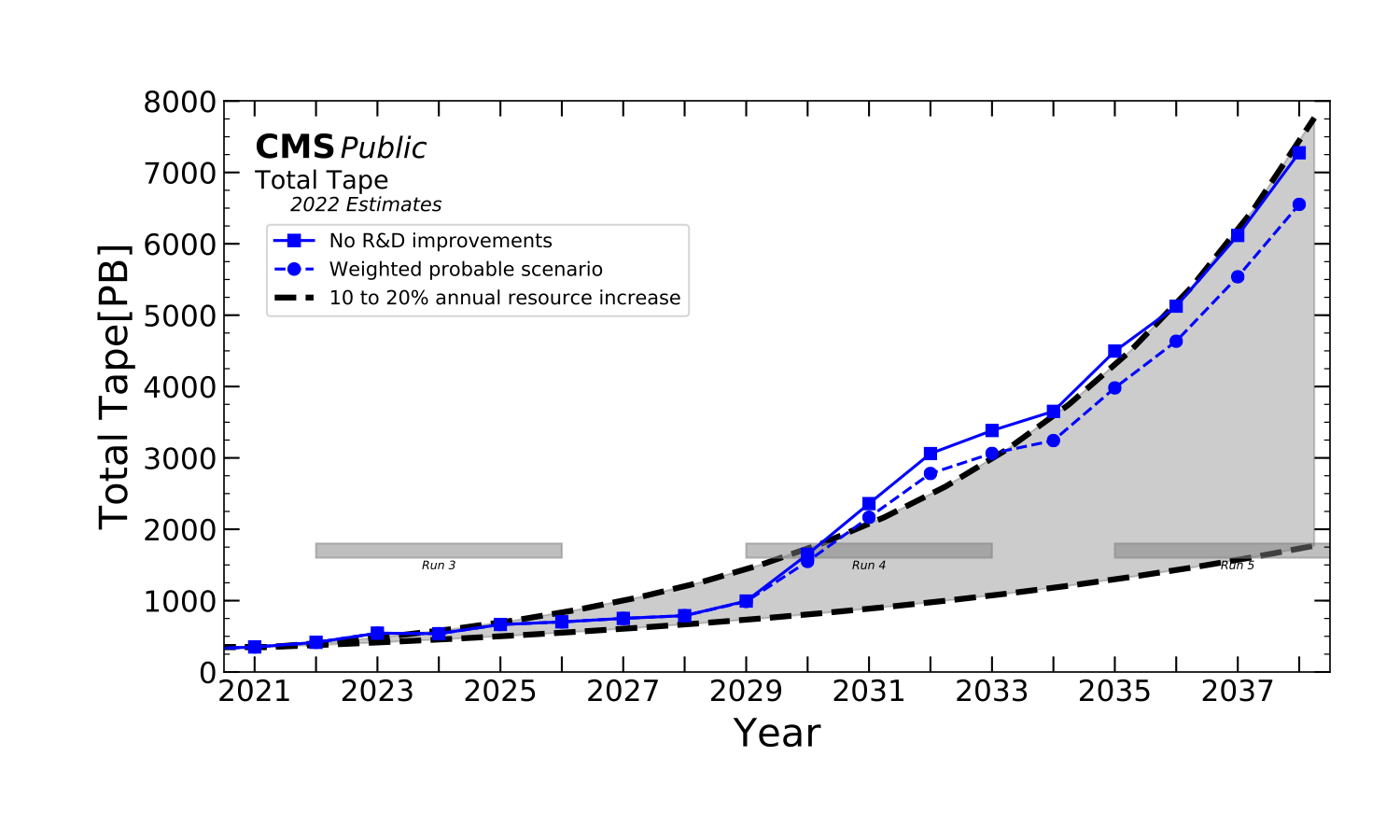}
\end{dunefigure}

\begin{dunefigure}
[CPU estimates]
{fig:est:cores}
{Estimated CPU needs for  various samples.  The units are kHS06-years (2020 vintage CPUs are $\sim$11 HS06~\cite{bib:HS06}) assuming 70\% efficiency. Left is through 2025, right is the same through 2040. CPU utilization in 2021 was lower than the model due to the absence of a yearly reconstruction pass.}
\includegraphics[width=0.49\textwidth]{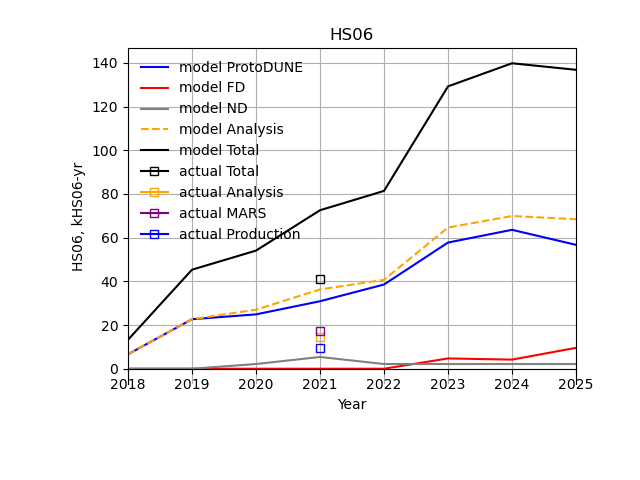}
\includegraphics[width=0.49\textwidth]{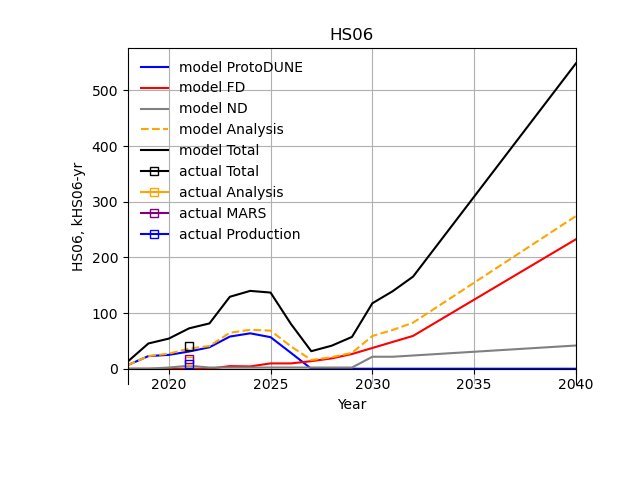}
\end{dunefigure}

\begin{dunefigure}
[CPU estimates compared to CMS]
{fig:est:cores-cms}
{Estimated CPU needs for DUNE compared to CMS.  The units are kHS06-years 
(2020 vintage CPU's are $\sim$11 HS06~\cite{bib:HS06} per core) assuming 70\% efficiency. CPU utilization in 2021 was lower than the model due to the absence of a yearly reconstruction pass.}
\includegraphics[width=0.49\textwidth, trim=0.0in 0.30in 0.0in 0.0in]{graphics/IntroFigures/compare2LHC/Parameters_2022-07-10-2040-HS06.png}
\includegraphics[width=0.49\textwidth]{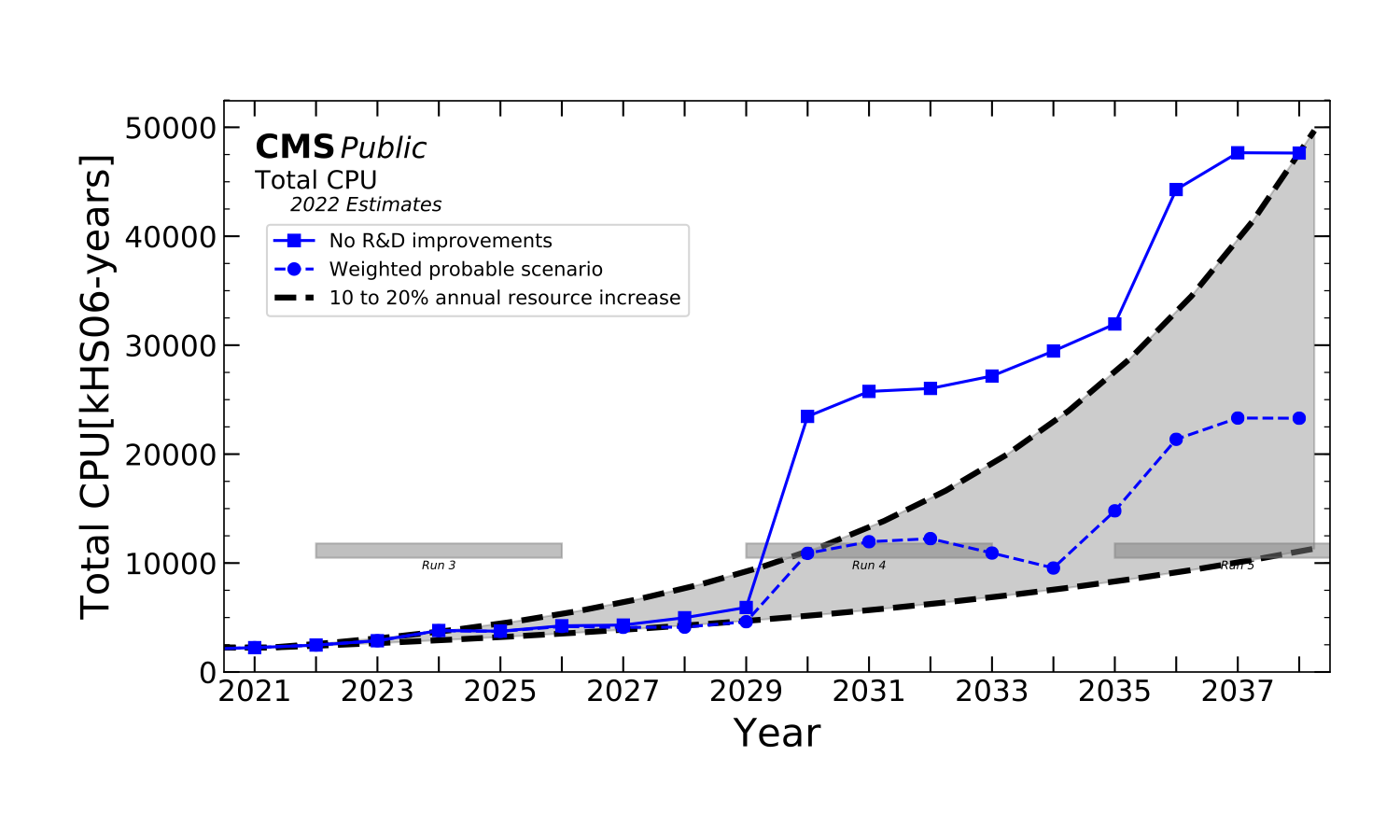}
\end{dunefigure}

\cleardoublepage

\chapter{Overview of the Computing Model \hideme{Anne comments addressed}}
\hideme{Anne looking again 30 sept}
\label{ch:cm}

\section{Introduction \hideme{JZ /anne comments 4/27}}\label{ch:cm:intro} %

The current DUNE global computing model is an organic extension of the \dword{fnal} \dword{fife}~\cite{box2014fife} computing model, used for smaller Intensity Frontier experiments, to the full global DUNE collaboration.  This effort  has relied heavily on global infrastructure such as \dword{osg} and the \dword{wlcg} and was successful for the first small-scale \dword{protodune} tests. However, it needs substantial enhancement to cope with  anticipated data and processing volumes.  This chapter describes the current situation and proposals for future improvements. 

\subsection{Global Resources}\label{model:global}

\dshort{dune} is a global collaboration  with contributions from institutions worldwide.   The long-term strategy for computing resources is for primary raw data storage to reside at the large host labs (the \dword{cern} and \dword{fnal}) and other national class facilities, with computing resources such as %
CPU and storage largely contributed by collaborating %
institutions. CPU contributions are provided by a large number of sites worldwide as shown in Figure~\ref{fig:prodsites}, while storage is concentrated at a few larger %
sites. Table~\ref{tab:coop:disk} shows the distribution of disk space pledges for 2021 and 2022, while Tables~\ref{tab:coop:sites} and~\ref{tab:coop:ussites} list the sites contributing CPU resources. 

Requests for resource pledges are made early in the calendar year, with pledges made before April for a computing year based on the UK fiscal year of April 1-March 31. This process is described in more detail in Section~\ref{sec:ccb}. The actual need can change over time, especially due to delays in experimental running and validation of new codes for production.  Our current experience is that \dword{cpu} needs have been met but storage pledges fall below the estimated need. For storage, shortfalls in pledges relative to %
requests are addressed by reducing the lifetime of second copies of some samples. For CPU, most usage is currently opportunistic and is running below the pledged levels.  Chapter~\ref{ch:est} gives a comparison of actual use in 2021 to the model used to make these requests.

\begin{dunefigure}
[Wall time distribution of production jobs FY22]
{fig:prodsites}
{Distribution of wall time for DUNE production jobs, October 2021 to March 2022. Inner ring: country. Outer ring: site. Over 50\% of wall hours have come from outside the United States in Fiscal Year 2022.}
{\includegraphics[height=3.5in,width=0.5\textwidth]{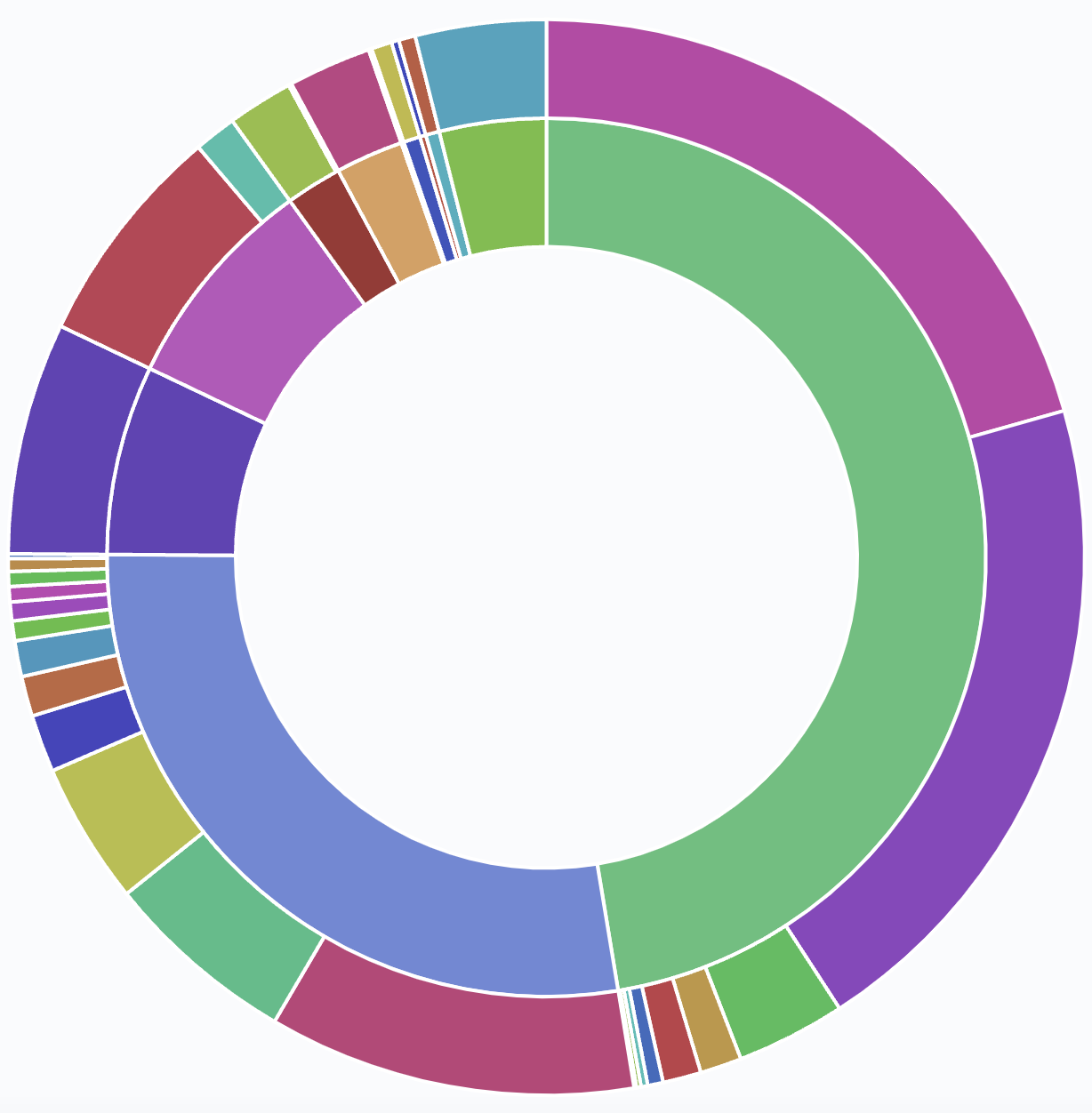}
\includegraphics[height=3.5in,width=0.12\textwidth]{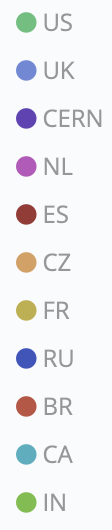}
\includegraphics[height=3.5in,width=0.15\textwidth]{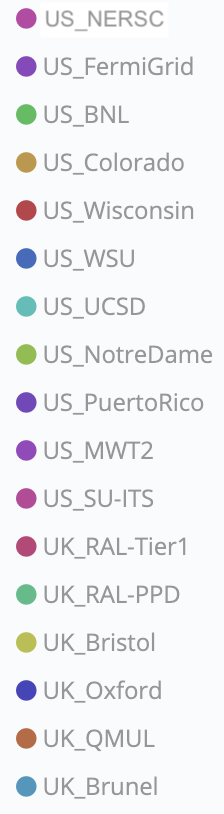}
\includegraphics[height=3.5in,width=0.15\textwidth]{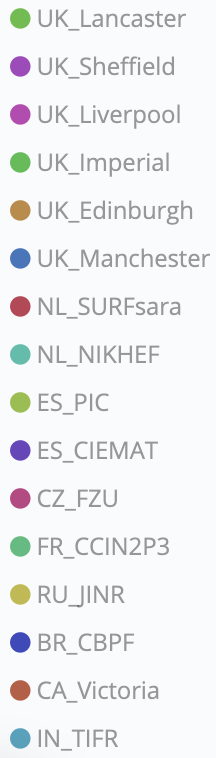}}
\end{dunefigure}

\begin{dunetable}
[National disk pledges]{llrr}{tab:coop:disk}
{Disk pledges in PB for 2021 and 2022.}
Country/Lab	&	Name	&	2021	&	2022	\\
\dword{fnal}	&	\dword{fnal}	&	2.2	&	7.6	\\
\dword{cern}	&	\dword{cern}	&	2.2	&	3.0	\\
\dword{bnl}	&	\dword{bnl}	&	0.5	&	0.5	\\
United Kingdom	&	GridPP	&	4.0	&	4.0	\\
France	&	CC-IN2P3	&	0.5	&	0.5	\\
Spain	&	PIC Tier-1	&	0.5	&	0.72	\\
Netherlands	&	NL/LHC Tier-1	&	1.9	&	1.8	\\
Czechia	&	CZ-Prague-T2	&	0.3	&	1.0	\\
India	&	TIFR	&	0.75&	0.75\\
Russian Federation	&	JINR	&	-	&	0.5	\\
\hline
Total pledge	&		&	12.85	&	18.97	\\
Total request & & 20.4 & 27.3 \\
\end{dunetable}

\begin{dunetable}
[List of DUNE compute sites]
{l l r l}{tab:coop:sites}
{List of non-US international DUNE compute sites as of December 2021.  Sites with substantial \dword{rucio} controlled disk are noted.}
Site name	&	RC Site	&	Disk	&	Country	\\
\hline
BR\_CBPF	&	BR\_CBPF	&		&	Brazil\\
BR\_UNICAMP	&BR\_UNICAMP	&		&	Brazil\\
CA\_Victoria	&	CA\_Victoria	&		&	Canada\\
CERN	&	CERN-PROD	&	Yes	&	Switzerland\\
CH\_UNIBE-LHEP	&	UNIBE-LHEP	&		&	Switzerland\\
CZ\_FZU	&	FZU	&	Yes	&	Czechia\\
ES\_CIEMAT	&	CIEMAT-LCG2	&		&	Spain\\
ES\_PIC	&	pic	&	Yes	&	Spain\\
FR\_CCIN2P3	&	IN2P3-CC	&	Yes	&	France\\
IN\_TIFR	&	IN\_TIFR	&	Yes	&	India\\
NL\_NIKHEF	&	NIKHEF-ELPROD	&		&	Netherlands\\
NL\_SURFsara	&	SURFsara	&	Yes	&	Netherlands\\
RU\_JINR	&	JINR\_CONDOR\_CE	&	Yes	&	Russian Federation\\
UK\_Bristol	&	UKI-SOUTHGRID-BRIS-HEP	&		&	United Kingdom\\
UK\_Brunel	&	UKI-LT2-Brunel	&		&	United Kingdom\\
UK\_Edinburgh	&	UKI-SCOTGRID-ECDF	&		&	United Kingdom\\
UK\_Imperial	&	UKI-LT2-IC-HEP	&		&	United Kingdom\\
UK\_Lancaster	&	UKI-NORTHGRID-LANCS-HEP	&	Yes	&	United Kingdom\\
UK\_Liverpool	&	UKI-NORTHGRID-LIV-HEP	&		&	United Kingdom\\
UK\_Manchester	&	UKI-NORTHGRID-MAN-HEP	&	Yes	&	United Kingdom\\
UK\_Oxford	&	UKI-SOUTHGRID-OX-HEP	&		&	United Kingdom\\
UK\_QMUL	&	UKI-LT2-QMUL	&	Yes	&	United Kingdom\\
UK\_RAL-PPD	&	UKI-SOUTHGRID-RALPP	&		&	United Kingdom\\
UK\_RAL-Tier1	&	RAL-LCG2	&	Yes	&	United Kingdom\\
UK\_Sheffield	&	UKI-NORTHGRID-SHEF-HEP	&		&	United Kingdom\\

\end{dunetable}

\begin{dunetable}
[List of US DUNE compute sites]
{l l r }{tab:coop:ussites}
{List of US DUNE compute sites as of December 2021.  Sites with substantial rucio controlled disk are noted.}
US\_UConn-HPC	&	UConn-HPC	&	Disk\\
US\_BNL	&	BNL-SDCC-CE01	&	Yes	\\
US\_Caltech	&	CIT\_CMS\_T2	&	\\
US\_Clemson	&	Clemson-Palmetto	&	\\
US\_Colorado	&	UColorado\_HEP	&	\\
US\_Florida	&	UFlorida-HPC	&	\\
US\_FNAL	&	GPGrid	&	Yes	\\
US\_KSU	&	BEOCAT-SLATE	&	\\
US\_Lincoln	&	Rhino	&	\\
US\_Michigan	&	AGLT2	&	\\
US\_MIT	&	MIT\_CMS	&		\\
US\_MWT2	&	MWT2	&		\\
US\_Nebraska	&	Nebraska	&	\\
US\_NERSC & NERSC &   \\
US\_NMSU-DISCOVERY	&	SLATE\_US\_NMSU\_DISCOVERY	&	\\
US\_NotreDame	&	NWICG\_NDCMS	&	\\
US\_Omaha	&	Crane	&	\\
US\_PuertoRico	&	UPRM-CMS	&	\\
US\_SU-ITS	&	SU-ITS-CE2	&	\\
US\_UChicago	&	MWT2	&	\\
US\_UCSD	&	UCSDT2	&	\\
US\_Wisconsin	&	GLOW	&	\\
US\_WSU	&	WSU - GRID\_ce2	&	\\
\end{dunetable}

Computational resources are currently dominated by conventional UNIX batch systems accessible via the \dword{osg} and \dword{wlcg} but HPC resource such as \dword{nersc} and commercial resources such as GPUs from Google Cloud\cite{Wang:2020fjr} %
are also being tested and used when %
needed and available.  These systems are believed to be well-suited to many DUNE workflows, but are very diverse in the hardware offered and require significant effort to access, see Section~\ref{sec:model:hpc} for further discussion.

\dword{dune}'s model is flatter and more network oriented than the original model for \dword{lhc} experiments.   As a result, it relies mainly on well connected data centers for storage and \dword{cpu} provision.  To date, we have found that most users prefer to do their work through the large national facilities available to them rather than building and using small local clusters at their home institutions.

\section{Current Performance}\label{ch:model:perf} %

\dshort{dune} has performed multiple simulation and reconstruction passes on the \dword{pdsp} data and is running significant simulation campaigns for the \dword{fd} and \dword{nd} design and physics studies. The \dword{poms} and \dword{sam} described in Chapters~\ref{ch:datamgmt} and~\ref{ch:wkflow}, are highly instrumented and allow assessments of the performance of the global computing system in near-real time.  There are four major data/CPU access patterns:
\begin{itemize}
    \item Simulation requires little input (mainly beam flux files and photon libraries),  has a large memory footprint, uses significant CPU resources and writes back a few large files.
    \item Hit reconstruction and pattern recognition both read in large files, have an intermediate memory footprint and use $\sim$10\,sec/MB of input data.  
    \item Data reduction reads the reconstructed data and produces small tuple outputs for further analysis.  Reduction uses $\sim0.1$\,sec/MB of input data and is generally I/O limited.
    \item Data analysis consists of repeated access to smaller tuple outputs for calibration and parameter estimation.
\end{itemize}

Each of these use cases is best suited to a different combination of data/CPU resources and the global compute model should be able to allocate resources appropriately.  Currently the default configuration for all \dword{htc} jobs is for data delivery to be handled through \dword{xrootd} streaming. To understand the impact of this choice on efficiency, we have used the \dword{sam} instrumentation to measure the \dword{xrootd} streaming performance for disk/CPU location combinations. 

\begin{dunefigure}
[Streaming speeds for reconstruction]
{fig:recospeed} 
{Streaming speeds for reconstruction jobs running at different locations. Raw data are stored at \dword{cern} and \dword{fnal}.  The histograms show the log$_{10}$ of the inferred streaming rate (wall time/file size) for reconstruction jobs running at selected sites with different data sources.}
\includegraphics[width=0.49 \textwidth]{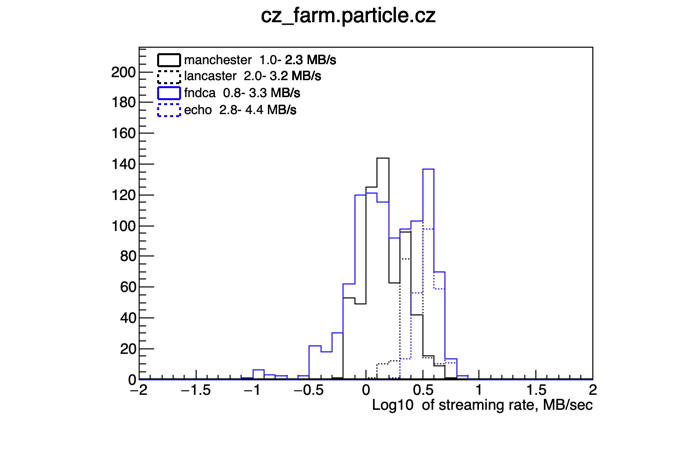}
\includegraphics[width=0.49 \textwidth]{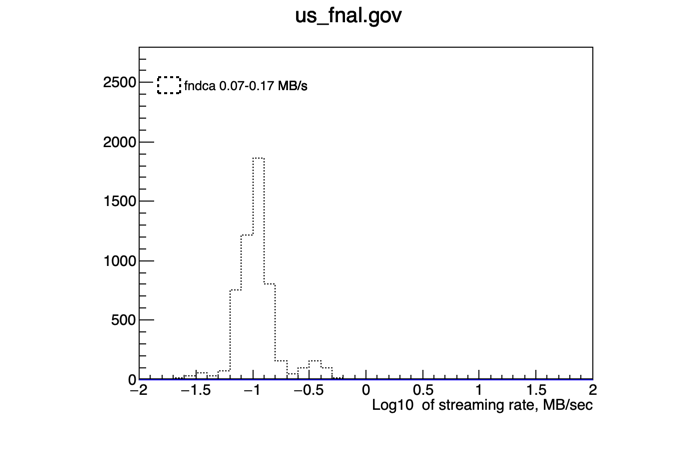}
\includegraphics[width=0.49 \textwidth]{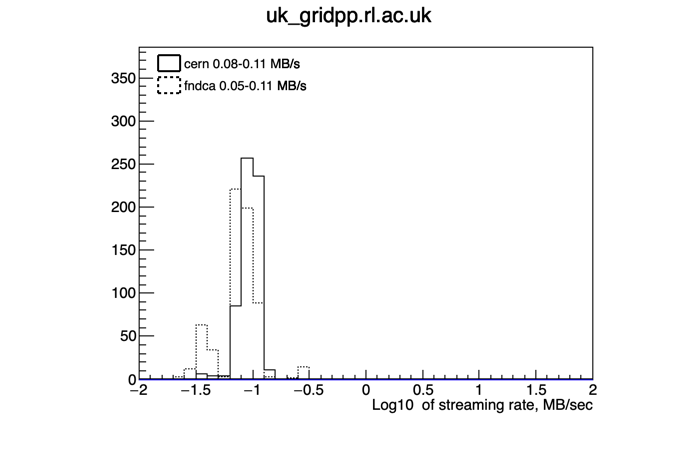}
\includegraphics[width=0.49 \textwidth]{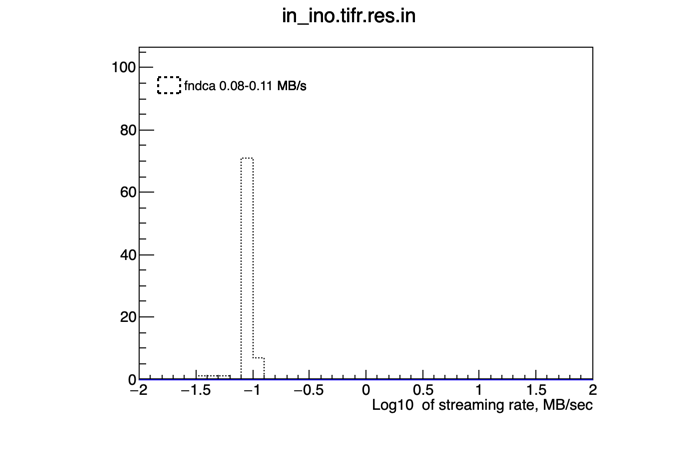}
\end{dunefigure}

\begin{dunefigure}
[Streaming speeds for tuple creation]
{fig:tuplespeed} 
{Streaming speeds for tuple creation jobs running at selected locations during a test in early January 2022. Reconstructed simulation files were located at sites in the UK and at \dword{fnal}.  The histograms show the log$_{10}$ of the inferred streaming rate (wall time/file size) for tuple creation jobs running at selected sites with different data sources.}
\includegraphics[width=0.49 \textwidth]{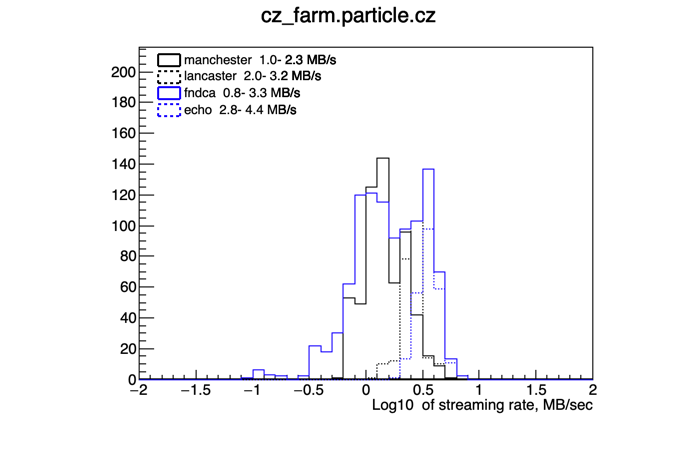}
\includegraphics[width=0.49 \textwidth]{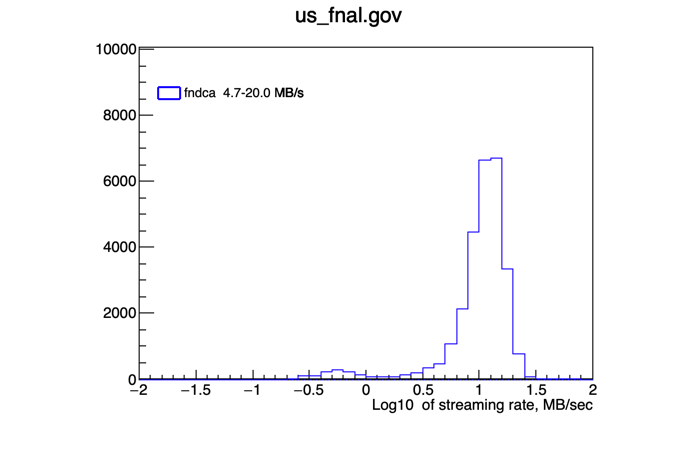}
\includegraphics[width=0.49 \textwidth]{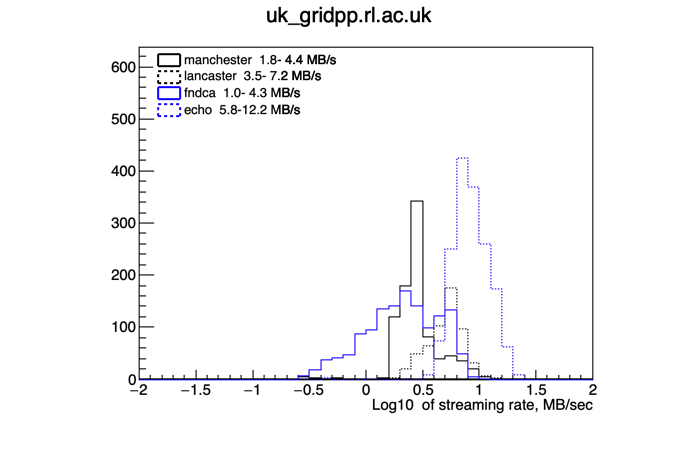}
\includegraphics[width=0.49 \textwidth]{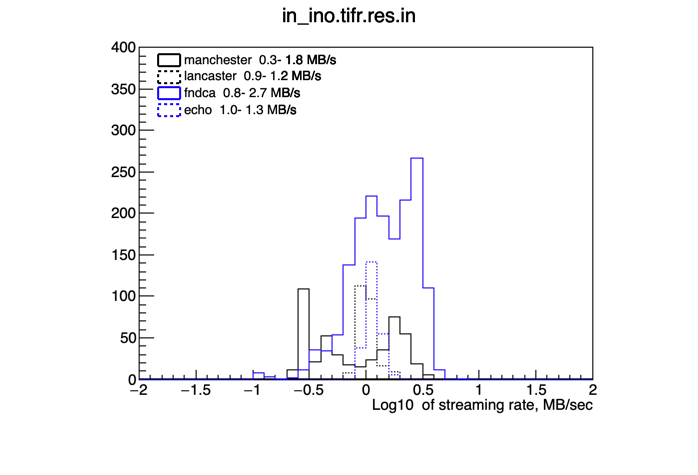}
\end{dunefigure}

\begin{dunefigure}
[Streaming speeds for tuple creation]
{fig:tuplespeedsummary} 
{Streaming speeds for tuple creation jobs running at multiple locations during a test in early January 2022. Reconstructed simulation was stored at sites in the UK and at \dword{fnal}.  The average estimated streaming rates are plotted as a function of disk location ($x$-axis) and compute site ($y$-axis). Jobs in the US were required to use \dword{fnal} disk but international sites were tested with multiple samples.}
\includegraphics[width=0.8 \textwidth]{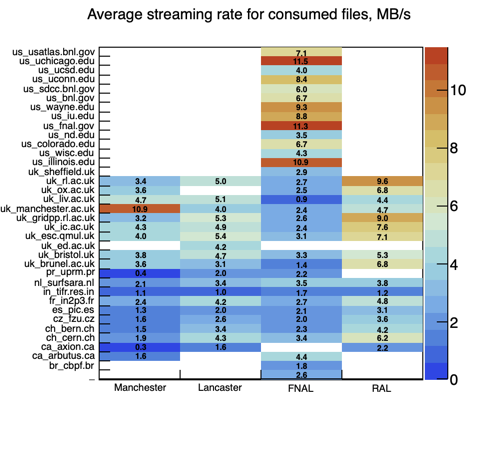}
\end{dunefigure}

\subsection{Implications for Data and Processing Placement}

The current study indicates that for CPU-dominated applications, notably reconstruction of raw data and simulation, the relative location of CPU resources is not critical. Wall time/MB   is similar regardless of location. For IO-dominated applications, proximity to the data is important.  Here there is a trade-off between the availability of resources and the efficiency with which they can be used. Intra-US processing, especially for locations near the national laboratories, is highly efficient, as is processing within the UK.  However, efficiency falls off as the CPU and disks become more separated.  

There are additional constraints imposed on our use of each site at scales greater than that of a single job. Due to local networking limitations, the number of jobs of a particular type we can run may need to be limited to avoid saturating the site's inbound or outbound capacity.

\section{Design Philosophy \hideme{McNab - draft}}
\label{sec:cm:philosophy}  %

The large \dword{lhc} experiments have historically relied on a tiered structure, with national Tier-1 centers %
and regional Tier-2 and Tier-3 centers.  The DUNE model builds on the emergence of faster networks to move to  a service-oriented model, where sites provide services -- disk, CPU, real memory/core and archival tape -- and projects are  distributed to them based on their capabilities and available networking.  For example, a site with large CPU and memory/core but slower networking would be ideal for simulation while small memory/core and fast access to large local disk stores would be ideal for high-level data analysis.  In the long run, this model will require a high-level view of data locations and job placement with continual monitoring for bottlenecks, but allows new sites to contribute in an optimal way. The intent is to lower the bar for contributions without overburdening the core computing operations of the experiment.  To date, we have successfully and quickly reconstructed and simulated the \dword{protodune} data  using this flatter model and were able to improve I/O bound data analysis processing speeds at European sites by a factor of two by tuning the \dword{sam} data locality tables based on the measurements shown in Section~\ref{ch:model:perf}.

We have the site information and monitoring tools to provide a workflow management system with the inputs it needs to do optimal placement of data and jobs to maximize efficiency. However our existing data management and workflow tools do not currently have a top-level request management overlay and  are not able to take full advantage of system-wide  information.  This motivates development of an improved workflow model that scales to \dword{dune}'s world-wide site model, in addition to replacement of the existing \dword{sam} functionality for simpler use cases.  %

In Chapters \ref{ch:datamgmt} and \ref{ch:wkflow}, we describe the present and future designs for data and workflow management. We are pursuing two complementary designs: (1) a Data Dispatcher replacement for the existing \dword{sam} file delivery system which retains the loose coupling between the job submission system and data management (Section  \ref{sec:dispatch})  and (2) a new, more tightly coupled, workflow system (Chapter~\ref{ch:wkflow}) which uses a global view of CPU and disk locality to further optimize efficient processing.  The existing \dword{sam} system is used by multiple \dword{if} experiments for interactive and batch use, mainly within the FNAL/OSG computing ecosystem, and those experiments will also benefit from the Data Dispatcher \dword{sam} replacement.  The new workflow system is being designed by DUNE-UK based on LHC experience and is intended to optimize %
processing efficiency based upon data location %
across the much wider DUNE computing landscape.  

\section{Sites and Services}\label{sec:cm:sites_and_services}
This section sets out our ``Sites and Services'' model for using sites for DUNE computing tasks, including sites that participate in \dword{osg} or \dword{wlcg} more generally. Since our requirements are not the same as the \dword{lhc} experiments, this %
requires implementing both a distinct naming scheme in
\dword{wlcg} and  mapping our scheme onto the \dword{wlcg} tier model.

At this stage, DUNE has chosen to express its requirements in terms of services provided by sites. Each site provides networking plus one or more approved DUNE services, which satisfy DUNE's minimum requirements for the service in terms of capacity, quality, and interfaces. This model does not rely on assumptions about how sites and federations of sites will be organized in the future, as the community evolves away from the strict \dword{wlcg} tiers model towards \dword{doma} and concepts such as %
\dwords{datalake}.

DUNE expects to be able to access services using broadly the same set of APIs as \dword{wlcg} (e.g., HTCondor-CE and \dword{xrootd}) and by using common cloud APIs (e.g., \dword{openstack} and \dword{s3}). 
For this reason, sites may be operated using conventional grid technologies, on-premises cloud systems, or commercial cloud services (where cost-effective). Nevertheless, sites do appear in the DUNE computing model, as the atomic unit for operations activity. %
For example, staff at a site can receive and process tickets, %
and may be required to have a representative at an operations meeting with the technical knowledge to comment on issues as they arise.

In terms of workflows and data management, DUNE does not impose or require any hierarchy or grouping of sites, and assumes that, in general, data may flow between services at any two sites. That said, DUNE expects to use network proximity and bandwidth information to guide the efficient transfers of data between services. The details of the different services within the DUNE Computing Model are described in detail in Section~\ref{sec:cm:types_of_service}.

\section{Sites, Federations, and Countries\hideme{McNab - draft}}
\label{sec:cm:federations}

As well as sites, there are two more administrative concepts: federations and nations. %
Federations are borrowed from \dword{wlcg} and represent one or more sites that together pledge a particular amount of capacity to DUNE and enter their pledges into a system such as \dword{cric}. %
Sites may choose to organize themselves this way as it allows more flexibility in how pledges are met against a background of planned upgrades at sites, unplanned outages, etc. 

Nations are represented directly or indirectly at the DUNE \dword{ccb}, and consist of one or more federations. Broadly, nations map to funding bodies and are 
the entity reviewed when evaluating the level of contribution to computing capacity relative to their number of DUNE members.   %

\section{Types of Service\hideme{McNab - draft}}
\label{sec:cm:types_of_service}

We %
have identified eight classes of service on which we will put requirements and request capacity:

\begin{enumerate}
    \item Network -- Section \ref{sec:cm:network}
    \item DUNE Computing Element (conventional) -- Section \ref{sec:cm:dce}
    \item DUNE Computing Element (HPC) -- Section \ref{sec:model:hpc}
    \item Data Cache -- Section \ref{sec:cm:data_cache}
    \item DUNE Storage Element -- Section \ref{sec:cm:dse}
    \item DUNE Data Archive -- Section \ref{subsec:data_archive}
    \item Interactive Analysis and Build Facility -- Section \ref{sec:interactive}
    \item DUNE Analysis Facility -- Section \ref{sec:af}
\end{enumerate}

Note that each class of service may have varying levels of %
resources provided within a service, and the details of those levels are discussed in the following subsections.

\subsection{Network\hideme{McNab - draft}}
\label{sec:cm:network}

Networking is needed at all sites, with basic requirements including IPv4 and \dword{ipv6}. %
All sites should be connected to the wide area network, typically via their \dword{nren}, with sufficient capacity to handle the data I/O commensurate with their fraction of the workload. In practice this means at least 40\,Gb/s for major sites with large amounts of storage (with 100\,Gb/s becoming the normal expectation for a shared site such as a \dword{wlcg} Tier-1) and at least 20\,Gb/s for smaller CPU-only sites (but with 40-100\,Gb/s becoming the norm in shared sites).  Sites with less network capability than this will be restricted to CPU-limited activities such as simulation.
Other DUNE-approved services may impose further requirements in terms of network capacity. %

\subsection{DUNE Computing Element (conventional) \hideme{McNab - draft}}
\label{sec:cm:dce}

A DUNE Computing Element is a service that provides access to conventional \dword{cpu} resources. %

We envisage %
three subclasses within the computing element services aimed at centrally managed data processing, at user or working group data analysis, and at detector simulation, with appropriate minimum standards for each case, involving the following criteria: 

\begin{itemize}
    \item the support level, in terms of whether tickets will be acted 24/7 or only during working hours; note that our data model envisions %
    dual copies of most user-facing data samples and a distributed processing system that is reasonably robust against single node or site failure. Services such as databases need to be at sites with 24/7 support.
    \item the total number of logical processors across the service;
    \item the interface used to submit jobs or create virtual machines;
    \item the operating system version for grid capacity;
    \item memory and scratch disk space per processor
    (Note that DUNE computing, due to the large data objects, often requires a large memory footprint.  Provision of high-memory  resources is very valuable.);
    \item incoming and outgoing network capacity per processor; and
    \item %
    for, I/O limited activities, access to DUNE Storage Elements or Data Caches (defined below), which allows data-intensive jobs to execute without an unacceptably low CPU efficiency.

\end{itemize}

\subsection{High-Performance Computing Elements (HPCs) \hideme{McNab - draft}}
\label{sec:model:hpc}

High-Performance Computing facilities form a special class of compute elements.  These are large multi-purpose national-scale systems, such as \dword{nersc}, that give access to large amounts of specialized, interconnected back-end hardware such as \dwords{gpu} or \dwords{fpga}.  There are several hurdles that stand in the way of \dword{dune} taking full advantage of these resources. An initial challenge is that the HPCs available to \dword{dune} are generally not controlled by DUNE, or even \dword{hep}, with access subject to  annual applications and restrictions. In addition to a mismatch between \dword{hep} experimental time scales and the center's annual allocation models, current \dword{hpc} centers have limited WAN connections at the node level making it difficult to move 100's of TB of data into, or out of processes. To address some of these challenges, \dword{dune} is in active dialog with \dword{nersc}, as a first step, to gain access to larger temporary storage and bandwidth to support large-scale DUNE processing in the late 2020's. Other large US and European centers, as well as commercial systems, are also being explored but, for the moment, each requires custom negotiations and code adaptions.
In addition, DUNE is taking an active role in the design and alignment discussions as part of the Integrated Research Infrastructure Architecture Blueprint Activity \footnote{An overview presented to the Open Science Grid Council can be found here https://indico.fnal.gov/event/52594/} that will define the landscape for DOE HPC in the next decades. 

Given the large compute resources available at these centers, and the match between DUNE's image-like data and \dword{hpc} hardware, we  anticipate making substantial use of these very diverse resources. To date, DUNE collaborators have made significant progress in exploring algorithms that utilize the capabilities of \dwords{gpu}, but those research and development efforts have not yet been incorporated into the production offline workflows. It should be noted that \dword{dune} analyzers have individually been taking advantage of opportunistic access to GPU resources on the \dword{osg}, local GPU computing clusters (e.g. Wilson Cluster at \dword{fnal}), university resources, and individual allocations at some \dword{hpc} sites. With the exception of the Google Cloud studies described in Section~\ref{sec:google}, \dword{dune} has not made use of the full capability of \dword{hpc} systems in production (e.g. via MPI or multi-node calculations), but we consider incorporating these algorithms an important part of the framework and workflow development in the future. The integration of GPU and accelerator-based algorithms, and matching those with the available resources as part our production workflows, will take substantial administrative and technical effort to achieve. Given that at this time GPU utilization is on an individual basis with  myriad algorithms and is not part of a centrally-managed workflow, \dword{dune} has not been making predictions within the current computing model about the amount of GPU resources that will be needed in the future. It is a near-term goal to begin estimating the \dword{gpu} resource needs and the effort needed to integrate %
those resources as production algorithms and workflows start to incorporate GPU-based processing. At this time, \dword{dune} has been able to meet the development and analysis needs of the collaboration %
with opportunistic access, limited \dword{nersc} allocations, and university-based hardware.

\subsection{Data Cache\hideme{McNab - draft}}
\label{sec:cm:data_cache}
A Data Cache is local storage used temporarily by DUNE jobs to optimizes data movement into and out of compute and storage sites. 
A data cache service provides transient storage that is not managed by DUNE; %
it improves the access speed for data %
located on remote storage %
that is accessed multiple times. Technologies such as XCache, StashCache, and ``Data Lake'' proposals may be able to fulfill this role. 

A data cache  is generally associated with CPU resources and requires:
\begin{itemize}
\item suitable networking,
\item sufficient resiliency against transient problems in order to prevent jobs from failing and losing the outputs from the work they have already done.
\end{itemize}

\dword{dune} may be able to achieve similar functionality using some of the storage that it does manage to create temporary copies of data files, but this will require further development of the Data Management system. %
Support for data caches in the DUNE computing model may still be beneficial, by bringing in resources from sites that are willing to operate a cache service but do not wish to support long-term managed storage for DUNE.

\subsection{DUNE Storage Element\hideme{McNab - draft}}
\label{sec:cm:dse}
A DUNE storage element is a large storage element dedicated to \dword{dune} and managed via \dword{rucio}. 
The concept of a DUNE Storage Element mirrors that of a DUNE Compute Element. It must be of sufficient capacity, measured in hundreds of TB or in PB, for the operational overheard to be worthwhile. It must have a suitable support level, depending on the degree to which its services are replicated elsewhere. %
There must be enough inbound and outbound networking capacity for global data placement operations, and for jobs to write data there or to consume the data already present. In particular, for I/O intensive applications, there must be %
a minimum amount of DUNE Compute Element capacity available nearby, on which DUNE jobs can access 
 appropriate  storage services without   unacceptably low CPU efficiency. For some jobs, for example full reconstruction and simulation, almost all storage services are appropriate while for analysis jobs, the only appropriate storage may be co-located at the same site. 

This formulation allows conventional grid sites %
to place CPU and disk storage in adjacent racks, and it also supports novel regional architectures such as data lakes %
with sufficient network capacity to link CPU and storage at different locations. At this stage of the project, DUNE does not want to prejudge what will be available at the start of \dword{fd} data taking, and does not want to discourage the exploration of new and more efficient ways of providing resources.

\subsection{DUNE Data Archive}
\label{subsec:data_archive}
Data archive is designed to be archival storage of \dword{dune} raw data, simulation, and any data derived from those files that must be permanently stored. Traditionally and as foreseen with \dword{dune}, this service has been provided by tape storage facilities. There are currently at least five institutions providing tape storage facilities with the largest and most important providers being \dword{fnal} and \dword{cern}. The volume and access patterns for these services is an ongoing area of research and development to both keep up with modern technology and to understand how it impacts \dword{dune}'s ability to accomplish physics.  A central part of our model  is prepositioning the samples most likely to be accessed on DUNE managed disk for extended periods, preferably in two locations, with tape access reserved for production processing of raw data. 

\subsection{Interactive Analysis and Build Facility} \label{sec:interactive}

Both \dword{fnal} and \dword{cern} provide centrally managed interactive computing facilities for users to perform small-scale data analysis and algorithm development.  Fermilab hosts 15 4-core  \dwords{gpvm} dedicated to DUNE %
that run Scientific Linux 7 with $\sim$2-3 GB of memory/processor.  \dword{cern} hosts the \dword{lxplus} facility with  machines with similar characteristics running CENTOS variants.  The CERN machines are shared across CERN experiments. These systems have access to local network attached disk and to the \dword{pnfs}(Fermilab) and \dword{eos}/\dword{castor}/\dword{cta}(CERN) storage systems respectively. The laboratory interactive systems and many institutional clusters    access the DUNE computing environment through the \dword{cvmfs} and can access data via \dword{xrootd} once authenticated to the DUNE \dword{vo}.  

The Fermilab system has around 700 DUNE user accounts of which about 300 have been active over the past six months.  It is harder to get statistics on interactive DUNE activity at CERN; it is substantially smaller but non-zero. 

Fermilab also provides two 16-core virtual machines dedicated to fast code builds.

\subsection{Dedicated Analysis Facilities}\label{sec:af}

Most user data analysis is currently being done using serial reads of \dword{larsoft} outputs or small root ntuples on the generic interactive systems described in  \ref{sec:interactive}.  DUNE is actively exploring using and provisioning dedicated user analysis facilities capable of columnar access to data and more sophisticated analysis techniques in addition to notebook based analysis.

\subsubsection{Compute Canada Prototype}

An interactive analysis facility has been prototyped on a cloud allocation provided by Compute Canada’s OpenStack platform on the Arbutus Cloud (https://docs.computecanada.ca/wiki/Cloud\_resources). This experimental allocation is dedicated to DUNE and currently includes 300 virtual CPUs, 2.2 TB of shared memory, 2 TB of disk storage, and 1 TB of CephFS storage that uses a shared storage protocol. More storage has been requested for the next resource allocation year (2022). 

Several computing nodes, each with eight CPUs and 90 GB of memory, have been formed out of this allocation. The facility relies on Jupyterhub to implement a web-based interactive experience and uses containerization technologies to maintain a dedicated workspace for each authenticated user. Kubernetes is the industry-standard platform for this use case and can be deployed at scale for up to 5000 nodes\cite{web:Kubernetes}. A user authenticates through CILogon and a single-user JupyterLab server on a Scientific Linux 7 Docker image is quickly deployed thereafter. 

The facility provides a few features to facilitate analysis. \dword{larsoft} and DUNE-specific libraries are enabled through \dword{cvmfs}. Users building event loop-based analysis routines will have the identical experience to the bash-based virtual machines. Data files can be streamed with xrootd once users authenticate with the \dword{voms} server. 

\subsubsection{Fermilab Elastic Analysis Facility}
We are testing similar functionality at Fermilab on the Fermilab Elastic Analysis Facility.  Among other interfaces, this facility hosts the Fermilab Analytics Hub, which provides access to DUNE code and data to Jupyter notebooks via \dword{cvmfs} and \dword{xrootd}. This system has already been used to test reconstruction algorithms and validate data from the \dword{protodune} \coldbox{}es. For the future we are exploring modern analysis frameworks such as \dword{coffea}.

\subsubsection{Coffea}
Coffea\cite{web:Coffea}  is a new analysis framework that could enable analyses on the fly with a quick turnaround time and a low barrier for a new analyzer. The framework was originally developed by LHC collaborators to enable columnar-based analyses of large volumes of LHC data in flat ROOT format using Python. Coffea employs uproot for file I/O and uses a NanoEvent class to parse TTree branches into user-defined physics objects for columnar computation. Users have the option to distribute their computations across different CPUs using a Dask\cite{web:Dask} cluster. 

A prototype NanoEvent for the \dword{protodune} PDSPAnalyzer flat analysis format is already available with the goal of demonstrating a full-chain analysis (event selection, background tuning, systematics). Small-scale tests of data streaming from \dword{fnal} using xrootd has also been successful. 

Longer-term goals  include token-based authentication to seamless bridge user login and grid authorization, understanding network limitations and developing means to circumvent them. 

\subsubsection{Analysis Facility Issues}
One of the major issues for a dedicated analysis facility is the ability to access and maintain data samples.  We expect that, with many different analysis users, many disparate data samples will need to be available at any given facility but they must also be on local storage to maintain optimal efficiency.  We are designing user \dword{rucio}-controlled samples into our new systems, but do not yet have experience with them in production. The analysis-user-facing \dword{rucio} services will be need to be designed, developed, and integrated into the future analysis facilities. The expectation is that the majority of the effort will be the development of DUNE specific modules, scripts, and services within the analysis facility and framework, and that there will be limited development in \dword{rucio} to meet the needs of analysis workflows. \dword{dune} expects to draw requirements and end-user analysis workflow examples from \dword{pdhd}.

\cleardoublepage

\chapter{Data Formats}
\label{ch:format}

\section{Data Format Overview}

DUNE data are stored and processed in a variety of forms and representations which  are tuned to the analysis that is performed on them.
In general, %
the data are characterized and classified by their \dword{datatier} (described in \ref{sec:dataflow}) and their data format within that tier.  The data tier corresponds to the stage of the data in its life cycle and its progress from the original detector acquisition data and simulations, towards its final analysis stage.  The data format refers to the low level representation and organization of the data as it takes on in either a persistent or transient form during the analysis workflows.  This classification of data by tier and format allows \dword{dune} to develop, catalog, and maintain its data ecosystem in the context of the larger computing model and computing resources.

The specific data format of each data tier can have a large impact on the performance of processing and accessing that data tier.
A review of data formats and their serialization performance with respect to LHC data, was performed  based on the state of major data formats in 2017\cite{Blomer:2018icl}.  While the actual results have since been superseded by improvements in the formats they examined, their study
highlighted the importance of the data format and access libraries to analysis turn-around times in \dword{hep}.

The low-level representation of the \dword{dune} data is keyed to the data tier(s) at which they exist and are accessed.  It is common for these representations to change over the data lifecycle to reflect organizational needs of the algorithms that are run on the data.  In particular in \dword{hep}, it is common for raw data coming from the detector subsystems and DAQ systems to have representations that are closely tied to the electronics and readout (and is often customized due to optimizations that are needed for high speed readouts), while later stages of processing use more generic formats and representations more amenable to use in algorithms and transforms.  In the \dword{hep} and neutrino community, there have been a number of different ``high level'' data representation and I/O systems that have been used.  Starting in the late 1990s the \dword{root} I/O system became one of the dominant systems for \dword{hep} data, and has continued to feature prominently for experiments which make heavy use of C++ in their software stacks.  

In the case of \dword{dune}, we have historically represented our data through a combination of custom and \dword{root} I/O based data formats.  In our initial \dword{protodune} run 1 data processing, and in our current simulation chains, this has been our strategy and is fully supported throughout our processing chain.  \dword{root} I/O provides data objects and methods that have been optimized for certain types of \dword{hep} applications and access methodologies.  The \dword{larsoft} framework, which is discussed in section~\ref{sec:framework:status}, is compatible with the \dword{root} I/O system and other data formats.

Many \dword{ml} toolkits, which have been developed by the computational sciences and computational industries, do not use the \dword{root} I/O data formats and systems. These general toolkits have been successfully tested in DUNE for event classification and particle identification.  Written in Python, these \dword{ml} workflows read and write data in formats commonly used in other scientific communities, such as \dword{hdf5}~\cite{bib:hdf5}.   Not all workflows work in this manner.  Some export image data from LArTPCs as image data in two-dimensional ROOT histograms, and even comma-separated value (CSV) format has been used as a communication format between {\it art}/LArSoft jobs and external \dword{ml} tools.  In the case of ROOT histograms, a DUNE-supplied read-in layer is provided.

DUNE is planning to use HDF5 storage format for the upcoming raw data from ProtoDUNE-HD, ProtoDUNE-VD and ICEBERG.  Dedicated I/O modules have been written to read the HDF5-formatted data into an art/LArSoft job, and as of this writing, they are in use reading HD and VD coldbox data.  The DUNE software stack also contains modules that export data from an art/LArSoft DUNE job in HDF5 format, so that they can be read by external AI/ML software.  

Not all data used for AI/ML applications is in the HDF5 format.  For example, section~\ref{sec:google} describes a use of a \dword{cnn} in a LArSoft job in which data are exchanged via TCP/IP network communication between compute elements in a format dedicated for the purpose.

A common solution to the compatibility issue in \dword{hep} experiments is to use an interface layer that can mitigate the choice of persistent data format with some performance cost.  The scikit-hep project\cite{Rodrigues:2019nct} provides a python ecosystem to read \dword{root} persistent data formats and convert them to native data science formats.

The \dword{hdf5} format is designed to save data as organized tables of fundamental data types along with links between tables which then allow for complex record structures similar to modern databases.  The strength of \dword{hdf5} is that its tabular structure allows
efficient ``columnar'' analysis of data (i.e. analysis of one or more variables from across a dataset).
The format also has native support for parallel data access, and in particular for parallel reading and writing of compressed data.  This support extends to highly parallel file systems, such as those found in leadership computing facilities, and has been tuned to scale extremely well using the MPI protocol in these environments.  
For the large \dword{dune} event data, which can be highly compressed, this support for parallel I/O is advantageous, especially in the near real-time data acquisition environment.  As a result of these features, and the use of the format in other segments of \dword{dune}, we expect to support the \dword{hdf5} format as part of our computing model.

In contrast to the \dword{hdf5} format,
the \dword{root} I/O format has support for the recording and retrieval of complex data structures in the  C\raisebox{1pt}{++} language, and can be used to serialize and deserialize  C\raisebox{1pt}{++} structures and classes.  It 
is naturally
integrated with the \dword{root} data analysis framework which, while domain specific to \dword{hep}, is nonetheless one of the largest and most robust toolkits for \dword{hep} data analysis.  We also expect that support for parallel I/O within the \dword{root} I/O systems will be expanded by the time of the \dword{dune} era, and that the parallel reading and writing of compressed data will be supported.  Similarly we expect that in the \dword{dune} era, new organization and representations of tree'd data structures (e.g. n-tuple or TTree like structures) will be available through the \dword{rntuple} system\cite{Blomer:2020usr, ROOTTeam:2020jal}. It is expected that \dword{rntuple} will provide columnar data access and performance as good, and potentially better, than \dword{hdf5}. Indeed, the \dword{lhc} experiments are expected to migrate to the \dword{rntuple} format on the timescale of \dword{dune}.
As a result of these features of the format, and the use of the format across \dword{hep}, we expect to support the \dword{root} format as part of our computing model. 

One concern that arose during our format evaluation was the availability of a streaming option for \dword{hdf5}. Our large-scale offline processing is most efficient with a mix of both streaming and direct copies of input data.   Streaming also allows any \dword{dune} site to access centrally stored data transparently.  For this reason \dword{hdf5} streaming is very desirable.  We have successfully adapted the \dword{dune} \dword{larsoft} framework to read  local \dword{hdf5} input and very recently have demonstrated streaming via \dword{xrootd} in test mode. However, due to increased latency, efficient streaming has to include caching strategies which have not yet been investigated and making HDF5 streaming work in production will require further, crucial developments.

In addition to the \dword{root} and \dword{hdf5} data formats, we expect that the \dword{dune} data representations will need to be flexible and adapt to changing technologies and evolutions in the data science and physics communities.  For this reason we expect to allow for the addition of other data representations and have placed requirements on the software frameworks used by \dword{dune}, as discussed in chapter~\ref{ch:fworks}, to support multiple data formats and I/O layers, as well as requirements on the data management and cataloging systems to support multiple data formats.

Given the vast variation in scale and data access patterns that \dword{dune} computing needs to support, it is likely that most if not all of these options, and additional as-yet-unknown options, will have a role in creating the most effective suite of solutions to support the physics program. Detailed studies of the \dword{protodune} I \& II datasets will be needed to make cost-benefit analyses of the various options and guide the analysis model.  A final consideration, and by no means the least important, will be the ease of use of the analysis model for physicists to maximize their analysis output.

\cleardoublepage

\chapter{Data Placement \hideme{draft 3/10 moved from end to right after data management } }
\label{ch:place}

\section{Current Status}

DUNE relies on a multi-level data placement strategy that has grown out of the Fermilab fixed-target program with elements from CERN fixed target experiments.  Figure \ref{fig:storage} illustrates the storage available to users. Reference \cite{bib:storage} provides a more detailed description.
Access methods and catalogs are described in detail in Chapter \ref{ch:datamgmt}.

There are two general use cases considered for storage systems in \dword{dune}: production operations and analysis by end users. Production operations are in general focused on utilization of large-scale storage elements. End users, however, utilize a wide variety of storage systems and typically log into unix systems, at Fermilab, CERN, or home institutions, which have disk mounts of varying size and level of backup.

\subsection{Fermilab Storage systems}

At present, most users use the interactive resources at Fermilab and/or submit their grid jobs via the Fermilab systems so those systems are described in enhanced detail.

\begin{description}

    \item{\bf Local disk -}  There are small   volumes on local physical disks, directly
mounted on the machine hosting the \dword{vm} with direct links to the /dev/ location.
These are mainly intended  as temporary storage for infrastructure services (e.g., /var, /tmp).
These areas must be used to  store secure items such as %
user authentication and authorization credentials to avoid exposing them over the network. Such secure items are saved to local disk automatically with owner-read permission and other permissions disabled.
These local areas are usually very small and should not be used for data file storage %
or for code development.
    
    \item{\bf Network Attached Storage}
    
     Users have access to \dword{posix} compliant network attached storage areas with varying sizes and levels of backup. Network attached disks are not safe for storage of certificates and tickets.

 \begin{itemize} 
   \item Users are provided with a network mounted home area %
   with $\sim$2 GB of quota.    A valid Kerberos ticket is needed   to access files in the home area. Periodic snapshots are taken and available to users in case there is a need to recover deleted files.
\item A \dword{nas} Application volume is intended for code development and is  limited to 100 GB/user. The area has periodic snapshots to allow for recovery of unintentional deletion of files.

\item  Additional \dword{nas} data volumes provide storage for fast analysis and code testing interactively. The expectation is approximately 1 TB of storage per user will be sufficient. Any larger samples should be cataloged and managed by \dword{rucio} or by physics groups. 
\end{itemize}
 
The \dword{nas} disks are not accessible  to grid worker nodes. 
 
\item{\bf dCache -} Several PB of \dword{dcache}\cite{Millar:2014cfa} storage are available for multiple purposes. \dword{dcache} storage is not fully \dword{posix} compliant but \dword{nfs} mounts with limited functionality are available on Fermilab interactive machines. The \dword{dcache} storage is readable and writable from grid machines via \dword{xrootd} and transfer mechanisms such as \dword{ifdh}. It is expected that this is the main storage element for the output of both production and analysis distributed computing jobs.

\begin{itemize}
    \item The \dword{dcache} scratch area is used mainly for files returning from grid jobs.  The total volume of the \dword{dcache} scratch volume is (2022) greater than 5 PB. It is currently shared with other experiments at FNAL. Files are automatically removed based upon a LFA policy with typical lifetimes of one 1 month.
     \item The \dword{dcache} persistent volume is a $\approx 800$ TB area for persistent storage of user files.  Files must be removed by their owner. Half of this volume is currently controlled with physics group based quotas, while the other half will soon be transitioned to quota-controlled usage.
      \item The \dword{dcache} tape-backed volume is a $\approx 3.2$ PB disk cache sitting in front of the \dword{enstore} tape system. Files in this area are cataloged and tracked by the \dword{sam} system and, in future, by \dword{rucio}. Files that have been flushed from the tape-backed cache need to be prestaged before they can be used.  At the moment, individual users can do this but we anticipate, once most samples have been moved to DUNE controlled disk, that users will need to request prestage to avoid drive contention. 
\end{itemize}

Remote users also have similar access to local disk at \dword{cern} and their own facilities and to collaboration samples located on the large \dword{dune} \dwords{rse}.
Much of the \dword{dcache} storage at Fermilab and at other DUNE sites is migrating to management via the \dword{rucio} data management system that is described in more detail in Chapter \ref{ch:datamgmt}. A small fraction will remain  managed by physics groups or directly available to general users. 

\item{\bf Tape -} At Fermilab, DUNE shares the Public instance (separate from a CMS instance) of the tape storage infrastructure.  The physical system is currently (March 2022) composed of two IBM TS4500 tape libraries with LTO8 drives (72 total) with over 100 PB capacity per library.  In addition there is a legacy Oracle SL8500 library being replaced with a 150 PB capacity library (vendor selection and purchase pending).  Current DUNE utilization of the tape complex is 21 PB of the total (non-CMS) active utilization of 168 PB.
The tape storage software used is \dword{enstore}.  Fermilab is actively pursuing a transition to using the  \dword{cta} software, and plans a multi-year transition.

Tape storage is also available at CERN through the \dword{castor} system.  The CERN system is mainly used to archive the raw data from ProtoDUNE.  We are also in the process of integrating tape archives in the UK and France for processed samples.

\item{\bf Distributed caching systems - } \dword{dune} makes use of the \dword{cvmfs} caching system to distribute code, flux and shower libraries and user grid executables. \dword{cvmfs} mounts are available on DUNE grid nodes.

\begin{itemize}
    \item {\tt /cvmfs/ } mounts of DUNE-specific code,  shared executables such as \dword{larsoft}, and general utilities are distributed via \dword{cvmfs}. Version control is provided by the Fermilab \dword{ups} system. 
    
    \item \dword{stashcache} a.k.a. XCache is used to deliver larger payloads such as flux files and shower libraries to grid jobs.
    
    \item A \dword{cvmfs}-based  {\tt dropbox} is also available to transfer user executables to grid jobs.

 \end{itemize}

\end{description}
\begin{dunefigure}
[Storage Schematic]
{fig:storage}
{Storage systems for \dword{fnal}-based DUNE computing.  Thick lines denote mounts while thinner lines denote transfer methods such as ifdc and xroot streaming. \dword{cern}-based computing shares access to the \dword{dcache} and \dword{cvmfs} systems but uses lxplus and eos for local interactive computing.}
\includegraphics[width=0.8\textwidth]{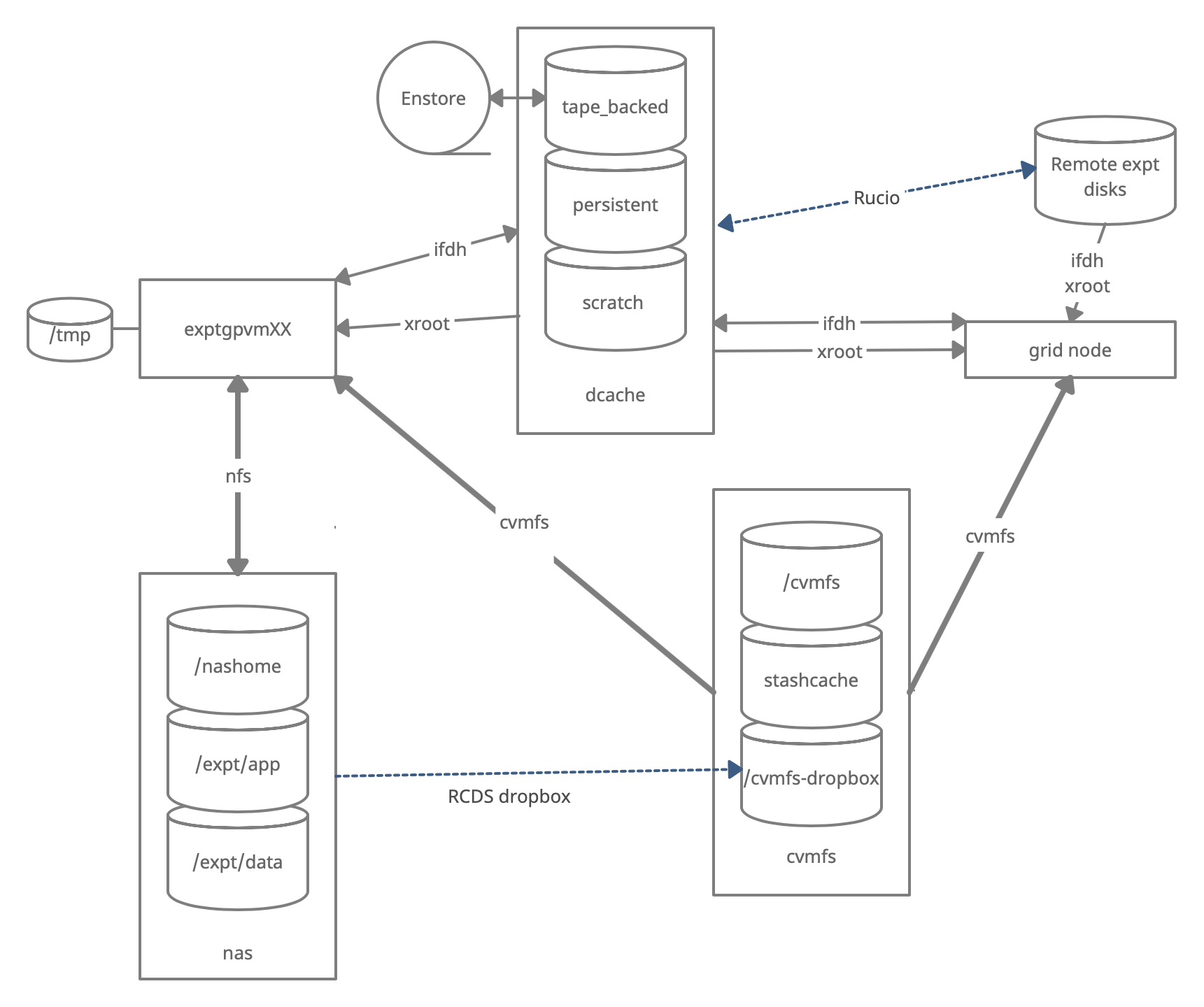}
\end{dunefigure}

\subsection{Storage systems at CERN}

Data at CERN are stored in the \dword{eos}, \dword{castor}, and \dword{cta} systems.  Some of this storage is under the control of the \dword{dune} \dword{rucio}
system, including a full copy of the \dword{protodune} raw data in \dword{castor}/\dword{cta}. Transfers out of \dword{cern} to storage elements elsewhere are made using FTS3 and standard transfer protocols. \dword{dune} is working to avoid Globus transfers in order to reduce cost and support effort. During \dword{protodune} runs, CERN provides a vital step in fast data transfer and the raw data on disk at CERN is very important for fast turnaround analysis. Along with archival storage, there is also substantial space for generic use by CERN-based users. The \dword{cvmfs} file systems are mounted at \dword{cern} and local users have access to the \dword{cern} batch resources through the normal \dword{dune} batch system which requires membership in the \dword{dune} \dword{vo} and through normal \dword{cern} submission systems. 

To date, most general users only use the CERN CPU and disk resources as they would use any other \dword{wlcg} site. The expectation is that access to files on storage elements at \dword{cern} will be through \dword{xrootd} transfers or streaming. 

\subsection{Collaboration disk stores}
DUNE collaborating institutions also contribute very substantial storage resources. These resources are populated and managed via \dword{rucio} and the  \dword{sam} catalog. As with the Fermilab \dword{dcache} stores they are available via streaming (xRootD) and grid-copy mechanisms (non-Globus) but are not mounted on interactive nodes.   These sites have come online over the past year (June 2021-2022) and contain copies of the most recent reconstructed and simulated samples.  Figure \ref{fig:rucioWorld} shows the distribution of disk storage across DUNE institutions as of July 2022. The status of storage elements from the \dword{dune} Data Management side will be monitored using Rucio tools, while the local monitoring by sites will be the responsibility of site administration.

\begin{dunefigure}
[Rucio RSE's]
{fig:rucioWorld}
{Summary of Rucio RSE's as of July 2022. Tape systems such as CASTOR are not shown.}
\includegraphics[width=0.99\textwidth]{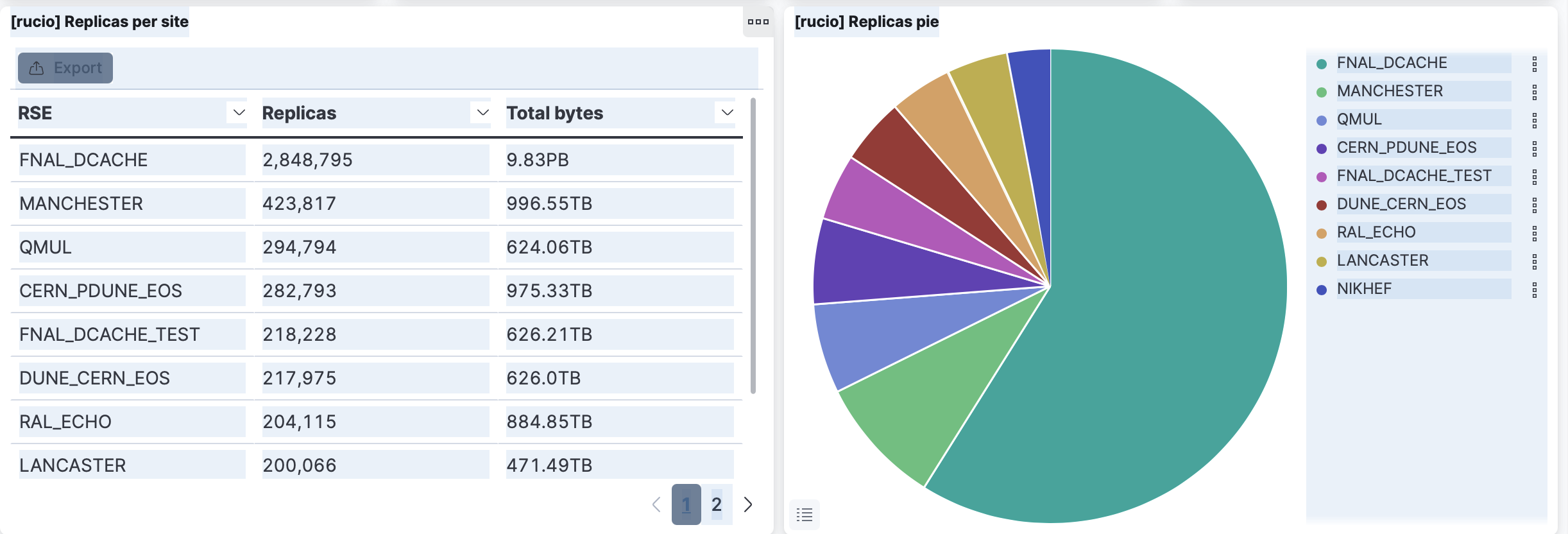}
\end{dunefigure}

\section{Data Placement Strategy}

\dword{dune} data has a large number of data access patterns based on the size, CPU/byte and frequency of access.   Major examples are:

\begin{description}
    \item{\bf Raw data -}  Raw data from the TPC detectors comes in 4-8 GB files which are generally run through reconstruction algorithms that take of order 5-10 sec/MB to process. Raw data is generally accessed a few times for calibration and reconstruction as part of an organized production effort.  
    \item{\bf Reconstructed data -}  Reconstructed data from \dword{pdsp} is around three times smaller  than the raw data once the raw TPC waveforms are dropped. It is accessed many times by end users for algorithm development, calibration and production of physics analysis samples.   Observed data access rates range from 1 to 40 MB/sec depending on the amount of reprocessing done to data. 
    \item{\bf Simulated data - } Simulated data are around twice as large as real data due to the large amount of simulation information that is kept.  Even after raw waveforms are dropped, records are still significantly larger. The access pattern for simulated data is very similar to that of reconstructed data.
    \item{\bf Analysis samples - } End-user analysis samples are significantly smaller.  For example the 1 GeV ProtoDUNE simulation sample consists of $\sim$80,000 files each 4 GB in size while the reduced analysis sample is around 8 GB total.  These samples are expected to be access multiple times a day for several months during the development and finalization of analyses.
    \item{\bf Flux files - } Simulation and cross section extraction require access to large libraries of particle fluxes.  These are currently delivered via \dword{stashcache}. These files are accessed with regularity based upon the production schedule of simulated samples. This access pattern lends itself to caching for increased efficiency to distributed grid jobs.
    
\end{description}

As described in the data volumes section \ref{ch:est}, raw data is kept in multiple tape copies but only kept on disk for a short period until calibration and reconstruction are complete.  If reprocessing is needed, the raw data must be prestaged to cache and work on optimizing this process is underway. Recent reconstructed data and simulation copies for users are kept on \dword{rucio}/\dword{sam} controlled disk, with one copy in the US and one in Europe if possible. Smaller final analysis samples are kept on user or group controlled persistent dcache or \dword{nas} areas. Rucio will perform replication and distribution of raw data and production datasets. 
 All datasets will have a well-defined data lifetime within the Rucio system. 
\cleardoublepage

\chapter{Data Lifetimes and Preservation: \hideme{coll-preservation.tex Anne's comments addressed but do we need comments on US/EU compliance?  }}
\label{ch:pres}

\section{Formal Data Management Policy}

This section describes policies and plans governing the lifecycle of scientific data from the DUNE family of detectors and derived data products from such.

The current authoritative version of the DUNE Data Management plan\cite{bib:pubdocdb23} %
forms the basis of the data management plans for institutions and sites within the collaboration.
The plan encompasses Raw, Analysis, and Scientific Results data tiers.  That version is the authoritative source and will be updated as needed.  This chapter summarizes the lifecycle elements.

DUNE and Fermilab data management policies are consistent with U.S. Department of Energy (DOE) policy "DOE Policy for Digital Research Data Management"\footnote{
\href{https://www.energy.gov/datamanagement/doe-policy-digital-research-data-management}{https://www.energy.gov/datamanagement/doe-policy-digital-research-data-management}}.
The host lab portions of the data management plan are consistent with the documented Fermilab "Data Management Practices and Policies for Fermilab Experiments"
\footnote{\href{https://computing.fnal.gov/atwork/data-management-practices-and-policies-for-fermilab-experiments/}{https://computing.fnal.gov/atwork/data-management-practices-and-policies-for-fermilab-experiments/}}.

The DOE policy references "DOE Policy for Digital Research Data Management: Suggested Elements for a Data Management Plan"\footnote{
\href{https://www.energy.gov/datamanagement/doe-policy-digital-research-data-management-suggested-elements-data-management-plan}{https://www.energy.gov/datamanagement/doe-policy-digital-research-data-management-suggested-elements-data-management-plan}}.
Following these suggested elements, the DUNE data management plan addresses the following data elements.
\subsection{
 Data Types and Sources}
The data lifecycle policies are applicable to the Raw instrument data from the DUNE detector elements as noted within the DUNE Data Management Plan.  The policies will also be applicable to certain Simulation, Analysis and Scientific Results data as deemed appropriate by the collaboration.  Policies are also applicable to the metadata needed to understand and catalog the above data types, along with configuration and calibration information.  In addition, to interpret and understand the data it is necessary to store and document software code bases used in reconstruction and analysis.

Test, commissioning, simulation, reconstruction and other generated data is expected to have a finite period of usefulness.  As such a data lifecycle plan must be followed to include lifetime information in the corresponding metadata at time of creation.  At the end of the lifetime a review process is instantiated to determine if the lifetime should be extended or the data may be deleted.

\subsection{Content and Format}
The scientific data contain raw values from the various detector elements and derived values from analysis of the raw information.  The data are stored in standard HEP formats (ROOT and HDF5) and various databases (\dword{postgres}), all of which may evolve over time.  Software and documentation will be stored in industry standard repositories (\dword{github}).

\subsection{ Data Sharing}
All scientific data are available to members of the DUNE collaboration and may be disseminated to collaborating institutions.  The data are globally accessible via the network to authenticated and authorized users.  The host labs,  FNAL and CERN, are the principle location for data, but data may be copied to other collaborating sites for expediency of access and redundancy.
Derived data that are to be directly linked to publications (for example tables) shall be granted the same access as the publication.  In the future meaningful and accessible data sets will be made publicly available.

\subsection{Data Preservation}
Fermilab, as the host laboratory, will maintain accessible copies of all raw data for a period dictated by lab and DOE policies, generally at least five years past the end of data taking.  A longer retention period may be negotiated.  Processed and derived data is retained for a lifetime deemed useful by the collaboration, which for ultimate processed results is also minimally five years past the end of data taking.  Published and other publicly accessible data sets will be retained as long as technically viable. The data will reside on commercially available  archive systems, requiring periodic migration to new technologies.

\subsection{Protection}
Raw and other valuable derived data will have a copy on tape at the host lab as the current solution for reliable storage.  The tape systems include periodic integrity tests.  Deleted data is only marked for removal, with a final review before tapes are destroyed or recycled.  Raw data and other highly valuable derived data will also have an additional redundant copy on external collaboration resources.  Tape and disk resident data is protected by storage system authentication and authorization controls.

\subsection{Rationale}
DUNE raw data and derived data products are unique and represent the return on an international investment in neutrino science.

\section{Policy Implementation}
Table~\ref{tab:est:retention} in Chapter~\ref{ch:est} describes the current retention policies, with simulated and reconstructed data samples retained on tape for 10 years. 

One of the major concerns for long-term preservation of data is the evolution of storage,  operating systems and data formats. Long-lifetime binary data may need to be migrated between storage technologies multiple times.   Our current experience is that data written in the \dword{root} format is readable for at least a decade.   

As operating systems and compilers evolve, old code may cease to work or yield consistent numeric results. %
We will not have the resources to perform continuous integration tests on all code versions, so it is likely that reviving old codes will require substantial effort. The priority will be ensuring that the raw data remain accessible. 

Long term, we anticipate that the smaller reduced samples, not the PB of reconstructed raw data or simulation, will form the legacy samples from the experiment.  Those formats have not yet stabilized, but will need to be carefully documented and duplicated for use past the formal end of the experiment.

\section{Data Releases}
Final public data releases are approved by the DUNE collaboration in conjunction with publications. Unfortunately, formal data portals, such as those provided by most astrophysics experiments are not currently available for the Intensity Frontier experiments.  It would be desirable to work with the host laboratories to set up a shared portal across the suite of neutrino experiments, and possibly the broader Intensity Frontier, with uniform data access methods that build on the experience of experiments such as the Sloan Digital and Dark Energy Sky Surveys in providing public access. At present, we are relying on the supplemental materials features on the arXiv and zenodo to make derived samples available while we negotiate a better technical solution.

\cleardoublepage

\chapter{Data Management \hideme{Timm, Mandrichenko - draft} }
\label{ch:datamgmt}

\section{Introduction \hideme{HMS update 3/9}}
\label{sec:datamgmt:xyz}  %

The Data Management subsystem brings data from the detectors to the archival storage facility 
and then distributes it to storage elements around the world.  The components include a data ingest manager to receive data from the detectors, a replica manager that knows the location of files and manages  transfers between storage elements, a metadata catalog that keeps track of the types and provenance of data, and  interfaces to deliver the appropriate files to workflow management systems and to interactive users.

The DUNE Data Management group is currently designing and deploying several new components in the Data Management
system to replace the legacy \dword{sam}~\cite{Illingworth:2014mba} system, which combined the functions of replica manager, metadata server, 
and file delivery.  Those functions are being separated with well-defined interfaces.  The goal is to have the new systems in place before beam operation begins on \dword{pdsp2}. %
As of the summer of 2022 all the new components of the data management system have been written and have been tested successfully at scale, but are not all in production as yet.
Figure \ref{fig:datamanagement} illustrates the old and proposed data management architectures with the legacy \dword{sam} system replaced by a new catalog and the \dword{rucio} storage management systems. 

\begin{dunefigure}
[SAM Data management architecture diagram]
{fig:datamanagement} 
{Top: Legacy \dword{sam} data management architecture.  The \dword{sam} system provides both catalog and data location information to processing nodes.
Bottom: Proposed data management architecture diagram.  The \dword{sam} functions are divided between the \dword{metacat} catalog, the \dword{rucio} data movement system and a \dword{datadis} that interfaces to processing and monitoring systems. }
\includegraphics[height=8cm]{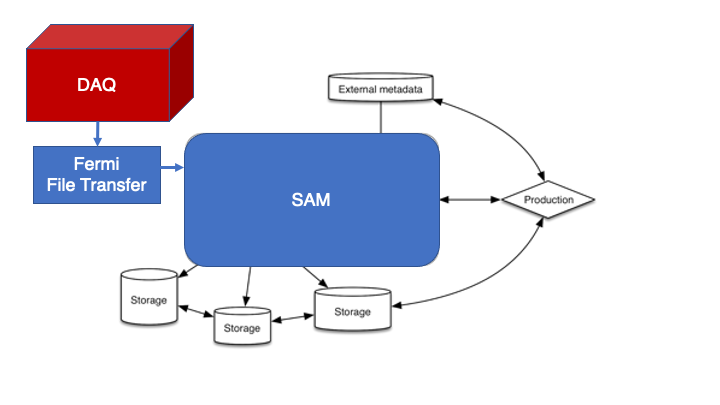}
\includegraphics[height=8cm]{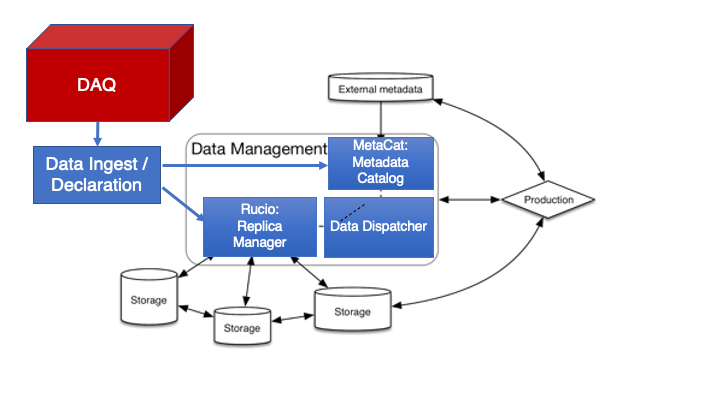}
\end{dunefigure}

The existing DUNE data catalog, \dword{sam}, was originally designed for the D0 and CDF high energy physics experiments at \dword{fnal}.  It is now used by most of the Intensity Frontier experiments at Fermilab. %

The most important objects cataloged in \dword{sam} are individual files and collections of files called
\dwords{samdataset}.
Data files themselves are not stored in \dword{sam}, their metadata is, and that metadata allows users to both identify and locate the physical files %

\dword{sam} was designed to ensure that large-scale data processing be done accurately and completely,  which led to  high standards of reproducibility and documentation in data analysis.
For example, at the time of the original design, the main storage medium was 8\,mm tapes using consumer-grade drives.  Drive and tape failure rates were $>1$\%.  Several \dword{sam} design concepts, notably luminosity blocks and parentage tracking, were introduced to allow accurate tracking of files and their associated normalization and provenance in a high error-rate environment. 

Unfortunately \dword{sam} is almost completely file-based and often duplicates run-level information for convenience.  The %
system has served the DUNE Collaboration through the first \dword{protodune} runs but going forward, a replacement is needed. %
In addition, replacement is also needed to provide better integration with external databases such as the conditions database, and to have data movement capacity for a distributed worldwide storage system.

\section{Requirements for Replacing SAM Functionality \hideme{HMS update 3/9}}

 The replacement for \dword{sam} should support use of run or trigger record level information, in addition to file information,  as appropriate. 
For example,  DUNE trigger records may span multiple files. The new model needs to generalize from ``events'' and files to data objects and collections across multiple scales. 

The functions of the existing \dword{sam} system that we wish to retain and extend are listed here. The first three functions that we need to retain %
relate to content and characteristics while the last four relate to data storage and processing tools.  We propose to separate these functions where appropriate.

\begin{enumerate}

\item	Describe the contents of individual files in a searchable manner to allow users to select well-defined data and simulation samples. 

\item	Create and document data collections ``datasets'' to allow later retrieval based on data characteristics.

 \item	Track object and collection parentage and describe processing transformations to document the full provenance of any data object and ensure accurate normalization.

\end{enumerate}
The next four functions of \dword{sam} relate to the physical location and delivery of files.
\begin{enumerate}
 \item	Store the physical  location of objects.

 \item	Track the processing of collections to allow reprocessing on failure and avoid double processing.

 \item	Provide methods for delivering and tracking collections in multi-process jobs.

 \item	Preserve data about processing/storage operations for debugging/reporting.

\end{enumerate}
 We propose to break the existing system up into functional sub-units specializing in cataloging, storage and delivery. 

\begin{description}
\item{\bf \dshort{metacat}}  The catalog function will be replaced by a combination of a file metadata catalog \dword{metacat} and a run configuration and conditions database that stores run level information.  
\item{\bf \dshort{rucio}}  The file location function is replaced by \dword{rucio}~\cite{Barisits:2019fyl}, the file tracking system originally developed by the ATLAS collaboration. \dword{rucio} provides a location catalog for files and rule-based transfers between sites. 
\item{\bf Data Dispatcher} The existing \dword{sam} system supports a station and \dword{samproject} ecosystem where \dwords{samdataset} are submitted for processing as \dwords{samproject}.  The \dword{samproject} is a server process that contains a list of files based on a \dword{samdataset} description  and, when asked, delivers location information for the next file in the list.  Request, delivery and processing are tracked and recorded. This allows resubmission on failure.  A website allows users to see the details of file delivery.

\end{description}

In the following subsections we describe the existing features that we wish to retain in more detail with the proposed replacement technologies in %
later sections. 

\subsection{Existing SAM  Features}

\subsubsection{Data Description Values and Parameters}

 \dword{sam} supports several types of data description fields.
\dword{sam}  Generic ``values'' such as data\_tier, run\_type and file\_size  are common to almost all  HEP experiments and are optimized for efficient queries.  These values are a limited set and it is possible to request a list of the values (of the ``values'') already in use. 

\dword{sam} also allows definition of free-form ``parameters'' as they are needed within each experiment's instance.  This allows the schema to be modified easily as needs arise. Unfortunately, a major problem is that it is  not possible to request a list  of the values for a given parameter and there is little protection against typographical errors in parameter names or their values.  This has, over time, led to considerable chaos. 
The new \dword{metacat} extends these concepts to make them easier to search and maintain.

\subsection{SAM Datasets and Projects}

\subsubsection{Datasets and Snapshots}

In addition to the files themselves, \dword{sam} allows %
definition of \dwords{samdataset}.
A  \dword{samdataset} is not a fixed list of files but a query against the \dword{sam} database. An example query would be ``run\_type protodune-sp and data\_tier full-reconstructed and run\_number 5141 and version v09\_41\_00'' which would be all files from protodune single phase run 5141 that are reconstructed data produced by version v09\_41\_00. The \dword{samdataset} is dynamic; if after this dataset is created, an additional file from run 5141 is reconstructed with v09\_41\_00, and satisfies the criteria,
 the \dword{samdataset} will grow to include it. There are also %
``\dwords{samsnapshot}'' that are derived from \dwords{samdataset} and capture the exact files in the \dword{samdataset} at the time the \dword{samsnapshot} was made. Both have their uses: \dwords{samsnapshot} are always reproducible while  \dwords{samdataset} are very valuable when processing data from a running experiment or simulation as they can incorporate new data without user intervention.

\subsubsection{SAM Projects}

\dword{sam} also supports access delivery and  tracking mechanisms called \dwords{samproject} and \dwords{samconsumer}. 

A \dword{samproject} is effectively a processing campaign across a \dword{samdataset} that is owned by the \dword{sam} system. At launch, a snapshot is generated after which the files in the snapshot are delivered to a set of \dwords{samconsumer}.  The \dword{samproject} maintains an internal record of the status of the files and \dwords{samconsumer}. Compute processes instantiate \dwords{samconsumer} attached to the \dword{samproject}.  Those \dwords{samconsumer} then request ``files'' from the \dword{samproject} and, when done processing, tell the \dword{samproject} of their status.  

The original \dword{sam} implementation actually delivered the files to local hard drives.  Modern \dword{sam} delivers the location information and expects the \dword{samconsumer} to find the optimal delivery method. This is a pull model, where the consuming process requests the next file rather than having the file assigned to it.  This makes the system more robust on distributed systems. 
There is also a web interface\footnote{\href{http://samweb.fnal.gov:8480/station_monitor/dune/stations/dune/projects}{http://samweb.fnal.gov:8480/station\_monitor/dune/stations/dune/projects}} that allows users to view the status of running \dwords{samproject}. 
This functionality will move to \dword{rucio}, the data dispatcher and the workflow systems.

\subsection{Data provenance and tracking}\label{sec:prov}
Because \dword{sam} stores the ancestry of files, both children and parents, complex queries such as "give me all the raw data files of type X that do not have children processed and stored by code type Y versions Z-AA" are possible.  Complex queries such as this can be used to pick up missing files in reprocessing without creating duplicates. 

An additional protection against duplicate files is requiring constrained filenames for output files based solely on the inputs and code characteristics. If variable fields such as timestamps are put into filenames, \dword{sam} will happily catalog duplicate instances of the same processing and the provenance information is then necessary to check for and remove duplicates.

\section{Future Components\hideme{Heidi added 3/9}}
We propose to replace  the \dword{sam} system  with %
the following set of components: 

\begin{itemize}
\item \dword{metacat} Metadata Catalog: stores file characteristics but not locations - described in section \ref{sec:metacat};    
    \item Data Ingest Manager: takes in data from detector sites, declares it to the catalog and transfers to the managed storage system - described in section \ref{sec:ingest};
    \item A \dword{rucio}-based Replica Manager: stores the physical location of files and provides tools to move them between storage elements -- described in section \ref{sec:rucio}; and
    \item Data Dispatcher: interacts with user and production processing systems %
    to provide file location information from the Replica Manager to clients and can track file processing -- described in section \ref{sec:dispatch}. 
\end{itemize}

The \dword{rucio} system is already being used for file placement.  We are testing use of \dword{metacat}/\dword{rucio} in hopes of putting them into production for \dword{pdsp2} runs late in 2022.

These components, collectively, will provide almost all of the existing \dword{sam} functionality while providing  significant enhancements in implementation and features.

\section{MetaCat Metadata Catalog \hideme{updated 3/9 may still be too long}}\label{sec:metacat}

The current \dword{sam} data catalog combines file description with location and delivery.  The new \dword{metacat} catalog   provides the permanent file description information, with location and delivery handled by \dword{rucio} and a data dispatch system. 
A prototype version of \dword{metacat} has been produced and is being tested as part of our 2022 data challenges. \dword{metacat} is described in Reference~\cite{Mandrichenko:2021spd}. 

DUNE file metadata includes mandatory and optional fields and can be represented to the API by a Python dictionary or a \dword{json} file. 
As with \dword{sam}, the metadata includes a description of the file itself and data about how it was created, including enough information to follow the full processing chain.

The DUNE data  management system operates on a collection of ``physical'' copies of files (or objects), moving them between storage elements and making them available to the data processing and analysis. 
A single  ``logical'' file can have multiple ``physical'' replicas across multiple storage elements. The \dword{metacat} package replaces the \dword{sam} file description function and generalizes it to objects.  
\dword{metacat} stores metadata about logical files and objects and makes it possible to select ``interesting'' logical files based on criteria expressed in terms of metadata parameters. 

The \dword{metacat} design builds on \dword{sam} experience, preserving the useful concepts while introducing additional flexibility and constraints where needed.   For example:

\begin{itemize}  
\item To enforce reproducibility, %
free-form datasets no longer exist and  must be explicitly modified.
\item Metadata values are more flexible.
\item Namespaces are introduced to separate different functions (e.g., user %
and production files)
\item Access to external databases (e.g., for run information) is available where needed.  In \dword{sam}, run characteristics were often duplicated in every associated file.
\item Modification permissions are more granular.

\item An Event Catalog is provided,
\dword{sam} currently does not contain a means of easily determining which file(s) contains a given event.  If a \dword{daq} system is writing multiple streams, an event from a given subrun could be in any stream.   Existing neutrino experiments are low enough rate that this has not been an issue but \dword{protodune} already needs this feature.
The \dword{sam} replacement will require the development of a true two-way map between objects (events) and collections. 
\end{itemize} 

Although the main target user of the project is \dword{protodune}/DUNE, \dword{metacat} is generic enough to be usable by %
other experiments, as well.

\subsection{MetaCat requirements}
\dword{metacat} has to satisfy the following general requirements: 

\begin{itemize} 
\item 
The Metadata representation must be powerful, flexible and abstract enough to accommodate a wide range of metadata data types and possibly complex metadata structures. 

\item
It needs to scale to several 100 million objects, based on Fermilab fixed-target experience, and likely to billions of objects by the end of \dword{dune} running. %

\item
The unit of operation of the catalog should be an abstract ``file'' or ``object'' with a  minimal but sufficient set of predefined   attributes and the ability to add user-defined   attributes in a flexible and convenient way. 

\item 
The file selection mechanism must be powerful and at the same time simple enough to be able to express a wide range of metadata selection criteria. 

\item 
The \dword{metacat} should not need knowledge of physical file replicas, aside from a checking mechanism to make certain that all physical objects have logical representations in the catalog. 

\end{itemize}

Based on our experience with \dword{sam}, an important improvement is to provide a mechanism to use data from external metadata sources, such as conditions or runs databases, as part of metadata queries without copying the external data into or making it appear as part of the metadata database.

\subsection{MetaCat implementation} 

In \dword{sam}, files were identified by a unique immutable name within a single namespace.  To support  multiple use cases (for example user and production spaces) \dword{metacat}'s objects %
can be identified  by names within namespaces. The name of an object is unique within its namespace. A namespace/name pair uniquely identifies  an object in \dword{metacat}. In addition to namespace/name, files are assigned unique text identifiers. These identifiers are primarily for internal use within the \dword{metacat} database, but are available to the user. At the time of the file declaration, the user can specify a file ID, which must be unique, otherwise \dword{metacat} will generate a unique file ID. After a file is declared to \dword{metacat} it can be renamed. Both namespace and name can be changed, provided the new namespace/name pair is unique. However, file ID can not be changed. 

As in \dword{sam}, \dword{metacat} stores metadata associated with files and datasets. %
A \dword{metacatdataset} is a collection of files. A \dword{metacatdataset} can have child \dword{metacatdataset}.  %
Unlike the active \dwords{samdataset} defined   in \dword{sam}, files must be explicitly added to and removed from a \dword{metacatdataset}. 
A file can belong to zero or more \dwords{metacatdataset}. \hideme{HMS - I think this may confuse the reviewers suggest leaving it out There is no requirement that the file namespace  be related in any way to the namespace(s) of the \dword{metacatdataset}(s) the file belongs to. If a file belongs to a child \dword{metacatdataset}, it does not %
necessarily belong to its parent \dword{metacatdataset}. However, the query language discussed below allows recursive inclusion of files from child \dwords{metacatdataset} into query results.}

As in \dword{sam} there can be a many-to-many provenance relationship between files. A file can have zero or more derived (child) files and zero or more parent files. The system makes sure the provenance relationship is not circular. 

Files do not necessarily have to have physical locations. Some use cases, for example those involving production and then merging of output files, may include a ``virtual'' file in the parentage relation between the large parent file and the merged child file. 

Metadata attributes are name-value pairs. Any file or \dword{metacatdataset} can have zero or more metadata attributes. Attribute names are alphanumeric words optionally combined with dots. Any \dword{json} structure can be an attribute value. Therefore, a file or a \dword{metacatdataset} attribute set is a \dword{json} dictionary.

Here is an example \dword{metacat} entry for a processed data file from \dword{pdsp}. 
\begin{verbatim}
checksums :    {'md5': 'ea8d1a009f23accf9582f9e2bd7f58fd',
                     'adler32': 'c689390f', 'enstore': '3896981774'}
children  :    ['52593351']
created_timestamp:  1616688097.41347
creator   :    dunepro
fid       :    52593202
name:    np04_raw_run005141_0006_dl1_reco1_42477998_0_20210324T231631Z.root
namespace :    pdsp_det_reco
parents.  :    ['6606706']
size      :    3599190934
metadata. :   {
    'DUNE.campaign': 'PDSPProd4',
    'DUNE_data.DAQConfigName',
    'np04_WibsReal_Ssps_BeamTrig_00021',
    'DUNE_data.acCouple': 0,
    'DUNE_data.calibpulsemode': 0,
    'DUNE_data.detector_config':  ...... 
    'DUNE_data.detector_config.object': ['cob2_rce01', 'cob2_rce02', .....
    'DUNE_data.feshapingtime': 2,
    'DUNE_data.inconsistent_hw_config': 0,
    'DUNE_data.is_fake_data': 0,
    'beam.momentum': 7,
    'data_quality.online_good_run_list': 1,
    'detector.hv_value': 180,
    ...... many other items not shown  .....
}


 \end{verbatim}

\subsection{Ownership and Permissions }

A \dword{metacat} user  is identified  by a unique username. A user can be a member of zero or more roles (groups). Namespaces and attribute categories have owners. A namespace or category owner can be either an individual user or a role. If the namespace or the category is owned by the role, %
that means it is automatically owned by all role members. 
The namespace owner automatically owns all the datasets and files in that namespace (either directly or via the role membership). Ownership does not automatically propagate along the dataset parent/child or file provenance relationships. 
Only the owner of a dataset %
can add or remove files from the dataset. Only the owner of a dataset can change its metadata attributes. The same is true for files in that only the owner of a file can change its metadata attributes. %

\subsection{Queries} 
One of the most important parts of the \dword{metacat} functionality is the ability to query the database for ``interesting'' files. Essentially, the query is a logical expression in terms of file and/or dataset metadata attributes, file provenance relationship, and dataset parent/child relationship. There are two types of queries in \dword{metacat}: file queries and dataset queries. File queries return a set of files (a list of file IDs) whereas dataset queries return a list of datasets. The file and dataset lists are not guaranteed to have a consistent order, so they are in fact ``sets'' rather than ``lists.'' 
A query is merely a formula specifying the selection criteria. \dword{metacat} does not save the results of the query, so re-running a query can produce different results as files are  added/removed from the system, to/from the datasets or their metadata attributes change.
However, there is an option to save results of the file query as a new \dword{metacatdataset} or add selected files to an existing \dword{metacatdataset}. 
A query can be saved into the database under a name within a namespace. This function can be used to publish complicated queries and make them reusable by other users. 

\subsection{MetaCat Query Language (MQL)}
The original \dword{sam} query language was produced in the late 1990s to provide a restricted command set and  avoid having users execute full SQL queries.

In \dword{metacat}  the query is written in a specialized query language, \dword{mql}. \dword{mql} allows the user to specify file/dataset metadata attribute criteria, use \dword{metacatdataset} parent/child relationships and file provenance relationships to select a set of file or a \dword{metacatdataset}. Further, simple queries can be combined into more complicated ones using logical operations like union, join, subtraction. Named queries can be referred to from within the \dword{mql} expression by their name. 

\subsection{External data sources }
\dword{metacat}   functionality includes the ability to access external metadata sources, such as conditions databases, and use the data stored there to filter file selection results. In order to make an external metadata source available to \dword{metacat} instance,  a Python plug-in module with a standard interface must be provided.  Once the module is plugged into a \dword{metacat} instance, it can be referred to in the \dword{mql} query as a named filter  and used to filter results of an intermediate query or queries within the \dword{mql} expression or even ``inject'' metadata from the external source into the query and make it available as a  selection criterion. 

\subsection{Architecture and Interfaces}
\dword{metacat} is a typical web-services database application. The underlying database is not exposed to end users. It can be accessed via a Python API, but the primary method of interacting with the system is through a web service or web GUI. Use of this REST-based web service 
improves the scalability and cacheability of the system and completely unties the server side from the client implementation. Any standard HTTP/HTTPS client can interact with the system either directly or through a standard HTTP proxy or cache.

The system publishes the following interfaces: 
\begin{itemize} 
\item a direct database access Python API, 

\item a Web services REST interface, 

\item a python client side API that communicates with the server via HTTP, and 

\item 
a command line interface with a basic set of commands to enter (modify) data into the database and to query the database. 
\end{itemize}

\subsection{File naming conventions}
In principle, a data catalog such as \dword{metacat} means that file names could be completely generic.  In practice random bitstrings makes it very difficult for humans to interpret and check what they are doing.  DUNE does not yet have a general file naming convention, with different subgroups defining their own. A general specification would be useful and will be produced. Similarly, datasets can be identified based on their characteristics via queries but should also have a naming convention.   Previous experiments have generally encoded the data format, data tier, run/subrun number and processing campaign in the file name for easy comprehension. 

\subsection{Current Status} 
\dword{metacat} has been successfully tested in the Data Challenge at full operational scale.  We are currently operating it in parallel with the \dword{sam} catalog.  A plan for switching from \dword{sam}/\dword{rucio} is being finalized with full deployment either during \dword{pdsp2} or for subsequent processing steps in 2022-2023.

\section{Data Ingest Manager \hideme{HMS moved here 3/9}} \label{sec:ingest}

The Data Ingest Manager will run at detector sites.  It consists of two components:  the Ingest Daemon and the Declaration Daemon.  The Ingest Daemon will 
run at each detector location, at CERN EHN1, SURF surface and Fermilab. Each instance will
detect new files in the data store, extract the metadata and add new metadata fields
if necessary, calculate the checksum of the files, initiate and monitor the transfers to the first managed 
storage element, and send a signal back to the \dword{daq} when the file has been transferred, properly stored, and can be deleted from the data store.
Files will be transferred to a drop box on the nearest managed storage element, using \dword{fts3} %
as the transport mechanism.  

The Declaration Daemon will run on or near a storage element that is managed by the replica manager. %
It will detect new files that have been added to the drop box, and declare them to the metadata catalog and the replica manager.  It will then instruct the replica manager to send these files to permanent tape-backed storage and monitor the transfer to be sure that this has been done, and then send a signal to remove them from the temporary drop box.

These daemons will be performing most of the functions of the current data transport system. They only 
require modifications to interface with the new replica manager and metadata catalog rather than the current ones.  These daemons were both successfully tested at full scale in the recent Data Challenge 4.  We expect to use them during the upcoming  \dword{pdhd} running.

\section{Rucio Replica Manager \hideme{HMS updated 3/9}} \label{sec:rucio}

The \dword{rucio}\cite{Baritsis:2019csbs} replica manager system is now used to track the physical location of  data files %
and ensure that they 
are moved from point to point as needed.   This system was originally developed by the ATLAS experiment and has now been deployed to 
a number of other HEP and astronomy experiments.  DUNE has had a \dword{rucio} server since 2019.  All raw data that was
taken in the fall 2018 \dword{pdsp} (\dword{np04}) run and since that time, as well as all raw data taken by the \dword{pddp} (\dword{np02}) run, are now tracked by \dword{rucio}.  All output of \dword{mc} simulation and reconstruction are also tracked by \dword{rucio}. \dword{dune} plans to take full advantage of the data lifetime features of \dword{rucio}.

\dword{rucio} is a rule-based system.  Files are sent to remote servers via a system of rules
that are created by the experiment data managers. \dword{rucio} provides tools to implement those rules.   Data can be accessed interactively or in batch jobs and
\dword{rucio} has a built-in system of delivering the URI of the closest replica to the running job for access via streaming.

DUNE has contributed several features to the \dword{rucio} base software.  These include a new method to map from the \texttt{scope:filename} logical file name to the path-based directory structure that we use on tape sites. We also have added a DUNE-specific hook that requires any new files declared to \dword{rucio} already be declared to the
\dword{metacat} system.   

We have implemented object store support, modularized the \dword{vo} customization code and added policy packages specific to the DUNE \dwords{vo}, started to develop a lightweight client, and factorized external dependencies in order to make integration into DUNE 
and other non-ATLAS \dword{vo}s easier. 
These contributions have been done in close cooperation with the CERN \dword{rucio} team and have been fed back to the main \dword{rucio} project code base. %

\dword{rucio} has an internal metadata functionality but our evaluation indicates that it doesn't provide the full functionality required by DUNE physics and offline processing, %
and thus DUNE has proceeded to develop its own metadata catalog. 

The development work needed on \dword{rucio} in the next few years includes completing the lightweight client, integration of this client into all DUNE workflows, adding quality of service support and proximity mapping and fully integrating the \dword{metacat}. In the longer term \dword{rucio} will constantly need development as issues are uncovered and new features are needed. The main work that remains to be done in implementing the \dword{rucio} server  for production is back-loading all relevant data from the legacy SAM system into \dword{rucio}. Currently about 50\% of the data by volume is known to \dword{rucio}.  In the Data Challenge 4 we used Rucio for all transport of data between sites and successfully tested it at the full expected data rate of \dword{pdhd}, which is also similar to the eventual data rate of the far detector.

\section{Data Dispatcher \hideme{HMS 3/9 updated}} \label{sec:dispatch}

The \dword{datadis} replaces the project management and file delivery functions that were previously done by the \dword{sam} system.  The \dword{datadis} functions include:
\begin{itemize}
    \item Creating projects to process collections of data,
    
    \item Delivering file handles of the files to consumers,
    
    \item Keeping track of the project progress, including consumer status and files consumed,
    
    \item Keeping record of projects, consumers, and file consumption for a specified amount of time,
    
    \item Providing project monitoring and control to a user or a client, and
    
    \item Organizing and coordinating data processing among data consumer processes.

\end{itemize}

A \dword{datadis} has been written and is now being interfaced to the general use Fermilab job submission system and tested for processing large quantities of data.  The exact boundary between the \dword{datadis} and the enhanced  Workflow System described in Chapter \ref{ch:wkflow} has not been finalized.  It is likely that other \dword{fnal} experiments will need the \dword{datadis} functionality if \dword{sam} is deprecated.

\section{Tools and Integration \hideme{Timm - needs more}}

The data management system relies on several sets of underlying tools.  The \dword{rucio} replica manager is dependent on the CERN File Transfer Service \dword{fts3}~\cite{Kiryanov:2015fts} to actually move files from one storage element to another, although it can also be configured to use other protocols as needed. 
The \dword{fts3} service in turn is dependent on the
GFAL2 (Grid File Access Library) set of utilities.

We have developed a suite of  data transfer monitoring tools and dashboards to track data placement and \dword{rucio} transfers. 
We have tested and  integrated the CERN CTA Tape service into DUNE data management and are now doing the same for the
RAL CTA based Antares Tape service.  As new sites come along each must be integrated as a \dword{rucio} Storage Element.
An unforeseen development was the need to deploy a StashCache instance in the UK to solve a low job efficiency problem due to slow data access.

\cleardoublepage

\chapter{Networking \hideme{Mike Kirby and Peter Clarke - in progress}}
\label{ch:netw}

With the DUNE experiment located at two geographically distant sites and having computing resources distributed around the globe, the design and operation of networking will be critical to the success of the DUNE experiment. The network requirements can in general be divided between three main areas: %
\dword{fd}, \dword{nd} and distributed computing. Assuring that interconnection and bandwidth requirements between each are met with appropriate uptime is necessary for operations and successful physics results. The requirements for each system are based upon current estimates of data rates for both the \dword{fd} and \dword{nd} \dword{daq}, technical networks, and slow control networks.

\section{SURF to Fermilab}

The \dword{fd} networking encompasses the \dword{lan} at \dword{surf}, the connection to the wide area network, and the network path back to storage services at \dword{fnal} in Batavia, IL. This network path between the \dword{fd} at \dword{surf} and the \dword{fnal} campus is the most important network path for the DUNE raw data. This path will be used both for transfers of raw data, and also for connectivity between \dword{surf} and operations centers at Fermilab and elsewhere. While operations traffic is expected to use this connection, it should be noted that safety systems will not rely upon the network connection.

The requirements for the network connection between \dword{surf} and \dword{fnal} were explored extensively in the High Energy Physics Network Requirements Review~\cite{bib:osti_1804717} . The primary bandwidth requirement is determined from both the steady-state yearly transfers from the \dword{daq}, and by burst 
transfer rates during calibration and extended-readout trigger records. The Consortium Interface document between the DUNE \dword{daq} Consortium and the DUNE Computing Consortium and the \dword{daq} specification~\cite{edms:dunedaq} states that no more than 30 PB/year of data will be output from the \dword{daq} for permanent storage. This translates into 8  Gb/s of steady-state data transfer from \dword{surf} to \dword{fnal}, and includes raw data from beam triggers, supernova candidates, cosmic triggers, and calibration runs. 

The most extreme burst requirement considered in the ESNet review was the handling of extended-time trigger records for %
\dword{snb} candidates. These trigger records, in contrast to the steady beam and cosmic ray rate,
are expected to occur one to two times/month and involve data volumes of up to 160\,TB/module of uncompressed \dword{fd} data. The ability to produce a reasonably fast pointing signal would be extremely valuable to optical astronomers doing follow-up, especially if the supernova were in a region where dust masks the primary optical signal. The need to be alert to %
\dwords{snb} and to quickly transfer and process these data imposes stringent requirements on triggering, data transfer, and reconstruction beyond those imposed by the more regular beam-based oscillation physics. For example, an uncompressed \dwords{snb} readout of the first two \dword{fd} modules will be on the order of 320\,TB in size and take a minimum of seven hours to transfer over a 100\,Gb/s network, and then take on the order of 90,000\,CPU-hrs for signal processing and pattern recognition at present speeds (See Chapter \ref{ch:use} for details). If processing takes the same time as transfer, a peak of 10-20,000 cores would be needed. In order to have initial analysis results of a \dwords{snb} trigger record within one day, the network between the  \dword{fd} at \dword{surf} and the \dword{fnal} campus is being designed for a capacity of 100\,Gb/s.

As the network will be used for both raw data transfer and for remote operations, there is a requirement to have both a primary and secondary network path between the \dword{fd} and \dword{fnal}. Working in conjunction with ESNet, the Core Computing Division at \dword{fnal} has developed a deployment plan for two geographically separate network paths. The map of the proposed paths is shown in Figure~\ref{fig:esnet_network_path_map} with the two paths merging between St. Cloud, MN and the \dword{fnal} campus. Having a single contracted path from MN to the \dword{fnal} campus is not considered a significant risk given the number of alternate paths between the two locations that are available. The primary path connects network switches at the top of the Ross Shaft (at \dword{surf} in Lead, SD) to \dword{fnal} via ESNet infrastructure. As of February 2022, the primary path is available from Lead, SD 
to \dword{fnal} at limited bandwidth of 10\,Gb/s. In order to complete this path, a vendor will be secured to provide infrastructure between Lead and the Ross Shaft. After completion, the primary path will provide 100\,Gb/s guaranteed bandwidth. The secondary path will provide 10\,Gb/s of network bandwidth and should be available in %
2022 or 2023, and discussions are underway to secure dedicated 100\,Gb/s of network bandwidth once detector operations have begun in 2028. \footnote{This information was updated at the beginning of 2022 from early estimates that had the secondary path at 10\,Gb/s.}  The secondary path will be from the Yates Shaft (also at \dword{surf}) through Rapid City and Sioux Falls, SD to \dword{fnal} via ESNet infrastructure. Additionally, a tertiary path that utilizes vLAN infrastructure from the South Dakota Higher Education Network (REED) and a southern path through Colorado and Kansas City, KS will provide 10\,Gb/s of bandwidth. During construction, the tertiary path will serve as the primary link to the DUNE \dword{fd} and then revert to a tertiary role once the dedicated primary path is complete. The networking is expected to be completed in stages with both paths capable of the full design bandwidth by the time that \dword{fd} physics operations commences. A detailed schedule for networking is shown in Figure~\ref{fig:esnet_network_timeline}. 

The network uptime is defined by the capability of the \dword{fd} \dword{daq} system to stage raw data locally in case all connectivity is lost. %
This \dword{daq} is designed to have capacity to store one week of raw data from the detector on local storage at \dword{surf}. %
Given the nominal rate of 7.98\,Gb/s from the \dword{daq}, and assuming that the network interface cards (NICs)
present in the local storage have effectively infinite bandwidth, the additional 92\,Gb/s of bandwidth available on the primary path means that 
were connectivity to be lost for an entire week, 
the backlog in raw data could be cleared within one day of reconnection. Given the presence of three geographically separate paths from \dword{surf} to \dword{fnal}, if all three networks have an uptime of > 90\%, then the expected networking live time (99.9\%) is more performant than the \dword{daq} required live time (98\%) by more than a factor of 10. If both the primary and secondary network paths were interrupted but the tertiary path operating, there would still be limited impact on primary physics data taking but operational adjustments would be necessary. As mentioned earlier in this paragraph, the annual data volume from the \dword{fd} requires only 7.98\,Gb/s of continuous bandwidth, and therefore could be actively transmitted via the tertiary path (10\,Gb/s) without need for caching at \dword{surf}. Any additional, data-intensive detector operations (e.g. calibrations or streaming of Supernova readouts) would require either local caching at \dword{surf} or be postponed until one of the primary or secondary paths are restored. Given the location of \dword{surf} and the potential for natural or man-made disasters, there is a real risk of weeks- or months-long repair times following physical damage to a network path (e.g. forest fire). The ability to continue normal physics operations via the tertiary path, and plans to adjust \dword{fd} operations during such a period, is an important part of the operational planning of \dword{dune} Computing.

\begin{dunefigure}
[ESNet map of network path]
{fig:esnet_network_path_map} 
{A map of the geographically separate primary (blue) and secondary (red) proposed network paths between \dword{surf} and \dword{fnal}. %
}
\includegraphics[width=0.9\columnwidth]{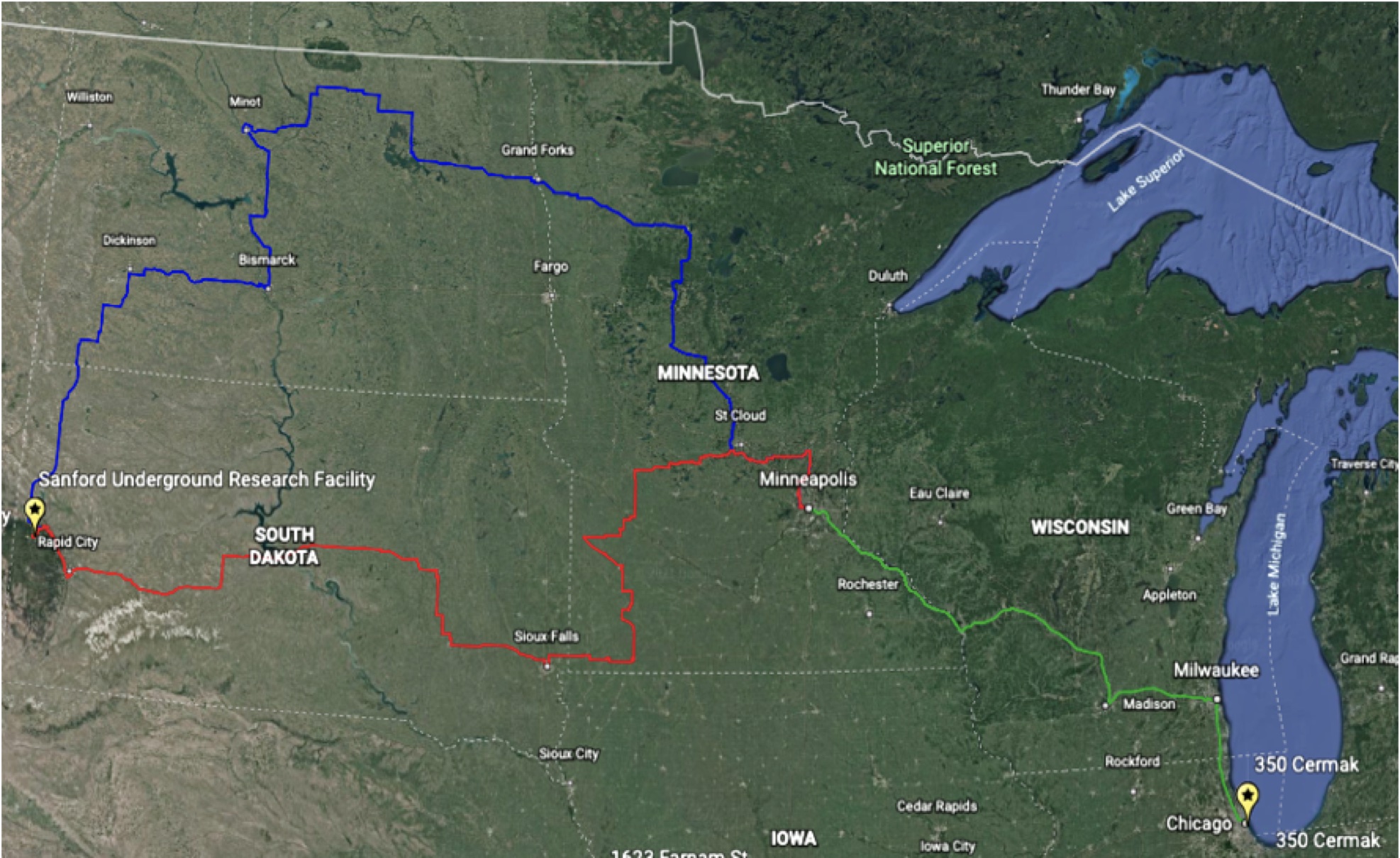}
\end{dunefigure}

\begin{dunefigure}
[Networking Timeline]
{fig:esnet_network_timeline} 
{The current timeline for the implementation of networking connection between \dword{surf} and \dword{fnal}. Note that while the secondary path bandwidth is listed as 10+\,Gb/s, as of Summer 2022 it is anticipated that there will be 100\,Gb/s of capacity once \dword{dune} FD is operational.}
\includegraphics[width=0.9\columnwidth]{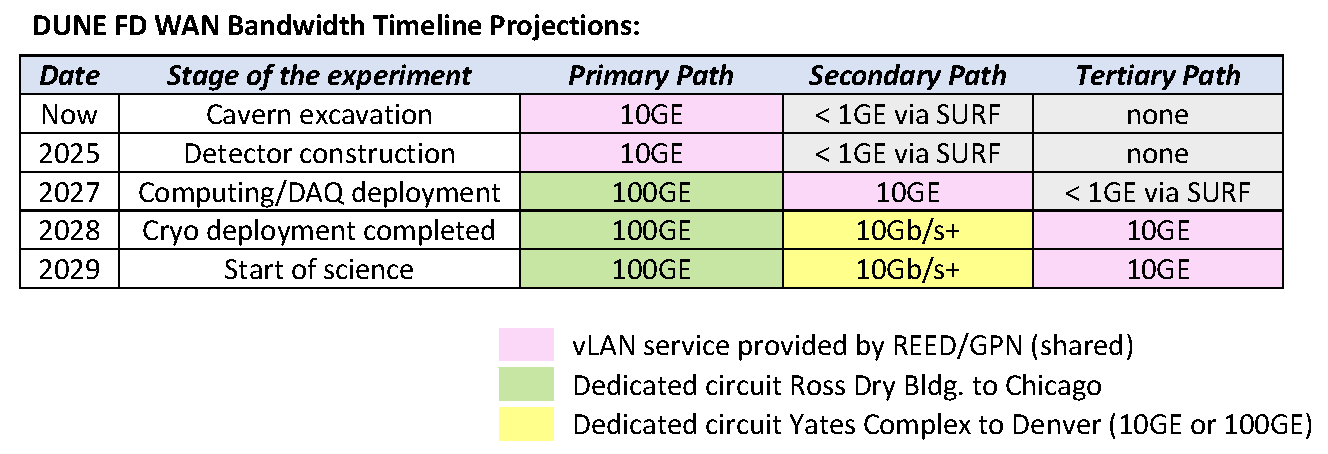}
\end{dunefigure}

\section{Far and Near Site Local Area Networks}

The \dword{lan} for the Far and Near Sites will be designed, installed, and commissioned as a collaboration between the \dword{dune} \dword{daq} %
consortium and the \dword{fnal} Core Computing Division (CCD). The \dword{daq} has provided requirements to CCD and preliminary designs of the LAN have begun. The installation of networking infrastructure at \dword{surf} will be paid for by the %
\dword{usproj} Project 
and is scheduled to start in %
2026. Once network installation has begun, there is a joint responsibility between \dword{surf} and CCD network experts to configure and maintain the interface with external high-capacity networking for data transfers to storage at \dword{fnal}. The installation of the \dword{lan} at the Near Site is planned to begin in %
2029, and follows a similar set of responsibilities as at \dword{surf}. The \dword{daq} group will provide requirements and CCD providing design, commissioning, and operational support for the Near Site LAN.

\section{Global Connectivity}
Connection to European sites is accomplished via the \dword{esnet},
the pan-European research network (\dword{geantnet}) and \dwords{nren}.
The current aggregate  transatlantic bandwidth is of the order of a Tbit/s, of which 400\,Gb/s is from  \dword{esnet} (Boston, New York and Washington DC)
to \dword{geantnet} (London, Amsterdam and Geneva (\dword{cern})).
\dword{geantnet} in turn peers with the \dword{nren}   in each participating country, where details
pertaining to each DUNE site vary. As an example, in the UK the
\dword{nren} is JISC-JANET, which at the time of writing has a 400\,Gb/s core and connects to the GridPP-RAL site redundantly at 100\,Gb/s.
Similarly the CC-IN2P3 site in France is connected to RENATER at 100\,Gb/s, and the FZU site in the Czech Republic is connected to CESNET
at 100\,Gb/s.  DUNE expects all participating countries to
ensure that as part of their pledges of CPU and storage capacity, %
their sites %
have commensurate network connections. %
No systematic issues are expected to arise, and none have done so in the
WLCG/LHC context. DUNE participates in the worldwide \dword{hep} Network
coordination body which meets twice per year  (the LHCOPN/LHCONE Group).

\dword{layer3} \dword{vrf} provision is now very prevalent in \dword{hep}. 
 \dwords{vrf} provide a logical routing overlay that can allow for traffic engineering to utilize high-capacity paths where needed. The \dword{lhc} community uses a \dword{vrf} called LHCONE, and this has also been used for DUNE traffic along with other non-\dword{lhc} experiments such as Belle II. 
At present DUNE is agnostic %
regarding the use of LHCONE, and since \dword{fnal} is connected to LHCONE it can easily accommodate sites with or without such provision.
Investigations are currently underway to determine the technical requirements for the creation of a separate DUNEONE \dword{vrf} were it to ever be required. It is, however, not currently foreseen, and does not form part of our baseline planning.
\cleardoublepage

\chapter{Workflow Management \hideme{McNab}}
\label{ch:wkflow}
\section{Introduction}
\label{sec:flow:intro} 

Efficiently matching CPU and data is a long-standing problem in \dword{hep} computing. DUNE proposes to use a relatively non-hierarchical system that uses improved knowledge of the computing properties of applications (I/O rate, memory needs, data size) and the network connections between \dwords{rse} and \dwords{cpu} to optimally match processing and data.
The \dword{sam}/\dword{jobsub} system described previously and in the first section \ref{sec:current} below, has served Fermilab experiments well for 2 decades but the complex international nature of \dword{dune}'s computing resources motivates the development of a new Workflow system  that uses more detailed information to define and optimize jobs running across  \dword{dune} global computing resources.  
This chapter describes the existing generic Fermilab production system which is being used  to process data from \dword{protodune}, and %
lays out the requirements for the future system and the prototype of it that has been produced. %

\section{Existing Production Submission Infrastructure \hideme{Herner - draft}}
\label{sec:current}

\subsection{Production Operations Management System}
\label{subsec:jobsub}

\begin{figure*}[htb]
\centering
\includegraphics[width=.9\textwidth]{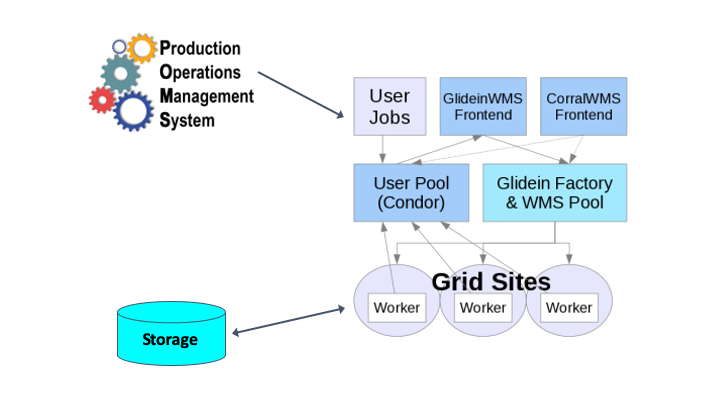}
\caption{Overview of the current DUNE Production workflow setup used also by the \dword{protodune} detectors for data reconstruction and simulation. Production group members interact with \dword{poms} to submit jobs, which uses the \dword{jobsub} tool to submit jobs to a HTCondor scheduler. \dword{glideinwms} provisions worker node resources and jobs match to the available worker node slots. DUNE jobs interact with storage elements  at \dword{fnal} and other sites both for input copy (or streaming for most production workflows) and output copyback.}
\label{fig:workflowPOMS}       %
\end{figure*}

Production and large-scale analysis jobs currently use \dword{fnal}'s \dword{poms}~\cite{Mengel:2020wev} for submission. \dword{poms} is used by a large number of experiments and provides both GUI and command-line options for job launches (both immediate and scheduled), recovery project setup, and  integrated monitoring with Fermilab's Landscape project. Job submission is typically performed via \dword{fnal}'s \dword{jobsub} tool~\cite{box2014fife}. \dword{jobsub} in
turn interfaces with the \dword{glideinwms} workflow management system~\cite{sfiligoi2009pilot} for resource provisioning and matchmaking to slots at \dword{fnal}, on dedicated DUNE resources at other sites, or opportunistic cycles on the \dword{osg}   and \dword{wlcg}. Figure~\ref{fig:workflowPOMS} illustrates the entire chain, including interaction with storage elements within the job.
The architecture can also provision resources on \dword{hpc} resources, such as Cori at the \dword{nersc}, within the HEPCloud~\cite{mhashilkar2019hepcloud} infrastructure. The submission mechanism is unchanged whether the jobs are \dword{htc} or \dword{hpc}; this seamless transition is key to efficiently utilizing available resources and also saves the job submitter significant effort by not requiring customized submission infrastructure for different resource types.

Choosing a Glidein-based system at this stage of the experiment had several advantages. DUNE was able to quickly leverage the existing FIFE~\cite{herner2019advances} toolset, including \dword{poms} and \dword{jobsub}, negating the
need for significant effort from the experiment in getting jobs running quickly. Since the system is in use by other neutrino experiments at \dword{fnal}, it is easy for new DUNE collaboration members coming from these experiments to 
begin submitting jobs quickly as they are working with a familiar system. The DUNE production workflows were also able to leverage the existing infrastructure support teams in place to serve other collaborations and consortia such as 
CMS and \dword{osg}. Finally, as \dword{glideinwms} is widely used in HEP, setting up new sites becomes extremely straightforward, especially if the site is already supporting another experiment that uses \dword{glideinwms}. Our integration times for new 
DUNE sites are typically less than one week and successful production jobs immediately after opening up the site are now the rule rather than the exception. This ease of setup has been a key enabler of DUNE's international expansion. International sites routinely deliver more than 50\% of \dword{dune}'s \dword{cpu} resources as illustrated by  the total DUNE Production wall hours for August 2021  shown in Figure~\ref{fig:country}. International sites regularly run the full suite of DUNE jobs, including \dword{protodune} data reconstruction and user analysis. %

\begin{dunefigure}
[Wall time distribution of Production jobs FY22]
{fig:country}
{Distribution of wall time for DUNE production jobs, October 2021 to March 2022. Inner ring: country. Outer ring: site. Over 50\% of wall hours have come from outside the United States in Fiscal Year 2022.}
{\includegraphics[height=3.5in,width=0.5\textwidth]{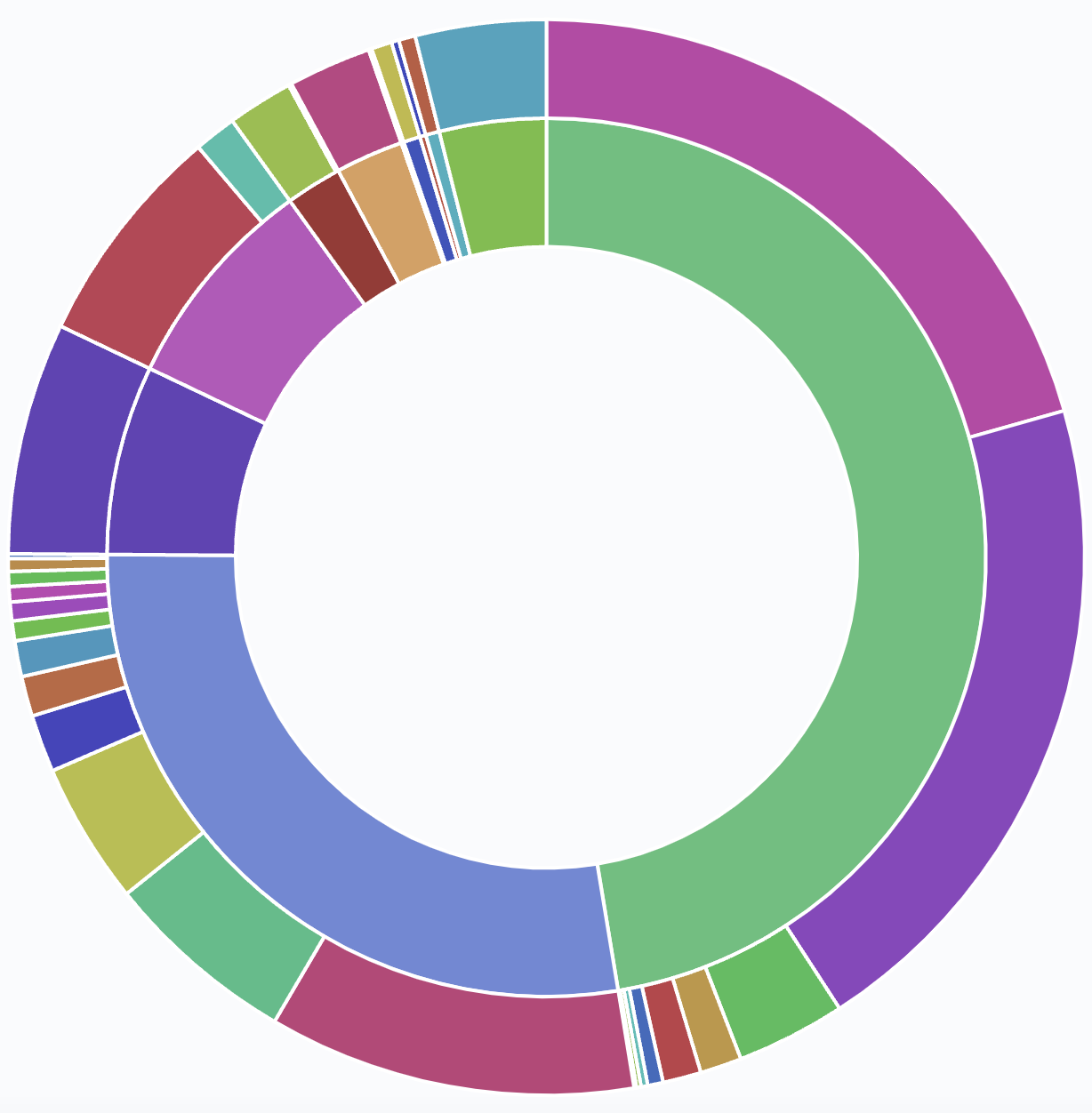}
\includegraphics[height=3.5in,width=0.12\textwidth]{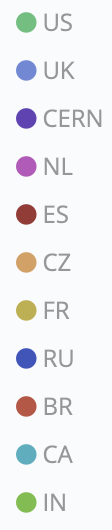}
\includegraphics[height=3.5in,width=0.15\textwidth]{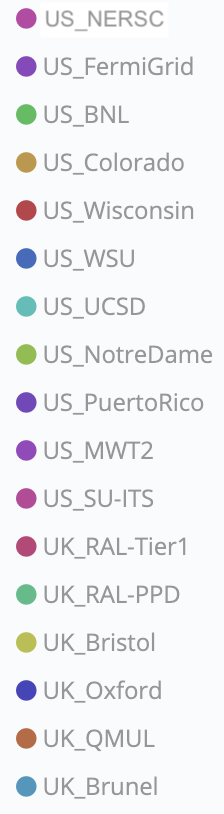}
\includegraphics[height=3.5in,width=0.15\textwidth]{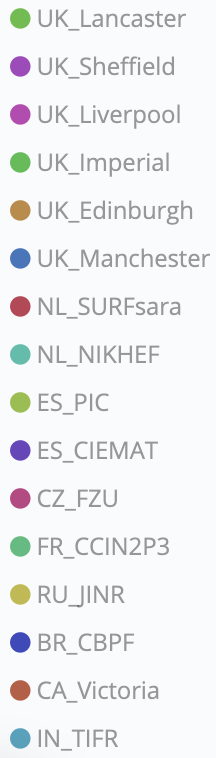}}
\end{dunefigure}

\subsection{%
Software, Input Data Distribution, and Output File Handling}
\label{subsec:io}
DUNE builds its software suite for  Scientific Linux. Since October 2019 DUNE jobs %
have been automatically run inside a Singularity\cite{osti_1627824} container at supported sites via a \dword{glideinwms} mechanism that requires no user knowledge of Singularity other than specifying the desired image. This %
reduces the possibility of errors and guarantees a homogeneous environment across all sites.

As described in Chapter \ref{ch:datamgmt},   \dword{rucio}~\cite{Barisits:2019fyl} now provides file location information to processes. Production jobs typically stream input data files via \dword{xrootd}, though in some cases they will copy a file to the worker node and directly read the local copy. The data source can be any storage system reachable from the worker
node and to which DUNE has access. This is frequently \dword{fnal} \dword{dcache}, but \dword{eos} at the \dword{cern} and other storage elements in Europe are also used (jobs run at \dword{cern} would get their inputs from \dword{cern} \dword{eos}, for example, while those run in the UK often use replicas sited in the UK). 

We have recently implemented minimal match optimization between \dwords{rse} and compute sites in \dword{sam} as many data samples used for user analysis have now been duplicated at European \dwords{rse}.  \dword{sam} now directs data from European sites to jobs running in Europe when possible.  This is done via a reasonably simple look-up table that provides a prioritized list of data locations for each compute site.  Even this simple system substantially raised the efficiency for high I/O applications at many sites. Section \ref{ch:model:perf} describes studies of RSE-to-CPU data rates that were used to implement the priority tables. 
The new workflow systems described below will support much more sophisticated matching algorithms.

Several DUNE workflows require one or more auxiliary input files %
such as calibration files or neutrino flux information, and other inputs necessary for \dword{mc} generation. Some simulation workflows randomly
choose several such input files for each job from a much larger set, so the file overlap between jobs is small. Additionally the files are typically tens of MB in size. These two attributes make these files poor candidates for placement in a standard \dword{cvmfs} repository. For such files we store them in a \dword{stashcache}~\cite{bib:stashcache} repository,
accessible in a POSIX-like fashion though a \dword{cvmfs} overlay. With this method there is still some level of shared caching on a worker node, but as files need to be copied in from the source (\dword{fnal} \dword{dcache}), it happens in a transparent way, meaning the user can simply access the files via a \dword{cvmfs} path in the {\tt dune.osgstorage.org} repository.

For output file handling nearly all workflows currently copy their outputs to \dword{fnal} \dword{dcache}, with a small minority (\dword{pddp} jobs) also copying to a storage element at IN2P3 in France. We use \dword{rucio} to manage file replication
to other sites. In the future DUNE will likely move to a more distributed copyback model: at sites with local storage, a job can simply copy its output to a local location, and then we can use \dword{rucio} for replication as is already done, rather than requiring everything first go through \dword{fnal}.

There is an exception for workflows run at NERSC; we read input auxiliary files from and copy output files back
to Cori's global scratch filesystem. For job outputs, a separate process performs bulk transfer back to \dword{fnal} from one of \dword{nersc}'s dedicated data transfer nodes. We expect that workflows on other future \dword{hpc} platforms
will follow a similar approach, especially at places without external connectivity on the worker nodes.

\section{Workflow System Requirements for Replacing SAM/JobSub Functionality}

As well as the data management functionality of \dword{sam} described earlier, it will be necessary to provide a replacement for \dword{sam}'s role in directing files to jobs and keeping track of what work has been done. We propose to take this opportunity to extend the \dword{sam} model further, by allowing the replacement Workflow system to determine which requests for work should be processed at a particular place and time, as well as determining which file to process next.

Discussions with partner projects reveal a variety of constraints imposed by the circumstances of different sites. Some sites have 
abundant network capacity and can readily support jobs that stream input data from elsewhere. Others have relatively limited networking
compared to the volume of computing power they make available, and
expect experiments to only send jobs that are either minimally I/O intensive or will process %
onsite data. A third class of sites has access to metropolitan or 
regional networks that can accommodate streaming from nearby sites only.

This landscape has led the computing consortium to look at  models where 
systems in which jobs can be accurately matched to sites where their input data is present, to deal with the most constrained class of sites. %
However, the system must also allow for a larger fraction of data access over the network for jobs running at %
less constrained sites. %

As noted above, \dword{sam} already has some region level matching capabilities, which have proven to be very useful, but significant effort will be needed to fully implement an adaptable application-aware matching system with site-level granularity.

To accommodate %
these constraints, we have designed and prototyped a generalization of the \dword{sam}/JobSub model with site-aware matching and an additional high level request interface.  This Workflow System gathers and tracks requests.  Using our Sites and Services model described in Section~\ref{sec:cm:sites_and_services}, generic jobs arrive at grid worker nodes on a Computing Element at a site and ask a central Workflow Allocator service what they are to work on and which file to process within that activity. This matching is based on the physical characteristics of the worker node job slot (such as memory, processors, and lifetime), the application characteristics and what data are unprocessed and suitable to be accessed from local or remote Storage Elements.

\section{Request Lifecycle}
\label{sec:flow:lifecycle}

The central concept of the proposed workflow system is a request %
that describes how some data processing activity is to be carried out. Requests are submitted by users (which may include members of a central production team) to the Workflow Database described below, where it progresses through %
several states, %
for example, draft $\rightarrow$ submitted $\rightarrow$  approved $\rightarrow$  running $\rightarrow$  paused $\rightarrow$  running $\rightarrow$  checking $\rightarrow$  completed $\rightarrow$  archived. Human intervention is needed for some transitions, e.g., from submitted to approved. Requests also have types and priorities. For example, simulation or user analysis, and high or low, respectively.

As part of its definition, a request may include one or more stages, each of which can apply a sequence of processing steps to the input or output files. %
Each stage specifies a bootstrap script used by generic jobs to run the relevant applications. The script specifies the requirements on the worker nodes %
(for example memory) and the maximum number of input files to be issued to %
the job executing that stage.

The %
request definition includes a MetaCat MQL query %
to generate a list of files to be processed in the first stage. This list of files is cached in the central Workflow Database, associated with the first stage of that request. All these files are set to the unallocated state. 

The request definition is an input to the Data Management placement agent, which transfers replicas of files to suitable sites as necessary. %
The location of the replicas of the files is also included in the database, cached from \dword{rucio}. 

Once the various agents have finished building the request, it can move to the running state and the bootstrap script associated with the first stage will begin execution. %

\begin{dunefigure}
[Workflow and data management architecture diagram]
{fig:workflow} 
{Workflow and data management architecture diagram.}
\includegraphics[width=0.95\textwidth]{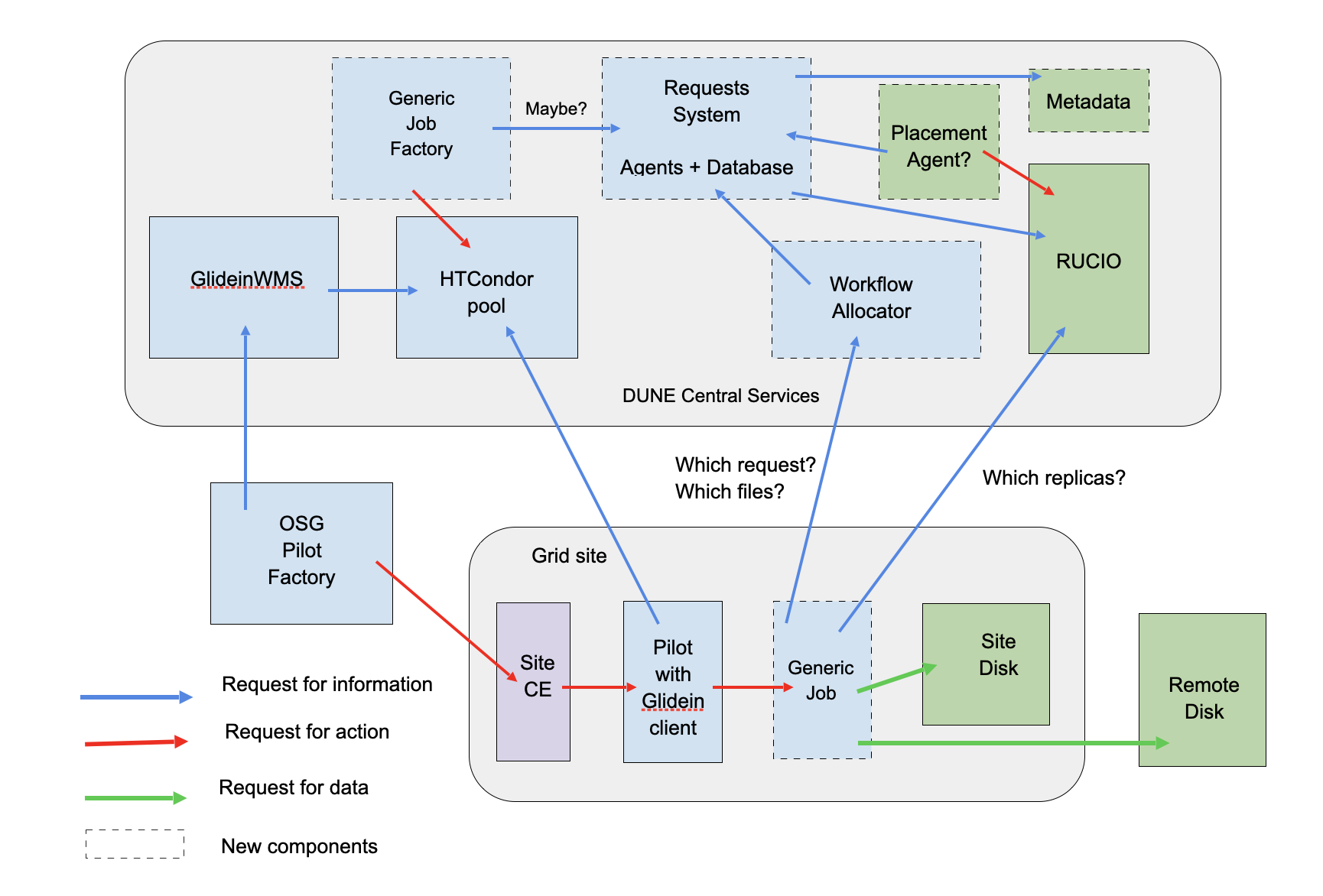}
\end{dunefigure}

\section{Grid Workload Systems}
\label{sec:flow:grid}

The workflow system makes use of the existing global grid infrastructures to deliver jobs for execution on worker nodes at sites. This allows DUNE's jobs to operate alongside those from other experiments in the \dword{wlcg} without placing additional requirements on sites.

\dword{fnal} operates a global HTCondor pool for DUNE which makes use of the existing \dword{osg} Pilot Factory service and HEPCloud to provision execution slots at sites. By using these existing systems we are able to use computing capacity presented with ARC CE or HTCondor CE grid interfaces, or on cloud and \dword{htc} services supported by HEPCloud.

\section{Generic Job Factory}
\label{sec:flow:factory}

The Generic Job Factory agent creates and submits HTCondor jobs, which are each assigned to a specific execution site. It uses a mixture of matching successes, site limits from DUNE \dword{cric}, and prioritization of sites to determine how many generic jobs to submit and have waiting at each site. It may also use inspection of the central workflow database to estimate whether more generic jobs should be submitted for a particular site. For ordinary jobs, the job factory must be able to prioritize use of pledged academic and lab sites over commercial cloud sites and specialized sites like \dword{nersc}, which are managed by HEPCloud, to allow those services to be used for specialized workflows. Inspection of files waiting to be allocated in the database may be more appropriate for HEPCloud managed sites, so that work requiring their features may be satisfied.

Once a generic job lands on a worker node, it contacts the Workflow Allocator described later which determines which unallocated files from one stage best match that worker node.

\section{Workflow Database}
\label{sec:flow:wfdb}

The central Workflow Database stores definitions of requests and their stages, and cached information about files, replicas, sites, and storage services. Together, all of this information is used by queries to determine what work to carry out where. It is implemented as an SQL relational database.

\section{Information Collector}
\label{sec:flow:collector}

The Information Collector agent runs periodically to obtain a list of eligible sites and storage services from \dword{cric} and other information sources, including any downtime notifications. This information includes a table recording whether data access from a particular site to a particular storage is classed as being at the same site, a ``nearby'' site, or is merely elsewhere on the grid but not blocked by firewalls or policies (for example at an \dword{hpc} site with no off site access.) The concept of ``nearby'' is defined as being accessible with an acceptable level of inefficiency.

\section{Finder}
\label{sec:flow:finder}

The Finder agent uses the input dataset definition in a request to construct a list of input files for the first stage of that request. Typically this involves making queries to MetaCat to obtain a list of files in the given dataset. This list is cached within the Workflow Database so it can be used as part of further SQL queries. As the files are identified, the location of their replicas are also obtained and cached from Rucio. Again, this allows the proximity of replicas of unprocessed files to be used in deciding what work a job should do.

\section{Archiver}
\label{sec:flow:archiver}

The Archiver agent has responsibility for removing information about requests, stage, files, and replicas once they are no longer needed, archiving whatever is needed for future reference to long term storage. It is intended that important information about how and where files were processed will be stored in the MetaCat database by the generic jobs themselves, but some higher level information about the operation of the workflow system and the processing of requests will be saved by the Archiver.

\section{Workflow Allocator}
\label{sec:flow:allocator}

Once a generic job arrives at a worker node, it contacts the Workflow Allocator service which determines which unallocated file from one stage best matches that worker node. This matching includes the characteristics of the worker node job slot (memory, time limit, etc.), and whether the site is eligible to access a replica of that data file. The matching takes into account that some stage definitions allow access to remote input files anywhere on the grid, and others require files to be at a ``nearby'' site. Replicas are prioritized based on whether the worker node and replica are at the same site, ``nearby'', or elsewhere but still eligible. 

  Details of the request and stage and the bootstrap script to be run are then provided to the generic job. The script can use these details to request a series of files to process with the applications the script invokes. Each input file successfully processed by an application is reported to the Workflow Allocator so that the input file’s status can be updated from Allocated to Processed. Unprocessed input files are returned to the unallocated state for processing in another job. 

If the stage is not the final stage for that request, each output data file is also inserted into the list of files associated with the next stage for that request, in the unallocated state. 

\section{User Commands}
\label{sec:flow:commands}

Command line tools are provided to operators and users to allow queries of Workflow Database contents and the creation of requests. These tools are envisaged to be of most use during testing of new workflows and for short workflows during analysis. 

\section{Workflow Dashboard}
\label{sec:flow:dashboard}

The same functionality as the command line tools outlined above is provided by a Workflow Dashboard web interface. This allows more sophisticated searches in the Workflow Database to monitor the progress of running requests, and allows members of the operations team to examine the state of the system at the level of individual files and replicas which are due to be processed, or have recently been processed to enable debugging of problems with sites or workflow definitions.

The Workflow Dashboard is also intended to be used for all large scale productions, and allows submitters to draft request definitions, circulate their proposal with colleagues for checking, before submitting them to any formal approval and checking procedure. The lifecycle of a workflow request accommodates manual approval and prioritization of large requests while smaller requests can be approved automatically according to preset limits. A library of bootstrap script templates and requests will be provided as part of the dashboard to help users make sensible choices for common types of workflow.

\section{Workflow System Prototype}
\label{sec:flow:prototype}

A prototype implementation has been produced, which includes the Workflow Database, Workflow Allocator service, command line tool, generic job factory, finder and information collector, and the monitoring aspects of the Workflow Dashboard. 

Using this prototype, we have created requests based on existing productions, populated them with lists of files and replicas held by Rucio, and then processed the files within generic jobs executing the bootstrap script included in the stage definitions, at sites, using matching performed by the Workflow Allocator, and registered the resulting files in MetaCat and Rucio.

\section{Implementation Plan \hideme{McNab/Timm/Herner needed}}
\label{sec:flow:implementation}

During 2022, we plan to continue improve the components of the prototype until it becomes a viable system which can be used for production processing of \dword{protodune} II data. This will allow further comparisons between the functionality requirements of the production team and users, and the Workflow System design.

\cleardoublepage

\part{Integration and Evolution}

\chapter{Services Overview \hideme{revisions for SK/Anne 5/8}}
\label{ch:serv}

\section{Introduction}
DUNE computing is dependent on a number of services that are not operated by DUNE itself.
Some of these are provided by the host laboratories (CERN and FNAL), such as experiment LAN's and primary raw data storage,  others are provided by the various remote sites where
computing and data storage are done, and still others are operated by commercial companies and hosted in the cloud.

\subsection{Computer Security}
Computer security has multiple layers, as we work with multiples sites, each with their own security requirements.
DUNE has a named computing security officer who is responsible as the contact person for reporting potential bad actors or security problems based on DUNE \dword{vo} distributed computing activity. DUNE will maintain traceability for each Distinguished Name (DN) in order to respond to security issues and take appropriate action. The security office is also responsible for ensuring custom DUNE software or services follows good security practices. For common shared tools, patches and security updates are propagated from central maintainers out to WLCG/OSG sites. At larger sites, DUNE relies upon the local admins to monitor, report, and address local security issues that are independent of common services or tools. Overall, this model is very comparable to the models used by other experiments operating on the WLCG/OSG. 

\section{Host Lab Provided Services}
DUNE relies on the host laboratories, \dword{fnal} and \dword{cern}, to provide a wide variety of central services.  These include web sites (for example, EDMS at CERN and docdb at Fermilab for document management),  databases and oversight of cyber security. Compute and storage services are more widely distributed across the collaboration, see \ref{model:global} for more detail.  The networking services are described in \ref{ch:netw}.  We summarize the other services briefly.

\subsection{Web Services}
The conference-scheduling service (currently Indico), SharePoint, the DUNE Document Database, the DUNE Wiki, and the main {\tt dunescience.org} web page, are all hosted at \dword{fnal}.  
\dword{fnal} also hosts authentication and authorization facilities such as \dword{voms}, and manages the business relationship with CILogon.org to provide X.509 certificates and WLCG \dword{jwt} tokens for batch authentication. %
\dword{fnal} and \dword{cern} both  provide  electronic logbook services and 
host the web sites for a number of monitoring services. 

\subsection{Database Services}
\dword{fnal} maintains the collaboration database, which tracks the membership of the DUNE collaboration, and runs the \dword{ferry} 
database, which records compute permissions for DUNE collaborators.  \dword{fnal} hosts the underlying databases for the data management
services \dword{rucio} and \dword{metacat}, as well as the legacy \dword{sam} workflow management. %

\subsection{Compute Support Services}
In addition to the storage and computer services provided by collaborating institutions through their national resources, central coordination and distributions support is needed. 
\dword{fnal} provides the \dword{jobsub} service for batch job submission, the \dword{poms} workflow service to submit campaigns, and the \dword{glideinwms}
and \dword{hepcloud} services to access remote sites including high performance computing and commercial clouds. It provides system administration and hardware maintenance of the various interactive and batch clusters (FermiGrid).  It also provides
continuous integration facilities (Jenkins), and build service machines.

\dword{fnal} currently provides the main instance for several distribution services, namely \dword{cvmfs} and its associated Rapid Code Distribution Facility for user code.  Streaming of large auxiliary data files is accomplished using the Open Science Data Federation (until recently known as 
\dword{stashcache}), which also uses the \dword{cvmfs} user interface.

\dword{fnal} also maintains
the monitoring and log retrieval services for the batch and storage systems, collectively known as \dword{fifemon}. 

\dword{fnal} is assisting DUNE and other experiments to transition to the \dword{spack}\footnote{Spack\copyright, \href{https://spack.io/}{https://spack.io/}} code packaging system.  Significant effort is also provided on managing the \dword{art} framework and \dword{larsoft} project.

\subsection{Storage Services}
\dword{fnal} and \dword{cern} provide the primary archival tape storage for raw data, currently delivered through the \dword{enstore} and \dword{cta}  tape library systems. A tape copy of reconstructed and simulated data is also currently maintained at \dword{fnal} and in future, at several collaborating sites in Europe. At \dword{fnal} the \dword{dcache} 
disk caching system is the front end to this system, and also provides some standalone disk.  \dword{fnal}, \dword{bnl} and \dword{cern} also provide a number of data management and data movement services, including CERN-FTS, Fermi-File Transfer Service, \dword{sam}, \dword{rucio}, \dword{metacat}, and the Data Dispatcher.

\section{Collaboration Contributed Services}
DUNE receives compute and storage resources from a number of sites around the world, not all of which are formally DUNE collaborators,  as detailed in the section on the Computing Contributions Board 
Section~\ref{sec:ccb}.

\dword{cern} provided services include: the CERN \dword{indico},  the \dword{edms} document
management system, the MONIT monitoring system,  \dword{cric} - which is the master list of all DUNE compute and storage resources as well as the means to track allocations,  and the \dword{etf} testing service which routinely tests all remote compute sites on DUNE's behalf. %

While \dword{cern} and \dword{fnal} remain the major providers of support services, some monitoring and control activities are moving to collaborating institutions, notably the \dword{rucio} monitoring, now hosted by Edinburgh and the new workload system at Manchester.  We anticipate a move to an even more distributed system as the experiment evolves. 

\section{Cloud-Hosted Services}
DUNE uses the GitHub service for its code management, the Slack and Microsoft Teams services for interactive communication, the Overleaf service for editing \LaTeX{} documents, and the Zoom teleconferencing application.  The cloud-hosted \dword{servicenow} application is used for communication between DUNE liaisons and the two
\dword{fnal} Computing Divisions about outages and changes to service and also for internal DUNE use in 
tracking of issues and workflow requests such as production requests and data movement requests.

\cleardoublepage

\chapter{Information Systems and Monitoring \hideme{5/4 HMS included comments from Anne, Doug and Steve K. }}
\label{ch:mon}

\FPset{MonEtfOpsPeople}{3.1} %
\FPset{MonEtfDevPeople}{1.0} %
\FPadd\MonEtfTotalPeople\MonEtfOpsPeople\MonEtfDevPeople %

This section %
describes the monitoring of services and resources on the grid.  The actual monitoring of individual jobs that are submitted by DUNE 
is provided as part of the Workflow System (see Chapter \ref{ch:wkflow}).

As the global infrastructure is broadly similar to that used by the current \dword{lhc} experiments, DUNE plans to reuse the relevant monitoring tools (\dword{etf}, \dword{perfsonar}) developed for this purpose by the \dword{wlcg} which we cover in this section.  %

Going forward, DUNE plans to be agile, keep up with upcoming developments in this field, and develop new tools as needed. For example, the widespread deployment of \dword{ipv6}~\cite{bib:ipv6TaskForce} networking may require re-evaluation of some of our computing tools.  The \dword{etf} and \dword{perfsonar} tools are already \dword{ipv6} compatible and can be deployed to test resources offering \dword{ipv6} connectivity in either pure or dual-stack mode.

\section{Tools}
\label{sec:mon:xyz}  %
\subsection{CRIC} %

DUNE intends to use \dword{cric} as a central source of information about sites and their compute and storage services. \dword{cric} is routinely used by ATLAS and CMS with \dword{osg} and other \dword{wlcg} resources, and is also familiar to site administrators. 

An evaluation of \dword{cric} is being done, with information about sites obtained from the configurations of the \dword{osg} pilot factories. This is browsable via the \dword{cric} web dashboard, and also used to generate XML files in the \dword{vo} Feed format required by the \dword{etf} testing framework used for monitoring.

\subsection{Experiment Test Framework (ETF)}

DUNE jobs run at a number of computer data centers across the grid. Checking the status of these jobs is one way of keeping an eye on whether a given site or resource is functioning normally. Given that issues can and usually do have multiple sources, it is essential to have an independent resource monitoring system to monitor and %
catch issues quickly. %

\dword{etf}\cite{bib:ETFDoc, bib:ETFStatus} is a system of tools  hosted by CERN that regularly test all the available resources for different experiments, i.e., \dwords{vo}. It has been developed and run on behalf of the \dword{wlcg} Virtual Organizations ( %
 \dwords{vo}) for well over a decade. %
The \dword{etf} framework runs tests customized to the \dword{vo}'s application and systems to provide a true picture of the  available resources and their reliability. The \dword{etf} tests run hourly and feed into the \dword{monit} 
framework, which keeps history information and the most recent test logs to enable debugging and metric measurement.

The \dword{etf} framework currently involves 
\begin{enumerate}
\item high-level functional testing of about 90 hosts defined in the \dword{glideinwms} configuration;
\item dashboard (checkmk) to show results; and
\item plugins conforming to Nagios\footnote{Nagios\textregistered, \url{https://www.nagios.com/}} standard. 
\end{enumerate}

For DUNE, customized tests are in place to verify %
a minimal simulation functionality on the worker nodes and to check that a given worker node has \dword{ipv6} networking. A customization to perform a lightweight test of access to  \dword{rucio} servers from the worker node is in progress.
Beyond this, DUNE-specific development will be required from time to time %
to synchronize with updates to the framework both from the \dword{etf} / \dword{monit} and the DUNE software ends. %

\subsection{PerfSonar}

Data distribution requires both high-performance networking between the sites and sophisticated debugging tools to identify and solve  problems that arise. The \dword{wlcg} network monitoring and debugging infrastructure that DUNE will adopt and deploy is \dword{perfsonar},
 an open source toolkit collaboratively developed by %
groups in Europe and the Americas that keeps a clean separation between network-related and other metrics. %
It is installed on dedicated hardware (\dword{perfsonar} boxes) at %
each resource site. %
The data from these \dword{perfsonar} boxes %
can be aggregated by a variety of tools and algorithms to %
visualize the connectivity in different ways, as described in  in various (kibana, grafana, checkmk, ...) ways, enumerated below.
\begin{enumerate}
    \item \url{https://psetf.opensciencegrid.org/etf/check_mk/index.py}
    \item \url{https://toolkitinfo.opensciencegrid.org/}
    \item \url{https://monit-grafana-open.cern.ch/}
    \item \url{https://atlas-kibana.mwt2.org/s/networking/app/dashboards}
    \item \url{https://perfsonar.uc.ssl-hep.org/}
    \item \url{https://sand-ci.org}
\end{enumerate}

These tools are currently well supported within \dword{wlcg} with %
expertise shared widely and regularly among administrators.
Given %
that networking is a core need for \dword{wlcg} as it is for DUNE, it is anticipated that \dword{perfsonar} %
will continue to be supported %
as long as DUNE requires it.

\section{FIFEMON}\label{is:fifemon}

DUNE relies on the \dword{fifemon}\cite{bib:fifemon}  family of monitoring tools which have been developed at Fermilab over the past
decade or so.  These dashboards show the CPU usage and efficiency and storage usage and traffic of
all the experiments on a user by user basis.

The dashboards are built on open-source technology, and use common ingest tools such as Apache Kafka, Logstash,
and Prometheus to collect data, the Graphite and ElasticSearch data stores to store it, and the Grafana and Kibana 
display utilities to display it.  Grafana is used for static displays, while Kibana has an active search function in which custom graphs can be made using a query language.  

There is information on batch usage by user and aggregate of the whole experiment.  CPU efficiency and memory usage are plotted. In addition storage total usage as well as the volume of reads and writes, as well as the age of files, are visualized.  We also have custom visualizations to display data movement from data center to data center and

\chapter{Authentication and Authorization \hideme{HMS 5/4 comments from Steve K. and Anne }}
\label{ch:auth}

\section{Obtaining Access to DUNE Computing}
\label{sec:auth:access}
Computing access for collaborators is a multi-step process.  New collaborators are first proposed by institutions (with senior collaborators requiring approval by the institutional board).  They are then added to the collaboration database.
Once this is done, they may apply for access to computing resources at \dword{fnal}, and the DUNE secretariat verifies that they
are members of DUNE.  %
The central repository %
\dword{ferry} %
is used to populate the user list for the 
\dword{fnal} interactive login machines and to determine if a user can obtain DUNE Grid X.509 credentials.
 When a collaborator's \dword{fnal} ID expires, or they leave the DUNE collaboration, their access to computing is automatically removed.  
 
 While the \dword{protodune} experiments \dword{np02} and \dword{np04} are running at the \dword{cern}, %
 collaborators %
 working on those activities may  apply %
 for interactive computer access at \dword{cern}, as well.

\section{Current State of Authentication and Authorization}
Interactive login to \dword{fnal} is strongly authenticated via Kerberos 5 credentials.  %
DUNE collaborators must be currently registered to have accounts at \dword{fnal}.

Authentication for batch submission and access to storage is currently done via X.509 certificates and proxies
on the %
\dword{osg} in the U.S. and \dword{wlcg} %
Computing Grid (\dword{wlcg}) elsewhere in the world.  %
Automated processes %
at \dword{fnal} %
dynamically generate 
X.509 certificates on behalf of the user at the time of batch submission, from the CILogon-Silver 
certificate authority\footnote{\url{https://www.cilogon.org/news/cilogonsilverenablesfederatedaccesstoopensciencegrid}}. This authentication is used to submit batch jobs both to the local
\dword{fnal} clusters known as FermiGrid and to the rest of the distributed computing sites.  The proxies are
also used to access files %
stored on the global storage elements around the world.  
X.509 certificates are issued by members of the International Global Trust Federation (IGTF).  They are then verified 
by the DUNE \dword{voms} server.  Certificates issued by other IGTF authorities can be associated with a user by request.

Access to DUNE web sites hosted at \dword{fnal} is controlled by the \dword{fnal} Single Sign-on (SSO) facility.  Some sites (e.g., the DUNE wiki) %
require a \dword{fnal} SSO.
 Some sites, e.g., %
document database (DocDB), require either a \dword{fnal} or \dword{cern} SSO or  %
an X.509 certificate. %
Both of these applications also currently allow read access via a group username and password.

\section{Planned Changes to Authentication Currently Under Way}

The \dword{osg} and \dword{wlcg} are both in the process of changing over to \dwords{jwt} %
for authentication of batch jobs and storage. This %
authentication protocol %
has significant adoption in industry. A common schema has been agreed on between the various issuing organizations. %
\dword{fnal} has set up a %
service to issue these tokens on behalf of DUNE and early tests have %
successfully demonstrated access to storage elements and batch job submission with these tokens. %
The token issuer is populated with information from the \dword{ferry} server. %
The tokens include a unique user identifier %
and a list of capabilities that the user is allowed to have, including which 
areas in the storage element the user is allowed to write.  The basic token infrastructure is available now but we expect a phased transition beginning in 2022, during which %
the X.509-based authentication infrastructure will continue to be available for some number of years.

Interactive web sites will continue to use %
\dword{fnal} SSO 
for the foreseeable future, while gradually allowing more Identity Providers. 
It is likely that \dword{jwt} tokens will be used to access various non-interactive web
services besides the compute and storage.

\section{Requirements for Authentication and Authorization}

DUNE is an international collaboration and it is important that all collaborating scientists be able access to the data, codes and documents that result from our common efforts.  To this end DUNE Computing Management has %
submitted to \dword{fnal} the following requirements:

\begin{itemize}\item  DUNE collaborators without a \dword{fnal} SSO must be able to see and edit internal DUNE web pages and submit to the collaboration document server.  
\item  Computing staff who work at collaborating institutions but who are not themselves DUNE collaborators must be allowed access to the computing documentation.
\end{itemize}

 There are a number of technical solutions for federated web identity, including the Edugain federation of identity providers.  For now, we are working to extend access to collaborators with \dword{cern} credentials.
\dword{fnal} is responding, and as of this writing, users with \dword{cern} credentials can now use those credentials to access the DUNE DocDB. \dword{fnal} is in the process of federating its SharePoint instances with \dword{cern}, and the DUNE wiki and Indico sites are next in line.

\cleardoublepage

\chapter{Code Management \hideme{comments from Anne /BV included 5/8}}
\label{ch:codemgmt}

\section{Liquid Argon TPC Code Management \hideme{Junk/Calcutt - needs update}}
\label{sec:codemgmt:dunetpc}  %

Software for simulation, data read-in, and reconstruction %
has a great deal of commonality across the %
liquid-argon time projection chamber detectors 
 planned to be deployed by the DUNE collaboration.  These  
  include the %
  horizontal- and vertical-drift far detector modules, the 35-ton prototype, ProtoDUNE-SP, ProtoDUNE-DP, ProtoDUNE-2-HD, ProtoDUNE-VD, ICEBERG, and the pixel \dword{nd} and its prototypes.
The same software stack used by DUNE and the protoDUNE detectors is used for  wire and pad readout, the photon detectors, and the cosmic-ray taggers used in some of the prototypes.  

Despite the commonalities, the interfaces to event generators, \dword{geant4}, \dword{art} framework and the data handling systems require effort to build and maintain, and data products need to be thought out carefully.  To reduce the development burden and expand the pool of expertise, DUNE has chosen to use the \dword{larsoft} toolkit, which is built on \dword{nutools} and \dword{art}.  

\dword{larsoft} is supported by \dword{fnal}'s Scientific Computing Division (SCD) and is shared among several other experiments, such as ArgoNeuT, MicroBooNE, ICARUS and SBND. The pool of potential developers is quite large,  with a variety
 of software experience and length of time commitment to maintenance.
Substantial effort goes into maintaining coherence across the project with new releases of \dword{larsoft}   made weekly.  DUNE's \dword{larsoft}-based code stack follows a similar release schedule.

DUNE's \dword{larsoft}-based software historically had been collected in a single git repository called dunetpc, starting in the early days of the collaboration.  This repository was hosted using \dword{fnal}'s Redmine service.  Compiling all of the code in this single repository became slower as more source files were added.  In January 2022, a split of the dunetpc repository into ten smaller repositories was deployed on GitHub, taking advantage of GitHub's superior performance, feature set, and openness compared with Redmine.  The old dunetpc repository is kept in a read-only state in Redmine for inspection purposes.  The new top-level product is called dunesw.  A \dword{ups} dependency graph made in January 2022 is shown in Figure~\ref{fig:dunetpcdeptree}, showing the dunesw stack, the \dword{larsoft} \dword{ups} products, and their dependencies.    The January 2022 split of the DUNE components of the stack helps speed build times for users who only are interested in modifying a small portion of DUNE software at a time.   The division of code into several repositories also helps define the boundaries of maintenance responsibility.

\begin{dunefigure}
[Dependency graph for the dunesw software stack, Jan 2022]
{fig:dunetpcdeptree}
{\dword{ups} dependency graph for the dunesw software stack, for version v09\_42\_00\_01, January 2022.  DUNE-specific products have ``dune'' in their names, and LArSoft product names begin with ``lar''.  External products, such as \dword{root}, \dword{geant4} and python are also shown.  This diagram is an aid to users and managers who need to find code, as well as check out and build portions of it while setting up the remainder from pre-installed images.}
\includegraphics[width=\textwidth]{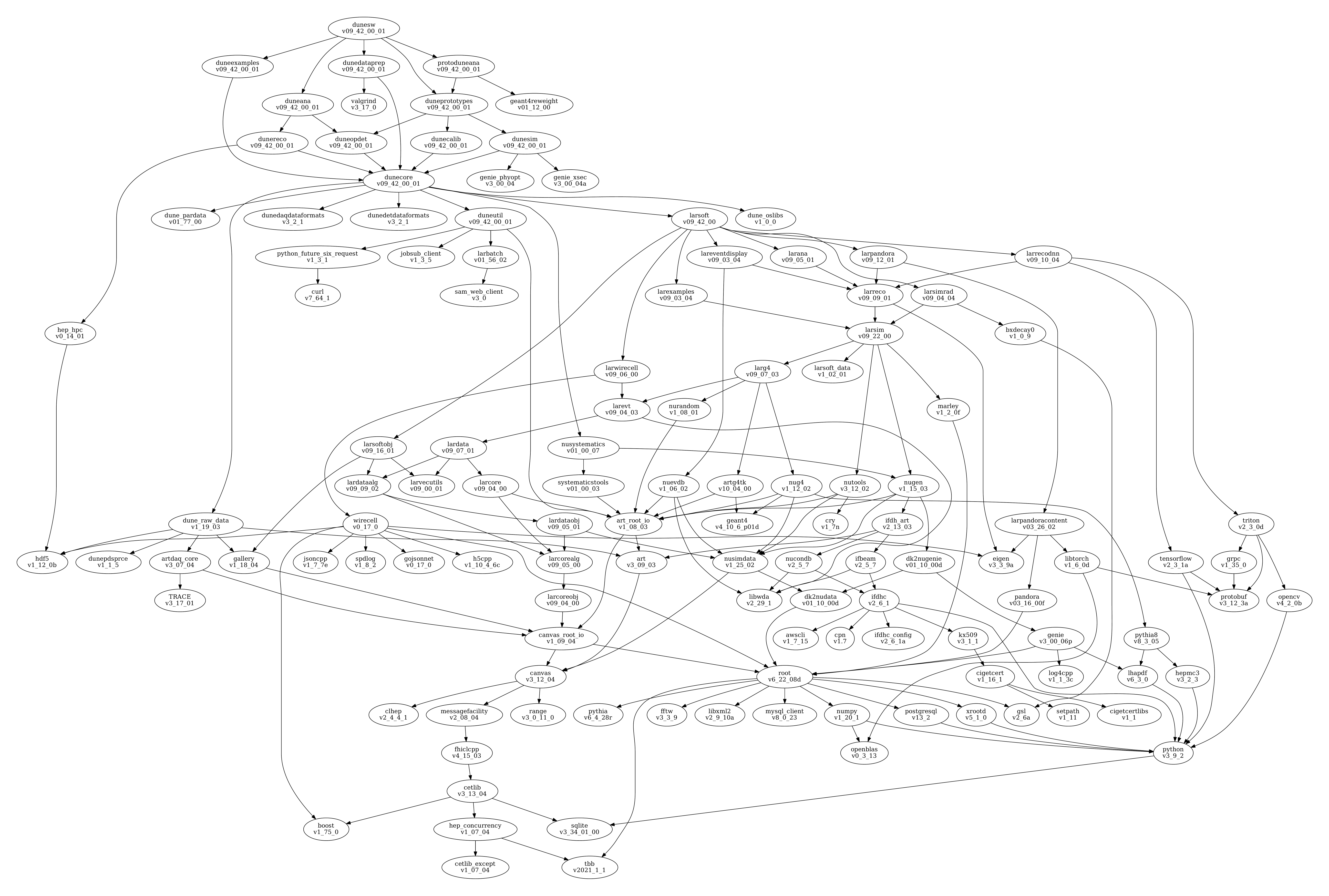}
\end{dunefigure}

Since the dunesw stack depends on \dword{larsoft} and sets it up via \dword{ups}, dunesw also has a weekly release schedule, and releases are tied to \dword{larsoft}'s releases and use the same version tag names. 
\dword{larsoft}'s pull-request system, based on the one used by the CMS experiment, performs automated checks that the proposed new code compiles and tests properly.  Once the tests are successful, a review of the proposed new code is required before it can be merged into the repository.  As of this writing, DUNE is considering moving to a pull-request model similar to that of \dword{larsoft}.  

Code maintenance is required, even for unmodified algorithms, when the external environment changes, such as a compiler or build-system upgrades.  DUNE software managers have maintained legacy software when this occurs, though improvements and additional features to be added to software require stakeholders to perform the work.  Code that is no longer needed by DUNE may be deleted in order to speed build times, reduce maintenance effort, and save space.  Deleted code is always present in previous installed releases and in the git repository in case it needs to be recovered or used as an example for writing new software.

The build system used is {\tt mrb}, which is provided by and supported by \dword{fnal}'s SCD.  Software is built in \dword{ups} products which are distributed to the collaboration in two ways.  The primary distribution mechanism is via a set of installed products in the \dword{cvmfs} which can be directly set up by end users of Scientific Linux 7 and CENTOS 8 without the need to install anything locally, except \dword{cvmfs} itself.  \dword{cvmfs} maintains a cache on the user's computer to store copies of the released software files for repeated local access.  \dword{cvmfs} is also used to distribute released software to batch worker nodes.  The second software distribution mechanism is via the {\tt scisoft.fnal.gov} web server.  When a release is made, the repositories are tagged and the release manager triggers \dword{fnal}'s Jenkins build servers to compile the software and install auxiliary files in \dword{ups} products.  The \dword{ups} products are then tarred up and the tarballs are uploaded to the scisoft web server, along with a manifest file that describes which tarballs need to be downloaded and installed to get a fully
functioning dunesw software stack.

In addition to code, some data files also are distributed via \dword{cvmfs}.  For example, photon lookup libraries are stored in several-hundred-megabyte \dword{root} files which are inconvenient to store in a git repository.  They are loaded into a special repository in \dword{cvmfs} and scisoft called {\tt dune\_pardata}.  Even larger files are stored in \dword{stashcache} which has a \dword{cvmfs} interface for the user, but which retrieves files out of a dedicated area in \dword{dcache}.

As operating systems have become more secure, some of the ways of distributing, setting up, and running HEP software, such as the use of the environment variables {\tt LD\_LIBRARY\_PATH} and \linebreak{\tt DYLD\_LIBRARY\_PATH}, have conflicted with security features of at least one operating system.  We are evaluating \dword{spack} as a replacement for {\tt mrb} and \dword{ups}.  \dword{spack} is a build system that also allows users to select from a list of installed versions, like \dword{ups} does. \dword{ups} is now approximately 25 years old, it is not an industry standard, and it is somewhat linked to \dword{fnal}.  \dword{spack} has a significant community using it outside of HEP, and thus there is a large volume of documentation available for it on the web.  The DAQ group is already using \dword{spack} for online systems.

\section{Near Detector Code Management \hideme{Muether/Cremonisi/Junk needs update}}
\label{sec:codemgmt:neardet}  %

When DUNE starts running with beam, the expected \dword{nd} configuration will consist of a \dword{lartpc} with pixel readout (\dword{ndlar}), a magnetized steel muon spectrometer (\dword{tms}), and an on-axis beam monitor, \dword{sand}.  A proposed future upgrade is replacement of the \dword{tms} with a subdetector that consists of a \dword{gartpc} with pixel readout, an \dword{ecal} and a muon system (\dword{ndgar}). The software structure reflects the detector configuration.  

The software %
contributions for each subdetector %
come from the groups that are designing and proposing to construct the subdetectors and their subsystems.  Below is a summary of the code management strategies used for each subdetector, %
and also for the shared tools. In general, official repositories are hosted in GitHub with executable binaries and the associated libraries installed in \dword{cvmfs} so that the software can easily be version controlled and run on the grid. These requirement must be met in order for software to be run at the production level and stored in group accessible areas. 

\subsection{ND-LAr Code Management}
\label{sec:codemgmt:ndlar}

The ND-LAr code, larnd-sim and ndlarcv, is hosted in GitHub. This code handles the standard reconstruction and detector response mock-up. The ML and Pandora based reconstruction code is currently under private development, but will be made available this spring prior to production integration.  

\subsection{TMS Code Management}
\label{sec:codemgmt:tms}

The TMS code, dune-tms, is hosted in GitHub. This code handles the reconstruction and detector response mock-up. 

\subsection{ND-GAr Code Management}
\label{sec:codemgmt:ndgar}

The software for simulating and reconstructing events in \dword{ndgar} is, at the time of writing, contained in two github repositories:  \dword{garsoft} and \dword{garana}.  \dword{garsoft} builds on the functionality of \dword{art} and NuTools, and it is maintained as the \dword{art} API changes and when the build system is upgraded.  Following \dword{larsoft}'s pattern, it is built with {\tt mrb} and set up with \dword{ups}.   %
\dword{garsoft} functionality is described in Sections~\ref{sec:usecases_ndgardetsim} and~\ref{sec:algo:reco:gartpc:pixels}.  \dword{garana} provides a facility for making an analysis ntuple from information stored in \dword{garsoft} data products.  Both repositories are hosted in GitHub.  Continuous integration has not yet been set up for \dword{garsoft} and \dword{garana}.  Executable binaries and the associated libraries are built using \dword{fnal}'s Jenkins build servers, and the build artifacts are installed in \dword{cvmfs} so that the software can easily be run on the grid.

\begin{dunefigure}
[Dependency graph for the \dword{garsoft} software stack. March 2022]
{fig:garsoftdeptree}
{Dependency graph for the \dword{garsoft} software stack, for version v02\_16\_00, current as of March 2022.}
\includegraphics[width=\textwidth]{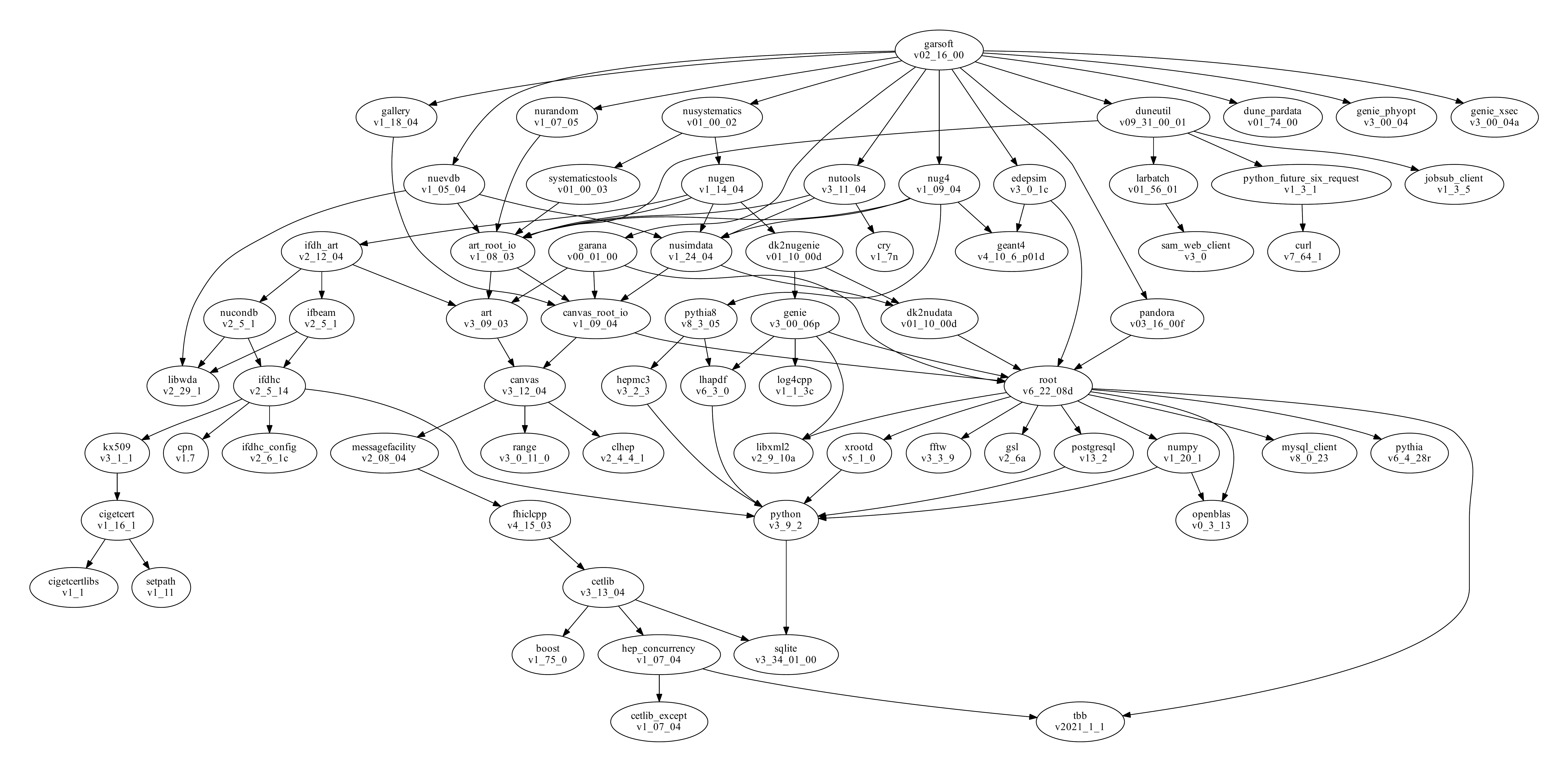}
\end{dunefigure}

\subsection{SAND Code Management}
\label{sec:codemgmt:sand}

The SAND code, ND-SAND-FastReco, is hosted in GitHub. This code handle the basic reconstruction and detector response mock-up. The full reconstruction and detector response code is currently under private development, but will be made available this spring prior to production integration.  

\subsection{Near Detector Common and Production Tools}
\label{sec:codemgmt:ndproduction}

A repository has been set up in GitHub to manage files that are needed for \dword{nd} production.  These include production scripts, detector geometry descriptions and flux specifications.  It is versioned and installed in \dword{cvmfs} like other DUNE repositories, so that production workflows can be reproduced at later dates. Additionally, common ND geometry code is stored in GitHub, dunendggd.

\section{Continuous Integration \hideme{Junk - draft}}
\label{sec:codemgmt:ci}

The stability of the functionality and performance of a large base of shared software requires attention to each change that is committed.  Software releases require validation and approval before being used in analyses intended for publication.  The rapid pace of software development early in the life cycle of the experiment, as well as increased activity as data are first collected and conferences come up, requires constant vigilance to ensure that bugs are not introduced and that software remains backwards-compatible.

To meet this requirement, an automated continuous integration (\dword{ci}) system is currently in operation for the \dword{larsoft}-based code base.  Similar systems, even using the same infrastructure, will be deployed for the \dword{nd} and beam simulations system.  The \dword{ci} system consists of a set of servers that monitor  commits pushed to the central code repositories.  On each commit,   a suitable delay of approximately 15 minutes allows aggregation of commits. %
The \dword{ci} system can also be triggered by user interaction, via an authenticated request.  Once triggered, the \dword{ci} system  compiles the changed code as well as any dependencies that are required.  Currently, the \dword{ci} system builds \dword{larsoft} and all experiment code from the head of the develop branch in the repositories. 

The status of the build is stored in a logfile and summarized on a web page.  If a commit causes the build to fail, software managers and the person who committed and/or pushed the commit are notified and the commit is blocked from being merged into the head of develop.  \dword{larsoft} currently implements a pull-request model, in which experiment-appointed Level-2 managers comment on and sign off on changes to the central code, and Level-1 managers perform the actual merging.  A proposed change will not even be sent out for approval unless the \dword{ci} system can build it and validation tests are run to compare output that ought not to have been changed by the new code.

This second step, physics output validation, requires a more lengthy run through both unit tests and integration tests that run simulation and reconstruction workflows.  These tests run on standard input files and have their random number seeds fixed to constants.  The outputs are compared with reference histograms and a web page summarizes the comparison of the output logfiles, physics histograms, as well as run times and memory consumption, all of which are available on a web site that monitors these tests.  History plots of variables such as run time and memory consumption are made available on the monitoring web site so that investigations of when the memory usage of a job jumped up can be done without laboriously checking out, building, and running the software in the suspected 
timeframe of interest to find  a particular change.

While the \dword{ci} system is designed to ensure basic software quality as it is developed and checked in,  the final validation of a software release must be performed and approved by the \dword{dune} collaboration.  This process will involve significant person-power and it cannot be fully automated.  At the time of writing, formal procedures are not yet in place, though it is expected that the physics coordinators and the relevant physics working groups will perform these tasks.  Reviews of the physics performance of software releases must be scheduled with enough lead time for problems to be fixed, calibrations recomputed, re-approvals performed as needed, and samples generated for conferences and publications.  The reviews must be open to the entire \dword{dune} collaboration.  The criteria for acceptance of a software release will depend on the target precision of the physics intended to be performed with it.  Approved releases should be accompanied with documentation describing the quality of the data modeling, calibrated efficiencies, and recommended systematic uncertainties. Sufficient resources need to be allocated to physics working groups performing validations and calibrations so that they can achieve the goals on the required timescales.
\cleardoublepage

\chapter{Training and Documentation \hideme{Addressed Anne's comments 4/29}}
\label{ch:train}
The DUNE collaboration has grown rapidly since its inception, and it currently has members from over 200 institutions.  Even if the size of the collaboration were to remain constant, new members will always be joining while others move on to other positions and projects.  The scale of the experiment demands a harmonized documentation effort, coupled with consistent training for both newcomers and existing users. Although treated separately in this section, the tasks of documentation and training are highly correlated. Both are crucial for the long-term success of DUNE. 

\section{Documentation}
The documentation related to the DUNE’s computing aspects will be accessible on a variety of platforms, each with specific goals and access policies.

One evolving issue is the move to tighter control of public-facing websites at Fermilab and elsewhere.  Public facing websites are now subject to review before posting and this has led to much of the DUNE documentation becoming non-public except via \dword{sso} access.   Fermilab and CERN are working to make both Fermilab and CERN trusted credentials provide access to DUNE documentation but useful tools, such as the ability to find documents using google searches are becoming less and less available as documentation moves to protected areas.  This has become an issue in the preparation of this document, as it is good practice to reference the many DUNE documents that support this report but are no longer outside of the \dword{sso}. 

\subsection{Wikis}
The existing DUNE wiki \cite{comp:wiki} %
provides the landing page and starting point for ``DUNE Computing." It acts as a portal gathering information and links about the DUNE Computing consortium groups, %
their activities, and related resources. The template %
was revamped in early 2021 with a ``block'' design for a more visual and compact layout. As with the rest of the DUNE wiki,  access is restricted; users need a \dword{fnal} or CERN, Single Sign-On (SSO) service domain since the wiki contains information about computing access, node names, and other sensitive information.%

The top level DUNE wiki has six main blocks: Organization, Computing Toolbox, Operations, Working Groups, Getting Started and Resources.
\begin{itemize}
    \item 
The Organization block covers the consortium, the calendar of meetings, and lists the associated collaborations. 
\item The Computing Toolbox collects links to data access, analysis tools and algorithm frameworks.
\item The Operations covers the operations groups.
\item The DUNE Computing working groups are listed on the Working Groups area, and conveners of each working group will maintain their page.
\item Getting Started has  pointers to the tutorials, training sessions, and how-tos as well as the DUNE Slack channels and \dword{servicenow} helpdesk.
\item Last but not least, the Resources block gathers links about data policies and preservation, information protection, and the documents in progress.
\end{itemize}

\subsection{Redmine}

Fermilab's \dword{redmine} service was historically used to host DUNE \dword{fd} and \dword{protodune} software, wikis, and issue tracking.  Prior to 2021, read access to DUNE \dword{redmine} was open to the public, but write access required authentication.  Starting in the summer of 2021, all access to \dword{redmine} requires authentication.  As of January 2022, the DUNE \dword{fd} and \dword{protodune} software repositories and associated issue tracking have been migrated to \dword{github}. 

Many of the other Fermilab Intensity Frontier Experiments and much of the general Fermilab software documentation still resides in \dword{redmine}.

\subsection{Code Documentation}
One of the challenges for any large experiment is documentation of the algorithms and software. Following the lead of the Belle2 experiment, DUNE plans to explore the implementation of the Sphynx\footnote{https://www.sphinx-doc.org/en/master/} package for the automatic generation of code documentation. The generation of in-code comments and dedicated text files utilized by Sphynx will be the responsibility of code librarians and contributors of each sub-repository within the \dword{dunesw} code stack.

\subsection{GitHub}
DUNE software repositories have migrated to the version-controlled platform \dword{github}.  \dword{github} offers a ticketing system to facilitate debugging, revisions and updates from the community of users, though DUNE does not yet use \dword{github}'s wikis extensively.  The \dword{art} and \dword{larsoft} teams have transitioned from the \dword{redmine} system to using \dword{github}'s ticketing system.  The work to make this transition for DUNE is underway and follows the guidance of the \dword{hsf} community. %
The goal is for code to be publicly readable but protected from modification. 
\subsection{Code standards}

DUNE does not yet have \dword{dune}-specific coding standards.  We use the \dword{larsoft} standards which are documented at:

\href{ 
https://cdcvs.fnal.gov/redmine/projects/larsoft/wiki/The_rules_and_guidelines}{https://cdcvs.fnal.gov/redmine/projects/larsoft/wiki/The\_rules\_and\_guidelines}

and a longer list, including documentation and plug-ins and error handling:

\href{https://cdcvs.fnal.gov/redmine/projects/larsoft/wiki/Developing_With_LArSoft#Guidelines}{https://cdcvs.fnal.gov/redmine/projects/larsoft/wiki/Developing\_With\_LArSoft\#Guidelines}

\subsection{Frequently Asked Questions}
We are actively looking for a reliable \dword{faq}system.  We are currently using Github issues and projects as a place holder while we identify and implement a permanent solution. %

\section{User support}

In addition to formal documentation, users often have questions, either how to do things or problems with systems.  \dword{dune} currently relies on two systems to support users, \dword{slack} and \dword{servicenow}.  

\subsection{Slack}
\dword{dune} uses the commercial \dword{slack} platform to communicate urgent user questions.  The \dword{slack} systems is staffed by other users, including collaboration experts, who can provide quick answers to simple questions. 

\subsection{Service Now}
\dword{servicenow} is the official \dword{fnal} issue reporting system. Problems, including operational issues that cannot be resolved with peer support through \dword{slack} are directed to the \dword{servicenow} system.

\section{Training \hideme{3/12 HMS update}}

\subsection{Goals of DUNE Training}
DUNE’s training aims to serve both newcomers and existing users, offering a smooth start for the former and continuous support for the latter. %
DUNE recognizes that the computing environment and tools used within a HEP experiment are unique and evolve over time, and thus they require specialized training. The goals are to teach the basics of the environment and software used for analysis, as well as best practices in programming and data management. The training is offered through various formats, tools and platforms, as well as partnering with other collaborations, all discussed below.

\subsection{Training Sessions}
The primary training for new DUNE members is done twice a year during  dedicated sessions that extend over several days. These sessions run before or after collaboration meeting weeks. Such timing secures a good attendance for both new collaborators and presenting experts. Since 2020, the tutorials have been given online via zoom over several 1/2 day sessions to accommodate multiple time zones.  In particular, DUNE has a significant population of young scientists from India who cannot participate in person.  

The format of the training is an alternation of lectures, where students can follow running commands themselves, with interleaved quizzes and hands-on sessions. Introductory instructions and homework exercises are sent to participants prior to the tutorials to ensure that trainees 
arrive able to access the computing resources and are prepared for the material. %

The goal of this training is to make certain that new people have access to DUNE computing resources (at either FNAL or CERN) and understand the basics of logging in, storage areas, running applications, making minor modifications to code and submitting batch jobs. 

 All materials and documentation from past tutorials are retained in github and prominently linked on the DUNE wiki to serve as a library of self-education material.

However, due to the vast suite of software used by the DUNE collaboration for the various analysis steps and sub-detectors, additional training is needed and is currently provided on an ad-hoc basis by physics and hardware groups.   %

\subsection{Training Tools}
Training materials are now hosted in \dword{github} using the Software Carpentries framework. \cite{github:training}.
GoogleDocs are used %
for anonymized real-time questions by the participants. %
Most participants use the Fermilab general purpose \dwords{gpvm}, although  materials for everything except the batch submission parts can be run at CERN (or even local clusters with \dword{cvmfs} access).

Experts from the training team  %
 provide written answers to the questions as they come in and the responses remain available from the Tutorial github page. %
The sessions are delivered over zoom and recorded, with edited videos posted to the github tutorials so that they can be reviewed later.

In the future, DUNE %
hopes to use Jupyter\footnote{Jupyter\textcopyright{} \url{https://jupyter.org/}} notebooks, %
which provide an interactive environment for the trainees to run the examples, complete the exercises, and tweak portions of code to deepen their understanding of the analysis software. %
These notebooks will be accessible through %
JupyterHub servers provided by \dword{fnal} and CERN, as well as by other hosting institutions equipped with computing resources and analysis facility tools. The notebooks will be archived and referenced on the github pages as self-training modules for newcomers as well as %
reference material for %
lecturers.

\subsection{Audience for training}
As part of the 2022 training, incoming participants were surveyed about their level of experience and needs.  35 out of 50 participants responded.
We found that:
\begin{itemize}
\item 50\% of participants are graduate students, 25\% are postdocs with the rest divided between undergraduates (8\%), senior researchers (8\%) and engineers and programmers. 
\item Almost all participants were already associated with some DUNE hardware effort.
\item 16/35 (46\%) were not yet associated with a physics analysis group. 
\item 31/35 (88\%) were familiar with C++
\item 29/35 (83\%) were familiar with ROOT
\item 28/35 (80\%) were familiar with python
\item 26/35 (74\%) were familiar with github
\item 19/35 (54\%) were familiar with Jupyter notebooks
\item a smaller number were familiar with specialized HEP codes such as geant4 and LArSoft.
\item 31\% had no experience with batch environments while another 29\% were not confident in their abilities.
\item 16 (46\%) were already using Fermilab computing resources while 6 (17\%) were using CERN facilities.
\item Unfortunately 6 (17\%) indicated that they did back up their code in response to a question about backup methods.
\end{itemize}

\subsection{Partnering with other Collaborations}
The \dword{hsf} has started a continuous effort of harmonizing their tutorials under a common template called ``training module'' %
to create an introductory curriculum %
with the basic set of software needed %
 to instill good programming practices right at the start. Several of their modules for beginners are offered by the Software Carpentry Foundation\footnote{https://software-carpentry.org}. The growing list of modules can be found on the \dword{hsf} website ``Towards a HEP Software Training Curriculum''~\footnote{https://hepsoftwarefoundation.org/training/curriculum.html}.

The DUNE Computing training group is working in collaboration with the \dword{hsf}. The DUNE training %
points to the \dword{hsf}'s curriculum page so that %
newcomers can learn the prerequisites before attending a %
DUNE training. This %
in turn provides an audience for the \dword{hsf} material, %
providing that organization valuable feedback. %

\subsection{From Trainee to Mentor, User and Lecturer}
A key aspect of the success of the \dword{hsf} workshops %
is its mentoring system. %
The mentors are %
alumni of %
one of the recent training sessions, typically young, easy-to-approach researchers. %
After a couple of training sessions, the mentors can be trained by experts to become lecturers, acquiring new skills with their favorite software. The DUNE Computing consortium plans to reproduce the \dword{hsf} mentoring and training scheme %
by issuing frequent calls for mentors within the collaboration.

\subsection{Future Formats}
The DUNE Computing training group is also committed to surveying the needs and %
demands for particular %
skills the members would like to learn. Recent surveys have shown that \dword{larsoft} and \dword{pandora} are popular topics, as are %
best practices for building analysis code, %
and machine learning techniques, such as Keras, Tensorflow and GPU libraries.  The breadth of the tutorial wish-list is too large to efficiently cover during one bi-annual DUNE training, but continued input from tutorial attendees allow the planning of future topic specific lectures and tutorials. Moreover, incoming students are each working 
on a specific aspect of the DUNE project. A 
 project-based approach such as a series of hackathons would be an ideal format for some of these more advanced topics. This would allow students to team up on an aspect 
 directly linked with their analysis, guided by an expert.

We anticipate that the computing project will continue to  provide infrastructure and support but not all content for algorithmic documentation by the physics groups.  %

\cleardoublepage

\part{Resources and Conclusions} %

\chapter{Resource Needs Summary\hideme{coll-org.tex Anne's comments addressed}}
\label{ch:resource}

\section{Hardware Resources}
Chapter~\ref{ch:est} describes the projected compute and storage needs through 2040; these resources are a shared collaboration responsibility and are requested and monitored via the \dword{ccb}, described in Section~\ref{sec:ccb}. Chapter~\ref{ch:netw} describes networking needs and requirements. Our best estimates, developed in Chapter~\ref{ch:est}, indicate that DUNE's hardware resource needs are similar in terms of specifications to CMS or Atlas over the same time scales, but of order 10\%  of their size.

\section{%
Software Development and Operations Resources} %
Building and managing a project of this complexity %
requires a large number of %
dedicated people. %
To reduce our development resource needs relative to the large LHC experiments, DUNE's strategy is to use common tools already developed for HEP wherever possible and to work collaboratively with other groups in development of new tools. Much infrastructure is already in place, but significant DUNE-specific effort will still be needed.
Section~\ref{org:internal} %
describes the top level organizational structure while Figure~\ref{fig:resources} shows a timeline for estimated human resource needs.   Efforts are currently spread across a large number of highly expert people working part-time on DUNE computing, adding up to around nine \dword{fte}/year, in addition to  the five dedicated \dword{doe}-funded postdocs.  We anticipate that these early-career people will move into leadership roles in HEP computing in the future and will need to be replaced as they move on. In addition to the core group of experts, a small number of collaboration volunteers assist with operational tasks such as setting up and running simulation and reconstruction campaigns. Our ability to make use general collaboration members in operations will grow as we move from development to stable operations. 

Although the current expert \dword{fte} %
number of $\sim$\,14 is close to %
the identified need for dedicated experts, there is some mismatch in skill sets. In particular, we anticipate needing several person-years of effort on framework development, after \dword{pd2} but well before the near and far detectors are operational, and this expertise is  largely missing from the current  mix of personnel. %
In addition, most of the existing expertise is either funded short-term through grants or is provided by professionals at facilities who also have other responsibilities. Long term, the project will require more stable funding and well defined institutional commitments.  

\begin{dunefigure}
[Development personnel needs]
{fig:resources}
{Estimated computing infrastructure personnel needs through 2030.  The dark colors show development areas where experts are needed and the lighter colors show operations tasks where non-experts can contribute. The dashed line shows the estimated effort currently formally allocated to the project, including additional effort from 2022-2024 from the three-year DOE grant and enhanced UK project funding.}
{\includegraphics[width=0.99\textwidth]{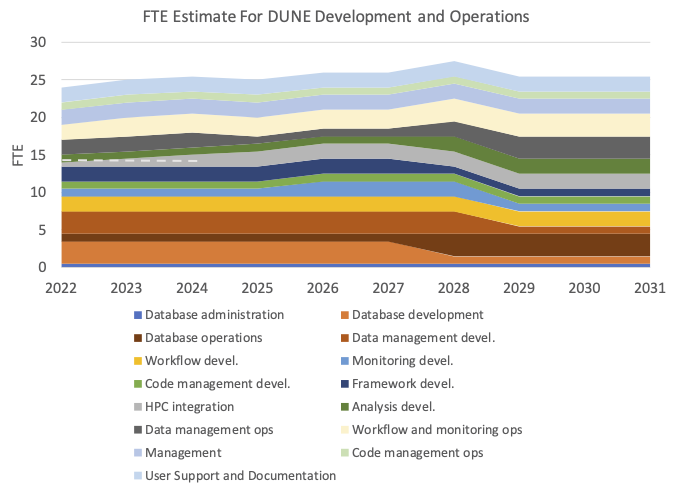}}
\end{dunefigure}

The roles listed in this chapter are limited to DUNE-specific projects and do not include the large number of personnel at multiple sites who support generic activities in areas such as storage and grid.

\subsection{Technical Roles}
\label{subsec:coll-org:Tech}

Technical roles require substantial computing expertise. The numbers listed below are the estimated \dword{fte}, not \dword{fte}-years, as the needs will be ongoing.  They appear as the darker areas in Figure~\ref{fig:resources}, adding up to 16-18 \dword{fte}. 

\subsubsection* {Database Design, Development and Management -- 3.5 FTE}

These roles include designing, developing, maintaining, and scaling databases for  
tasks within DUNE. Considerable effort is needed to interface with the large number of physics, calibration, data acquisition, and hardware tasks. People are needed with expertise in databases, data acquisition, and project management. Here we include 0.5 \dword{fte} of database administration effort for direct database interventions and operations as it requires specialized skills and access to restricted resources. %
The majority of the development database roles are expected to evolve into corresponding database operations roles (see Section~\ref{subsec:coll-org:Ops}) as a way of retaining expertise and allowing future upgrades of the database infrastructure\footnote{It is noted that many experiments have reported difficulties in retaining database expertise, which can result in greatly increased risk over time as technologies evolve.}.

\subsubsection*{Distributed Data Management Development -- 3.0 FTE}

These roles include oversight of all software engineering and development activities for packages needed to operate distributed storage resources. The roles require a good understanding of the distributed computing infrastructure used by DUNE and collaborating sites, as well as the DUNE computing model, including:

\begin{itemize} 
\item{Storage management, integration and monitoring -- 2.5 FTE}

These involve commissioning of new storage as \dword{rucio} storage endpoints (RSE's), data placement, construction and development of storage-usage monitoring and data-transfer monitoring, testing, commissioning of tape store services as they evolve, and participating in data transfer challenges.

\item {Core \dshort{rucio} Development -- 0.5 FTE}

\dword{rucio} development will continue in order to meet the emerging DUNE requirements. This ongoing work includes
further development and integration of the lightweight client, proximity mapping for optimizing job placement, quality-of-service support to differentiate between disk, tape-backed disk, and tape, and further developments that will certainly emerge after  the experience of \dword{protodune}.

\end{itemize}

\subsubsection*{Workflow Management Development 2.0  FTE}

This involves ongoing design of the Workflow System needed by DUNE for distributed processing of data and simulation; and the implementation and testing of that design, both as systems developed by DUNE or adapted from elsewhere, and the creation of any necessary interfaces for particular classes of computational resources including %
\dword{hpc} machines. 

DUNE's computing model is designed to be less hierarchical than the classic LHC systems, which motivates the creation of systems that are application and network aware and  take full advantage of the \dword{rucio}-based data management systems.

\subsubsection* {Monitoring Development -- 1.0 FTE}

This %
work includes oversight of all software engineering and development activities for packages needed to monitor distributed disk and compute resources. The role requires a good understanding of the distributed computing infrastructure used by DUNE as well as the DUNE computing model.

\subsubsection* {Code Management Development -- 1.0 FTE}

The code managers provide infrastructure to support applications for  data processing, simulation, and analysis, and also to coordinate activities in the areas of development, release preparation, and %
deployment of software package releases needed by DUNE. They organize the overall setup of software packages needed for releases.

The application managers need to keep up with evolution in operating systems, build systems, and compilers, so this role has a significant development component.

\subsubsection*{Core simulation and  reconstruction  framework development -- 2.0 FTE}

This role requires substantial expertise in the design and deployment of sophisticated frameworks for HEP algorithmic workflows. %
It involves coordinating %
significant additional effort from common projects (\dword{larsoft}, \dword{art} ...) and from collaboration algorithm development. 

\subsubsection*{HPC integration -- 2.0 FTE}

Integrating DUNE's code to run on the diverse \dword{hpc} systems available to us will require substantial administrative and technical development to negotiate resources and to adapt DUNE codes to run in very particular environments.

\subsubsection*{Analysis development -- 1.0 FTE}

This %
involves working with physics and detector groups on the development of specialized systems for efficient analysis of reconstructed data samples. 

\subsection{Operational  Roles}
\label{subsec:coll-org:Ops}
In addition to the development activities listed above, %
people are needed to manage and supervise collaborators in operations tasks. These roles are more fluid, based on the status of the experiment and %
many do not require computing expertise; they can be filled by other collaborators after some training.

\subsubsection* {Distributed Data Management  -- 2.0 FTE}

The distributed data managers are responsible for operational interactions with distributed computing disk and tape resources. The role includes but is not limited to helping to establish new storage areas, and to manage %
data replication, deletion, and movement.

\subsubsection* {Computing Coordination -- 2.0 FTE}

Coordinators %
manage the computing projects, %
and oversee the projection and distribution of resources needed to accomplish these projects' tasks. %
These roles serve largely as liaisons %
between the detector, physics, and computing resources of the experiment.
\begin{itemize}
\item{Computing Consortium Leads}

This group includes the consortium leads as well as the liaisons %
to the DUNE detector and physics groups and %
to external entities such as \dword{osg} and  \dword{wlcg}. The group's responsibilities include overall coordination of computing projects and negotiations regarding funding and resources with both agencies and institutions. 

\item {Resource Board Chair}

This role is responsible for chairing biannual meetings of the Computing Resource Board, which includes representatives from the %
various national funding agencies that support DUNE, to discuss %
funding for and delivery of the computing resources required for successful processing and exploitation of DUNE data.
\end{itemize}

\subsubsection*{Workload Management operations -- 2.5 FTE}
\begin{itemize}

\item{Distributed Workload Manager -- 1.0 FTE}

The distributed workload managers are responsible for operational interactions with distributed computing resources. %
They are responsible for helping to establish grid and cloud sites. They are also responsible  for the setup, launch, monitoring, and %
completion of processing campaigns (e.g., data processing, \dword{mc} simulation, and working group productions), executed on distributed computing resources for the experiment. %

\item {Computing Shift Leaders -- 1.0 FTE}

The shift leader is %
responsible for the experiment's distributed computing operations for %
week-long periods that run Monday through Sunday.  %
Shift leaders chair regular operations meetings during their week and attend general DUNE operations meetings as appropriate. %

\item {Distributed Computing Resource Contacts -- 0.5 FTE}

Distributed computing resource contacts are the primary contacts for the DUNE distributed computing operations team and for the operators of large (Tier-1) sites and regional federations. They interact directly with the computing shift leaders at operations meetings. 
\end{itemize}

\subsubsection*{Code Management Operations -- 1.0 FTE}

In addition to development associated with changes in systems, code librarians and application managers must continually prepare releases and %
deploy  software packages needed by DUNE.  Builds are currently done on a weekly basis.

\subsubsection*{User Support -- 2.0 FTE}

User support (software infrastructure, applications, and distributed computing) underpins all user activities of the DUNE computing project. 
User support %
personnel %
respond to questions from users on mailing lists, Slack-style chat systems, and/or ticketing systems, %
and are responsible for documenting solutions in knowledge bases, FAQs and wikis.

\cleardoublepage

\chapter{Summary}
\label{ch:conclusions}

This document has outlined the major use cases and challenges identified for \dword{dune} computing for time scales ranging from %
the upcoming year, e.g., \dword{pd2}, to long-term operations of the full \dword{dune} %
\dword{fd} and \dword{nd}. 
Substantial preliminary studies and design work have also been outlined. 

\section{Review of Challenges}

In the introduction to this document (Section~\ref{intro:challenges}), the following challenges were %
introduced---some unique to DUNE, some more general, but all of them significant.

\begin{description}
\item{\bf Large memory footprints -}  DUNE events, with multiple data objects consisting of thousands of channels and thousands of time samples,   present formidable challenges for reconstruction on typical \dword{hep} processing systems. %
This report describes the use cases (Chapter~\ref{ch:use}), framework (Chapter~\ref{ch:fworks}) and compute requirements (Chapter~\ref{ch:cm}) to address these use cases.

\item{\bf Storing and processing data on heterogeneous international  resources -} 
DUNE depends on the combined resources of the collaboration for large-scale storage and processing of data.   Tools for using shared resources ranging from small-scale clusters to dedicated \dword{hpc} systems are being designed but will need to be further developed and maintained.  
Chapters~\ref{ch:est}, \ref{ch:cm} and~\ref{ch:datamgmt} describe the data volumes, computing model, and data management plans. 

\item{
\bf Machine learning - }  Use of \dword{ml} techniques can greatly improve simulation, reconstruction and analysis of data. However, integration of \dword{ml} techniques into a software ecosystem %
the size and complexity of a large \dword{hep} experiment requires substantial effort beyond proof-of-concept and small-scale demonstration. 
Chapters~\ref{ch:use} and~\ref{ch:codemgmt} discuss some known applications and their management, but substantial effort will be needed to keep up with this rapidly evolving field. 

\item{\bf Efficient and sustainable use of resources -} %
The proposed global computing model allows \dword{dune} to make flexible use of resources worldwide.  Our workflow systems are being designed to ensure that CPU, network, and storage resources are well matched to achieve high efficiency. We are working with the HEP community on improved documentation and training in best practices to ensure that our user community can do their work accurately and efficiently. 

\item{\bf Keeping it all going -}  %
Many computing activities, both novel and mundane, need to continue over the full lifetime of the experiment. The chapters on databases(\ref{ch:db}), data management(\ref{ch:datamgmt}, \ref{ch:place}), workflow(\ref{ch:wkflow}), 
services (\ref{ch:serv}), authentication (\ref{ch:auth}), code management (\ref{ch:codemgmt}) and training and documentation (\ref{ch:train})
describe many of the systems that must be designed and then maintained over this time period. %

In addition, while our computing challenges are not on the same scale as those of the large \dword{lhc} experiments, substantial human effort, CPU resources, dedicated storage, and fast networks will need to be acquired and made easily available to the collaboration. 
And whereas \dword{dune} will not be as resource intensive as the \dword{lhc} experiments, it comes with %
unique computing challenges %
related to \dwords{snb}, calibration, and other time-variable trigger records. Chapter~\ref{ch:cm} provides estimates of our needs while Sections ~\ref{sec:ccb}, \ref{ch:cm} and~\ref{ch:resource} describe our existing collaborative resources and future needs. The timetable shown in Figure~\ref{fig:timeline} gives a rough estimate of the timeline for development, commissioning, and operations of different components of the DUNE Offline Computing and their relation to the experiment timeline.
\end{description}

\begin{dunefigure}
[Offline computing timeline]
{fig:timeline}
{DUNE Computing Consortium timeline for development, commissioning, and operations of various services and tasks within DUNE Offline Computing.}
{\includegraphics[width=0.9\textwidth]{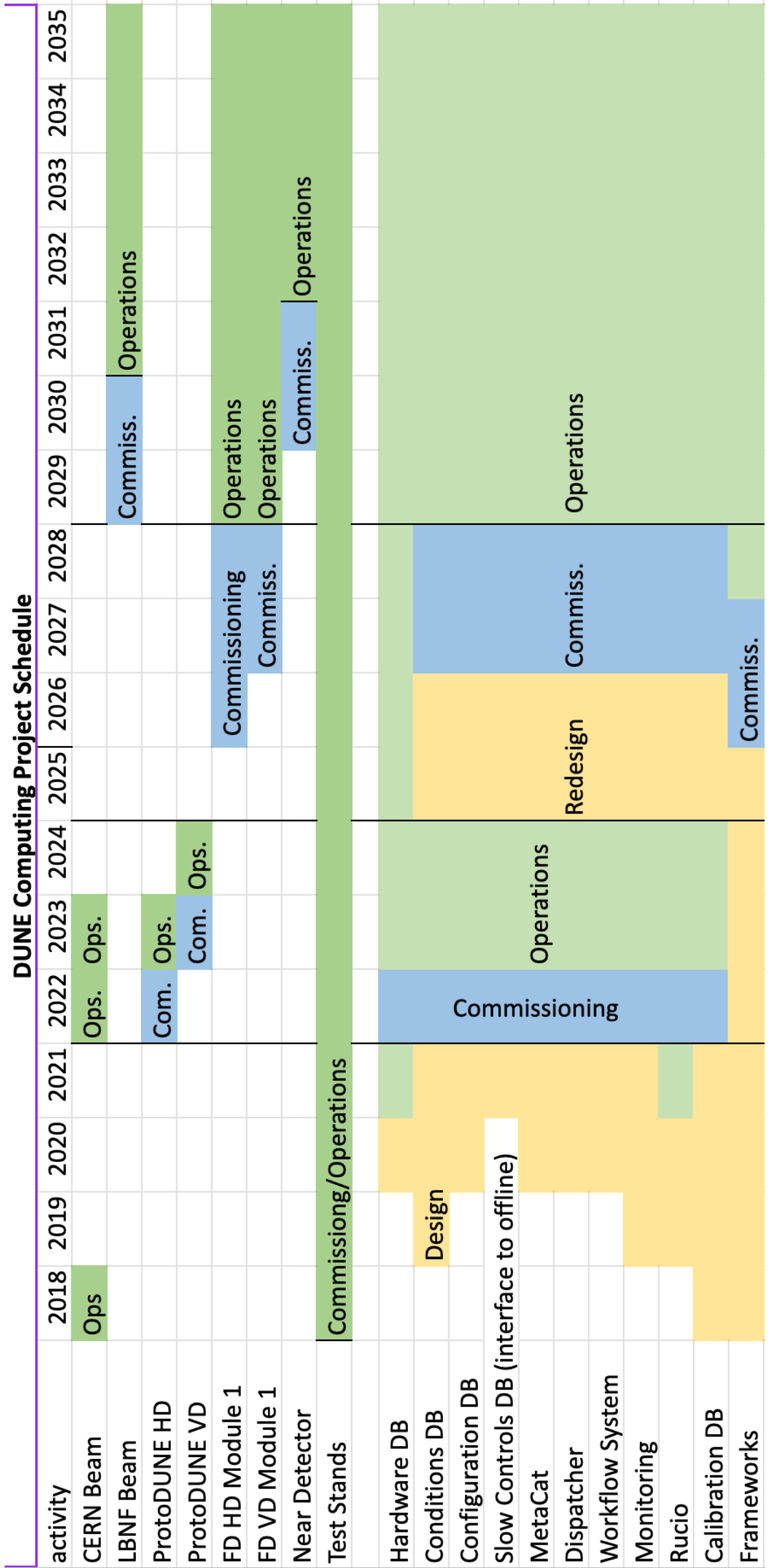}}
\end{dunefigure}

\section{Conclusion}

In conclusion, we have identified the major challenges facing \dword{dune} computing and have begun the process of designing and deploying solutions. We have preliminary estimates of resource needs and have broad connections to collaborating institutions and countries that have already provided substantial resources on a voluntary basis. %
This document lays out our understanding of the resources and activities that will be required to support timely data reconstruction and analysis as \dword{dune} grows, and ultimately to allow the \dword{dune} experiment to fully exploit its physics potential.

\cleardoublepage

\printglossaries

\cleardoublepage
\cleardoublepage
\renewcommand{\bibname}{References}
\renewcommand\bibpreamble{Some references to internal DUNE technical documents may be password protected. Please contact \href{mailto:dune-computing-cdr@fnal.gov}{dune-computing-cdr@fnal.gov} to request access.}
\bibliographystyle{utphys} 
\bibliography{tdr-citedb}
\end{document}